\documentclass[a4paper, english, 10pt, twoside]{memoir}

\usepackage[T1]{fontenc}
\usepackage[utf8]{inputenc}
\usepackage[english]{babel}

\usepackage{amsmath, amssymb}

\usepackage{color}

\usepackage{float}

\usepackage{graphicx}

\usepackage{parskip}

\usepackage[colorlinks=true, urlcolor=blue, citecolor=blue, linkcolor=black]{hyperref}

\usepackage[backend=bibtex, backref=true, style=numeric, sorting=none, minnames=4, maxnames=4, minbibnames=4, maxbibnames=4, mincitenames=4, maxcitenames=4]{biblatex}
\setsecnumdepth{subsubsection}
\maxtocdepth{subsubsection}

\usepackage{lmodern}
\usepackage{calligra}

\usepackage{eso-pic}

\usepackage{caption}
\usepackage[top=1.3in, bottom=1.3in, left=1.in, right=1.in]{geometry}

\usepackage{fancyhdr}
\DeclareMathOperator{\sign}{sgn}

\newcommand{\GeVc}{GeV c$ ^{-1} $}
\newcommand{\um}{$ \mu $m}
\newcommand{\us}{$ \mu $s}
\newcommand{\pT}{$ p_T $}
\newcommand{\pZ}{$ p_Z $}
\newcommand{\axis}[1]{#1}

\title{Development and Study of Different Muon Track Reconstruction Algorithms for the Level-1 Trigger for the CMS Muon Upgrade with GEM Detectors}
\author{Thomas LENZI}
\date{May 2013}

\begin{document}

\frontmatter

{ \pagestyle{empty}

\AddToShipoutPicture*{
	\put(\LenToUnit{73px}, \LenToUnit{220px}){
		\includegraphics[width=\textwidth]{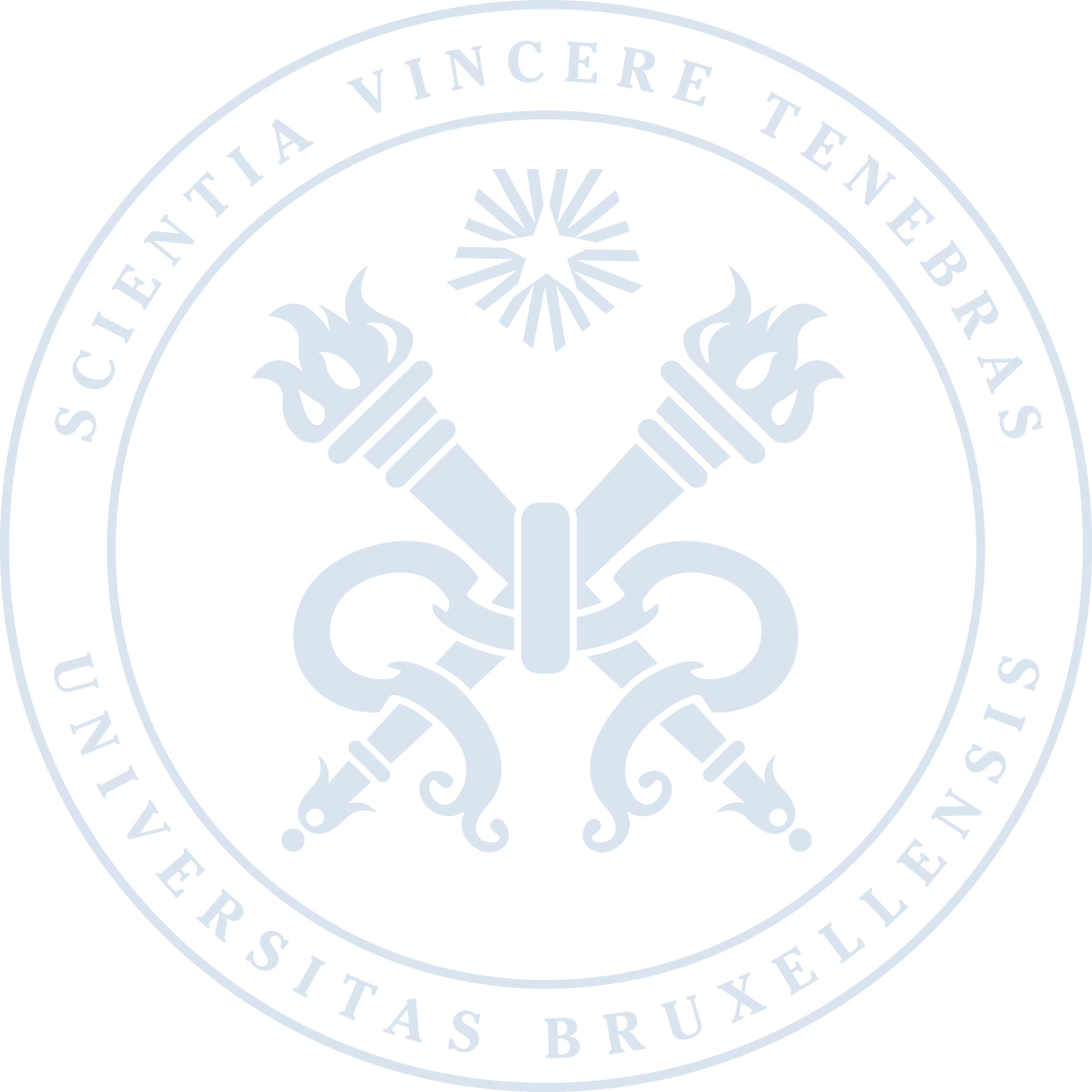}
	}
}

\begin{center} 
	\normalsize 
	\textsc{Université Libre de Bruxelles} \\
	\textsc{Faculté des Sciences - Département de Physique} \\
	\textsc{IIHE - Interuniversity Institute for High Energies (ULB-VUB)} 
\end{center} 

\vfill

\begin{center}

	{ \huge \textbf{\parbox[b]{400pt}{ \center Development and Study of Different Muon Track Reconstruction Algorithms for the Level-1 Trigger for the CMS Muon Upgrade with GEM Detectors}} } \\ [1.2cm]
 	{ \large Master thesis } \\ [0.2cm]
 	{ \LARGE Thomas LENZI } \\ [1.cm]
	{ \footnotesize Under the supervision of } \\ [0.2cm] 
	{ Dr. Gilles De Lentdecker } \\ [1.1cm]

\end{center}

\vfill

\begin{center}

	{\large May 2013} \\ [1.1cm]

	\textsc{\footnotesize Mémoire présenté en vue de l'obtention du diplôme de Master en Sciences Physiques}

\end{center}
\cleardoublepage

\vspace*{2cm}

\begin{flushright}
{ \Large \calligra À mes parents \\*[0.4cm] et \\*[0.4cm] À Arwen \& Violette }
\end{flushright}
\cleardoublepage	

}

\setcounter{page}{1}

{ \pagestyle{plain}

\addcontentsline{toc}{chapter}{Abstract}

\begin{abstract}

	In the upcoming years, with the upgrade of the LHC to higher luminosities, CMS will be subject to an increasing flux of particles, especially in its most forward region, at $ | \eta | $ > 1.6. To consolidate the muon spectrometer of CMS and increase redundancy, the CMS GEM collaboration proposes to instrument the 1.6 < $ | \eta | $ < 2.1 region, originally foreseen to be equipped with RPC, with Triple-GEM detectors. This technology has proven to remain efficient under high fluxes and meets the requirements of CMS. \\

	This work, using simulations, studies three track reconstruction algorithms intended to be installed in the first selection stage of CMS for Triple-GEM detectors: a Least Squares fit, a standard Kalman filter, and a modified Kalman filter. A comparative analysis is performed between the results obtained with Triple-GEM detectors and those yielded by the actual system in order to quantity the improvements made to the CMS muon spectrometer and to the CMS trigger system. \\

\end{abstract}

\renewcommand{\abstractname}{Résumé}

\begin{abstract}
	
	Dans les années à venir, avec la mise à niveau du LHC pour de plus hautes luminosités, CMS sera soumis à un flux croissant de particules, surtout dans la région avant, à $ | \eta | $ > 1.6. Afin de consolider le spectromètre à muons de CMS et d'augmenter la redondance, la collaboration CMS GEM propose d'installer des détecteurs Triple-GEM dans la région 1.6 < $ | \eta | $ < 2.1, qui devait initialement être équipée de RPC. Les Triples-GEMs peuvent resister à des flux intenses sans perdre de leur efficacité et ainsi satisfaire aux exigences de CMS. \\

	Ce travaille étudie, à l'aide de simulations, trois algorithmes de reconstruction de traces destinés à être installés dans le premier niveau de déclenchement de CMS: un Moindres Carrés, un filtre de Kalman standard et un filtre de Kalman modifié. Une analyse comparative est faite entre les résultats obtenus en utilisant les Triple-GEMs à ceux résultant du système actuel afin de caractériser l'impact des différents algorithmes sur le spectromètre à muons et sur le système de déclenchement de CMS.

\end{abstract}

\vfill
\cleardoublepage

\chapter*{Acknowledgment}
\addcontentsline{toc}{chapter}{Acknowledgment}
\label{chap:acknowledgment}

	Je tiens avant tout à remercier le Dr. Gilles De Lentdecker qui m'a offert l'opportunité de réaliser mon mémoire sur un sujet qui m'intéresse et me correspond. Je le remercie de m'avoir accompagné et encadré tout au long de l'année. Ce travail ne serait pas le même sans ses conseils avisés. \\
	 
	I would like to thank the members of the GEM group at the IIHE led by Dr. Gilles De Lentdecker: Erik, Florian, Geoffrey, Patrizia, Thierry, and Yifan, for their time and useful advice, and all the members of the IIHE for their warm welcome. A special thanks to Dr. Pascal Vanlaer and Dr. Kael Hanson for their help. \\
	 
	Je remercie également Gwen, Nicolas et Thibault, pour les bons moments passés, et ceux à venir et qui ont partagé avec moi ces années d’études. Je leur souhaite le meilleur pour la suite. \\
	 
	I would like to thank my Dad, Marlène, and Greg for the rereading of this work and apologize for the headaches it might have caused. \\

	Finalement, je remercie mes parents de m’avoir toujours soutenu et encouragé dans mes choix et d'avoir toujours été là pour moi. Ils m'ont donné envie d'apprendre et de grandir. Sans eux je ne serais pas le même et pour cela je leur suis infiniment reconnaissant. Ce mémoire leur est dédié.  
\cleardoublepage

	\tableofcontents
	\cleardoublepage 

}

\mainmatter

\chapter*[Introduction]{Introduction}
\addcontentsline{toc}{chapter}{Introduction}
\label{chap:introduction}

	Before its temporary shutdown early 2013, the \emph{Large Hadron Collider} (LHC) at the \emph{European Organization for Nuclear Research} (CERN) collided protons at energies up to 8 TeV in the center of mass reference frame with an instantaneous luminosity above 10$ ^{33} $ cm$ ^{-2} $ s$ ^{-1} $. Two \emph{Long Shutdowns} (LS) are planned during the first operation phase of the LHC: the first one in 2013-2014 to make the necessary adjustments to reach the nominal energy of 14 TeV in the center of mass reference frame, the second one in 2018-2019 to increase the luminosity beyond the initially foreseen 10$ ^{34} $ cm$ ^{-2} $ s$ ^{-1} $. These maintenance periods offer the possibility to the experiments recording the LHC beam collisions to maintain and upgrade their detectors, and prepare for the high luminosity phase after LS2. \\

	The LHC is designed to collide protons at a frequency of 40 MHz. We do not yet have the technology to handle and store all the produced data at this rate. Inside the \emph{Compact Muon Solenoid} (CMS), one of the four LHC's experiments, each event produces approximatively 1 Megabytes of analyzed data, while the detector generates several Terabytes every second. The maximum amount of data that CMS can store every day is of the order of the Terabytes, yielding a rate of accepted events of 100 Hz, requiring the total rate to be divided by a factor of 4 10$ ^{5} $. \\

	This requires the installation of a trigger system which handles data in real-time, coupled with a complex data acquisition system. The first stage of the CMS's trigger system, called the \emph{Level1 Trigger} (L1 Trigger), analyzes the information from the calorimeters and the muon chambers, using algorithms programmed on dedicated electronics, and performs a first selection of interesting events. It is important to ensure that the system has the capability to recognize the signature of interesting physical processes, while rejecting the other 99.99975\% of the events forever. \\

	The forward region of the CMS muon spectrometer is equipped with two different technologies of gaseous detectors: \emph{Cathode Strip Chambers} (CSC) which yield a good spatial resolution of the order of 100 \um{} and a time resolution of 5 ns, and \emph{Resistive Plate Chambers} (RPC) which offer a lower spatial resolution around 1 mm but an excellent time resolution down to 1 ns. In the most forward region of CMS, the RCPs have not been installed and the L1 Trigger relies on CSCs only. Currently, CMS has the least redundancy, trigger capability, and reconstruction efficiency in the most challenging region for muon detection. High background fluxes and shorter tracks in the transverse plane constitute a challenge when trying to identify muons. \\

	The presence of muons in the final state is a signature of many interesting processes such as the decay of the Brout-Englert-Higgs boson or new physics like super-symmetry. High energy muons often constitue the \emph{golden channel} due to their high detection and reconstruction efficiency. At the higher luminosity at which the LHC will run after LS2, the selection of muons will suffer from an increased background generating coincidences in the detectors and confusing the trigger. With only breadcrumbs of data available, the efficiency of the L1 Trigger will quickly diminish, degrading the performances of the CMS muon spectrometer. \\

	The standard RPCs are not designed to operate at the high rates of particles that will be reached after LS2 and will loose efficiency. New \emph{Gas Electron Multiplier} (GEM) already used in other experiments present the opportunity to equip the vacant region with detectors that have proven to maintain a spatial resolution of the order of 100 \um{}, a time resolution below 5 ns, and a detection efficiency above 98\% even at elevated fluxes. The objective of the CMS GEM collaboration is to instrument the most forward region of the CMS muon spectrometer with Triple-GEM detectors during LS2. \\

	Taking advantage of the improvements made on the performances of dedicated electronics and of the strengths of Triple-GEM detectors, we intend to develop new and more complex algorithms for the L1 Trigger to perform track reconstruction. \\

	The aims of this thesis are the development and the study of three muon track reconstruction algorithms to be run possibly at the L1 Trigger: a Least Squares fit, a standard Kalman filter, and a modified Kalman Filter. To analyze the performance of these algorithms, we have used a self-developed Fast Simulation environment and the official simulation framework of CMS. \\

	Chapter \ref{chap:lhc_and_cms} provides the reader with a general overview of the LHC and CMS, while Chapter \ref{chap:muon_chambers} focuses on the muon spectrometer of the latter. This chapter also reviews the theory behind gas detectors and states the difficulties that will arise once entering the high luminosity phase of the LHC. The GEM technology proposed to be installed as an upgrade in the most forward region of CMS to address these issues is detailed in Chapter \ref{chap:gas_electron_multiplier_detectors}. The Trigger system of CMS and the algorithms currently used to reconstruct and select events are described in Chapter \ref{chap:trigger_system_and_reconstruction_algorithms} which moreover presents various track reconstruction algorithms. \\

	Chapter \ref{chap:simulation_environment} introduces the simulation environments we developed and used to test our reconstruction algorithms. The first tested method, a Least Squares fit, is presented in Chapter \ref{chap:least_squares_fit} along with the obtained results and a thorough analysis of various observable effects. Increasing the complexity of the algorithms, Chapter \ref{chap:kalman_filter} describes the standard Kalman filter and the modified version we implemented. Chapter \ref{chap:algorithms_performances_timing_prospects} compares the algorithms and their impact on the trigger system, and gives some inside on the implementation of the algorithms on a programmable electronics.
	\cleardoublepage


\chapter{LHC and CMS}
\label{chap:lhc_and_cms}

	The first part of this chapter is dedicated to the Large Hadron Collider (LHC). After a short historical introduction about the LHC's construction, we characterize the particle beam and the machine. Further on, we give an overview of the current and future operation plan of the LHC. The second part of this chapter reviews the Compact Muon Solenoid (CMS), one of the experiments recording the LHC beam collisions. The structure of CMS is described, along with the technologies used to track particles. For each of them, a detailed description is provided.

	\section{Large Hadron Collider}
		\label{sec:lhc_and_cms__large_hadron_collider}

		The \emph{Large Hadron Collider} (LHC) \Cite{LHC_Machine} is state of the art in particle accelerators and colliders engineering. Built at and by the \emph{European Organization for Nuclear Research} (CERN), it is located 100 m beneath the Franco-Swiss border in the 27 km long tunnel that previously hosted its predecessor the \emph{Large Electron Positron} collider (LEP). Using smaller accelerators as injectors, it is designed to accelerate and collide protons or heavy ions at energies up to 14 TeV in the center of mass reference frame. These collisions take place at four different locations where the ALICE \Cite{ALICE_at_LHC}, ATLAS \Cite{ATLAS_at_LHC}, CMS \Cite{CMS_at_LHC}, and LHCb \Cite{LHCb_at_LHC} experiments collect and analyze data.
		
		\subsection{Timeline}
		\label{sec:lhc_and_cms__timeline}

		The LHC's concept dates back to the 80's. Although LEP was not yet running, physicists were already thinking about its successor. Multiple groups started to consider the possible measurements and discoveries that could be made with the future hadron collider. The project was approved by the CERN Council in December 1994 and construction began in 1998. During the first five years, excavation crews dug service tunnels and caverns which would host the detectors. Meanwhile, various parts of the machines were being developed and tested for later assembly. \\

		In 2003, the first section of the accelerator was assembled inside the tunnel. The construction continued until 2008 when the last piece was mounted and the LHC was complete and ready to produce its first collisions. After a successful preliminary run in August an incident occurred shutting down the machine for nearly one year. \\

		Since then, the LHC is slowly building up in energy, reaching 8 TeV in the center of mass reference frame before the \emph{Long Shutdown-1} (LS1) which started at the beginning of 2013. This 20 months long period will be used to perform maintenance and to upgrade the LHC as well as the experiments, and to prepare them to run at nominal energy and luminosity.
		
		\subsection{Injection Chain}
		\label{sec:lhc_and_cms__injection_chain}

\begin{figure}[h!]
			\centering
			\includegraphics[width = 12cm]{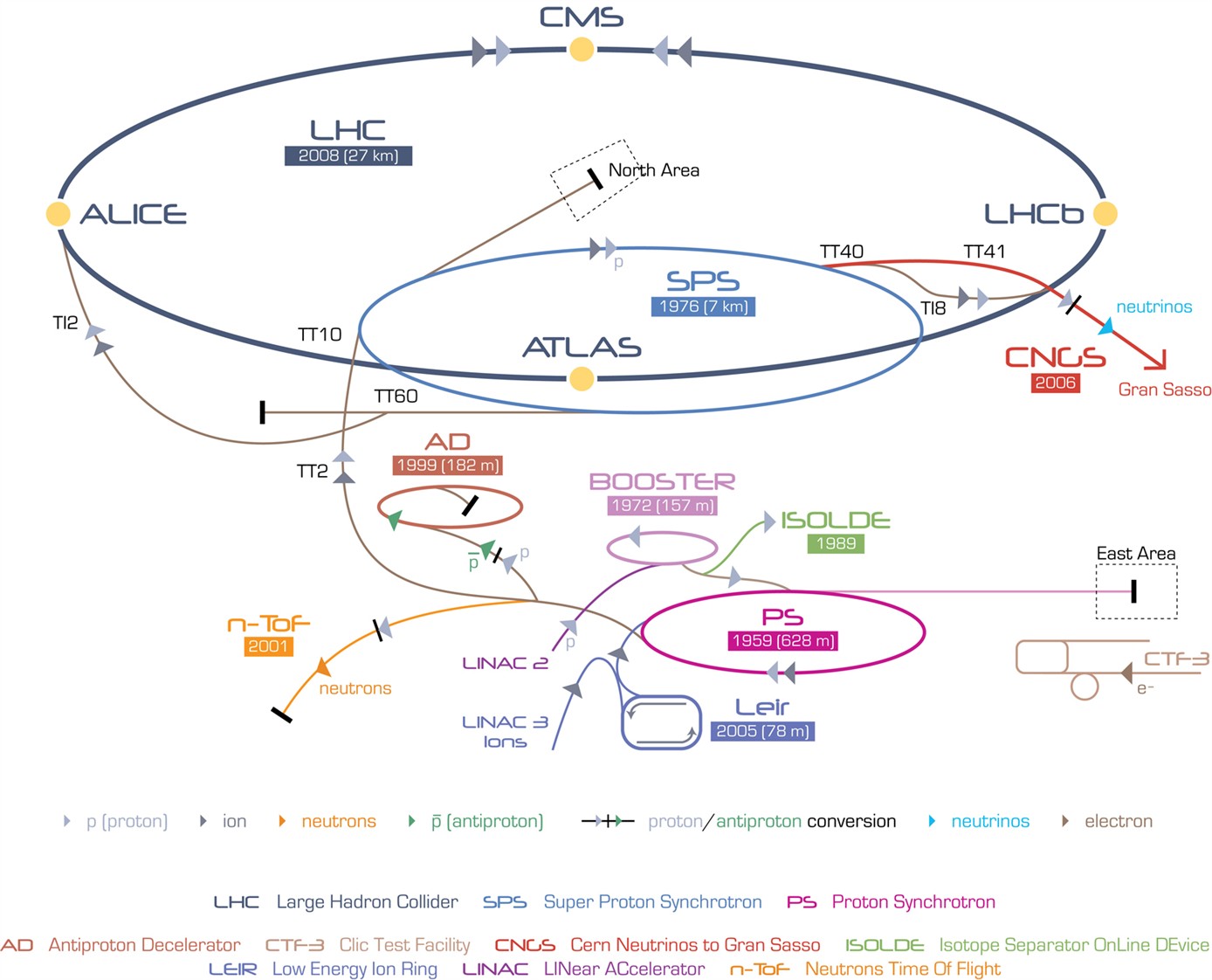}
			\caption{Schematic representation of the LHC's injection chain composed of multiple smaller accelerators \Cite{Fig_LHC_Chain}.}
			\label{fig:lhc_and_cms__lhc_injection_chain}
		\end{figure}

		Before being accelerated and collided by the LHC, the protons and ions are produced and sped up by multiple accelerators. Figure \ref{fig:lhc_and_cms__lhc_injection_chain} depicts the injection chain of the LHC. The first element in this chain is the \emph{Linac2}, which produces protons by ionizing gaseous hydrogen. These protons are accelerated and regrouped into bunches by a set of magnets before being transferred to the \emph{Proton Synchrotron Booster} (Booster), \emph{Proton Synchrotron} (PS), and \emph{Super Proton Synchrotron} (SPS), which furthermore increase the bunches' energy. Finally, the particles enter the LHC through two different tubes traveling in opposite directions.

		\subsection{Beam Structure and Luminosity}
		\label{sec:lhc_and_cms__beam_structure_and_luminosity}

		As previously mentioned, the particle beam is not a continuous flux of protons but rather a series of bunches that travel along the tube and are accelerated, regrouped, and focused by a set of magnets. Once they achieve the desired energy they are collided at a frequency of 40 MHz at the four crossing sites. \\ 

		The LHC also delivers a high instantaneous luminosity $ \mathcal{L} $ which is essential to detect rare processes as it yields the apparition's frequency of events in a given interaction process
		\begin{equation}
			f_{process} = \mathcal{L} \sigma_{process} \ ,
		\end{equation}
		where $ f_{process} $ is the number of expected events per second, and $ \sigma_{process} $ is the interaction cross-section of the process. For a circular collider, the instantaneous luminosity is defined as
		\begin{equation} 
			\mathcal{L} = \frac{N^2_b n_b f_{rev} \gamma}{4 \pi \epsilon_n \beta^*} F \ ,
			\label{eq:lhc_and_cms__luminosity}
		\end{equation}
		where $ N_b $ is the number of protons or ions per bunch, $ n_b $ is the number of bunches per beam, $ f_{rev} $ is the revolution frequency, $ \gamma $ is the Lorentz factor, $ \epsilon_n $ is the beam emittance, $ \beta^* $ is the beta function at the \emph{Interaction Point} (IP), and $ F $ is a function of the crossing angle between the beams at the IP. The $ \epsilon_n $ and $ F $ parameters are related to the bunches' structure and more specifically to their spatial spreading. These parameters change during the machine's operation as the number of protons per bunch decreases, and the bunches spread out. We can integrate $ \mathcal{L} $ over a long period of time in order to get the integrated luminosity
		\begin{equation}
			L = \int \mathcal{L} \ dt \ ,
		\end{equation}		
		which results in the number of events we can expect for a given interaction process
		\begin{equation}
			N_{process} = L \sigma_{process} \ .
			\label{eq:lhc_and_cms__luminosity_to_N}
		\end{equation}

		\subsection{Future Operations and Performances}
		\label{sec:lhc_and_cms__future_operations_and_performances}

			The LHC's operation plan \Cite{CMS_Upgrades} is divided in two phases: phase 1 during which the machine will slowly reach its nominal capabilities, and phase 2 where the machine will run at even higher luminosity after undergoing a major upgrade. \\

			Phase 1 extends from 2010 to about 2020 and is divided into three shorter periods separated by two Long Shutdowns. Table \ref{tab:lhc_and_cms__lhc_performances} shows the energy and luminosity at which the LHC will be running after both maintenances (LS1 and LS2). While the machine is shut down, physicists will have access to the detectors and will be able to perform repairs and upgrades. \\
			
			\begin{table}[h!]
				\centering
				\begin{tabular}{l|c|c}
					Period & Energy & Luminosity \\ \hline
					2010-2012 & 7-8 TeV & 0.5 10$ ^{34} $ cm$ ^{-2} $ s$ ^{-1} $ \\
					Long Shutdown 1 (LS1) & - & - \\
					2015-2017 & 13-14 TeV & 10$ ^{34} $ cm$ ^{-2} $ s$ ^{-1} $ \\
					Long Shutdown 2 (LS2) & - & - \\
					2019-2021 & 14 TeV & 2 10$ ^{34} $ cm$ ^{-2} $ s$ ^{-1} $
				\end{tabular}
				\caption{Energy and luminosity of the LHC during the different periods of phase 1 \Cite{LHC_Machine}.}
				\label{tab:lhc_and_cms__lhc_performances}
			\end{table}

			After LS1, the energy will be increased by using the magnets at full capability, and the luminosity will be multiplied by a factor of two by bringing down the time between two collisions to 25 ns instead of 50 ns. The luminosity will further be increased after LS2 when the machine enters the \emph{High Luminosity LHC} (HL-LHC) era. As reviewed in Equation \ref{eq:lhc_and_cms__luminosity}, various parameters can be tuned in order to achieve that goal. The main possibilities are to \Cite{LHC_LS1, LHC_Upgrade_Scenarios}
			\begin{enumerate}
				\item increase the number of bunches $ \sim n_b $;
				\item increase the number of protons per bunch by making them longer $ \sim N_b $;
				\item increase the collisions' frequency $ \sim f_{rev} $;
				\item decrease the bunches' spread by improving the efficiency of the focusing magnets $ \sim \epsilon_n $;
				\item decrease the collisions' angle by adding magnets near the IPs $ \sim F $. \\
			\end{enumerate}

			Phase 2 involves a major upgrade of the LHC and of the injection chain which would occur during LS3 after 2021. The objective is to increase the luminosity by a factor of 10 to reach 10$ ^{35} $ cm$ ^{-2} $ s$ ^{-1} $.
	
	\section{Compact Muon Solenoid}
	\label{sec:lhc_and_cms__compact_muon_solenoid}

\begin{figure}[h!]
			\centering
			\includegraphics[width = 16.5cm]{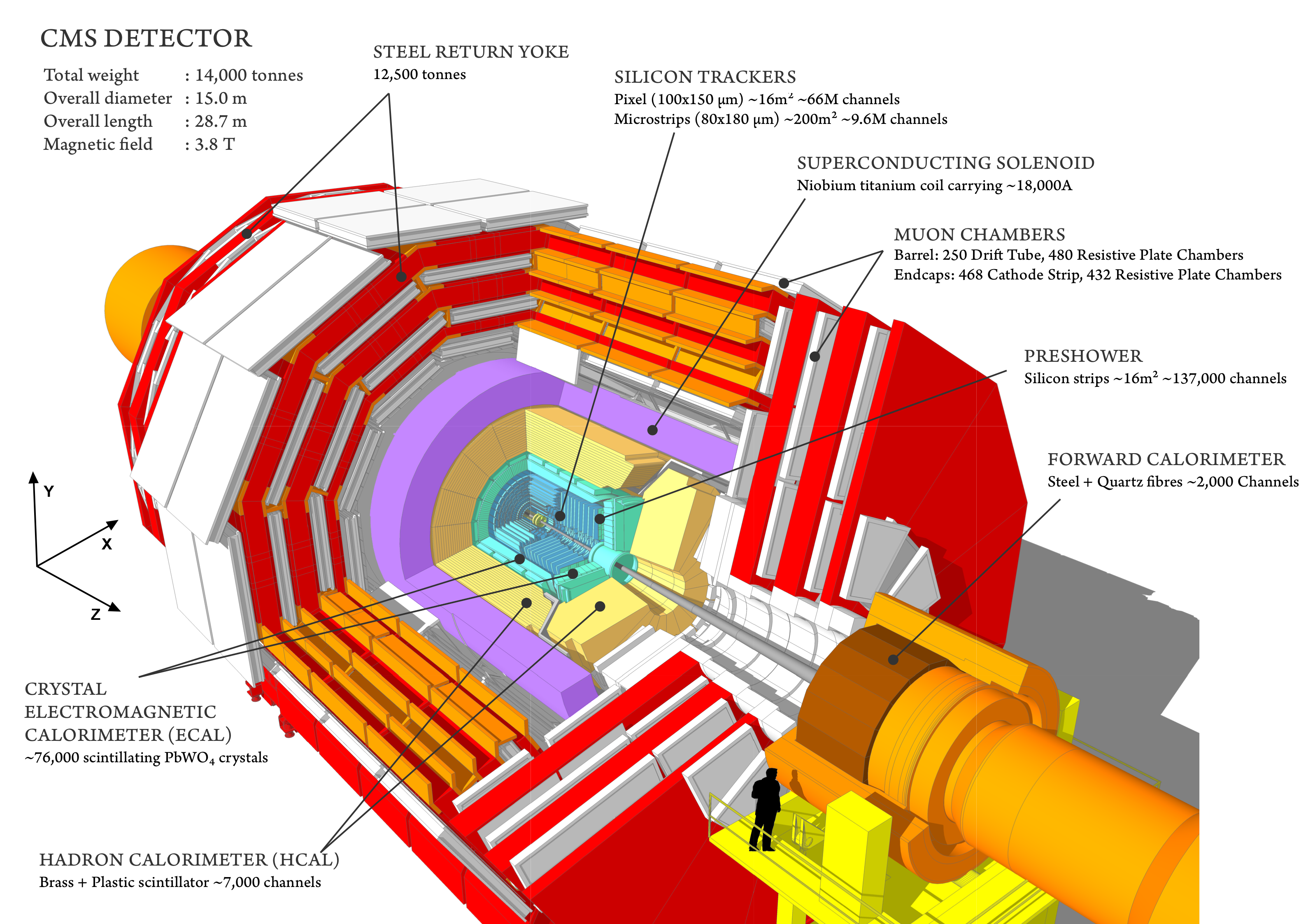}
			\caption{Representation of CMS and its different parts: the silicon tracker (blue), the electromagnetic calorimeter (green-blue), the hadronic calorimeter (orange), the magnet (purple), and the muon chambers (white) \Cite{Fig_CMS_Detector}.}
			\label{fig:lhc_and_cms__cms_global_view}
		\end{figure}

		The \emph{Compact Muon Solenoid} (CMS) \Cite{CMS_at_LHC} is, along with ATLAS, ALICE, and LHCb, one of the four main experiments recording the LHC beam collisions. Its structure and components are depicted in Figure \ref{fig:lhc_and_cms__cms_global_view}. As represented, it is composed of five main parts: the \emph{silicon tracker} (TK) in blue, the \emph{electromagnetic calorimeter} (ECAL) in green-blue, the \emph{hadronic calorimeter} (HCAL) in orange, the \emph{magnet} in purple, and the \emph{muon chambers} in white. \\
		
		CMS is divided into two regions: the barrel where the detectors are laid out cylindrically around the beam, and the endcaps where the detectors are placed perpendicularly to the beam. Although CMS has the ability to detect a wide range of interaction channels, it is characterized by its effective trigger system for muons and its strong magnetic field, which gave its name to the detector.
		
		\subsection{Appropriate Coordinates}
			\label{sec:lhc_and_cms__appropriated_coordinates}

\begin{figure}[h!]
				\centering
				\includegraphics[width = 11cm]{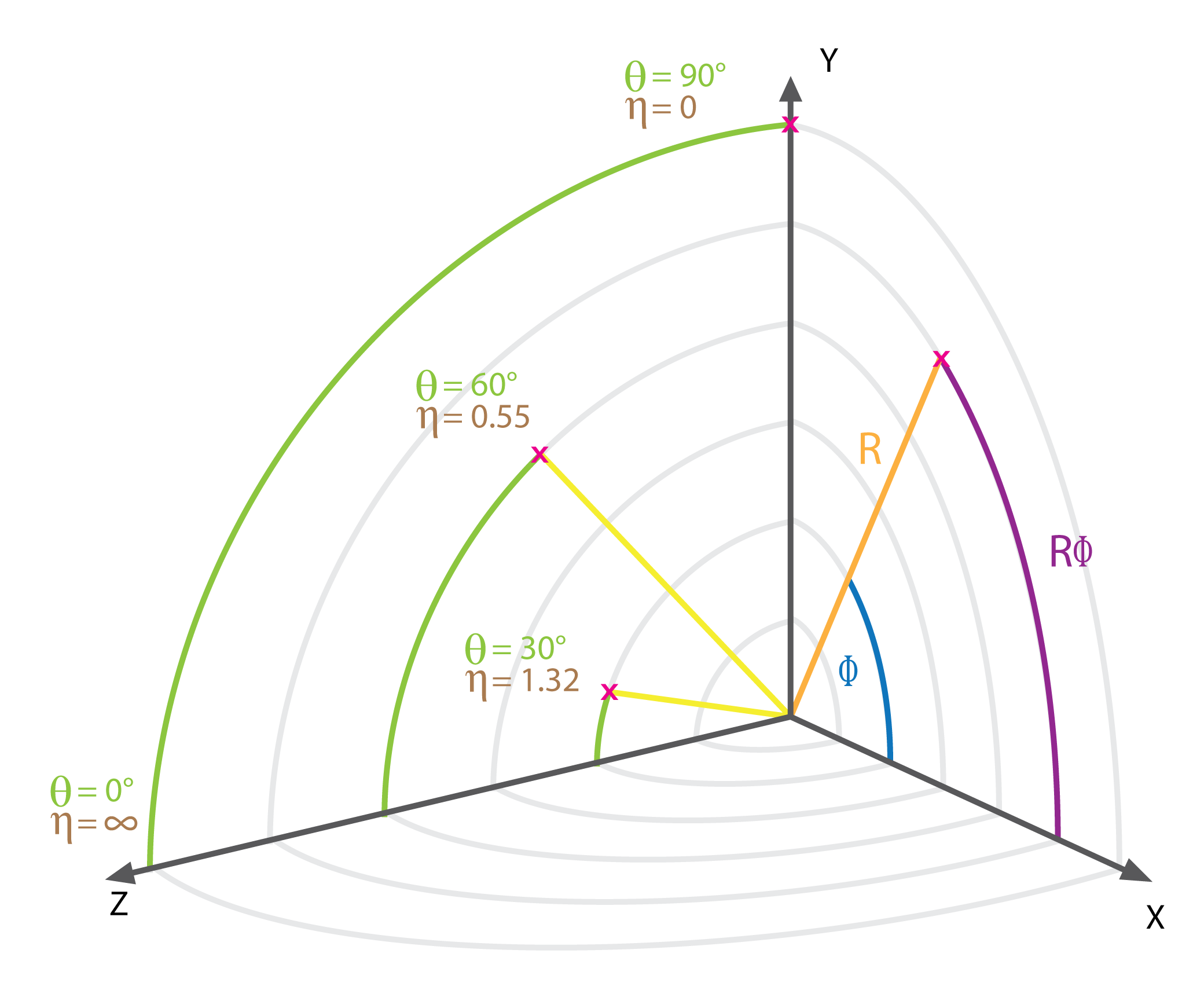}
				\caption{Pseudo-rapidity $ \eta $ and azimuthal angle $ \phi $ used to track particles inside CMS.}
				\label{fig:lhc_and_cms__cms_coordinates}
			\end{figure}

			As CMS is of the cylindrical type, appropriate coordinates are defined to track particles. The Cartesian coordinates \axis{X}, \axis{Y}, and \axis{Z} are first set: \axis{X} points to the center of the accelerator, \axis{Y} to the surface, and \axis{Z} in the direction of the beam, as illustrated in Figure \ref{fig:lhc_and_cms__cms_coordinates}. The \axis{XY} plane is also referred to as the transverse plane. The problem with this set of coordinates is that the physics is not symmetrical under these, which would be a beneficial feature. Therefore, the rapidity $ y $, a Lorentz invariant over which the particles in the final state are equally distributed, is used. The rapidity is defined as 
			\begin{equation}
				y = \frac{1}{2} \ln \left( \frac{E + p_Z}{E - p_Z} \right) \ ,
			\end{equation}
			which for highly relativistic particles can be approximated by the pseudo-rapidity
			\begin{equation}
				\eta = - \ln \left[ \tan \left( \frac{\theta}{2} \right) \right] \ ,
			\end{equation}
			where $ \theta $ is the polar angle (angle between the particle and the \axis{Z} axis). The other coordinates are the azimuth $ \phi $ (angle in the \axis{XY} plane), and the distance to the beam in the transverse plane $ R $ for tracks the barrel, and the \axis{Z} coordinate for tracks the endcaps. \\

			As previously stated, the particles in the final state are equally distributed over $ \eta $. This means that the particles fluxes are lower in the barrel than in the endcaps. This impacts the detectors' geometry.
	
		\subsection{Silicon Tracker}
		\label{sec:lhc_and_cms__tracker}

\begin{figure}[h!]
				\centering
				\includegraphics[width = 13cm]{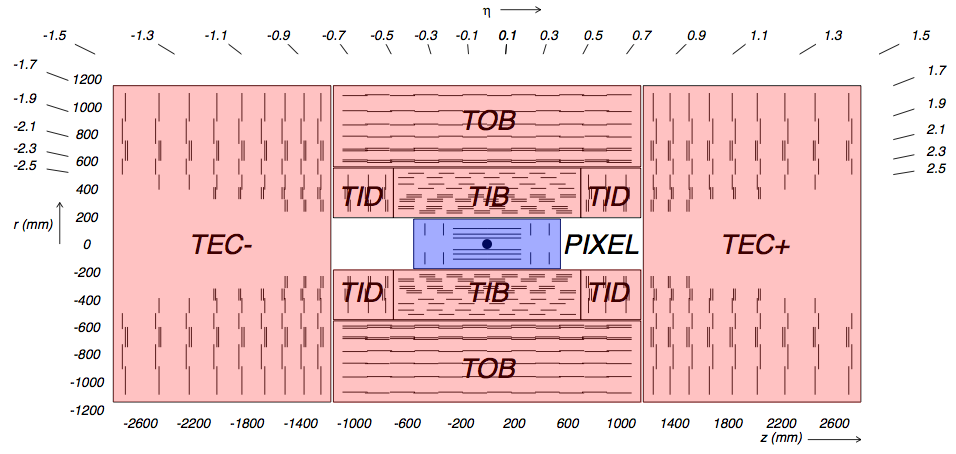}
				\caption{Disposition of the different detectors in the silicon tracker. PIXEL (blue) refers to silicon pixel detectors while TIB, TID, TOB and TEC (red) all refer to silicon strip detectors \Cite{CMS_at_LHC}.}
				\label{fig:lhc_and_cms__cms_tracker}
			\end{figure}

			The silicon tracker is the detector closest to the IP. It is composed of two different types of semiconductor detectors: \emph{silicon pixels} and \emph{silicon strips}. These detectors have an excellent spatial resolution (down to 25 \um{}) which yields excellent momentum reconstruction capabilities (resolution of the order of 1\% at low transverse momenta). The disposition of the different technologies is represented in Figure \ref{fig:lhc_and_cms__cms_tracker}. The silicon pixels are represented in blue, while TIB, TID, TOB, and TEC refer to different regions of the silicon strip detector. \\

			Semiconductor detectors are made out of two pieces of silicon, one negatively doped containing more unbounded electrons, and one positively doped containing more unbounded holes (absence of electrons), put together to form a n-p junction. At the interface, the electrons and the holes diffuse in the opposite region and recombine with the particles of opposite charge, creating an unbalance in charge: the n region and p region close to the junction become, respectively, positively and negatively charged. From this, an electric field is formed which slows the diffusion down until the system reaches equilibrium. When a charged particle passes through this region and losses energy, electrons switch from non-conductive to conductive bands creating electrons/holes pairs. Under the action of the electric field, they migrate towards the n or p regions and form the signal on the readout electronics. Unfortunately, the number of unbounded charges is still high compared to the formed signal and the active region is small. To increase the detection efficiency, a voltage difference is applied to the semiconductor further diffusing the electrons and holes through the junction. Typically, for a voltage difference of 100 V, the size of the active region is of the order of 300 \um{}.
			
			\subsubsection{Silicon Pixel Detectors} 
			\label{sec:lhc_and_cms__silicon_pixel_detectors}

\begin{figure}[h!]
					\centering
					\includegraphics[width = 8cm]{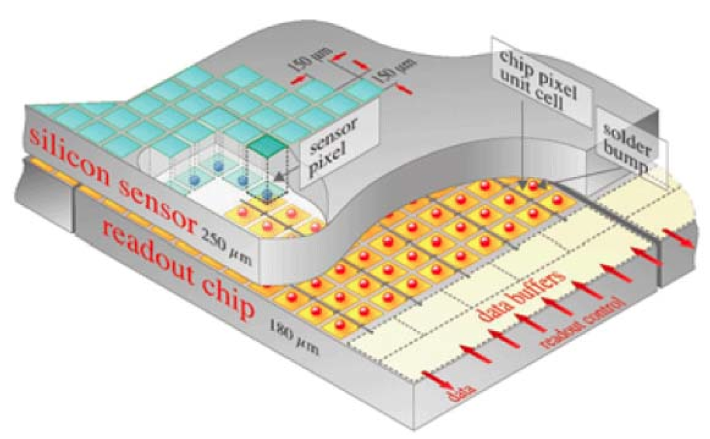}
					\caption{Disposition of the readout pixels (orange and red) on the same detection block (gray) inside a silicon pixel detector \Cite{CMS_Tracker_Construction}.}
					\label{fig:lhc_and_cms__cms_pixel_detector}
				\end{figure}

				The silicon pixels detectors, represented in Figure \ref{fig:lhc_and_cms__cms_pixel_detector}, are composed of small 100 \um{} x 150 \um{} rectangles of readout material disposed on a block of detection medium formed by a n-p region. The electrons are formed in that region and migrate towards the silicon pixels. The challenge arising is that each pixel needs its own readout electronics which takes a significant amount of space and requires output cables. These cables prevent the placing of detectors which creates dead-zones. Physicists and engineers must find the right balance between the number of pixels (granularity), the size of the electronic, and the detectors' resolution. \\

				Nevertheless, the pixel detectors are the most precise tracking technology in CMS with a spatial resolution of 15 to 20 \um{}. This value is smaller than the size of the pixels because of \emph{charge sharing}. All the pixels sharing the same detection block also share the energy deposited by a particle. By reading the total deposited charge on each of the pixels, we can find the \emph{Center of Gravity} (COG). The COG method is based on the assumption that there exists a linear relation between the induced pulse's height on a pixel and the distance between its center and the particle's hit, so that each pixel is assigned a weight proportional to the deposited charge. The reconstructed coordinate $ \mathbf{x}_{COG} $ of the cluster is then given by 
				\begin{equation}
					\mathbf{x}_{COG} = \frac{\sum_i \mathbf{x}_i q_i}{\sum_i q_i} \ ,
					\label{eq:lhc_and_cms__charge_sharing}
				\end{equation}
				where $ q_i $ is the individual pixel signal in the cluster, and $ \mathbf{x}_i $ is the positions of the pixel in the defined coordinate system. The same technique can be applied to other detectors as long as the charge is read out analogically and not digitally\footnote{One can consider using the same method with digital readouts but will not be able to achieve the same resolutions.}.			
			
			\subsubsection{Silicon Strip Detectors}  
			\label{sec:lhc_and_cms__silicon_strip_detectors}

				The most outer layers of the tracker cannot have a granularity as high as the pixel detectors, for both financial and technical reasons. The amount of data that would have to be read out is considerable, and the technology to do so is not yet available. A way to reduce the granularity of the detectors is to measure only one coordinate by using silicon strips instead of pixels. The strips are separated by 80 to 122 \um{} which gives a resolution between 23 \um{} and 53 \um{} in the direction perpendicular to the strips. Unfortunately, as expected, there is a larger error on the other coordinate (along the strip) corresponding to the size of the detection cell. To improve global precision, some of the cells have two strip detectors placed with a small stereo angle, typically 100 mrad, allowing them to measure both coordinates. However, this set up generates \emph{ghosts} because of the ambiguity created when more than one particle hits the detector at the same time.
			
			\subsubsection{System Performances}  
			\label{sec:lhc_and_cms__tracker_system_performances}		

				Due to the magnetic field generated by the solenoid, the trajectories charged particles are bent inside the tracker. The relation between the bending radius of the track $ R $ , the transverse momentum $ p_T $, and the intensity of the magnetic field $ B $ is
				\begin{equation}
					R[\mbox{m}] = \frac{p_T[\mbox{GeV c}^{-1}]}{0.3 B[\mbox{T}]} \ .
					\label{eq:lhc_and_cms__radius_to_momentum_relation}
				\end{equation}
				By measuring the bending radius of the track and inverting the previous relation, the transverse momentum of the particles can be obtained. Tracks created by high energy particles will be straighter than those left by low energy particles and therefore more difficult to reconstruct. \\

\begin{figure}[h!]
					\centering
					\includegraphics[width = 6.3cm]{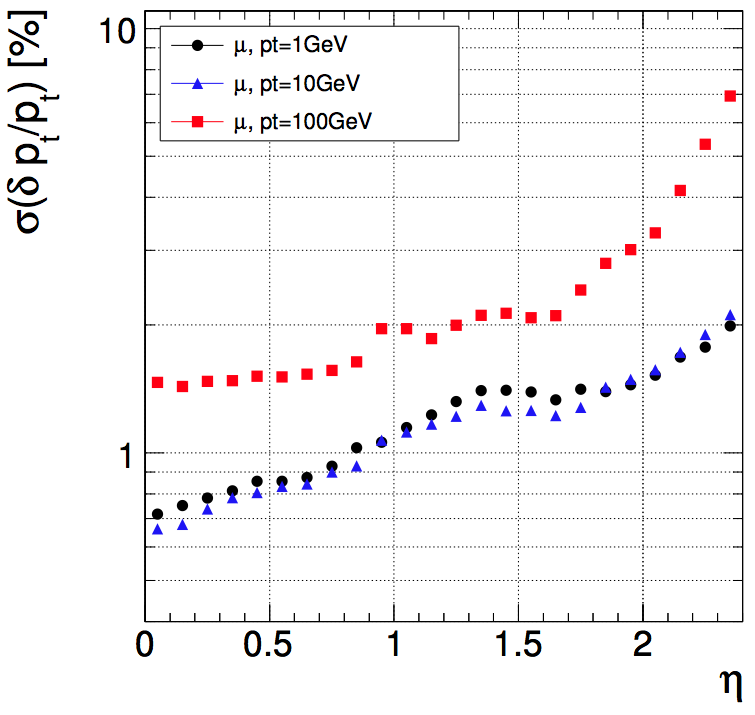}
					\includegraphics[width = 6.3cm]{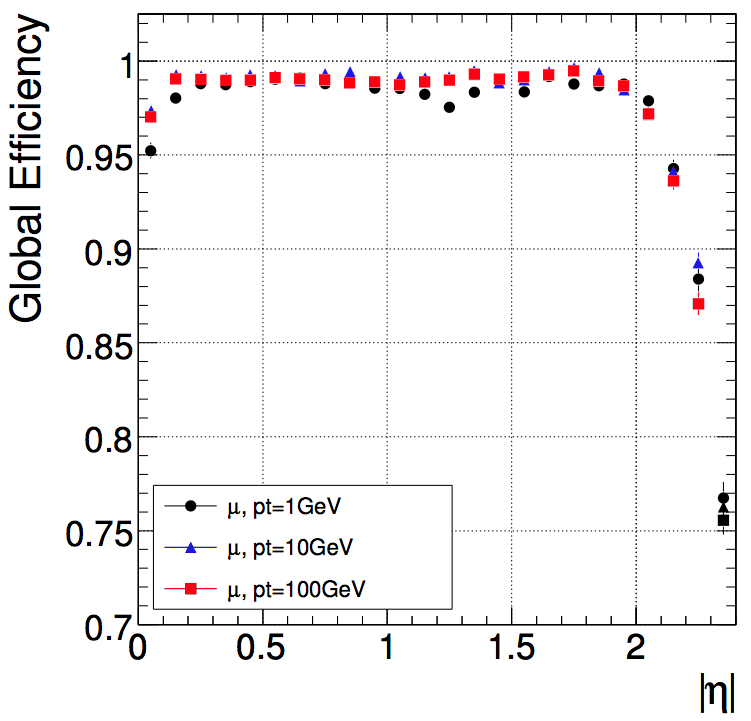}
					\caption{Resolution on the transverse momentum \pT{} (left) and detection efficiency (right) of the tracker as a function of the pseudo-rapidity $ \eta $ for muons of transverse momenta \pT{} of 1, 10, and 100 \GeVc{} \Cite{CMS_at_LHC}.}
					\label{fig:lhc_and_cms__cms_tracker_performances}
				\end{figure}

				As previously stated, the tracker offers an excellent resolution on the position of the particles and therefore gives precise measurements of the particles' momentum. Figure \ref{fig:lhc_and_cms__cms_tracker_performances} shows the resolution on the transverse momentum \pT{} (left) and detection efficiency (right) of the tracker as a function of the pseudo-rapidity $ \eta $ for muons of transverse momenta \pT{} of 1, 10, and 100 \GeVc{}. The resolution is less than 1\% for muons of 1 and 10 \GeVc{} in the barrel ($ \eta $ < 1) and quickly rises in the most forward region. The same effect is observed for the detection efficiency which is close to 100\% in the barrel but significantly diminishes at higher pseudo-rapidities. Precision decreases in the most forward region where the strip pitch is greater and more material is present, causing more scatterings.

		\subsection{Calorimeters}
		\label{sec:lhc_and_cms__calorimeters}

			Calorimeters are devices that absorb the full kinetic energy of a particle by creating particle cascades, and provide a signal that is proportional to that deposited energy. \\

			Two types of calorimeters are present in CMS: the electromagnetic to detect electrons and photons, and the hadronic to detect hadrons. Each of them corresponding to two of the elementary interactions through which particles can interact with the medium: electromagnetic interaction and strong interaction. \\

			The parameters characterizing particles showers are the radiation length $ \lambda_R $, and the Molière radius $ R_M $ for the electromagnetic calorimeter, and the absorption length $ \lambda_a $ and the interaction length $ \lambda_I $ for the hadronic calorimeter, all depending upon the atomic properties of the material.	
			\begin{enumerate}
				\item[] $ \lambda_R $ is the distance that an electron or photon has to travel inside the calorimeter to, respectively, emit a photon or create an electron/positron pair.
				\item[] $ R_M $ gives the radius of the cylinder in which 90\% of the electromagnetic shower is contained.
				\item[] $ \lambda_a $ is the average distance that a hadron has to travel before undergoing an inelastic interaction with the medium.
				\item[] $ \lambda_I $ yields the distance after which a hadron will have scattered inelastically and also gives the radius of the cylinder in which 95\% of the hadronic shower is contained.
			\end{enumerate}
			Those parameters result in the spatial extension of the cascades giving an idea of the granularity needed to correctly distinguish showers. Because the interaction length $ \lambda_I $ of hadrons is much larger than the radiation length $ \lambda_R $ of electrons and photons, the ECAL is placed first. This also implies that hadrons can pass through the ECAL without interacting, or at least not significantly, even if it is multiple $ \lambda_R $ long. \\

			The energy resolution of calorimeters depends upon the number of particles in the cascade hence the energy of the particle
			\begin{equation}
				\left( \frac{\sigma_E}{E} \right)^2 = \left( \frac{a}{\sqrt{E}} \right)^2 + \left( b \right)^2 + \left( \frac{c}{E} \right)^2 \ ,
			\end{equation}
			where $ a $ is the stochastic term depending upon the development of the shower and the detector's response, $ b $ is the constant term determined by the calibration and the uniformity of the crystal, and $ c $ is the noise term from the electronics. Unlike the tracker, the calorimeters' resolution increases with the energy, offering the best resolution at high energies.
			
			\subsubsection{Electromagnetic Calorimeter}
			\label{sec:lhc_and_cms__electromagnetic_calorimeter}

				The two main processes allowing the detection of electrons and photons are respectively Brëmsstrahlung and pair creation. These occur as long as the resulting particles (electrons and photons) have enough energy to repeat the process, creating an electromagnetic cascade inside the material. The size of the cascade hence the number of photons emitted by the scintillator is proportional to the energy of the incident particle. Muons do not significantly interact with the ECAL because the radiative processes are greatly suppressed. Indeed, Brëmsstrahlung is proportional to m$ ^{-2} $ (inverse-squared mass) and is therefore only significant for electrons. \\

				\begin{figure}[h!]
					\centering
					\includegraphics[height = 4cm]{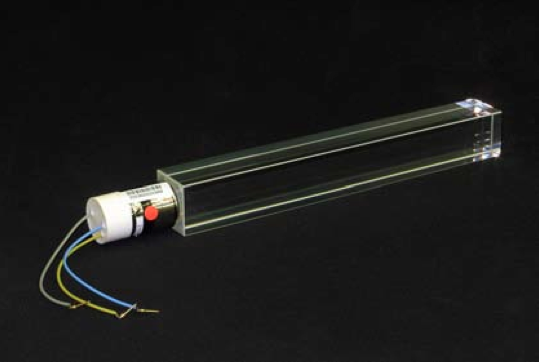}
					\includegraphics[height = 4cm]{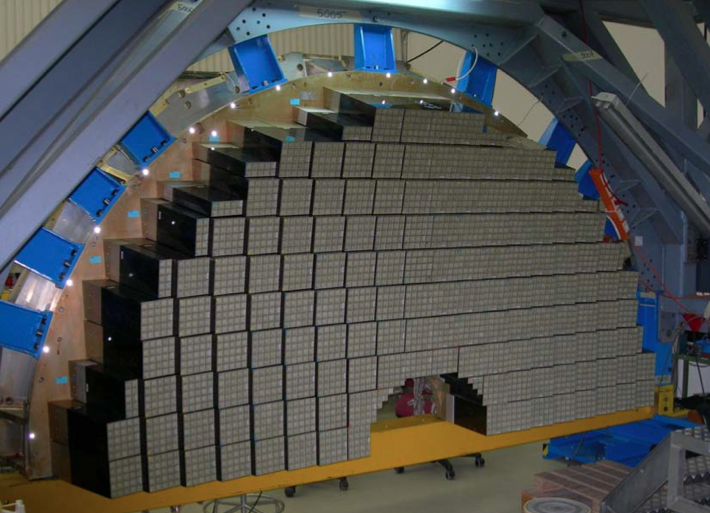}
					\caption{Picture of a PbWO$ _4 $ crystal (left) used in the ECAL with its photomultiplier, and of the endcap ECAL (right) showing the crates in which the crystals are placed \Cite{CMS_at_LHC}.}
					\label{fig:lhc_and_cms__cms_ecal_view}
				\end{figure}

				In CMS, the ECAL is composed of PbWO$ _4 $ crystals, acting both as interaction media and as scintillators, attached to photomultipliers to amplify the relatively small amount of photons they emit. The crystals measure 2.2 cm x 2.2 cm, which is equivalent to one Molière radius $ R_M $, by 23 cm, which corresponds to several radiation lengths $ X_0 $. Figure \ref{fig:lhc_and_cms__cms_ecal_view} shows one of these crystals (left), the crates that hold them, and their disposition in the endcap (right). The number of photons collected is proportional to the energy deposited in the calorimeter modulo a correction factor due to the aging of the material. The ambient radiation causes the crystals to become opaque and release less photons which in turn implies a constant need for recalibration of the detectors.

			\subsubsection{Hadronic Calorimeter}
			\label{sec:lhc_and_cms__hadronic_calorimeter}

				Where the ECAL relies on radiative processes to detect particles, the HCAL uses strong interactions between the hadrons and the material to create hadronic cascades. These are much longer than electromagnetic showers, requiring longer detectors. The most created particles are pions as they are the lightest hadrons. This induces an electromagnetic component as the $ \pi^0 $ principal decay channel is $ \pi^0 \rightarrow \gamma \gamma $. This creates a problem, as the response of the material can be different for the hadronic and electromagnetic component. \\

\begin{figure}[h!]
					\centering
					\includegraphics[height = 5cm]{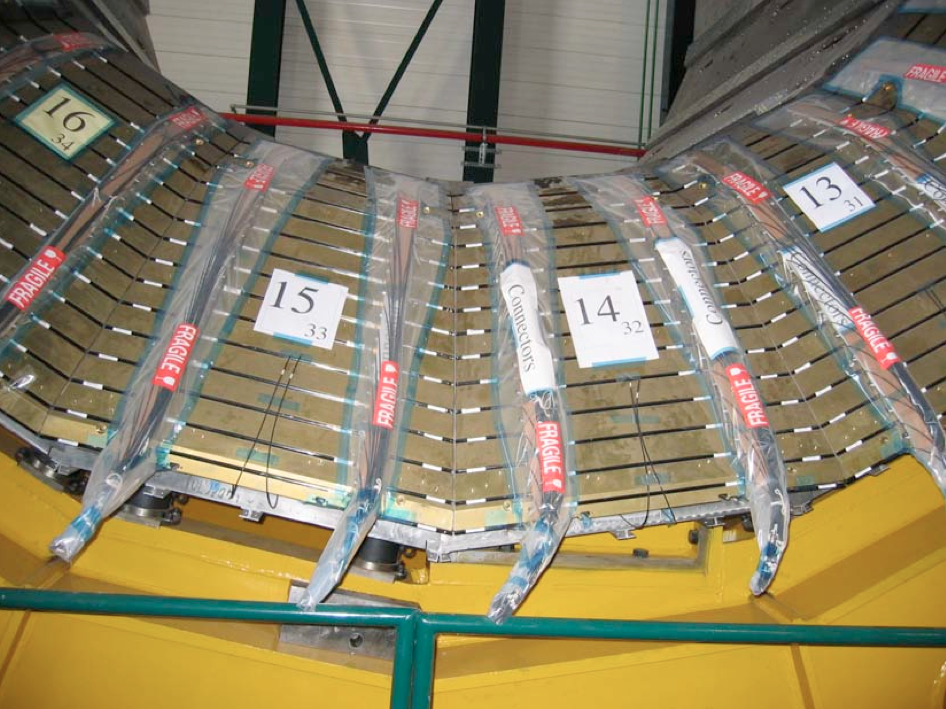}
					\includegraphics[height = 5cm]{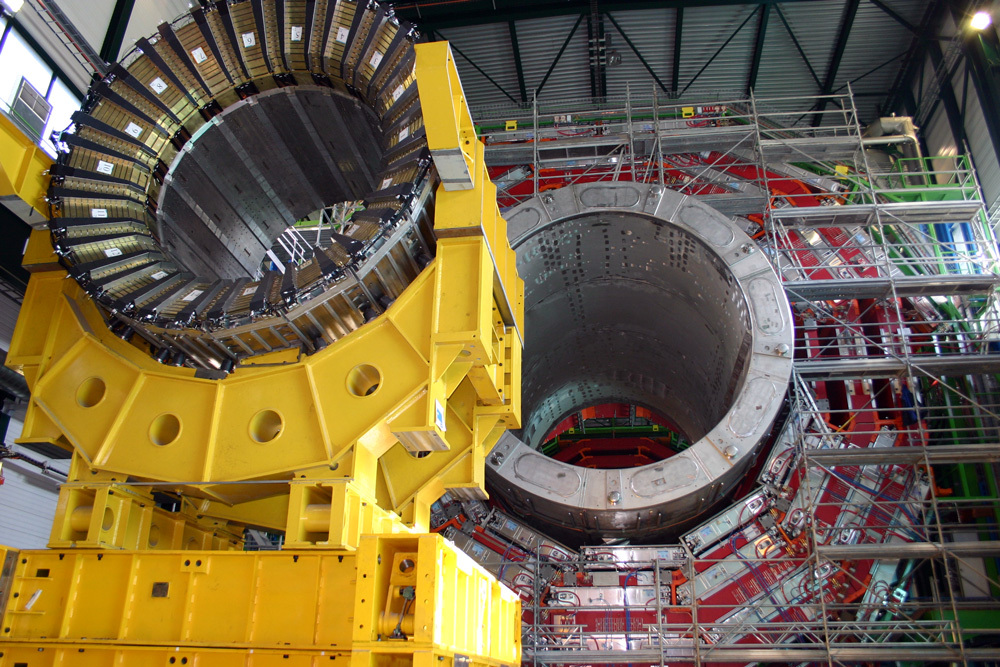}
					\caption{Picture of the barrel HCAL composed of several dense absorber (golden plates) and smaller scintillators placed in between (left) \Cite{CMS_at_LHC}, and installation in CMS of the barrel HCAL (right) \Cite{CMS_HCAL_Install}}
					\label{fig:lhc_and_cms__cms_hcal_view}
				\end{figure}

				Figures \ref{fig:lhc_and_cms__cms_hcal_view} are a picture of a section of the barrel HCAL representing the absorber (golden plates) with the scintillator in between and of the installation of the barrel HCAL in CMS. The HCAL is composed of an alternation of 16 layers of absorbers, made out of 40 to 70 mm thick steel plates and 50 to 56 mm thick 70\% Cu and 30\% Zn alloy plates, and 3.7 to 9 mm thick plastic scintillators. When particles hit the detectors perpendicularly, they have to travel through 79 cm of matter equivalent to 5.82 interaction lengths $ \lambda_I $. The barrel HCAL is divided into 72 segments in $ \phi $ and 16 $ \eta $ sectors while the endcap HCAL has 36 and 72 $ \phi $ segments for the inners and outers rings respectively, and 14 $ \eta $ sectors.

			\subsubsection{System Performances}  
			\label{sec:lhc_and_cms__calorimeters_system_performances}	

				The CMS ECAL's energy resolution is \Cite{CMS_Performances}
				\begin{equation}
					\left( \frac{\sigma_E}{E} \right)^2 = \left( \frac{2.8\%}{\sqrt{E}} \right)^2 + \left( 0.30\% \right)^2 + \left( \frac{0.12}{E} \right)^2 \ ,
				\end{equation}		
				where $ E $ is given in GeV. The CMS HCAL's energy resolution is
				\begin{equation}
					\left( \frac{\sigma_E}{E} \right)^2 = \left( \frac{120\%}{\sqrt{E}} \right)^2 + \left( 6.9\% \right)^2 \ .
				\end{equation}							
			
		\subsection{Solenoid}
		\label{sec:lhc_and_cms__solenoid}

\begin{figure}[h!]
				\centering
				\includegraphics[width = 12cm]{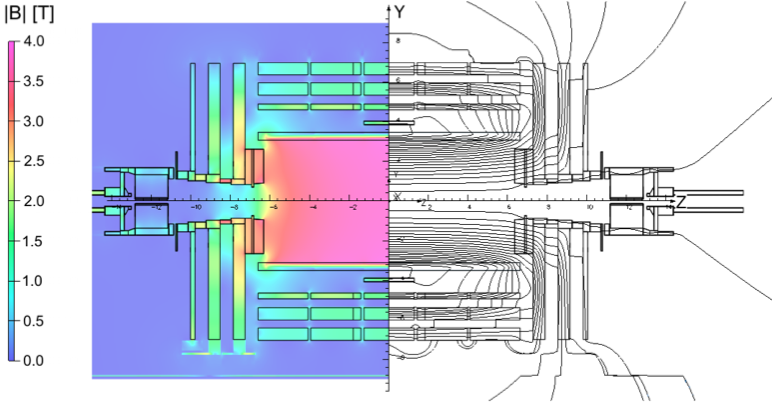}
				\caption{Field map of the magnetic field of CMS measured using cosmic rays \Cite{CMS_B_Field}.}
				\label{fig:lhc_and_cms__cms_magnetic_field}
			\end{figure}

			The intense magnetic field of CMS is created by cooling a solenoid down to 4.5 K, temperature at which the metal becomes supra-conductive, and by passing strong currents through it. The resulting field is uniform inside the solenoid but more complex outside, as shown in Figure \ref{fig:lhc_and_cms__cms_magnetic_field} which represents the measured magnetic field. The constant and strong field in which the tracker is placed allows it to measure the particles' transverse momentum with high-precision (resolution of less than 1\% in the tracker as seen in Figure \ref{fig:lhc_and_cms__cms_tracker_performances}). The intensity of the field is of 3.8 T inside the solenoid and typically 2 T outside the solenoid. \\

			Having the calorimeters inside the magnet improves the energy resolution as particles have less matter to travel through before reaching them, but increases the size of the solenoid. Due to the technical difficulty to build large magnets, the muon chambers are placed on the outside. This layout has the advantage to use the magnet as barrier for most particles escaping the calorimeters, ensuring that only muons will be detected by the muon system.
		
		\subsection{Muon Chambers}
		\label{sec:lhc_and_cms__muon_chambers}

			The muons system is composed of several types of gaseous detectors which physics and functioning are exposed in the next chapter. They are placed in the outer section of CMS because muons are the only detectable particles left at this stage as they do not or slightly interact with the calorimeters.
		
	\cleardoublepage


\chapter{Muon Chambers}
\label{chap:muon_chambers}

	This chapter reviews the theory behind gaseous detectors and their use inside CMS. We first describe the interactions between the particles and the detectors, which create small energy depositions. We then detail how these are amplified by the chambers, and the signals that they generate on the readout electronics. We also define parameters of the detectors that will be of importance inside CMS. Finally, we look at the functioning of the technologies used inside the CMS muon spectrometer and the challenges they will face in the future.
	
	\section{Particle Detection in Gas Detectors} 
	\label{sec:muon_chambers__particle_detection_in_gas_detectors}

		Multiple processes are involved between the interactions of the particle with the detectors, and the formation of the signals at the output of the readout electronics. All these steps will be detailed in the following.
		
		\subsection{Energy Losses}
		\label{sec:muon_chambers__energy_loss}

\begin{figure}[h!]
				\centering
				\includegraphics[width = 12cm]{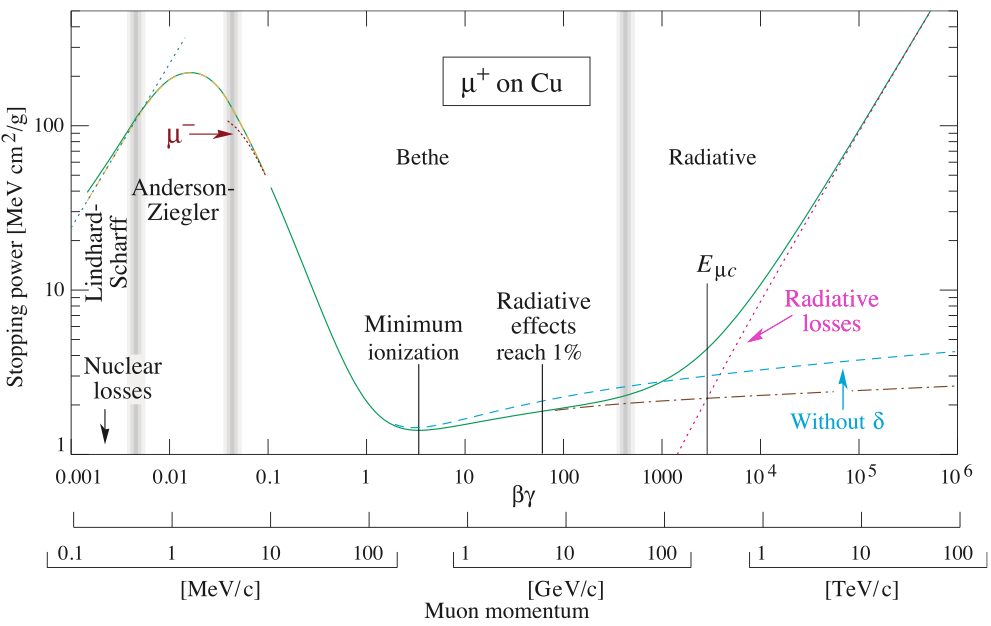}
				\caption{Mean energy loss per unit of length of traversed material, normalized to the density of the medium $ - \left\langle \frac{d E}{dX} \right\rangle \frac{1}{\rho} $ as a function of the $ \beta \gamma $ parameter measured using muons passing through copper \Cite{PDG_Particle_Review}.}
				\label{fig:muon_chambers__energy_lost}
			\end{figure}

			Particles passing through matter \Cite{Muons_Drift_Chambers} can interact with the medium through multiple processes. For instance, electrons lose energy through radiative processes, and photons above 1 MeV undergo $ e^+ / e^- $ pair creation, as reviewed in Section \ref{sec:lhc_and_cms__calorimeters}. Muons' energy losses, on the other hand, are dominated by the Coulomb interaction with the electrons of the medium. Figure \ref{fig:muon_chambers__energy_lost} shows the mean energy loss per unit of length of traversed material, normalized to the density of the medium $ - \left\langle \frac{d E}{dX} \right\rangle \frac{1}{\rho} $, as a function of the $ \beta \gamma $ parameter for muons passing through copper. \\

			For particles in the $ 0.1 \lesssim \beta \gamma \lesssim 1000 $ region, the energy losses have been quantified by Bethe as follows \Cite{PDG_Particle_Review}
			\begin{equation} 
				- \left\langle \frac{d E}{dX} \right\rangle = K z^2 \frac{Z}{A} \frac{1}{\beta^2} \left[ \frac{1}{2} \ln \frac{2 m_e c^2 \beta^2 \gamma^2 T_{max}}{I^2} - \beta^2 - \frac{\delta(\beta \gamma)}{2} \right] \ ,
				\label{eq:muon_chambers__bethe}
			\end{equation}
			where $ z $ is the charge number of the particle, $ Z $ the atomic number of the medium, $ A $ the atomic mass of the medium, $ m_e $ the mass of the electron, $ r_e $ the classical electron radius, $ T_{max} $ the maximum energy that can be transfered during a collision, $ I $ the mean excitation energy of the medium, $ \delta(\beta \gamma) $ an ultra-relativistic correction, and $ K = 4 \pi N_A r_e^2 m_e c^2 $. We can outline that Equation \ref{eq:muon_chambers__bethe} only involves the $ \beta \gamma $ parameter and not the mass of the incident particles. \\

\begin{figure}[h!]
				\centering
				\includegraphics[width = 9cm]{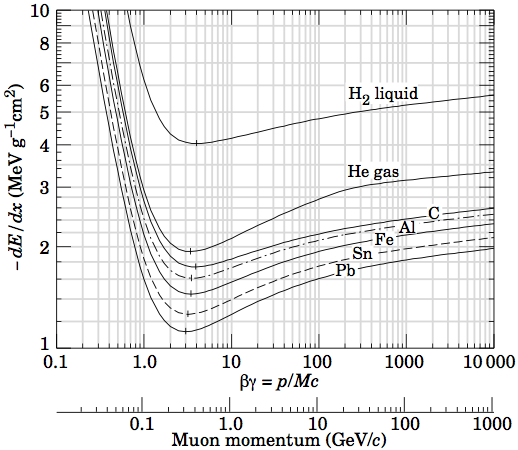}
				\caption{Mean energy loss per unit of length of traversed material, normalized to the density of the medium as a function of the $ \beta \gamma $ parameter for multiple materials and gases near the MIP \Cite{PDG_Particle_Review}.}
				\label{fig:muon_chambers__pdg_bethe}
			\end{figure}

			The plot in Figure \ref{fig:muon_chambers__pdg_bethe} depicts the evolution of the mean energy loss per unit of length of traversed material, normalized to the density of the medium as a function of the $ \beta \gamma $ parameter for different types of media. Both ends of the curve are highly $ Z $ dependent. At low energies ($ \beta \gamma < 1 $), the speed of the particles is comparable to the speed of the atomic electrons, significantly increasing the stopping power of the material. The opposite effect, the \emph{density effect} $ \delta(\beta \gamma) $, arises at high energies ($ \beta \gamma > 10 $). When particles become ultra-relativistic, they create a temporary polarization of the medium screening the interactions with the electrons and reducing the energy losses. In the $ 1 < \beta \gamma < 10 $ region, energy losses are minimal, giving particles of such energies a specific name: \emph{Minimum Ionizing Particles} (MIP). Detectors must be tested in this specific region to ensure their detection capabilities and to calibrate their response. \\

			It is important to emphasize that energy losses are stochastic processes. Therefore, particles do not leave behind a constant trail of ionization, but rather localized energy depositions, resulting in the detection of multiple hits inside one detector. The deposited energy follows a Landau distribution where the most probable value is given by Bethe's formula review in Equation \ref{eq:muon_chambers__bethe}.

		\subsection{Primary and Secondary Ionizations}
		\label{sec:muon_chambers__primary_and_secondary_ionization}

			Energy losses for muons result in an energy transfer from the muons to the electrons of the medium. If the energy losses are greater than the \emph{ionization potential} $ I $ of the medium, atomic electrons can be ejected from their atom, creating the primary ionization
			\begin{equation}
				A + \gamma \rightarrow A^+ + e^- \ .
			\end{equation}
			If these electrons have a kinetic energy $ K $ so that $ K > I $, they can in turn ionize the medium. This is the secondary ionization. \\

			More complex processes can occur, such as the Auger effect, during which an electron from higher orbitals takes the place of the ejected electron, emitting a photon. This photon can either escape and remain undetected or, in most cases, be absorbed by another electron causing a double ionization
			\begin{equation}
				A + \gamma \rightarrow A^{+*} + e^- \rightarrow A^{++} + e^- + e^- \ .
			\end{equation} \\

			In fact, not all the energy is used to ionize the medium. The average number of ionizations (primary and secondary) by unit of length $ n $ is given by 
			\begin{equation}
				n = \frac{1}{W} \left\langle \frac{dE}{dX} \right\rangle \ ,
			\end{equation}
			where $ W $ is the mean energy needed to ionize the medium. Table \ref{tab:muon_chambers__number_of_electrons} lists the values of $ W $ and $ n $ for different gases. Considering that the detectors are only a few millimeters thick, the number of electrons created is relatively small and has to be amplified as will be described in the next section.

			\begin{table}[h!]
				\centering
				\begin{tabular}{l|c|c}
					Gas & W (eV) & n (cm$ ^{-1} $) \\ \hline
					Ar & 26 & 97 \\
					CH$ _4 $ & 30 & 54 \\
					CO$ _2 $ & 34 & 100 \\
					CF$ _4 $ & 54 & 120  
				\end{tabular}
				\caption{Mean energy required to ionize the medium $ W $ and average number of ionizations by unit of length $ n $ for various gases \Cite{PDG_Particle_Review}.}
				\label{tab:muon_chambers__number_of_electrons}
			\end{table}			
		
		\subsection{Signal Amplification}
		\label{sec:muon_chambers__signal_amplification}

			An amplification stage is required as the readout electronics generates a noise around the 1000 $ e^- $ \emph{Equivalent Noise Charge} (ENC), which would entirely mask the signal (at maximum a few hundreds electrons). Therefore, the signal is amplified inside the chambers using strong electric fields. In the regions where the field is intense (several kV cm$ ^{-1} $), electrons gain enough energy to ionize the medium and create avalanches.
			
			\subsubsection{Formation of the Avalanche}
			\label{sec:muon_chambers__formation_of_the_avalanche}			

\begin{figure}[h!]
					\centering
					\includegraphics[width = 11cm]{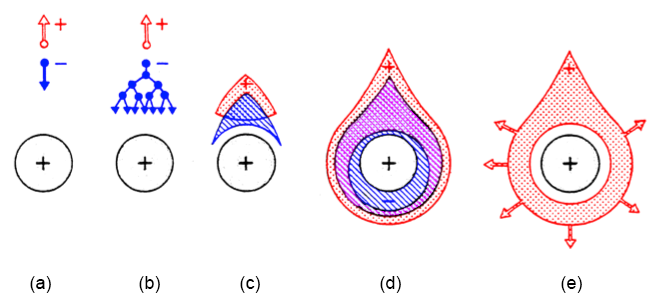}
					\caption{Development of an avalanche near an anode wire \Cite{Muons_Avalanches}.}
					\label{fig:muon_chambers__avalanches}
				\end{figure}

				Once the avalanche begins, it develops exponentially as represented in Figures \ref{fig:muon_chambers__avalanches} (a), (b), and (c). The electrons moving faster than the ions, they accumulate at the front of the avalanche, while the ions slowly drift towards the cathode. This process continues until the electric field generated by the electrons and ions compensates the external field. At that moment, typically for gains of $ 10^8 $, the charges start to recombine, preventing further multiplication. \\

				Figures \ref{fig:muon_chambers__avalanches} (d) and (e) illustrate the case of an avalanche surrounding an anode wire. Once the electrons are absorbed by the electrode, the repulsion between the ions and the anode is no longer screened, causing the ions to drift away. Meanwhile, the wire is said to be dead, as no other signal than the ions' one can be formed. This means that no other particle can be detected by this particular wire as long as the ions have not been evacuated.
			
			\subsubsection{Gain of the Avalanche}
			\label{sec:muon_chambers__gain_of_the_avalanche}

				In order to quantify the gain of the avalanche $ G $, it is defined as
				\begin{equation}
					G = \frac{N}{N_0} \ ,
				\end{equation}
				where $ N $ is the number of electrons in the final state, and $ N_0 $ is the number of electrons in the initial state. $ N $ will depend upon the width of the region where the avalanche occurs $ L $, and upon the first Townsend coefficient $ \alpha(x) $. $ \alpha $ is an empiric parameter varying from gas to gas according to the cross-section of the ionization, the intensity of the field, and other, more complex, parameters. It cannot be computed and has to be experimentally measured for each mixture. \\

				$ \alpha $ yields the variation of the number of electrons $ dN $ along a path $ dx $ through the following relation
				\begin{equation}
					dN = N \alpha(x) dx \ .
				\end{equation}
				Once this equation is integrated, it can be determined that the gain for a certain distance $ L $ is given by
				\begin{equation}
					G = \exp\left( \int_0^L \alpha(x) dx \right) \ .
				\end{equation} \\

				Even if $ \alpha(x) $ cannot be predicted, certain tendencies appear. Mono-atomic gases have a larger Townsend coefficient than poly-atomic gases as they have less degrees of freedom (rotation, vibration, etc) that can absorb energy and prevent ionization. On the other hand, they emit highly energetic photons that can damage the components of the detectors. To absorb this radiation, a gas with multiple degrees of freedom called a \emph{quencher} is added. This complex element will also prevent the formation of avalanches caused photons emitted by the recombination of electrons and ions. \\

		\subsection{Electrons and Ions Drift}
		\label{sec:muon_chambers__electron_and_ion_drift}

			In order to illustrate the fact that electrons drift faster than ions, let us consider a gas at thermodynamical equilibrium. The energy of the particles is given by the Maxwell-Boltzmann distribution
			\begin{equation}
				\langle \epsilon \rangle = \frac{1}{2} m \langle v^2 \rangle = \frac{3}{2} k T \ .
			\end{equation}
			This simple relation shows that due to the ions' higher mass, their drift will be slower. A complete derivation in the presence of a magnetic and electric field will not be presented in this thesis, but can be found in reference \Cite{Muons_Drift_Chambers}. The following example provides an order of magnitude for the ions' drift speed: [CO$ _2 $]$ ^+ $ ions in a Ar gas in an electric field of 100 V cm$ ^{-1} $ drift at a speed of 172 cm s$ ^{-1} $, meaning they cross a 1 cm wide volume (roughly the width of the hereafter considered detector) in 5.8 ms. In the same gas and electric field, the electrons' drift speed reaches up to 2 10$ ^5 $ cm s$ ^{-1} $, meaning they cross a 1 cm wide volume in 5 \us{}. \\

			While drifting, electrons and ions scatter with the gas, spreading the primary and secondary ionizations along a Gaussian distribution. The diffusion's amplitude mainly depends upon the composition of the gas and is much more significant for electrons than for ions. For example, electrons in a CO$ _2 $ gas in an electric field of 100 V cm$ ^{-1} $ will have a diffusion coefficient $ D $ of the order of 230 \um{} cm$ ^{-0.5} $ so that the transverse spread $ \sigma_{xy} $ after a traveled distance $ z $ is given by
			\begin{equation}
				\sigma^2_{xy} = D^2 \ z \ . 
			\end{equation}
			Ions' diffusion is typically two orders of magnitude smaller than for the electrons.
		
		\subsection{Importance of the Geometry}
		\label{sec:muon_chambers__importance_of_the_geometry}

\begin{figure}[h!]
				\centering
				\includegraphics[width = 11cm]{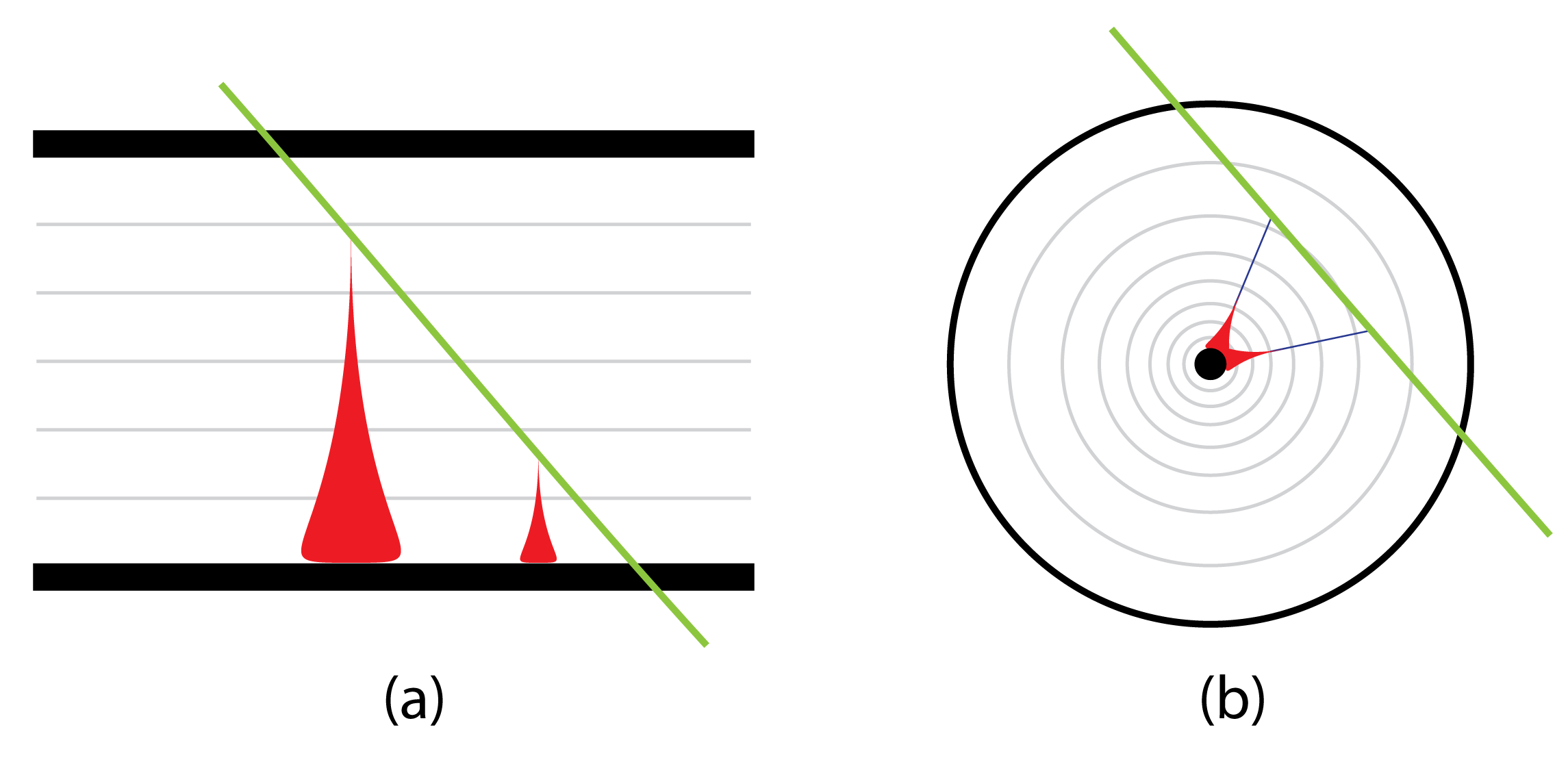}
				\caption{Schematic representations of two electrode planes (a) and a wire chamber (b) and the creation of the avalanches.}
				\label{fig:muon_chambers__geometries}
			\end{figure}

			The geometry of the detector, including the shape of the electrodes and their placing, significantly influences the formation of the avalanches and their shape. Figure \ref{fig:muon_chambers__geometries} depicts the two examples that will be reviewed: two electrode planes facing each other (a) and an anode wire placed inside a cathode tube (b).

			\paragraph*{Electrode Planes}
			\label{sec:muon_chambers__electrode_planes}	

				The electric field generated by the two facing electrodes is constant in all the volume. This results in the creation of avalanches anywhere in the chamber. In consequence, the size of the avalanche will not depend upon the amount of energy deposited by the particle, but rather upon the distance at which the avalanche occurred.
			
			\paragraph*{Wire Chamber}
			\label{sec:muon_chambers__wire_chamber}				

				In a wire chamber, the electric field goes as $ r^{-1} $. Therefore, the avalanches will only initiate when the electrons have reached a certain distance to the wire where the electric field becomes more intense. All the signals will be amplified equally, meaning that their intensity will be related to the size of the energy deposition. \\
			
			It is also important to take into account the ions' evacuation time. As described in Section \ref{sec:muon_chambers__formation_of_the_avalanche}, parts of the chambers will remain unusable as long as the ions disturb the signal formation. The faster they reach the cathode, the more events the chambers will be able to detect. In both previously described cases, the ions must drift a long distance before reaching the cathode, yielding a long dead time.

		\subsection{Signal Formation}
		\label{sec:muon_chambers__signal_formation}

			The signal is formed by the induced currents created on the electrodes by the electrons' and ions' drift, and not by the collection of the charges. Indeed, the collection time is too short to generate a visible current. Therefore, the longer the drift time is, the more intense the signal will be. \\

			\begin{figure}[h!]
				\centering
				\includegraphics[width = 9cm]{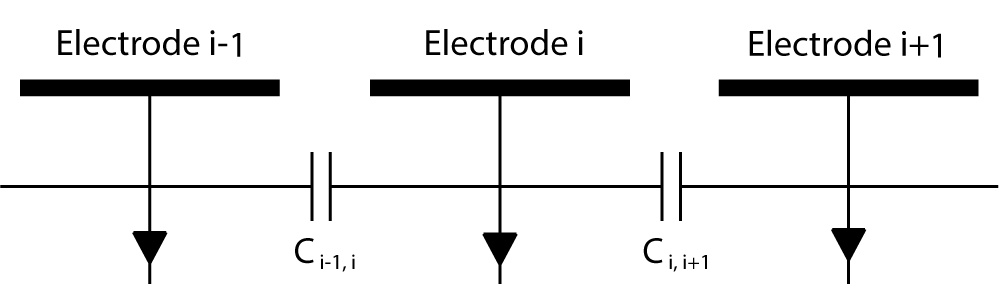}
				\caption{Simplistic view of a series of electrodes connected to each other by capacitors.}
				\label{fig:muon_chambers__electrodes}
			\end{figure}	

			In order to compute the current induced on a set of electrodes connected to each other by readout electronics, as represented in Figure \ref{fig:muon_chambers__electrodes}, we apply the Shockley-Ramo theorem \Cite{Misc_Shockley, Misc_Ramo}. When considering $ n $ electrodes and aiming to compute the induced current on the $ i $th one, the weighting field $ \bar{\mathbf{E}}_i $ created by placing that electrode at a potential $ V_i $ and all other to the ground must first be determined. The induced current on electrode $ i $ is then given by
			\begin{equation}
				I_i(t) = - \frac{q}{V_i} \bar{\mathbf{E}}_i(\bar{\mathbf{x}}(t)) \bar{\mathbf{v}}(t) \ ,
			\end{equation}
			where $ q $ is the charge of the particle $ \bar{\mathbf{x}}(t) $ the position of the charge and $ \bar{\mathbf{v}}(t) $ the velocity of the charge. These currents are then used as ideal current sources in a circuit where all the electronic components are present. By taking into account all the weighting fields, the capacitances between the electrodes $ C_{i, i+1} $ can be compute and yield the measured current. In order to calculate the current induced by multiple charges, the individually induced currents must be summed up.

		\subsection{Readout Electronics}
		\label{sec:muon_chambers__read_out_electronic}

\begin{figure}[h!]
				\centering
				\includegraphics[width = 11cm]{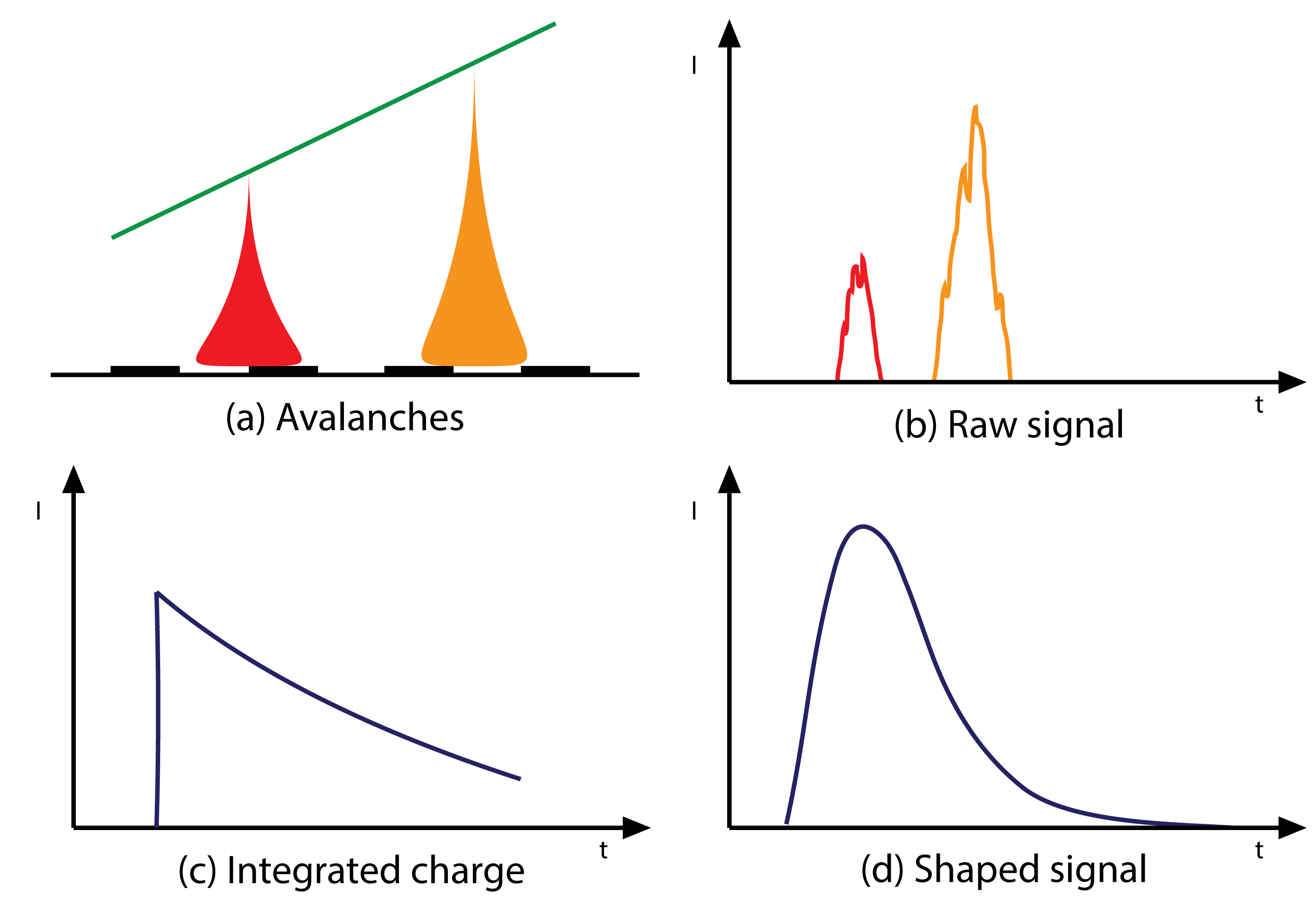}
				\caption{Signal treatment from the creation of the avalanches (a), the formation of the signal on the electrodes (b), the integration of the raw signal (c), to the shaping (d).}
				\label{fig:muon_chambers__shaping}
			\end{figure}

			The current induced on the electrodes depends upon the number of charges hence the energy lost by the particle. More precisely, it varies with the energy's distribution inside the chamber. If, for example, a particle created two primary ionization clusters inside the chamber, two peaks in the current distribution should be observe (if they are not overlapping) as represented in Figures \ref{fig:muon_chambers__shaping} (a) and (b). This makes the raw signals unusable as their shape cannot predict. Therefore, the readout electronics should first integrate the total deposited charge (c) and then shape it (d). Defining the impulsion response $ T(t) $ as the response of the readout electronics to a delta function like input pulse, for an arbitrary input signal $ S_{in}(t) $, the output signal is given by
			\begin{equation} 
				S_{out}(t) = (T * S_{in})(t) = \int_{-\infty}^\infty T(t - \tau) \ S_{in}(\tau) \ d\tau \ .
			\end{equation}		
			All the signals will have the same distribution, but will differ in amplitude, making these usable to perform the event's analysis. \\

\begin{figure}[h!]
				\centering
				\includegraphics[width = 8cm]{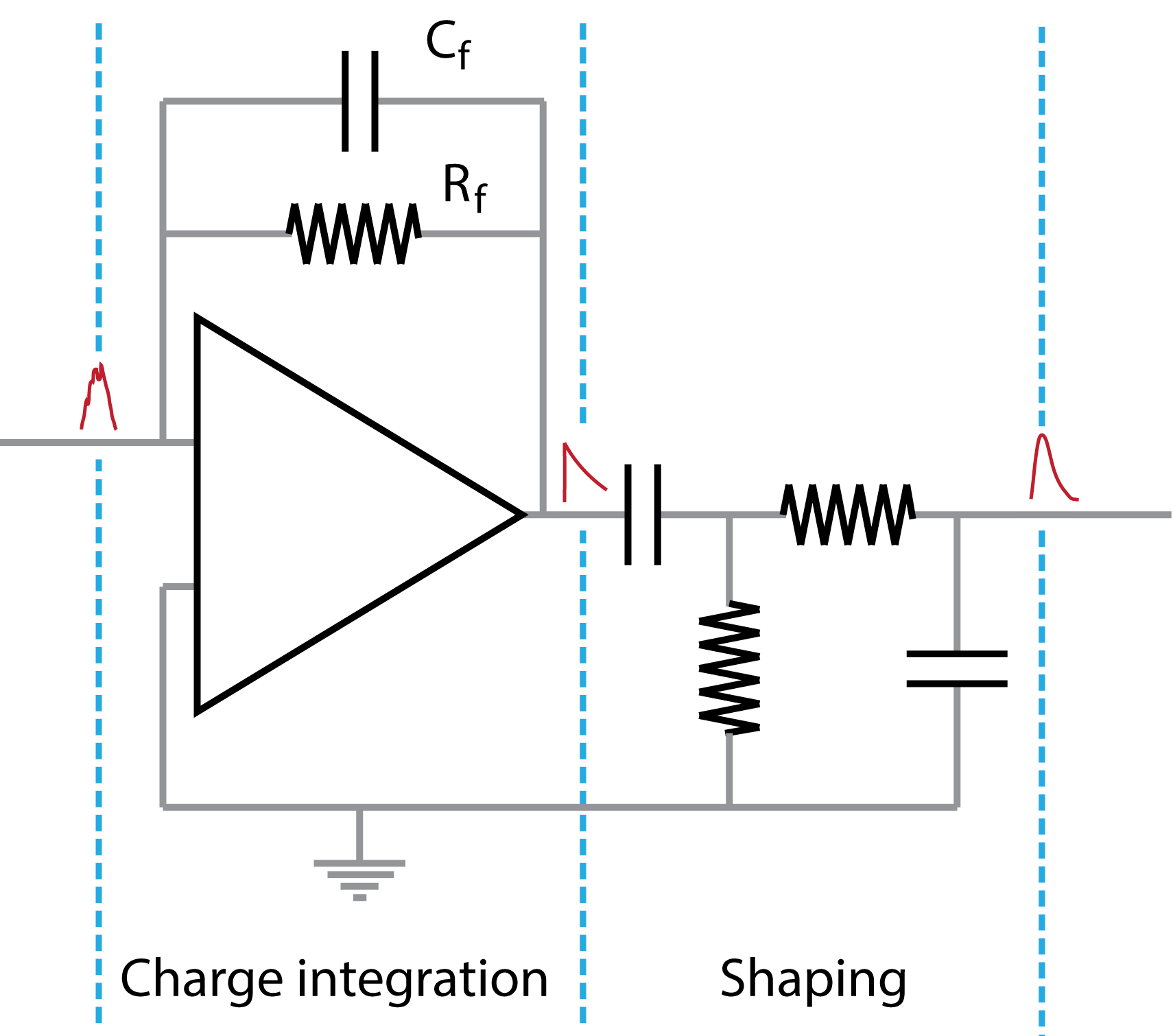}
				\caption{Schematic view of a simplistic circuit performing charge integration and shaping of the raw signal.}
				\label{fig:muon_chambers__shaping_electronic}
			\end{figure}

			Figure \ref{fig:muon_chambers__shaping_electronic} illustrates a typical electronic circuit performing charge integration and shaping. The feedback capacitor $ C_f $ is rapidly charged by the current arriving from the electrodes and slowly discharged through the feedback resistor $ R_f $. This prevents the capacitor to continuously charge and reach its maximum capacity. High and low frequency filters then cut the tails of the distribution to shorten the length of the signal.

	\section{Multiple Scattering}
	\label{sec:muon_chambers__multiple_scattering}

		While passing through matter, particles scatter with the atoms of the medium, slightly deviating from their original trajectory. This is called \emph{multiple scattering}. \\

		\begin{figure}[h!]
			\centering
			\includegraphics[width = 9cm]{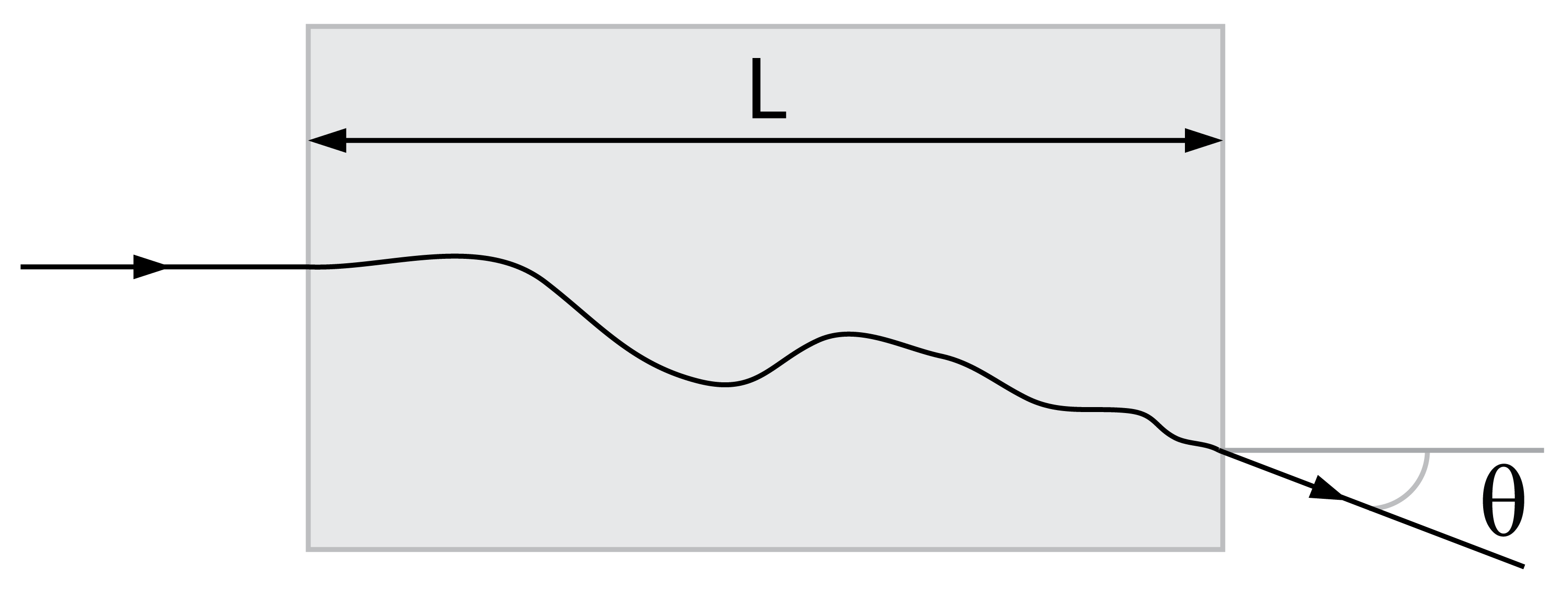}
			\caption{Scattering of a particles that passes through matter and deflects from its original trajectory.}
			\label{fig:muons_chambers_multiple_scattering}
		\end{figure}			

		The root mean square of the deviation angle $ \theta $ for a travel distance $ L $, as depicted in Figure \ref{fig:muons_chambers_multiple_scattering}, is given by 
		\begin{equation}
			\theta_{RMS} = \sqrt{2} \frac{13.6[MeV]}{\beta c p[MeV c^{-1}]} z \sqrt{\frac{L}{X_0}} \ ,
			\label{eq:muon_chambers__multiple_scattering_RMS}
		\end{equation}
		where $ c $ is the speed of light, $ p $ is the magnitude of the momentum, $ z $ is the charge number of the particle, and $ X_0 $ is the radiation length of the medium. 98\% of $ \theta $'s distribution is Gaussian leaving a 2\% non-Gaussian component for the tails. This effect plays an important role at low momenta but is negligible at higher energies.
	
	\section{Characterizing Parameters}
	\label{sec:muon_chambers__characterizing_parameters}

		Four parameters characterizing the detectors are of importance in CMS: the spatial resolution, the time resolution, the detection efficiency, and the rate capability, each of which plays an important role in the reconstruction of the events.
		
		\subsection{Spatial Resolution} 
		\label{sec:muon_chambers__spatial_resolution}

			The spatial resolution is the error on the position's measurement made by the detectors. It yields the resolution on the momentum when reconstructing the track, as hits are used to measure the bending radius of the trajectories. The momentum being the key quantity used to analyze events, it is important to ensure a good spatial resolution on the hits' position, of the order of 250 to 500 \um{} in the CMS muon spectrometer. For gaseous detectors, the spatial resolution will vary with the placement and spacing of the electrodes. As reviewed in Section \ref{sec:lhc_and_cms__silicon_strip_detectors}, some detectors only measure one direction, neglecting the other. This will often be the case in muon chambers as the area to cover is quiet large, and costs must stay reasonable. 
		
		\subsection{Time Resolution} 
		\label{sec:muon_chambers__time_resolution}

			The time resolution is important in order to correctly assign particles to a certain \emph{Bunch Crossing} (BX). With the high frequency at which the LHC runs, it is important to determine during which collision a certain particle was created. For detectors close to the IP, the \emph{Time of Flight} (TOF) is relatively short. However, muons reaching the muon system can have a TOF up to 40 ns. This means that particles from one interaction reach the outer detectors while another collision already took place. To unambiguously assign an event to a BX, the time resolution must be less than to 5 ns.
		
		\subsection{Detection Efficiency} 
		\label{sec:muon_chambers__detection_efficiency}

			The detection efficiency $ \epsilon $ is the percentage of particles passing through the chamber and which are detected. CMS requires all detectors to have a minimum efficiency of 95\%. 	

		\subsection{Rate Capability}
		\label{sec:muon_chambers__acceptance_rate}

			The rate is the maximal flux of particles under which the detection efficiency remains above 95\%. As noted in Sections \ref{sec:muon_chambers__formation_of_the_avalanche} and \ref{sec:muon_chambers__importance_of_the_geometry}, parts of the gaseous detectors remain unusable for a certain amount of time after the avalanche process. In order to increase the system's efficiency, the rate capability should be of the order of the particles' flux traversing the detectors. Detectors in the most forward region of CMS should be able to sustain rates of the order of 1 kHz cm$ ^{-2} $.

	\section{CMS Muon System}
	\label{sec:muon_chambers__cms_muon_system}

		Currently, the CMS muon system \Cite{CMS_at_LHC, CMS_Performances} is composed of three different types of gaseous detectors: \emph{Drift Tube} (DT), \emph{Cathode Strip Chamber} (CSC), and \emph{Resistive Plate Chamber} (RPC).

		\subsection{Disposition of the Detectors}
		\label{sec:muon_chambers__disposition_of_the_detectors}

\begin{figure}[h!]
				\centering
				\includegraphics[width = 13cm]{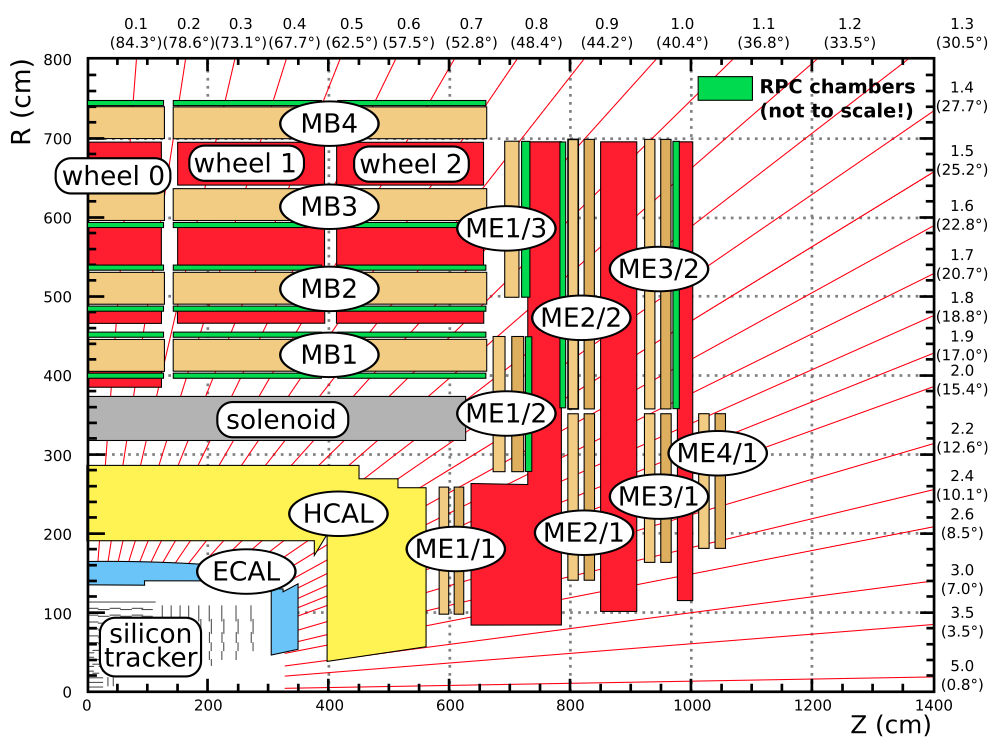}
				\caption{Disposition of the muon chambers inside CMS. MBn refer to DTs, MEn to CSCs and the green lines to RPCs \Cite{CMS_Upgrades}.}
				\label{fig:muon_chambers__placement}
			\end{figure}

			Like all the CMS detectors, the muon system is divided into two regions: the barrel ($ | \eta | $ < 1) and the endcaps (1 < $ | \eta | $ < 2.4). The chambers are regrouped into stations attached to the wheels of CMS. The barrel stations contain DTs (identified by MBn) and RPCs while the endcaps stations hold CSCs (identified by MEx/y) and RPCs (identified by REn), as represented in Figure \ref{fig:muon_chambers__placement}. For financial reasons, the RPCs were not installed for the LHC's start-up in the 1.6 < $ | \eta | $ < 2.4 region where only CSCs are present. 

\begin{figure}[p!]
				\centering
				\includegraphics[width = 16.5cm]{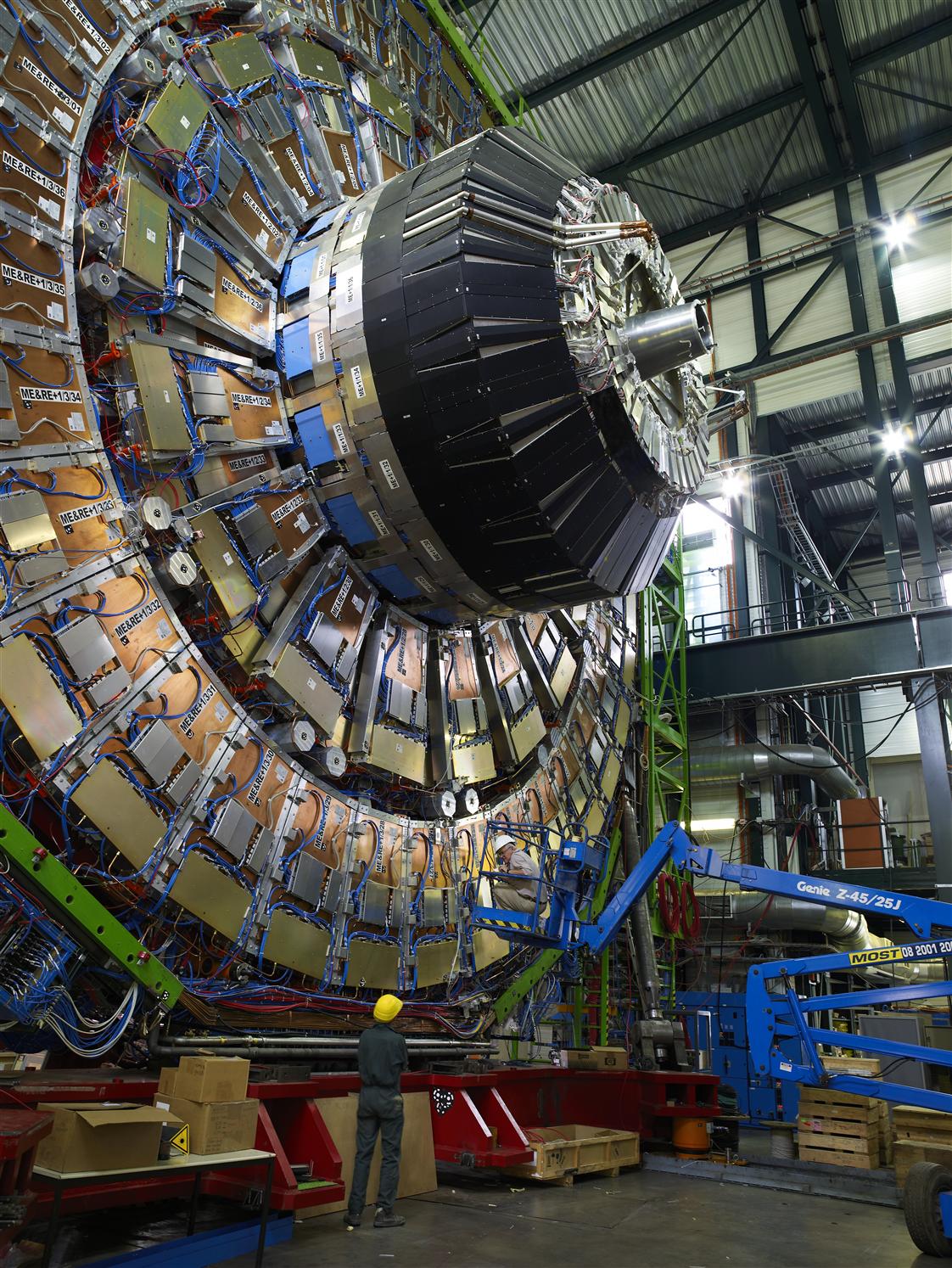}
				\caption{Picture of one of the endcaps' yokes. We can observe the two outer rings, ME1/2 and ME1/3. Chambers of the inner ring, ME1/1, are hidden inside the \emph{nose}, in black \Cite{Fig_CMS_Endcap}.}
				\label{fig:muon_chambers__cms_endcap}
			\end{figure}

			The barrel is composed of 5 wheels on which 4 layers of detectors are attached, each divided into 12 stations along $ \phi $. The endcaps have 4 layers of detectors divided into 1, 2 or 3 rings partitioned into 36 or 72 stations that overlap to ensure maximum efficiency. Figure \ref{fig:muon_chambers__cms_endcap} shows the first station of the muon endcap, ME1. The inner ring, called ME1/1 is hidden by the so-called \emph{nose}, in black. The two outer rings, ME1/2 and ME1/3 are well visible. In ME1/2, we can observe the overlap between the chambers. \\

			The use of two different kinds of detectors in each station ensures that the system meets the required detection efficiency for muons imposed by CMS. This redundancy is crucial to select and reconstruct events with high momentum muons in the final state, signature of the Brout-Englert-Higgs boson's decay and of many processes of new physics, including super-symmetry.
		
		\subsection{Drift Tubes}
		\label{sec:muon_chambers__drift_tubes}

\begin{figure}[h!]
				\centering
				\includegraphics[width = 8cm]{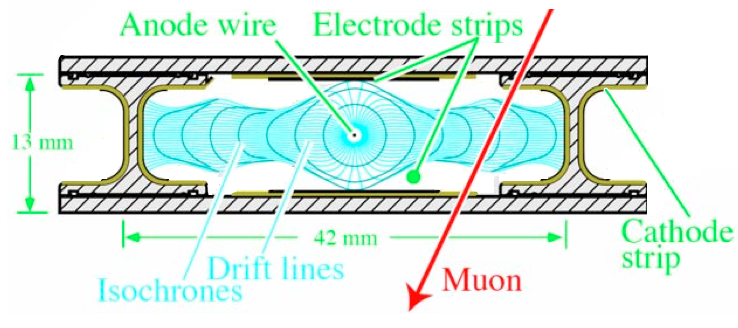}
				\caption{Schematic view of a drift cell along with the electric field line \Cite{CMS_at_LHC}.}
				\label{fig:muon_chambers__dt}
			\end{figure}

			DTs are rectangular parallelepiped detectors composed of an anode wire stretched between two cathode strips as represented in Figure \ref{fig:muon_chambers__dt}. The chambers are 2.4 m long by 13 mm height by 42 mm wide. A strong electric field (of the order of 1.5 kV cm$ ^{-1} $) is formed by applying a high voltage difference between the electrodes, causing the electrons and ions to drift into the gas, and provoking avalanches near the anode. The two electrodes placed near the anode help flatten the electric field and improve the charges' drift. \\

			Four DTs are assembled to create a \emph{Super Layer} (SL), and two or three SLs compose a DT module. Each SL has a spatial resolution of 100 \um{} in the direction perpendicular to the wire. To improve global precision, two SLs are used to measure the $ \phi $ coordinate and sometimes one additional SL is used to measure $ \eta $. DT modules have a time resolution of 3 ns. Their rather large size limits their rate capabilities, explaining why they are only present in the barrel where particles' fluxes are lower (< 10 Hz cm$ ^{-2} $).

		\subsection{Cathode Strip Chambers}
		\label{sec:muon_chambers__cathode_strip_chambers}

\begin{figure}[h!]
				\centering
				\includegraphics[height = 4.5cm]{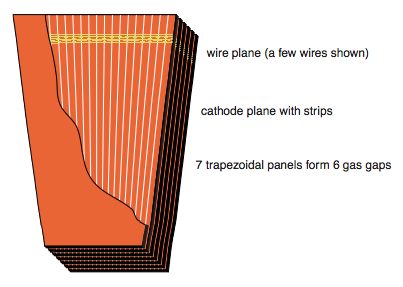}
				\caption{A representation of a CSC with its wires and strips \Cite{CMS_Performances}}
				\label{fig:muon_chambers__csc}
			\end{figure}

			CSCs are trapezoidal multiwire proportional chambers placed in the endcaps of CMS. Multiple anode wires (about 1000 spaced by 3.2 mm) are stretched radially in the chamber above perpendicularly placed cathode strips (typically 80 separated by a pitch of 8.4 mm on the narrow side and 16 mm on the large side) as depicted in Figure \ref{fig:muon_chambers__csc}. As for the DTs, an electric field is formed between the wires and the strips, accelerating the electrons and forming the avalanches near the anodes. By reading-out both electrodes, the CSCs provide a measurement of both coordinates.

			One CSC module is made out of six chambers put together (7 cathode planes and 6 wire planes). Due to the large number of readout channels in these modules, the spatial resolution is as good as 33 \um{} for ME1/1 and ME1/2, and 80 \um{} for the other stations. The time resolution for one cathode plane is 11 ns that can be brought down to the order of 5 ns when combining the measurements of all the planes. The largest CSC modules, ME2/2 and ME3/2, are 3.4 m by 1.5 m. \\ 

\begin{figure}[h!]
				\centering
				\includegraphics[width = 8cm]{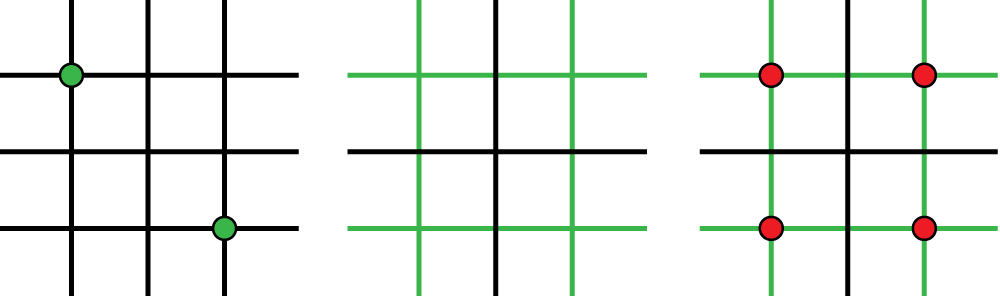}
				\caption{Ambiguities arise when more than one particle hit the chamber at the same time.}
				\label{fig:muon_chambers__ghosts}
			\end{figure}

			Note that the two dimensional readout configuration can create ambiguities called \emph{ghosts particles} as shown in Figure \ref{fig:muon_chambers__ghosts}. When two particles hit the chamber (left), four possibilities arise when reading the output signal (right). Two of them correspond to real particles, and the two others to ghost particles. It is easy to see that for $ n $ particles interacting in the chamber, $ n^2 $ particles can be reconstructed. This limits the rates at which the CSCs can function to 1 kHz cm$ ^{-2} $. 

		\subsection{Resistive Plate Chambers}
		\label{sec:muon_chambers__resistive_plate_chambers}

\begin{figure}[h!]
				\centering
				\includegraphics[width = 10cm]{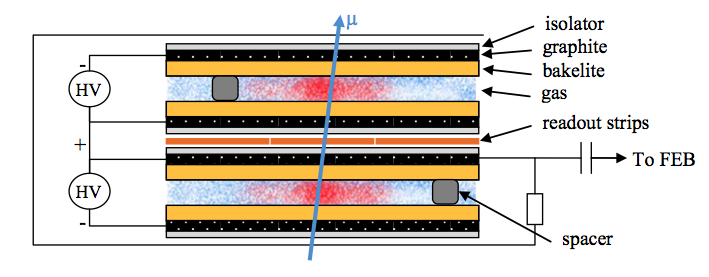}
				\caption{Representation of an RPC with two gas gaps for one readout strip plane \Cite{These_Karol}.}
				\label{fig:muon_chambers__rpc}
			\end{figure}

			RPCs, represented in Figure \ref{fig:muon_chambers__rpc}, are gaseous parallel plate detectors. They consist of two parallel plates, made out of bakelite with a high resistivity (10$ ^{10} $ to 10$ ^{11} $ $ \Omega $ cm) separated by a gas gap of a few millimeters. The outer surfaces of the resistive materials are coated with conductive graphite to form the HV and ground electrodes. Due to the fact that ions and electrons never come in contact with the electrodes, the evacuation time can be of importance if too many charges are produced. Therefore, the gain of the detectors are reduced and most of the amplification is done by the readout electronics and not by avalanches. This allows the RPCs to run at rates up to 1 kHz cm$ ^{-2} $, while maintaining an excellent time resolution down to 1 ns. On the other hand, they have a poor spatial resolution of the order of 1 mm. \\

			Since the RPCs can operate at high hit rate, they are used in both the barrel and the endcaps as trigger system. In the barrel, RPCs are rectangular chambers covering the DTs, while in the endcaps, they have a trapezoidal shape like the CSCs.
		
		\subsection{System Performances}
		\label{sec:muon_chambers__system_performances}

\begin{figure}[h!]
				\centering
				\includegraphics[width = 13cm]{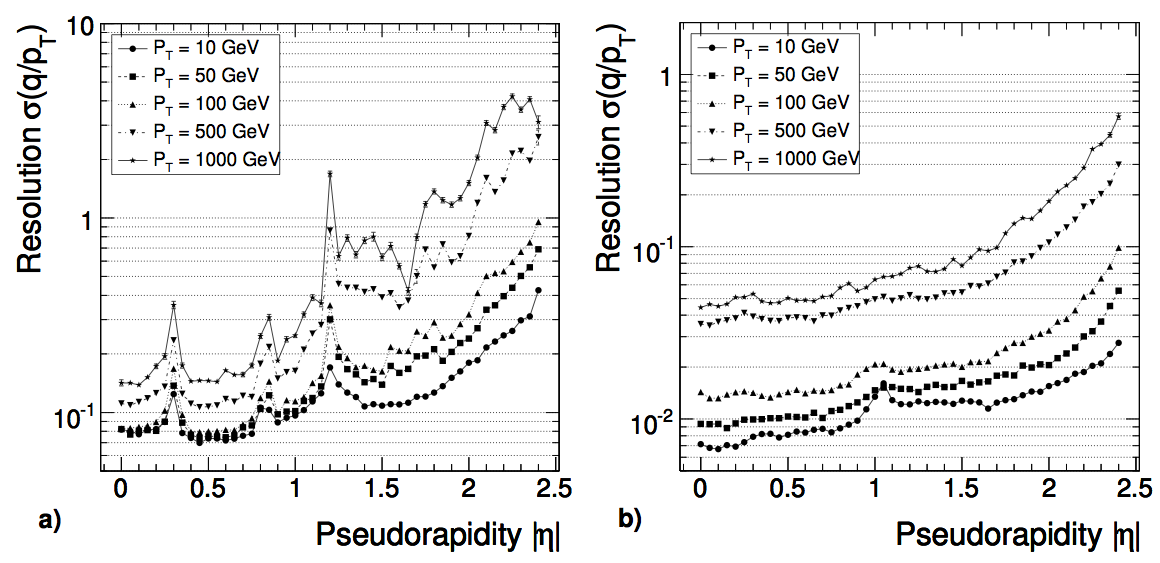}
				\caption{Resolution on the transverse momentum $ p_T $ of muons as a function of the pseudo-rapidity $ \eta $ for the muon system in standalone (left) and combined with the tracker's data (right) \Cite{CMS_Performances}.}
				\label{fig:muon_chambers__performances}
			\end{figure}

			Figure \ref{fig:muon_chambers__performances} represents the resolution on the transverse momentum $ p_T $ of muons as a function of the pseudo-rapidity $ \eta $ for the muon system in standalone (left) and combined with the tracker's data (right). The standalone system suffers from discontinuities in $ \eta $ when transitioning from the barrel to the endcaps ($ | \eta | $ = 1) or between stations in the endcaps where particles are not detected. These imprecisions can be removed by considering data from the trackers, which also improves the overall precision by a factor of 10. For a muon with a transverse momentum of 10 \GeVc{}, the resolution goes down from about 10\% in the standalone reconstruction to about 1\% when considering the tracker.

	\section{LS2 Upgrade and Challenges}
	\label{sec:muon_chambers__ls2_upgrade_and_challenges}

		After the LS2 upgrade, when the HL-LHC starts running, the particles fluxes inside CMS will increase, creating an even harder environment for detectors to operate in. Multiple upgrades of CMS \Cite{CMS_Upgrades} are planned in order to maintain the detection efficiency. Regarding the muon system, it is proposed to complete the CSC and RPC systems in the endcaps by installing the ME4/2 and RE4/2 stations. The readout electronics of the DTs and CSCs will be changed to sustain the radiation and increase the bandwidth and readout speed of the modules. It has not yet been decided if the 1.6 < $ | \eta | $ < 2.1 region will be instrumented and if so, which technology would be used. 

		\begin{table}[h!]
			\centering
			\begin{tabular}{l|c|c}
				 & LS2 requirement & RPC capabilities \\ \hline
				Spatial resolution & 250-500 \um{} & 1 mm \\
				Time resolution & 4-5 ns & 1 ns \\
				Rate capability & > 10 kHz cm$ ^{-2} $ & 1 kHz cm$ ^{-2} $ \\ 
				Efficiency & 95\% & 98\%
			\end{tabular}
			\caption{LS2 requirements and RPC capabilities regarding spatial and time resolution, rate capabilities and efficiency \Cite{GEM_Thierry_Pres}.}
			\label{tab:muon_chambers__ls1_requirements}
		\end{table}	

		The actual RPCs' design is not able to meet the LS2 requirements listed in Table \ref{tab:muon_chambers__ls1_requirements}. New glass RPCs are under study and offer an alternative. Meanwhile, the CMS GEM Collaboration proposes to instrument the 1.6 < $ | \eta | $ < 2.1 by installing new detectors: \emph{Gaseous Electron Multiplier} (GEM). A detailed description of this technology and its benefits is done in the next chapter.

	\cleardoublepage

\chapter{Gas Electron Multiplier Detectors}
\label{chap:gas_electron_multiplier_detectors}

	In this chapter, we review the functioning and the mechanical design of GEM detectors, and present the performances obtained with GEM prototypes. We then state the problems that the actual CMS muon spectrometer will face after the LS1 and LS2 upgrades of the LHC, and motivate the proposition to install GEM detectors in the forward region of CMS instead of RPCs.
	
	\section{Triple-GEM Detectors}
	\label{sec:gas_electron_multiplier_detectors__triple_gem_detectors}

		The following describes the detectors' layout and functioning for an installation in the CMS muon spectrometer.

		\subsection{Chamber Mechanical Design}
		\label{sec:gas_electron_multiplier_detectors__chamber_mechanical_design}

\begin{figure}[h!]
				\centering
				\includegraphics[height = 4cm]{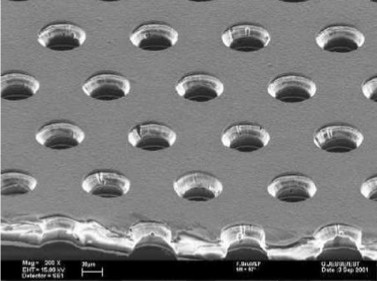}
				\includegraphics[height = 4cm]{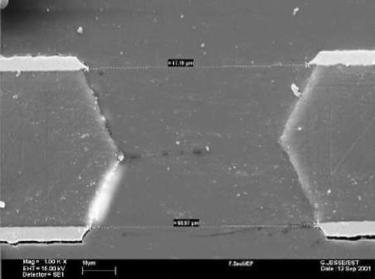}
				\caption{Electron microscope view of the honeycomb pattern of holes in a GEM foil \Cite{GEM_Construction}.}
				\label{fig:gas_electron_multiplier_detectors__construction}
			\end{figure}

			GEMs \Cite{GEM_Technical_Proposal, GEM_Construction, GEM_Construction_and_Performance} are made out of a 50 \um{} thick kapton foil coated with a 5 \um{} copper layer on each side, that is chemically drilled to create a honeycomb pattern of holes, as represented in Figure \ref{fig:gas_electron_multiplier_detectors__construction}. The holes have a diameter of 70 \um{} on both ends and a diameter of 50 \um{} in the middle. They are separated by 140 \um{}. \\

\begin{figure}[h!]
				\centering
				\includegraphics[width = 9cm]{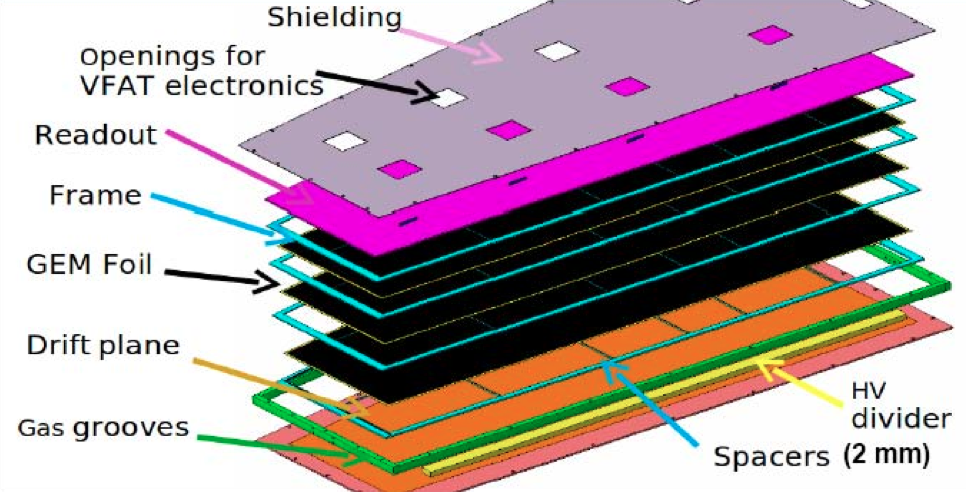}
				\caption{Schematic view of the constitutive layers of a Triple-GEM detector \Cite{GEM_Test_in_Beam_1}.}
				\label{fig:gas_electron_multiplier_detectors__structure}
			\end{figure}

			Three GEM foils are stretched and spaced by a few millimeters\footnote{The final design has not yet been chosen, leaving room for changes.} (1 to 3 mm) using a frame to form a Triple-GEM detector. A cathode plane is placed on one side and a series of anode strips (384 strips by 10$ ^\circ $ in $ \phi $) for the readout on the other. Figure \ref{fig:gas_electron_multiplier_detectors__structure} depicts the structure and the layers that compose a chamber. GEMs\footnote{In the rest of this work, \emph{GEM detectors} or simply \emph{GEMs} will refer to \emph{Triple-GEM detectors}.} have the same trapezoidal shape as CSCs as they would be installed in the endcaps. This means that even if the anodes are always spaced by 100 \um{}, the pitch between them (distance from middle to middle) is not constant, but varies between 600 \um{} on the narrow side and 1.2 mm on the long side. \\

			Each chamber covers 10$ ^\circ $ in $ \phi $ and is divided into 3 segments in $ \phi $ and 6, 8 or 10 segments in $ \eta $. This is done to
			\begin{enumerate}
				\item regroup a set of anodes into a single electronic readout chip; 
				\item increase the resolution in $ \eta $ as only the $ \phi $ coordinate is measured by the strips which leave a larger incertitude on the other coordinate;
				\item allow the detector to work even if a discharge occurs in one of the regions, requiring time to bring the voltage back up (the chamber's segmentation is not identical to the HV segmentation, but the argument remains valid).
			\end{enumerate}

\begin{figure}[h!]
				\centering
				\includegraphics[height = 5cm]{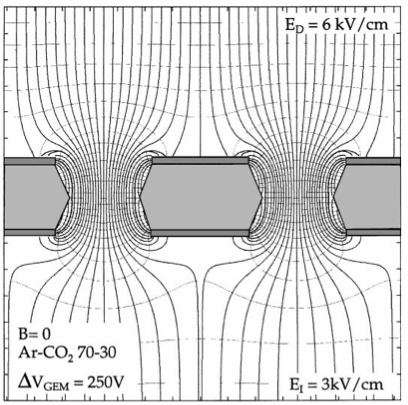}
				\includegraphics[height = 5cm]{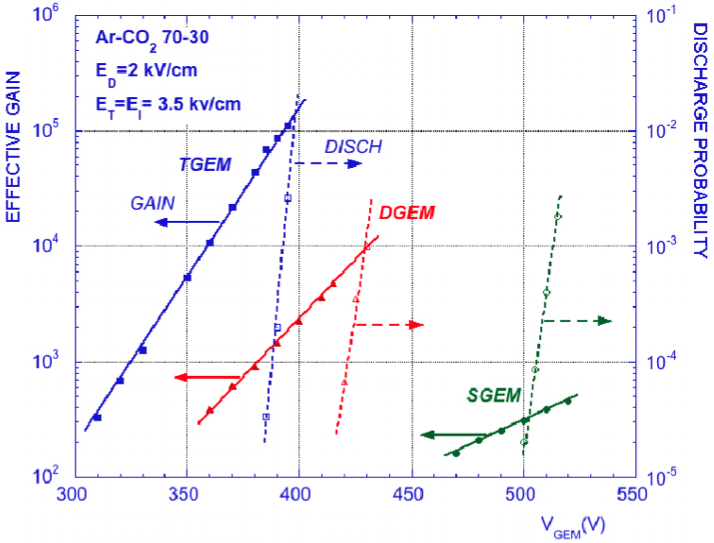}
				\caption{Representation of the electric field created inside the holes (left) \Cite{GEM_Electric_Field}; Gain for different numbers of GEM foils according to the applied voltage difference (right) \Cite{Thesis_Laura}.}
				\label{fig:gas_electron_multiplier_detectors__field_gain}
			\end{figure}

			Once the detector is assembled, a gas mixture of Ar : CO$ _2 $ : CF$ _4 $ (45\% : 15\% : 40\%) is injected in the chamber. A voltage difference is applied between the two copper layers of each GEM foil to create a strong electric field inside the holes that act as multipliers. As represented on the left in Figure \ref{fig:gas_electron_multiplier_detectors__field_gain}, the field also channels the electrons towards the regions where they will be accelerated and create avalanches. Due to the geometry of the GEM foils, even a small voltage difference of the order of 400 V generates the intense fields (> kV cm$ ^{-1} $) required to initiate avalanches. Nevertheless, one amplification stage would not be enough to create detectable signals, justifying the presence of three layers. A graph representing the amplification according to the applied voltage and the number of layers is shown on the right in Figure \ref{fig:gas_electron_multiplier_detectors__field_gain}. This image depicts that increasing the number of GEM layers (SGEM stands for Singe-GEM, DGEM for Double-GEM, and TGEM for Triple-GEM) allows the system to be ran at lower voltage and achieve higher amplifications (continuous lines). The dotted lines illustrate the probability that a discharge occurs in the chambers. When using more GEM layers, the voltage range for which the probability is quasi null increases.

			To improve the detection performances, two chambers are mounted back-to-back at each site, forming a \emph{super-chamber}.
		
		\subsection{Readout Electronics}
		\label{sec:gas_electron_multiplier_detectors__read-out_electronics}	

\begin{figure}[h!]
				\centering
				\includegraphics[width = 10cm]{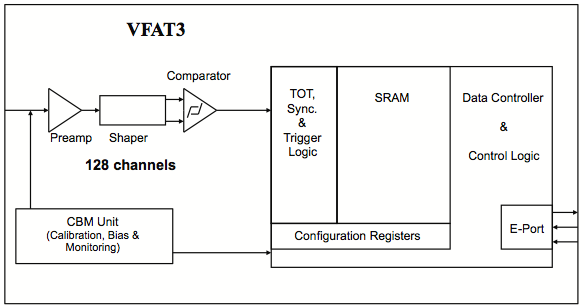}
				\caption{Simplified view of the on-board electronics performing the readout, amplification (preamp), shaping (shaper), and digitization (comparator) of the strips' signal \Cite{GEM_Technical_Proposal}.}
				\label{fig:gas_electron_multiplier_detectors__vfat3}
			\end{figure}

			The detectors will be equipped with VFAT3 electronics which architecture (still under design) is depicted in Figure \ref{fig:gas_electron_multiplier_detectors__vfat3}. These modules have 128 analogical channel readouts (one per strip) that amplify (preamp) and shape the signal (shaper). The data is then digitized (comparator) and stored in memory (SRAM) before being sent through optical links (E-Port) to off-detector electronics. \\

			The off-detector electronics serves as an interface between the on-board chip and the \emph{Data Acquisition} (DAQ) system that collects the output of all the detectors. It is placed in a cavern next to the detector to minimize the information's travel time and to protect it from the radiations. This system also applies a first series of filters on the events and takes the decision to keep them or drop them for ever.
		
		\subsection{Functioning}
		\label{sec:gas_electron_multiplier_detectors__functioning}

\begin{figure}[h!]
					\centering
					\includegraphics[width = 9cm]{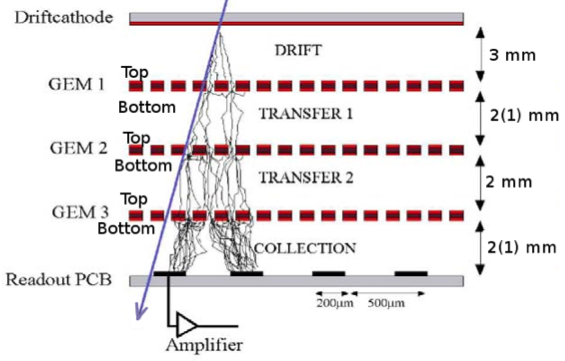}
					\caption{Signal amplification for a Triple-GEM detector \Cite{Thesis_Laura}.}
					\label{fig:gas_electron_multiplier_detectors__foils}
				\end{figure}

				As reviewed in Section \ref{sec:gas_electron_multiplier_detectors__chamber_mechanical_design}, three foils are used in the GEM detectors' design, creating four gaps, as represented in Figure \ref{fig:gas_electron_multiplier_detectors__foils}.

\begin{figure}[h!]
					\centering
					\includegraphics[width = 12cm]{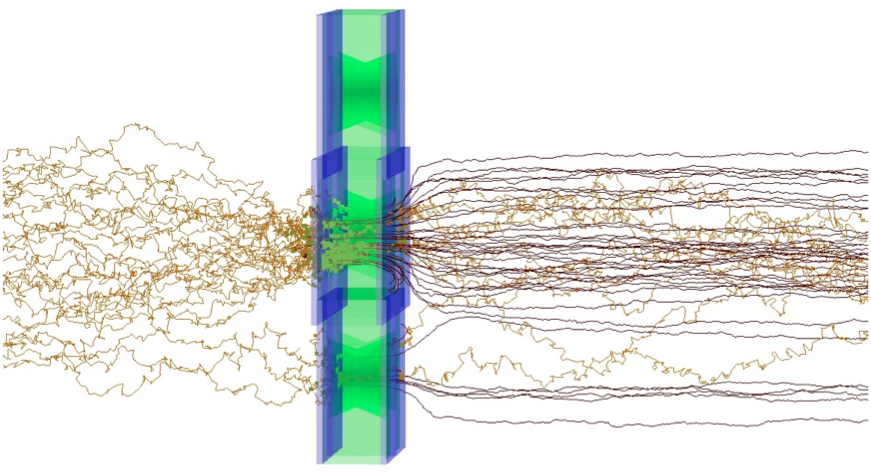}
					\caption{Simulation of the electrons' and ions' drift and multiplication for a Single-GEM foil \Cite{These_Geoffrey}.}
					\label{fig:gas_electron_multiplier_detectors__amplification}
				\end{figure}

				Before describing the role of each gap, it is interesting to note that they can be seen as independent from one another, as the voltage difference applied on each side of the foils makes them look like electrode planes. Therefore, we can separate the function of GEMs into two parts: the electrons' multiplication that occurs between each gap, and the charges' drift that takes place inside a gap. Figure \ref{fig:gas_electron_multiplier_detectors__amplification} shows a simulation of the amplification performed by a Single-GEM. Electrons (orange) arrive from the right and are amplified inside the holes. The resulting electrons then continue their drift towards the left, scattering with the gas, while ions (red) are attracted by the previous foil's cathode.

				Each gap plays a different role according to its width and to its order inside the chamber. The first gap is called the \emph{Drift gap}. It is where most of the detectable signals will originate from. It is larger than the other regions to increase the number of ionization clusters created by the particles. Electrons are then amplified by the first and second foil, and, respectively, enter the \emph{Transfer 1} and \emph{Transfer 2 gap}. Finally, after being amplified one last time, the electrons enter the \emph{Induction gap} where they will produce the readout signal. Due to the electrons' long drift distance in the last gap (1 or 2 mm), they are able to form a visible signal on the anodes, in opposition to the ions whose signal is a thousand times weaker as it is screened by the foils. \\

\begin{figure}[h!]
					\centering
					\includegraphics[width = 10cm]{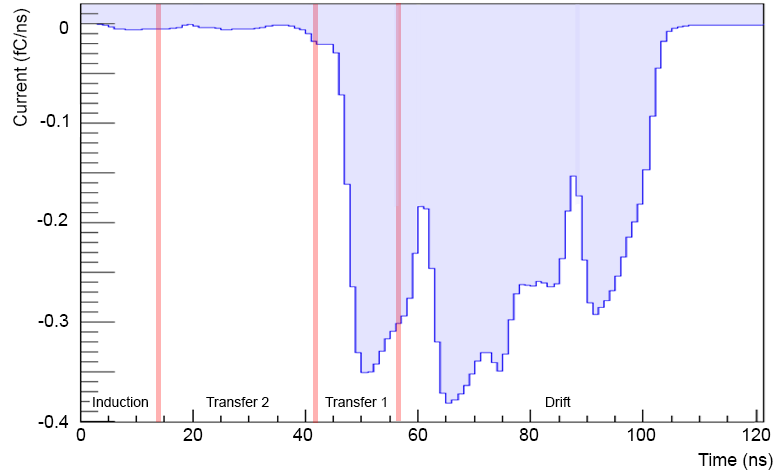}
					\caption{Simulation of the signal formed on the anodes by a single 1 GeV muon passing through the chamber perpendicularly to the readout plane \Cite{These_Geoffrey}.}
					\label{fig:gas_electron_multiplier_detectors__raw_signal}
				\end{figure}

				We stated that only energy deposited in the Drift gap can be detected, which is not entirely true. If an energy loss occurs in the Transfer 1 gap and leaves behind a sufficient amount of energy, it can create a strong visible signal. With the same logic, multiple peaks can be observed in the Drift gap, corresponding to multiple energy losses. Table \ref{tab:gas_electron_multiplier_detectors__signal_timing} lists the arrival time ranges for the electrons originating from the different gaps. The mean drift speed of the electrons is of the order of 74.6 \um{} ns$ ^{-1} $. Figure \ref{fig:gas_electron_multiplier_detectors__raw_signal} represents the induced current on an anode inside a GEM detector filled with a gas mixture of Ar : CO$ _2 $ (70\% : 30\%) for a simulated muon of 1 \GeVc{} passing through the chamber perpendicularly to the readout plane. Each region defined with a red line corresponds to the signal created in a specific gap.

				\begin{table}[h!]
					\centering
					\begin{tabular}{l|l}
						Gap & Timing \\ \hline
						Induction & 0 - 14 ns \\
						Transfer 2 & 14 - 42 ns \\
						Transfer 1 & 42 - 56 ns \\
						Drift & 56 - 98 ns 
					\end{tabular}
					\caption{Timing of the signals formed by the avalanches originating from the different gaps inside a Triple-GEM detector \Cite{GEM_Thierry_Pres}.}
					\label{tab:gas_electron_multiplier_detectors__signal_timing}
				\end{table}		

				The raw output signal is then shaped by the VFAT3 electronics which impulse response $ T(t) $, as defined in Section \ref{sec:muon_chambers__read_out_electronic}, is given by
				\begin{equation} 
					T(t) = \left( \frac{t}{\tau} \right)^2 \exp\left(- 2 \frac{t}{\tau} \right) \ ,
					\label{eq:gas_electron_multiplier_detectors__transfert_function}
				\end{equation}
				where $ \tau $ is the shaping time which can be programmed and take different values: 20, 50, 100, 250, or 500 ns. The shaping time has to be long enough to integrate the various peaks and always yield a well defined output signal function. 	
		
		\subsection{Performances}
		\label{sec:gas_electron_multiplier_detectors__performances}

			Small- and full-scale GEM prototypes have been tested using 150 \GeVc{} $ \pi / \mu $ beams \Cite{GEM_Construction_and_Performance, GEM_Test_in_Beam_1, GEM_Test_in_Beam_2, GEM_Test_in_Beam_3, GEM_Test_in_Beam_4}. Some of those results are presented in the following.
			
			\subsubsection{Spatial Resolution} 
			\label{sec:gas_electron_multiplier_detectors__spatial_resolution}

				When the avalanche's charge hits only one strip, the spatial resolution is given by the pitch size of the strips divided by $ \sqrt{12} $, which is the variance of a uniform distribution. 
				\begin{equation} 
					\sigma_{\phi} = \frac{pitch}{\sqrt{12}} \approx 170 - 340 \, \mu m
				\end{equation}
				This theoretical value has been confirmed by the beam tests. \\

				Due to limitations in the amount of data that can be transfered from the detector to the storage units, it is possible that several strips will be regrouped into super-strips, or that the strips will be read in a binary mode (hit or not) instead of analogically (induced current). The granularity of the detectors is defined as the number of super-strips for each segment. It can be of 128, 64, 32, 16, or 8. The first one meaning that each strip acts like a super-strip, and the last one meaning that strips are regrouped in sets of 8. The lower the granularity, the higher the pitch, yielding a decrease of the spatial resolution.

			\subsubsection{Time Resolution} 
			\label{sec:gas_electron_multiplier_detectors__time_resolution}

				No algorithm to compute the resolution on BX assignment has yet been defined for the new VFAT3 electronics that will equip the CMS GEM detectors. However, a technique called \emph{Time Over Threshold} (TOT) is under study. Preliminary results show that this method is able to give the relative time at which a particle passed through the detector with a resolution of 4.44 ns \Cite{GEM_Thierry_Pres}. \\

\begin{figure}[h!]
					\centering
					\includegraphics[width = 12.8cm]{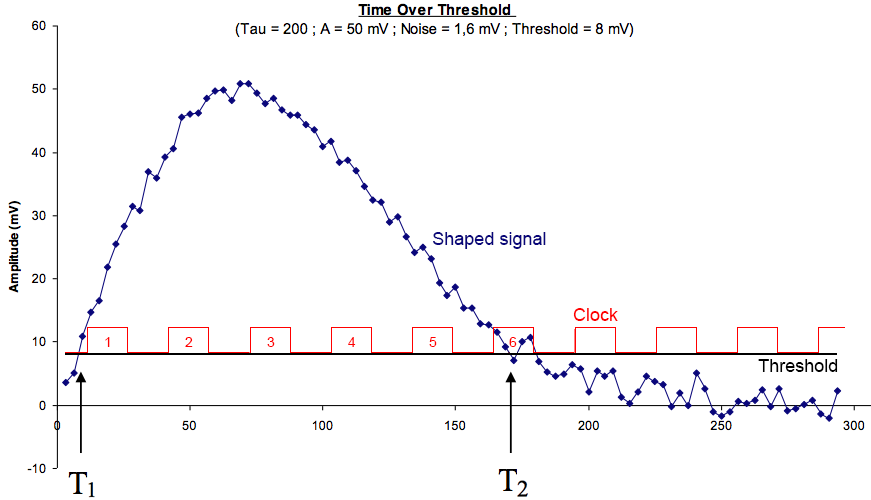}
					\caption{Time Over Threshold (TOT) algorithm for Bunche Crossing (BX) assignment \Cite{GEM_Thierry_Pres}.}
					\label{fig:gas_electron_multiplier_detectors__tot}
				\end{figure}

				Figure \ref{fig:gas_electron_multiplier_detectors__tot} illustrates the TOT method by representing a shaped signal (blue) for the electronics, a clock pulse (red), and the defined threshold (black). Using the shaped signal, $ T_1 $ and $ T_2 $ are respectively define as the moment that the signal passes over, and below the defined threshold. The TOT, given by
				\begin{equation}
					TOT = T_{2} - T_{1} \ ,
				\end{equation}
				is measured with a clock running at a predefined frequency, yielding the number of clock cycles where the signal was above the threshold. The fact that the exact shape of the output signal (which only varies in amplitude) is known allows to run simulations that associate a specific arrival time to each value of the TOT. Those values are stored in a \emph{Look Up Table} (LUT). When a particle generates a signal, the number of clock cycles is measured and the arrival time is fetched in the LUT. If the shaping time of the electronics, the threshold value, or the clock speed are changed, a different LUT has to be generated. \\

				The standard deviation of the Gaussian fit on the difference between the simulated arrival time and the value returned by the LUT, gives the time resolution of the detector.
			
			\subsubsection{Detection Efficiency} 
			\label{sec:gas_electron_multiplier_detectors__detection_efficiency}

				Using test beam results, a maximal efficiency of 98\% is achieved for a single GEM detector. When using a super-chamber, the inefficiency of the system drops bellow 0.04\%.

			\subsubsection{Rate Capability}
			\label{sec:gas_electron_multiplier_detectors__acceptance_rate}

				Earlier work \Cite{GEM_Rates} demonstrated that GEMs can sustain rates up to 10 MHz cm$ ^{-2} $ before loosing detection efficiency. This is a strong argument in favor of the installation of GEMs in the high $ \eta $ regions of CMS where the rates are expected to be higher than 10 kHz cm$ ^{-2} $.   
	
	\section{Upgrade of the CMS Muon System}
	\label{sec:gas_electron_multiplier_detectors__upgrade_of_the_cms_muon_system}		

		Events with high energy muons in their final state are a signature of many interesting processes such as the decay of the Brout-Englert-Higgs boson or possible new physics, like super-symmetry. Unfortunately, these events are rare and require a high detection efficiency and reconstruction capability. Even more so in the forward region of CMS where the tracks' projections in the transverse plane are shorter due to the high momentum of the muons, degrading the resolution on the reconstructed trajectory. \\

		As reviewed in Section \ref{sec:muon_chambers__disposition_of_the_detectors}, the muon system of CMS is not complete in the 1.6 < $ | \eta | $ < 2.4 region. Lack of redundancy in this region will become critical during the HL-LHC phase after LS2. The higher rates will confuse the muon spectrometer as more random coincidences will happen inside the detectors. Specifically CSCs will suffer due to the increase in the number of ghost particles they will reconstruct. Performances of standard RPCs will also drop as the rates will exceed those for which they were designed. \\

		To increase redundancy, improve the muon spectrometer's performances, and make use of the free space, the installation of GEM detectors has been proposed by the CMS GEM Collaboration. \\

		Initially foreseen to instrument the 1.6 < $ | \eta | $ < 2.4 region near stations ME1/1 and ME2/1 with super-chambers, elevated costs have forced the collaboration to revise their plans. As for today, the objective is to install a full ring of super-chamber (GE1/1) in the ME1/1 station of both endcaps covering the 1.6 < $ | \eta | $ < 2.1 region during LS2. \\

		Studies are being led in order to send data collected by GEMs to CSCs in the ME1/1 station to improve their detection efficiency \Cite{GEM_CSC_Trigger}. By matching hits in both detectors, CSCs should be able to decrease their number of ghosts. Most of the CSCs' resolution is given by ME1/1 where the magnetic field is still uniform and constant, and multiple scattering is at its least. Coupling GE1/1 and ME1/1 would significantly increase the system's efficiency.
	
	\section{LS1 Prototypes}
	\label{sec:gas_electron_multiplier_detectors__ls1_prototypes}	

\begin{figure}[h!]
			\centering
			\includegraphics[height = 5.3cm]{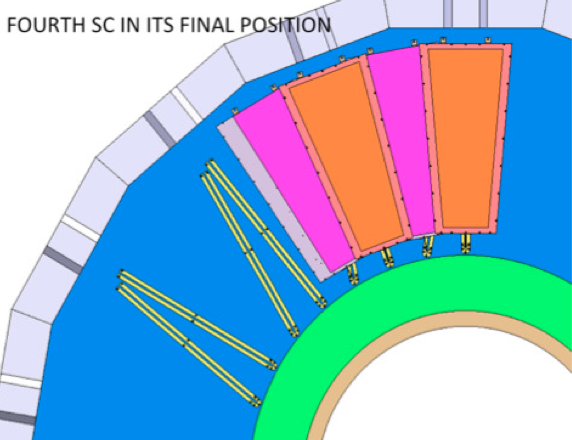}
			\includegraphics[height = 5.3cm]{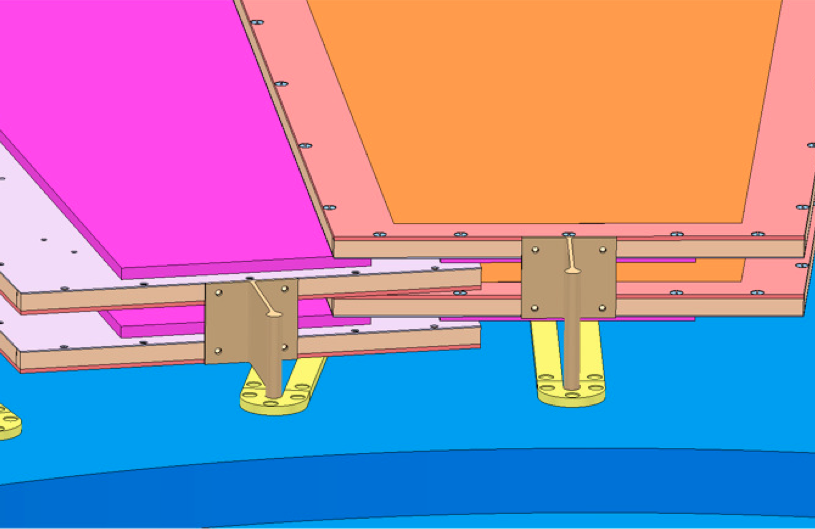}
			\caption{Disposition of the GEM prototypes on the yoke of CMS \Cite{GEM_Technical_Proposal}.}
			\label{fig:gas_electron_multiplier_detectors__install}
		\end{figure}

		Even though GEMs are proposed to be installed during LS2, the CMS GEM Collaboration has been allowed to install four prototypes in CMS during march 2014, at the end of LS1, in order to test the mechanical feasibility of the installation. Figure \ref{fig:gas_electron_multiplier_detectors__install} illustrates the disposition of the super-chambers on the yoke. Fully operational prototypes should be installed during a long service stop in 2016. Using those, the CMS GEM Collaboration intends to perform preliminary studies of the improvements that GEMs could bring to the CSCs' event selection. This is important as only a limited amount of events can be selected due to the limited bandwidth of the DAQ. The selection is done by the trigger system of CMS which is the topic of the next chapter.

	\cleardoublepage

\chapter{CMS Trigger System and Reconstruction Algorithms}
\label{chap:trigger_system_and_reconstruction_algorithms}

	The first part of this chapter reviews the CMS trigger system. It describes the event selection performed by CMS and the algorithms used by the DTs, CSCs, and RPCs to reconstruct particle tracks. It also presents the performances of the system and the difficulties it encounters. The second part is dedicated to reconstruction algorithms. We present the theory behind four methods that are used in track reconstruction: pattern recognition, the sagita method, the Least Squares fit, and the Kalman filter.
	
	\section{CMS Trigger System}
	\label{sec:trigger_system_and_reconstruction_algorithms__cms_trigger_system}

		With the LHC running at a rate of 40,000,000 collisions per second, the amount of data produced by CMS is considerable ($ \sim $ 40 TB s$ ^{-1} $). We do not yet have the technology to transfer nor handle all this information. Therefore, a selection of interesting events has to be done in order to reduce the transfer's rate. The decision to keep or drop an event is taken by the CMS trigger system which includes two stages: the \emph{Level-1 Trigger} (L1 Trigger) and the \emph{High Level Trigger} (HLT) \Cite{CMS_at_LHC}. \\

\begin{figure}[h!]
			\centering
			\includegraphics[width = 10cm]{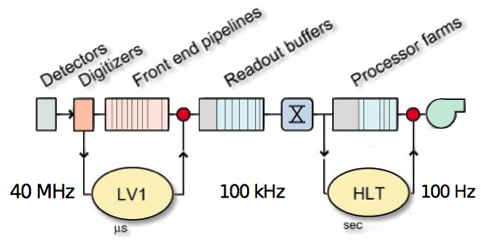}
			\caption{Data rates at each stage between the detectors and the data storage center. \Cite{CMS_Trigger_System}}
			\label{fig:trigger_system_and_reconstruction_algorithms__rates}
		\end{figure}

		Figure \ref{fig:trigger_system_and_reconstruction_algorithms__rates} depicts the different stages of the trigger and the maximal rate of events kept by each one of them, starting at 40 MHz and ending at 100 Hz. The kept events are sent and stored in multiple locations around the world (called \emph{Tiers}) where physicists can analyze them.

		\subsection{Level-1 Trigger}
		\label{sec:trigger_system_and_reconstruction_algorithms__level_1_trigger}

\begin{figure}[h!]
				\centering
				\includegraphics[width = 10cm]{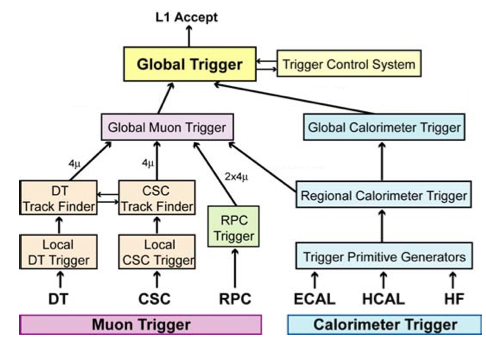}
				\caption{L1 Trigger decision flow of CMS before data is being transfered to the DAQ \Cite{CMS_at_LHC}.}
				\label{fig:trigger_system_and_reconstruction_algorithms__l1}
			\end{figure}

			The L1 Trigger is the first stage of selection of CMS and has to be able to handle all events successively. Therefore, it has to take a "keep or drop" decision every 25 ns (time between two BXs). The system is composed of dedicated electronic chips for each detector placed either on CMS or in the service caverns next to it, to protect them from radiations. A diagram of the L1 Trigger's decision flow is shown in Figure \ref{fig:trigger_system_and_reconstruction_algorithms__l1}. The decision is first taken locally by small groups of muon chambers and calorimeters before being sent to the \emph{Global Muon Trigger} (GMT) \Cite{Trigger_Muon} and \emph{Global Calorimeter Trigger} (GCT) respectively, that analyze the event over all the regions. Finally, the GMT and GCT send their keep/drop signal to the \emph{Global Trigger} (GT). As represented, the tracker is not involved in this first stage selection due to the time needed to reconstruct and transfer the output data. \\

			When a collision occurs, every 25 ns, the system reads out the response of every detector and stores it in a buffer that holds the last 128 events. Consequently, the algorithms have a maximum of 3.2 \us{} to process each event and return their decision. This also allows the decision taking process to be differed between the different triggers (GMT and GCT) as the particles take a certain time to travel from the IP to the various detectors. Once an algorithm has made a decision, it sends a one bit signal to the GMT or GCT. When all the detectors have responded, the GT either drops the event or tells the DAQ system to transfer it to the HLT. \\


			As the L1 Trigger only relies on the calorimeters and on the muon system, the events' selection is done according to the signature and transverse energy left in the calorimeters, and to the transverse momentum reconstructed by the muon system. Only the transverse components of the energy and the momentum are considered as they reflect the physics of the event. Indeed, when the protons collide, the fraction of energy put at play in the interaction is not the same. Therefore, the produced particles will be boosted along \axis{Z} according to these differences, while the total transverse momentum should remain null as the collisions are head to head.
		
		\subsection{High Level Trigger}
		\label{sec:trigger_system_and_reconstruction_algorithms__high_level_trigger}

			The HLT is composed of a farm of computers running reconstruction software that can perform complex calculations. Due to the filtering made by the L1 Trigger, the incoming data rate is lower (100 kHz), allowing for a longer processing time, of the order of 1 s. If an event passes through the multiple filters and is accepted, it is send to the storage unit and made available for analysis. Since this work aims to study track reconstruction performed at L1, the HLT will not be further reviewed.
		
		\subsection{Muon System L1 Trigger}
		\label{sec:trigger_system_and_reconstruction_algorithms__muon_system_l1_trigger}		 

			The different muon chambers have their own trigger system which benefits from the detectors strengths. DTs and CSCs have excellent spatial resolution (respectively of the order of 100 and 80 \um{}) and will therefore be used to filter the events according to their transverse momentum. RPCs on the other hand have great timing capabilities (down to 1 ns) which yields good BX assignment. By combining the DTs and RPCs in the barrel, and the CSCs and RPCs in the endcaps, the GMT can reconstruct event with a multitude of muons in the final state.
			
			\subsubsection{Drift Tubes and Cathode Strip Chambers}
			\label{sec:trigger_system_and_reconstruction_algorithms__dt_and_csc}

				DTs and CSCs use the same trigger system, the \emph{Track-Finder} (TF) \Cite{Track_Finder}, which relies on the measurement of the particles' bending angle when they pass through the detectors. The system is divided into a local and a global trigger. The local trigger validates hits in DT and CSC modules, before transmitting the information to the global trigger which tries to reconstruct tracks over all the stations.
				
				\paragraph*{Local Trigger}
				\label{sec:trigger_system_and_reconstruction_algorithms__local_trigger}	

\begin{figure}[h!]
						\centering
						\includegraphics[height = 3.5cm]{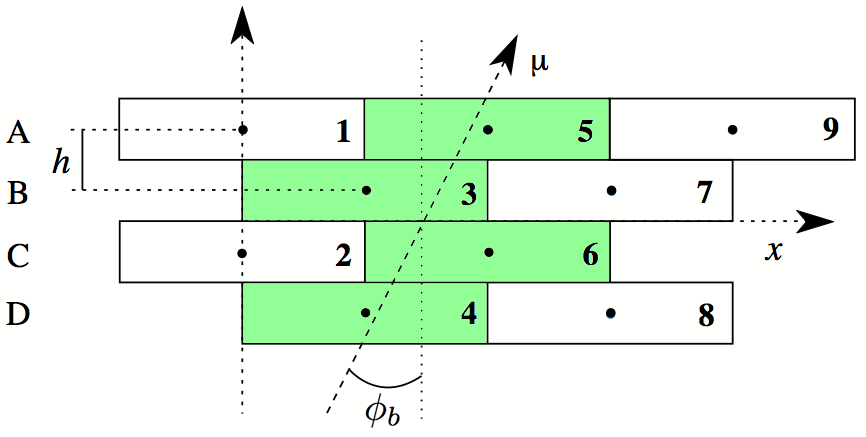}
						\includegraphics[height = 3.5cm]{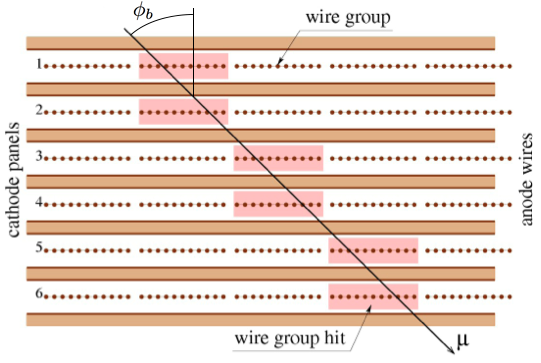}
						\caption{DTs' local trigger measuring the particle's incident angle using the ions' drift time (left) \Cite{CMS_at_LHC}; CSCs' local trigger relying on the multiple layers to measure the particle's incident angle (right) \Cite{Trigger_Muon}.}
						\label{fig:trigger_system_and_reconstruction_algorithms__dt_local}
					\end{figure}

					DT modules are divided into smaller segments as represented on the left in Figure \ref{fig:trigger_system_and_reconstruction_algorithms__dt_local}. Each couple of layers (AB, AC, AD, etc) is used to compute the position $ \mathbf{x} $ and the angle $ \phi_b $ of the track by measuring the arrival time of the signals to the anode. The further away a particle passes from the wire, the longer the drift time is, hence the time at which the signals are detected. If the values match for several couples, the segment is considered to represent a valid track and marked as such. The number of couples that return the same value defines the quality of the track. If an ambiguity appears, the parameters with the best quality are selected. \\		

					The same is done in CSC modules using the six planes of anode wires and seven planes of cathode strips, as seen on the right in Figure \ref{fig:trigger_system_and_reconstruction_algorithms__dt_local}. Due to the short amount of time available to run the reconstruction, anode wires are grouped by 5 to 16 by performing a logical \emph{OR} of the binary readout result.
				
				\paragraph*{Global Trigger}
				\label{sec:trigger_system_and_reconstruction_algorithms__global_trigger}	

\begin{figure}[h!]
						\centering
						\includegraphics[width = 12cm]{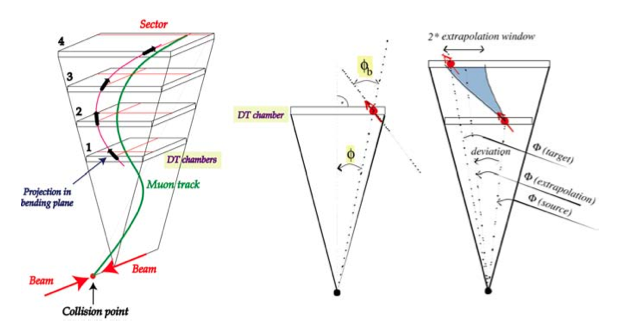}
						\caption{Reconstructed trajectory by the Track-Finder using pairwise matching between the stations \Cite{CMS_at_LHC}.}
						\label{fig:trigger_system_and_reconstruction_algorithms__dt_global}
					\end{figure}

					Once tracks have been reconstructed locally, the TF matches the different stations by comparing their hits. Figure \ref{fig:trigger_system_and_reconstruction_algorithms__dt_global} shows the pairwise matching between stations (left), the muon track (left; green), and the reconstructed track in the transverse plane (left; red). First, using the local reconstructed angle $ \phi_b $ of the track (middle), the parameters are extrapolated between layers (right) by using predefined parameters stored in LUTs for all the possible matching segments
					\begin{equation}
						\phi_{extrapolation} = \phi_b + \phi_{deviation} \ ,
					\end{equation}
					where $ \phi_{extrapolation} $ is the extrapolated parameter, and $ \phi_{deviation} $ is the deviation in $ \phi $ between two detection planes. If the extrapolation is close to the measurement, within a predefined range (right; blue), the site is added to the track and the propagation continues towards the next station. This method can result in more than one reconstructed track, which is why only the four tracks with the highest quality (best match between extrapolation and measurements, most matches, etc) are kept. Finally, an estimation of the transverse momentum in function of the bending angle is done.
					
			\subsubsection{Resistive Plate Chambers}
			\label{sec:trigger_system_and_reconstruction_algorithms__rpc_l1_trigger}

				RPCs use a \emph{Pattern Comparator} (PAC) algorithm \Cite{These_Karol} to retrieve the transverse momentum of the tracks. Each station is divided into segments that are considered to be active if they have been hit, or inactive otherwise. The state of the segments is read out by multiple chips that try to match the hits against a list of preloaded patterns they hold in a LUT. In order to fill the LUT, simulations are ran for a finite number of transverse momenta $ p_T $, associating a code, $ p_T^{Code} $, and a numerous amount of possible patterns to each one of them. For each $ p_T^{Code} $, a ranking is done according to the importance of the patterns 
				\begin{equation}
					E(pattern, \; p_T^{Code}) = \frac{N_0(pattern, \; p_T^{Code})}{N(p_T^{Code})} \ ,
				\end{equation}
				where $ N_0 $ is the number of identical patterns given by the same $ p_T^{Code} $, and $ N $ is the total number of patterns given by the $ p_T^{Code} $. Patterns are added to the LUT, starting with those with the highest $ E $, until 90 to 95\% of the tracks for each $ p_T^{Code} $ are present. \\

				\begin{figure}[h!]
					\centering
					\includegraphics[width = 6cm]{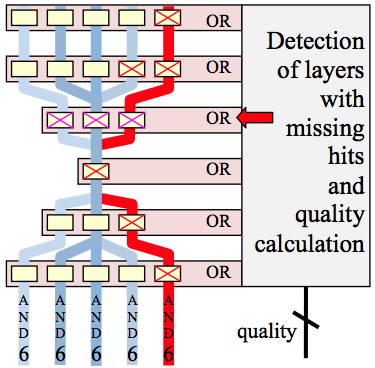}
					\caption{Pattern Comparator (PAC) algorithm matching hits against a multitude of predefined patterns \Cite{These_Karol}.}
					\label{fig:trigger_system_and_reconstruction_algorithms__rpc}
				\end{figure}

				Figure \ref{fig:trigger_system_and_reconstruction_algorithms__rpc} shows a couple of patterns that are being tested against measurements. If a layer has not been hit, all the segments are set to active, but the quality of the resulting track will be degraded. The four best reconstructed tracks in the barrel and the endcaps are kept and sent to the GMT.
			
			\subsubsection{Global Muon Trigger}
			\label{sec:trigger_system_and_reconstruction_algorithms__global_muon_trigger}

				For each event, the GMT receives the four best matches from the TF and from the PAC for the barrel and the endcaps. Those are compared and the best candidates are combined to increase the precision. 
			
			\subsubsection{System Performances}
			\label{sec:trigger_system_and_reconstruction_algorithms__system_performances}

\begin{figure}[h!]
					\centering
					\includegraphics[width = 8cm]{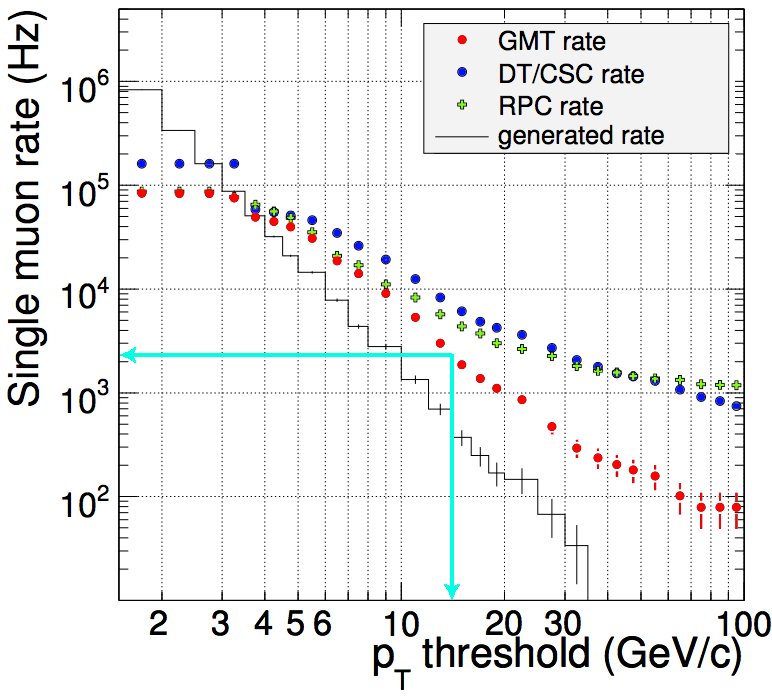}
					\caption{Generated and accepted rate of events for the L1 Trigger according to their transverse momentum $ p_T $ \Cite{CMS_Performances}.}
					\label{fig:trigger_system_and_reconstruction_algorithms__acceptance}
				\end{figure}

				An important parameter of the L1 Trigger is the applied threshold or cut on the transverse momentum above which all events are accepted. Figure \ref{fig:trigger_system_and_reconstruction_algorithms__acceptance} shows the generated (produced inside CMS; black line) and accepted (reconstructed and accepted by the L1 Trigger; red dots) rate of events according to the applied cut on the transverse momentum $ p_{T; threshold} $ for events with a single muon in the final state. Ideally, the dotted curves should match the continuous line, meaning that the system is able to perfectly identify the muons. Unfortunately, numerous low energy muons are reconstructed with a much higher energy, misleading the trigger. Improving those results would lead to a lower cut as the rates at high energies would drop. The current threshold is at 14 \GeVc{} (blue arrow) which yields a rate of the order of 2 kHz. The selection on a single muon is only one among several tens of L1 Trigger filters. The total bandwidth of the L1 Trigger has to be shared between all the muons, the electrons, the photons, and the jets triggers. Therefore, the maximum event rate allowed to the single muon trigger is limited to 2 kHz. \\

\begin{figure}[h!]
					\centering
					\includegraphics[width = 12cm]{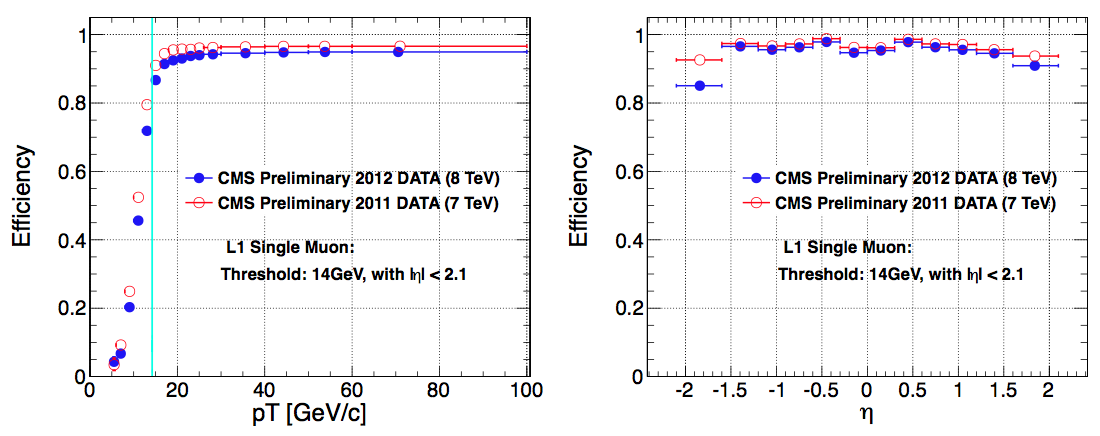}
					\caption{L1 Trigger's reconstruction efficiency as a function of the transverse momentum $ p_T $ (left) and pseudo-rapidity $ \eta $ (right) of single muons \Cite{CMS_L1_Efficiency}.}
					\label{fig:trigger_system_and_reconstruction_algorithms__l1_cut}
				\end{figure}

				Figure \ref{fig:trigger_system_and_reconstruction_algorithms__l1_cut} presents the L1 Trigger's efficiency as a function of the transverse momentum $ p_T $ (left) and pseudo-rapidity $ \eta $ (right) for a single muon. The plot on the left is called a \emph{turn-on} plot and shows the acceptance for various $ p_T $ for a defined threshold (14 \GeVc{}). Ideally, the curves should be equal to 0 below the cut and to 1 above the cut. Instead, because of the finite transverse momentum resolution in the muon track reconstruction, the trigger starts to accept events already above 5 \GeVc{}, and only keeps around 95 \% of them above the threshold. The second plot emphasizes what has been reviewed in Section \ref{sec:muon_chambers__ls2_upgrade_and_challenges}, namely the fact that efficiency drops at high $ | \eta | $ due to the lack of redundancy in the muon system.

	\section{Reconstruction Algorithms}
	\label{sec:trigger_system_and_reconstruction_algorithms__reconstruction_algorithms}

		Various algorithms can be used to perform track reconstruction or transverse momentum estimation. Four of them are presented in this section: pattern recognition, the sagita method, the Least Squares fit, and the Kalman fit.
		
		\subsection{Pattern Recognition}
		\label{sec:trigger_system_and_reconstruction_algorithms__pattern_recognition}	

			Pattern recognition algorithms compare the position of the hits inside the detectors against predefined entries stored in memory, as reviewed in Section \ref{sec:trigger_system_and_reconstruction_algorithms__rpc_l1_trigger}. Due to the extremely large number of possible trajectories, this method requires large amounts of memory and does not return the precise transverse momentum's value, but rather the closest match in a defined set. Nevertheless, the fact that patterns can be tested simultaneously makes this method fast and well suited for the L1 Trigger.
		
		\subsection{Sagita}
		\label{sec:trigger_system_and_reconstruction_algorithms__sagita}	

			As mentioned in Section \ref{sec:lhc_and_cms__tracker}, charged particles moving in the presence of a magnetic field describe helices along the axis of the field. The projection of those trajectories in the transverse plane are circles which radius is directly related to the transverse momentum of the particles. \\

			\begin{figure}[h!]
				\centering
				\includegraphics[width = 8cm]{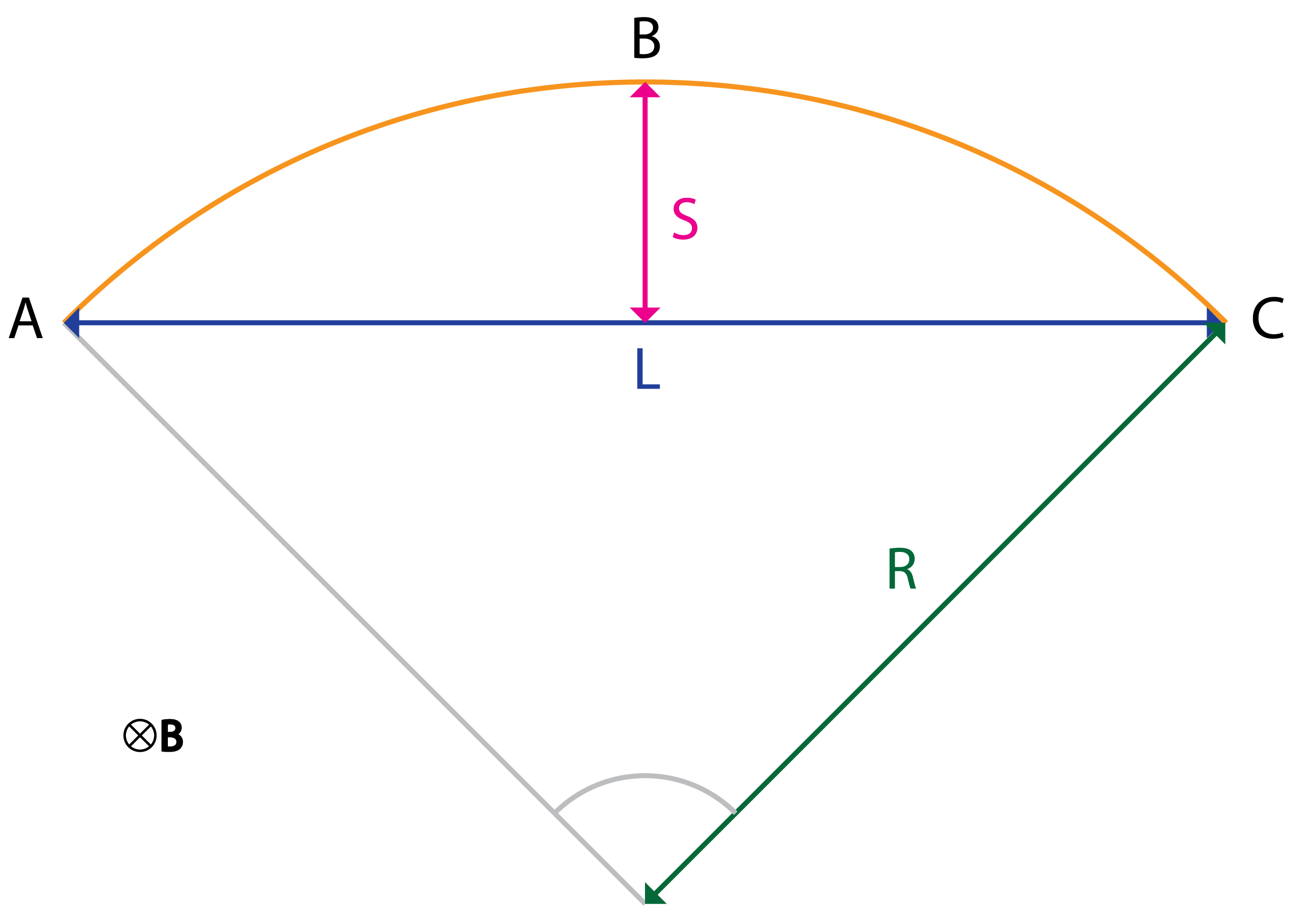}
				\caption{Illustration of the Sagita method using three equidistant hits (A, B, and C) to reconstruct the particle's trajectory (orange).}
				\label{fig:trigger_system_and_reconstruction_algorithms__sagita}
			\end{figure}			

			Using the parameters shown in Figure \ref{fig:trigger_system_and_reconstruction_algorithms__sagita}, and Pythagoras' theorem, we find that \Cite{Muons_Drift_Chambers}
			\begin{equation}
				\left( \frac{L}{2} \right)^2 + (R - S)^2 = R^2 \ .
			\end{equation}
			Assuming the sagita $ S $ is much smaller than $ L $, we have
			\begin{equation}
				R \approx \frac{L^2}{8 S} \ .
			\end{equation}
			Inserting this result in Equation \ref{eq:lhc_and_cms__radius_to_momentum_relation}, we find the following relation for the transverse momentum
			\begin{equation}
				p_T \approx \frac{0.3 B L^2}{8 S} \ .
			\end{equation}
			More than three points can be used, in which case, the error on the momentum is given by (neglecting multiple scattering)
			\begin{equation}
				\left| \frac{\sigma_p}{p} \right| = \frac{a_n p_T[GeV c^{-1}]}{0.3 B L^2} \sigma_x \ ,
			\end{equation}
			where $ a_n = \sqrt{\frac{720}{n + 4}} $, and $ \sigma_x $ is the detector's resolution. This method only yields good results if detectors are approximatively placed at equidistant points (A, B, and C), and if the magnetic field is constant. 
		
		\subsection{Least Squares}
		\label{sec:trigger_system_and_reconstruction_algorithms__least_squares}		

			The Least Squares \Cite{Fit_Least_Squares} method is the best known fitting algorithm due to its simplicity in both understanding and implementation. The principle is to model a set of points $ (\mathbf{x}_i, y_i) $ by a function $ f(\mathbf{x}_i, \mathbf{\beta}) $ where $ \mathbf{\beta} $ are the parameters that will be estimated. To obtain the best estimation, the sum of the squared residuals
			\begin{equation} 
				\chi^2(\mathbf{\beta}) = \sum_i \left[ y_i - f(\mathbf{x}_i, \mathbf{\beta}) \right]^2 \ ,
			\end{equation}
			hence the distance between the data and the model has to be minimized by the parameters, which gives the following condition
			\begin{equation}
				\frac{\partial \chi^2(\mathbf{\beta})}{\partial \beta_j} = 0 \ .
				\label{eq:trigger_system_and_reconstruction_algorithms__least_squares}
			\end{equation}
			One can also defined a weighted Least Squares fit where each point is weighted according to its variance $ \sigma_i $, in which case, we have the following equation
			\begin{equation} 
				\chi^2(\mathbf{\beta}) = \sum_i \frac{1}{\sigma^2_i} \left[ y_i - f(\mathbf{x}_i, \mathbf{\beta}) \right]^2 \ .
			\end{equation}
			If we want to compute the $ \chi^2 $ increment of measurement $ i $, it is given by the squared residuals
			\begin{equation} 		
				\chi^2_i = \left[ y_i - f(\mathbf{x}_i, \mathbf{\beta}) \right]^2 \ .
			\end{equation} 		
			
			\subsubsection{Linear Least Squares}
			\label{sec:trigger_system_and_reconstruction_algorithms__linear_least_squares}	

				In the \emph{Linear Least Squares} (LLSQ) method, the model function $ f(\mathbf{x}, \mathbf{\beta}) $ is a linear combination of the parameters
				\begin{equation} 
					f(\mathbf{x}_i, \mathbf{\beta}) = \sum_j g_j(\mathbf{x}_i) \beta_j \ .
				\end{equation}
				By injecting this in Equation \ref{eq:trigger_system_and_reconstruction_algorithms__least_squares}, we obtain
				\begin{equation} 
					\frac{\partial \chi^2(\mathbf{\beta})}{\partial \beta_j} = \sum_i \left[ y_i - f(\mathbf{x}_i, \mathbf{\beta}) \right] g_j(\mathbf{x}_i) = 0 \ ,
				\end{equation}					
				which can be expressed as 
				\begin{equation} 
					\sum_i \sum_k g_k(\mathbf{x}_i) \beta_k g_j(\mathbf{x}_i) = \sum_i y_i g_j(\mathbf{x}_i) \ .
				\end{equation}
				If we define 
				\begin{equation} 
					g_j(\mathbf{x}_i) = \mathbf{G}_{ij} \ ,
				\end{equation}				
				we find an equation system
				\begin{equation} 
					\left( \mathbf{G}^\intercal \mathbf{G} \right) \mathbf{\beta} = \mathbf{G}^\intercal \mathbf{y} \ .
				\end{equation}						
				that can be solved by inverting the matrix and finding the solutions for $ \mathbf{\beta} $
				\begin{equation} 
					\mathbf{\beta} = \left( \mathbf{G}^\intercal \mathbf{G} \right)^{-1} \mathbf{G}^\intercal \mathbf{y} \ .
				\end{equation}
				The estimated values are obtained by fitting a set of functions to all the points at the same time and only once.
			
			\subsubsection{Non-Linear Least Squares}
			\label{sec:trigger_system_and_reconstruction_algorithms__non_linear_least_squares}	

				When the model function is not linear in the parameters, we have to solve a \emph{Non-Linear Least Squares} (NLLSQ) problem. We approximate $ f $ by its Taylor expansion around an initial set of parameters $ \mathbf{\beta}^0 $
				\begin{equation} 
					f(\mathbf{x}_i, \mathbf{\beta}) = f(\mathbf{x}_i, \mathbf{\beta}^0) + \sum_j \frac{\partial f(\mathbf{x}_i, \mathbf{\beta}^0)}{\partial \beta_j} \left( \beta_j - \beta_j^0 \right) \ .
				\end{equation}	
				Using this in Equation \ref{eq:trigger_system_and_reconstruction_algorithms__least_squares}, we obtain			
				\begin{equation} 
					\frac{\partial \chi^2(\mathbf{\beta})}{\partial \beta_j} = \sum_i \left[ y_i - f(\mathbf{x}_i, \mathbf{\beta}) \right] \frac{\partial f(\mathbf{x}_i, \mathbf{\beta}^0)}{\partial \beta_j} = 0 \ .
				\end{equation}					
				If we define 
				\begin{equation} 
					\frac{\partial f(\mathbf{x}_i, \mathbf{\beta}^0)}{\partial \beta_j} = J_{ij} \ ,
				\end{equation}
				\begin{equation} 
					\beta_j - \beta_j^0 = \Delta \beta_j \ ,
				\end{equation}
				and
				\begin{equation} 
					y_i - f(\mathbf{x}_i, \mathbf{\beta}^0) = \Delta y_i \ ,
				\end{equation}
				we find the following equation
				\begin{equation} 
					\left( \mathbf{J}^\intercal \mathbf{J} \right) \Delta \mathbf{\beta} = \mathbf{J}^\intercal \Delta \mathbf{y} \ .
				\end{equation}	
				Unlike with the LLSQ, we cannot find an analytical solution to this equation, but have to solve it using numerical methods.
		
		\subsection{Kalman Filter}
		\label{sec:trigger_system_and_reconstruction_algorithms__kalman_filter}

\begin{figure}[h!]
				\centering
				\includegraphics[width = 8cm]{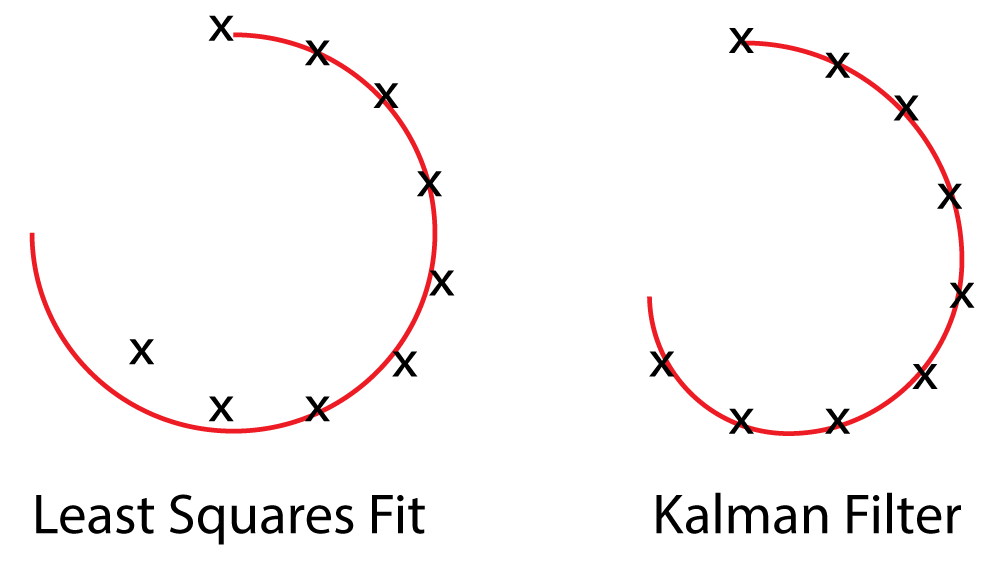}
				\caption{Reconstructed trajectory of a particle loosing energy in a magnetic field using two different algorithms: the Least Squares method and the Kalman filter.}
				\label{fig:trigger_system_and_reconstruction_algorithms__circle_fit}
			\end{figure}

			One of the drawbacks of the least squares fit is that it does not take into account multiple scattering or energy losses or at least the implementation of these features is not straightforward. This may result in bad estimations as shown in Figure \ref{fig:trigger_system_and_reconstruction_algorithms__circle_fit} where we try to fit a circle on the transverse trajectory of a particle loosing energy in a constant magnetic field. The least squares fit searches for solution using a constant bending radius, while the Kalman filter, a more complex fitting algorithm, is able to reconstruct the trajectory with varying parameters. \\

			The Kalman filter \Cite{Fit_Kalman, Fit_Kalman_Particles} is an iterative fitting algorithm that, at each measurement site, updates the parameters of the model by using the measurements themselves and the propagation of the parameters from the previous iteration. This allows the filter to take into account both the error made by the detectors on the measurements and the deviations induced by physical processes on the trajectories, which therefore differs from the ideal case. \\

\begin{figure}[h!]
				\centering
				\includegraphics[width = 9cm]{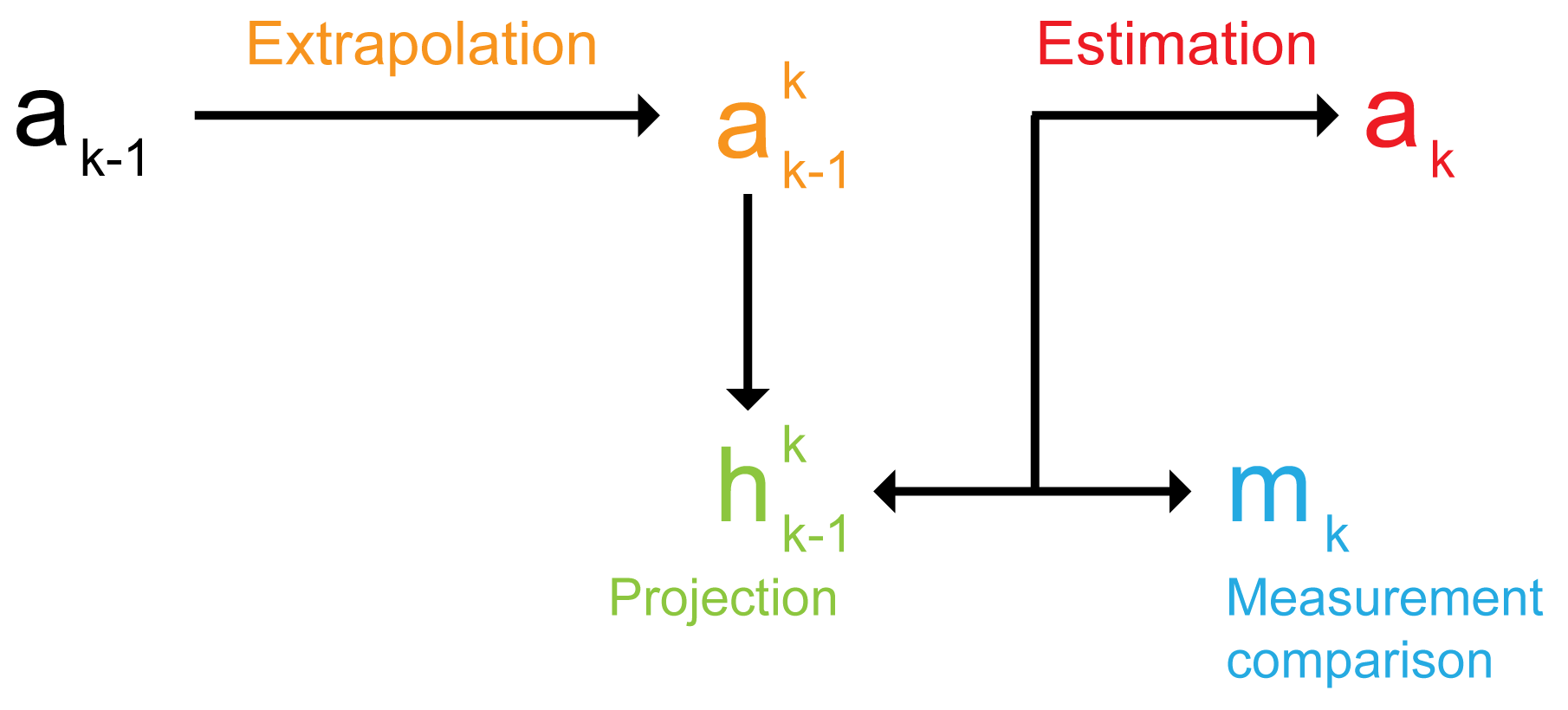}
				\caption{The Kalman filter iterative process flow.}
				\label{fig:trigger_system_and_reconstruction_algorithms__kalman_filter}
			\end{figure}

			The algorithm can be decomposed into four steps: extrapolation, projection, measurement comparison and estimation, as shown in Figure \ref{fig:trigger_system_and_reconstruction_algorithms__kalman_filter}. The last three steps are regrouped into the filtering process. First, the parameters are propagated from the old ($ \mathbf{a}_{k-1} $) to the new ($ \mathbf{a}^k_{k-1} $) measurement site. After this, the extrapolated parameters are projected onto the measurement space ($ \mathbf{h}^k_{k-1} $) which reflects the quantities the detectors measure ($ \phi $, $ \eta $, $ Z $, etc). Then they are compared to the actual measurements taken by the detectors ($ \mathbf{m}_k $) in order to obtain an appreciation of the parameters quality. Finally, a new estimation of the parameters ($ \mathbf{a}_k $) is done by weighting both the projection and the measurements.

			\subsubsection{Classical Kalman Filter}
			\label{sec:trigger_system_and_reconstruction_algorithms__kalman_filter_classical_kalman_filter}			

				Let us write $ \mathbf{a}_{k-1} $ the state vector or track parameters at measurement site $ (k - 1) $ and $ \mathbf{C}_{k-1} $ the covariance matrix of $ \mathbf{a}_{k-1} $. To obtain the new track parameters at site $ (k) $, we first have to extrapolate the state vector from site $ (k - 1) $ to site $ (k) $
				\begin{equation}
					\mathbf{a}^k_{k-1} = \mathbf{F}_{k-1} \mathbf{a}_{k-1} \ ,
					\label{eq:trigger_system_and_reconstruction_algorithms__propagator_matrix}
				\end{equation}
				where $ \mathbf{F}_{k-1} $ describes the evolution of the parameters when moving from one site to another, and $ \mathbf{a}^k_{k-1} $ is the extrapolated state parameters. The propagation being subject to error due to multiple scattering and energy losses, we define $ \mathbf{Q}_{k-1} $ as the propagation noise occurring between site $ (k - 1) $ and site $ (k) $. The covariance therefore becomes 
				\begin{equation}
					\mathbf{C}^k_{k-1} = \mathbf{F}_{k-1} \mathbf{C}_{k-1} \mathbf{F}^\intercal_{k-1} + \mathbf{Q}_{k-1} \ ,
					\label{eq:trigger_system_and_reconstruction_algorithms__propagated_covariance}
				\end{equation} 
				where $ \mathbf{C}^k_{k-1} $ is the extrapolated covariance. \\

				The extrapolated parameters are then projected onto the measurement space using the projector matrix $ \mathbf{H}_k $
				\begin{equation}
					\mathbf{h}^k_{k-1} = \mathbf{H}_k \mathbf{a}^k_{k-1} \ ,
				\end{equation}				
				where $ \mathbf{h}^k_{k-1} $ are the resulting measurements. These are compared to the measured values given by the detectors $ \mathbf{m}_k $ which covariance is notated $ \mathbf{V}_k = \mathbf{G}^{-1}_k $. \\

				Using these different quantities, we calculate the Kalman gain matrix
				\begin{equation}
					\mathbf{K}_k = \mathbf{C}^{k-1}_k \mathbf{H}^\intercal_k \left( \mathbf{V}_k + \mathbf{H}_k \mathbf{C}^{k-1}_k \mathbf{H}^\intercal_k \right)^{-1}
				\end{equation}				
				which describes the gain in precision that the measurement brings to the the extrapolation. Finally, the estimated state and covariance can be obtained through
				\begin{equation}
					\mathbf{a}_k = \mathbf{a}^{k-1}_k + \mathbf{K}_k \left( \mathbf{m}_k - \mathbf{h}^{k-1}_k \right)
				\end{equation}					
				and
				\begin{equation}
					\mathbf{C}_k = \left( \mathbf{I} - \mathbf{K}_k \mathbf{H}_k \right) \mathbf{C}^{k-1}_k \ .
				\end{equation}	
				The $ \chi^2 $ increment of site $ (k) $ is given by
				\begin{equation}
					\chi^2_k = \left( \mathbf{m}_k - \mathbf{h}^{k-1}_k \right)^\intercal \mathbf{R} \left( \mathbf{m}_k - \mathbf{h}^{k-1}_k \right) \ ,
				\end{equation}
				where 
				\begin{equation}
					\mathbf{R} = \left( \mathbf{I} - \mathbf{H}_k \mathbf{K}_k \right) \mathbf{V}_k \ .
				\end{equation} \\

\begin{figure}[h!]
					\centering
					\includegraphics[width = 9cm]{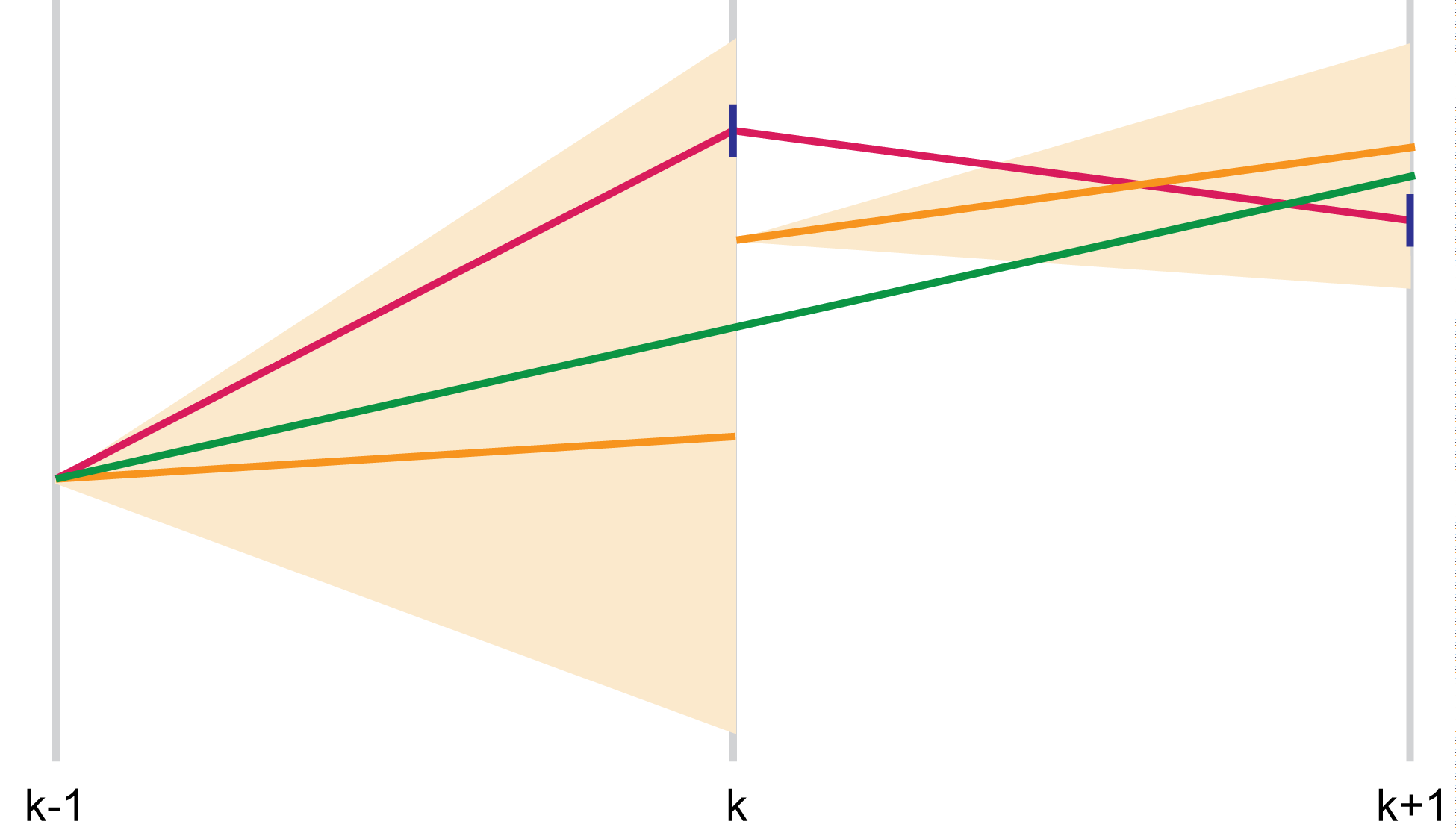}
					\caption{Illustration of the Kalman filter extrapolation and estimation processes (orange) for an ideal straight track (green) and the corresponding real track subject to multiple scattering (red).}
					\label{fig:trigger_system_and_reconstruction_algorithms__track_parameters}
				\end{figure}

				Figure \ref{fig:trigger_system_and_reconstruction_algorithms__track_parameters} illustrates the evolution of the parameters from site to site. The ideal track left by the particle if it was not subject to multiple scattering is represented in green, while the real trajectory is in red. At site $ (k - 1) $, we start with a rough extrapolation in orange towards site $ (k) $. Using the measurement, we update the state and get new parameters closer to the ideal track. At each iteration, the covariance on the estimated parameters decreases resulting in an improvement of the fit at each measurement site.

			\subsubsection{Extended Kalman Filter}
			\label{sec:trigger_system_and_reconstruction_algorithms__kalman_filter_extended_kalman_filter}	

				In the previous section, we made the assumption that the new track parameters are a linear combination of the old parameters. This is not always true and requires the kalman filter to be extended \Cite{Fit_Extended_Kalman}. The idea remains the same but both the propagation matrix $ \mathbf{F}_k $ and the projector matrix $ \mathbf{H}_k $ have to be derived from the system's equations. If the propagated track parameters are a function of the old parameters
				\begin{equation}
					\mathbf{a}^k_{k-1} = \mathbf{f}_{k-1}\left( \mathbf{a}_{k-1} \right) \ ,
				\end{equation}	
				we develop $ \mathbf{f}_{k-1} $ around $ \mathbf{a}_{k-1} $
				\begin{equation}
					\mathbf{f}_{k-1}\left( \mathbf{a}_{k-1} \right) = \frac{\partial \mathbf{f}_{k-1}\left( \mathbf{a}_{k-1} \right)}{\partial \mathbf{a}_{k-1}} \mathbf{a}_{k-1} = \frac{\partial \mathbf{a}^k_{k-1} }{\partial \mathbf{a}_{k-1}} \mathbf{a}_{k-1} \ .
				\end{equation}
				Comparing this to Equation \ref{eq:trigger_system_and_reconstruction_algorithms__propagator_matrix}, we find that
				\begin{equation}
					\mathbf{F}_{k-1} = \frac{\partial \mathbf{f}_{k-1}\left( \mathbf{a}_{k-1} \right)}{\partial \mathbf{a}_{k-1}} = \frac{\partial \mathbf{a}^k_{k-1}}{\partial \mathbf{a}_{k-1}} \ .
					\label{eq:trigger_system_and_reconstruction_algorithms__kalman_propagator}
				\end{equation}
				The same is done with the projection matrix. If we consider the following relation 
				\begin{equation}
					\mathbf{h}^k_{k-1} = \mathbf{m}_k \left( \mathbf{a}^k_{k-1} \right) \ ,
				\end{equation}
				and apply the same technique, we get
				\begin{equation}
					\mathbf{H}_k = \frac{\partial \mathbf{m}_k \left( \mathbf{a}^k_{k-1} \right)}{\partial \mathbf{a}^k_{k-1}} = \frac{\partial \mathbf{h}^k_{k-1}}{\partial \mathbf{a}^k_{k-1}} \ .
				\end{equation}	
		
		\subsection{Implementation}
		\label{sec:trigger_system_and_reconstruction_algorithms__implementation}

			Some of the here-above presented algorithms, namely a Least Square fit and two Kalman filters were implemented in C++ and tested in multiple simulation environments. Those are presented in the following chapter which reviews the simulation process and generated data.				

	\cleardoublepage

\chapter{Simulation Environment}
\label{chap:simulation_environment}

	To develop and study algorithms for the L1 Trigger of GEM detectors, we worked in two different simulation environments: a simplified simulation framework that we will call \emph{Fast Simulation} (FastSim) and a more advanced simulation based on the official CMS simulation software \emph{CMSSW} \Cite{CMSSW_Site}. Using both frameworks, we simulated the response of CSCs and GEMs in CMS for single muon events. This chapter reviews both tools, their functioning and the data they generate.

	\section{Simulation Steps}
	\label{sec:simulation_environment__simulation_steps}

		Both frameworks use the same simulation steps. First, a muon is generated at the IP with a given transverse momentum $ p_T $ and uniformly distributed in pseudo-rapidity $ \eta $ in the range 1.55 < $ \eta $ < 2.15 and in azimuthal angle $ \phi $ in the range 0 < $ \phi $ < 2 $ \pi $. It is then propagated inside the magnetic field towards the muon chambers. The program hereafter computes the interaction between the particles and the chambers, resulting in one or more \emph{Simulated Hits} (SimHits) which are the positions at which the particle deposited energy in the detectors. Then, the response of the detectors is simulated. Using the SimHits, the software generates \emph{Digitized Hits} (Digis) which correspond to the signals read by the electronics of the chambers. Finally, from the Digis, by for example using the COG method reviewed in Section \ref{sec:lhc_and_cms__silicon_pixel_detectors}, the \emph{Reconstructed Hits} (RecHits), which are the measured positions of the hit inside the detectors, are computed.

	\section{Fast Simulation}
	\label{sec:simulation_environment__fast_simulation}

		Due to the early stage of development of the GEM upgrade project within the CMS Collaboration, GEM detectors were not yet implemented in the official CMS simulation tool (CMSSW) at the beginning of our work. We, therefore, wrote our own simplified simulation framework (FastSim) using the same simulation steps as CMSSW. In this environment, physical processes like energy losses and multiple scattering are ignored, considerably increasing the speed at which events are generated. This framework is also very helpful to develop and debug the track reconstruction algorithms. \\

		Moreover, the FastSim differs from CMSSW due to the simplified geometry of the detectors it considers. Figure \ref{fig:gas_electron_multiplier_detectors__install} in Section \ref{sec:gas_electron_multiplier_detectors__ls1_prototypes} shows that the GEM detectors as they would be installed in CMS overlap and do not share the same \axis{Z} coordinates. In the FastSim, detectors are represented by full rings with a given inner and outer radius, and a given \axis{Z} coordinate, simplifying the structure of CMS. Furthermore, in CMSSW, the six planes of readout of the CSCs are implemented, while in the FastSim, only one ring is simulated for each station. \\

		Two versions of FastSim were developed with different CMS magnetic fields configurations: one using a constant 3.8 T field along \axis{Z}, and one using measurements of the actual field of CMS. The first solution allows us to validate algorithms in well known conditions without having to take into account effects caused by the non-uniformity of the field.

		\subsection{Muon Trajectory Generation and Propagation}
		\label{sec:simulation_environment__fs_generation_and_propagation}

			In presence of a magnetic field $ \mathbf{B} $, the particle's trajectory is described by the following equations of motion
			\begin{equation} 
				\left\{  \begin{split}
					\frac{d \mathbf{p}}{dt} & = \frac{q}{m \gamma} \mathbf{p} \times \mathbf{B} \\
					\frac{d \mathbf{x}}{dt} & = \frac{\mathbf{p}}{m \gamma}
				\end{split} \right. \ ,
				\label{eq:simulation_environment__eq_motion}
			\end{equation}	
			where $ \mathbf{p} $ is the momentum of the particle, $ \mathbf{x} $ is the position of the particle, $ q $ is the charge of the particle, and $ \gamma $ is the Lorentz factor. When considering the constant magnetic field, charged particles will describe helices along the direction of the field, in this case \axis{Z}
			\begin{equation} 
				\left\{  \begin{split}
					x & = \frac{p_T}{qB} \left(1 - \cos\left( \frac{qB}{m \gamma} t \right) \right) \\
					y & = \frac{p_T}{qB} \sin\left( \frac{qB}{m \gamma} t \right) \\ 
					z & = \frac{p_T}{m \gamma} \sinh\left( \eta \right) t 
				\end{split} \right. \ .
				\label{eq:simulation_environment__helix_equation}
			\end{equation} 
			To solve Equations \ref{eq:simulation_environment__eq_motion} and propagate particles in a non-uniform magnetic field, we used a numerical method called the 4th order Runge-Kutta method \Cite{Numerical_Recipes} which is one of the most stable and powerful differential equation solver. It is a more complex version of Euler's method which simply approximates 
			\begin{equation}
				\mathbf{p}_{i+1} = \mathbf{p}_i + \Delta t \; f(\mathbf{x}_i, \mathbf{p}_i) 
			\end{equation}
			where $ \Delta t $ is the time increment and 
			\begin{equation} 
				\frac{d \mathbf{p}}{dt} = f(\mathbf{x}, \mathbf{p}) \ . 
			\end{equation}	
			Instead, it defines Runge-Kutta coefficients
			\begin{equation} 
				\left\{  \begin{split}
					\mathbf{k}_1 & = f\left(\mathbf{x}, \mathbf{p} \right) \\
					\mathbf{k}_2 & = f\left(\mathbf{x}, \mathbf{p} + \frac{\Delta t}{2} \mathbf{k}_1 \right) \\
					\mathbf{k}_3 & = f\left(\mathbf{x}, \mathbf{p} + \frac{\Delta t}{2} \mathbf{k}_2 \right) \\
					\mathbf{k}_4 & = f\left(\mathbf{x}, \mathbf{p} + \Delta t \ \mathbf{k}_3 \right) \\
				\end{split} \right. \ ,
			\end{equation}		
			that are used to update the position and momentum
			\begin{equation} 
				\left\{  \begin{split}
					\mathbf{x}_{i+1} & = \mathbf{x}_i + \frac{\Delta t}{m \gamma} \mathbf{p}_i + \frac{\Delta t^2}{6} \left( \mathbf{k}_1 + \mathbf{k}_2 + \mathbf{k}_3 \right) \\ 
					\mathbf{p}_{i+1} & = \mathbf{p}_i + \frac{\Delta t}{6} \left( \mathbf{k}_1 + 2 \ \mathbf{k}_2 + 2 \ \mathbf{k}_3 + \mathbf{k}_4 \right)
				\end{split} \right. \ .
			\end{equation}
			Due to the small number of measurements given by the field map (one measurement every 5 cm in \axis{R} and \axis{Z}), we assume that the magnetic field stays constant during each step ($ | \Delta \mathbf{x} | \sim $ 10$ ^{-4} $). \\
			
			The simulated detectors being placed in the endcaps at a given \axis{Z}, we repeated the iteration until the desired \axis{Z} coordinate is reached. Due to the discrete nature of the propagation, a linear extrapolation is done between the two closest matches of the requested position. 
		
		\subsection{SimHits}
		\label{sec:simulation_environment__simhits}

			For each detector, one SimHit is created, corresponding to the intersection between the particle's trajectory and the chamber. No energy losses or secondary avalanches are computed in order to simplify the simulations.
		
		\subsection{Digis}
		\label{sec:simulation_environment__digis}

			The digitization step emulates the response of the detector to the interaction between the incident particles and the sensitive volume of the detector as well as the response of the readout electronics. Although such emulation already exists for GEMs \Cite{These_Geoffrey}, the electronic design is not yet available. In addition, this level of detail is not yet required at this stage of our study.

		\subsection{RecHits}
		\label{sec:simulation_environment__rechits}

			GEM RecHits are obtained from the SimHits by applying a gaussian noise to the $ \phi $ coordinate with a standard deviation corresponding to the resolution of the detectors (170 to 340 \um{}), as reviewed in Section \ref{sec:gas_electron_multiplier_detectors__spatial_resolution}. The $ \eta $ coordinate being unknown, we arbitrarily take the value at the middle of the segment. \\

			For CSC RecHits, a gaussian noise is applied to both coordinates according to the pitch between wire groups and between strips
			\begin{equation}
				\sigma = \frac{pitch}{\sqrt{12}} \ .
			\end{equation}

	\section{CMSSW Software}
	\label{sec:simulation_environment__cms_software}

		CMSSW is the official simulation and data analysis framework of CMS. It includes the complete geometry of all the detectors of CMS, described with the GEANT4 software \Cite{Simulation_GEANT4}, which allows to take into account multiple scattering and energy losses when propagating particles. \\

		CMSSW SimHits are more complex than FastSim SimHits, as GEANT4 computes primary and secondary ionizations inside the muon chambers resulting in up to 50 hits per chamber. Those hits are grouped to form one SimHit which position is equal to the mean position of the hits. \\

		When we started this work, neither the Digis nor the RecHits were implemented inside CMSSW for GEMs. We therefore decided to use the same process as the FastSim to generate the RecHits from the SimHits.

	\section{Validation Plots}
	\label{sec:simulation_environment__validation_plots}		

\begin{figure}[h!]
			\centering
			\includegraphics[width = 0.8 \textwidth]{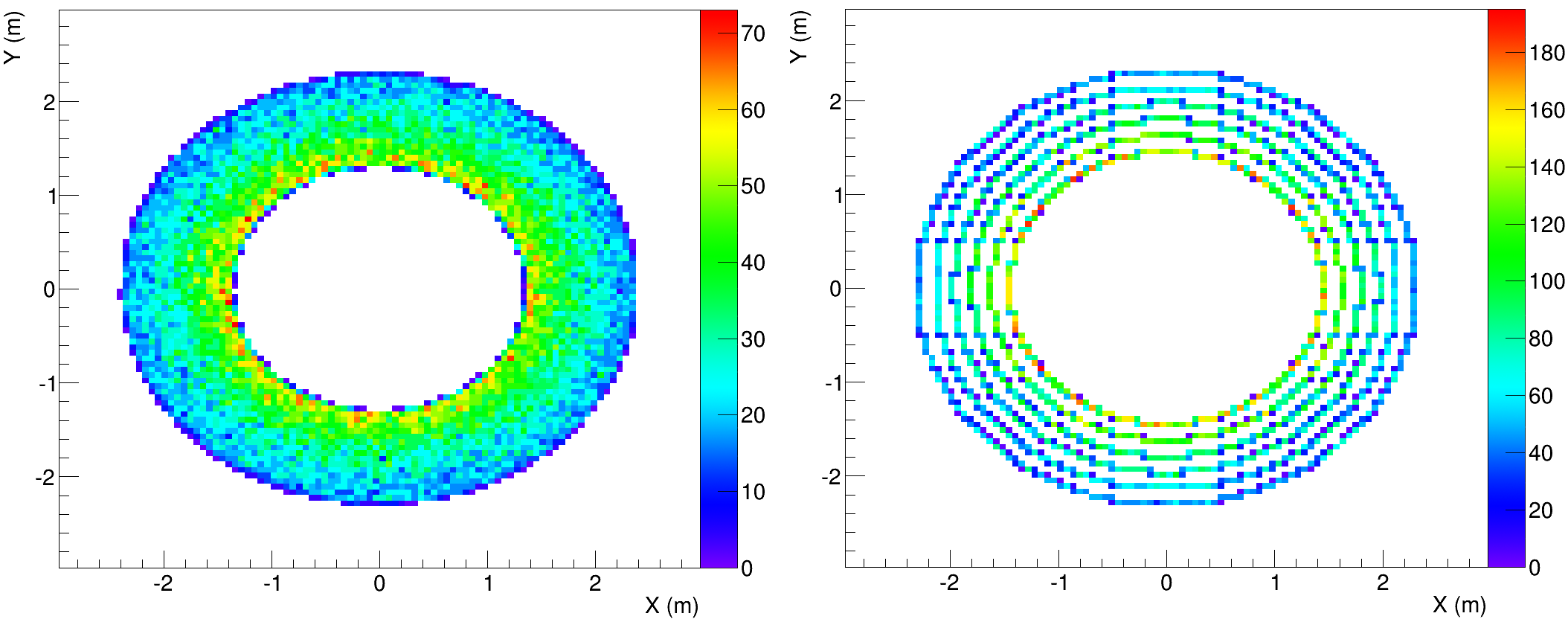}
			\caption{Occupancy in the transverse plane of the SimHits (left) and the RecHits (right) in the first GEM detector in GE1/1 for muon tracks generated with the FastSim in the constant magnetic field with a simulated \pT{} of 20 \GeVc{}.}
			\label{fig:simulation_environment__simhits_rechits}
		\end{figure}

		Validation plots are used to verify that the simulations produce the correct output and that the muons hits are correctly located inside CMS. Figure \ref{fig:simulation_environment__simhits_rechits} represents the occupancy in the transverse plane of the SimHits (left) and the RecHits (right) in the first GEM detector in GE1/1 for muon tracks generated with the FastSim in the constant magnetic field with a simulated \pT{} of 20 \GeVc{}. The $ \eta $ discretization is clearly visible in the image on the right where 6 consecutive rings appear after segmentation.

	\section{Geometry of the CMS Muon Endcap System}
	\label{sec:simulation_environment__geometry_of_the_cms_muon_endcap_system}

		We simulated one endcap of CMS by considering hits in the following CSCs: ME1/1, ME2/1, ME2/2, ME3/1, ME3/2, ME4/1, and ME4/2, and in the two GEMs of the GE1/1 super-chambers: GE1/1a and GE1/1b. \\

\begin{figure}[h!]
			\centering
			\includegraphics[width = 0.55 \textwidth]{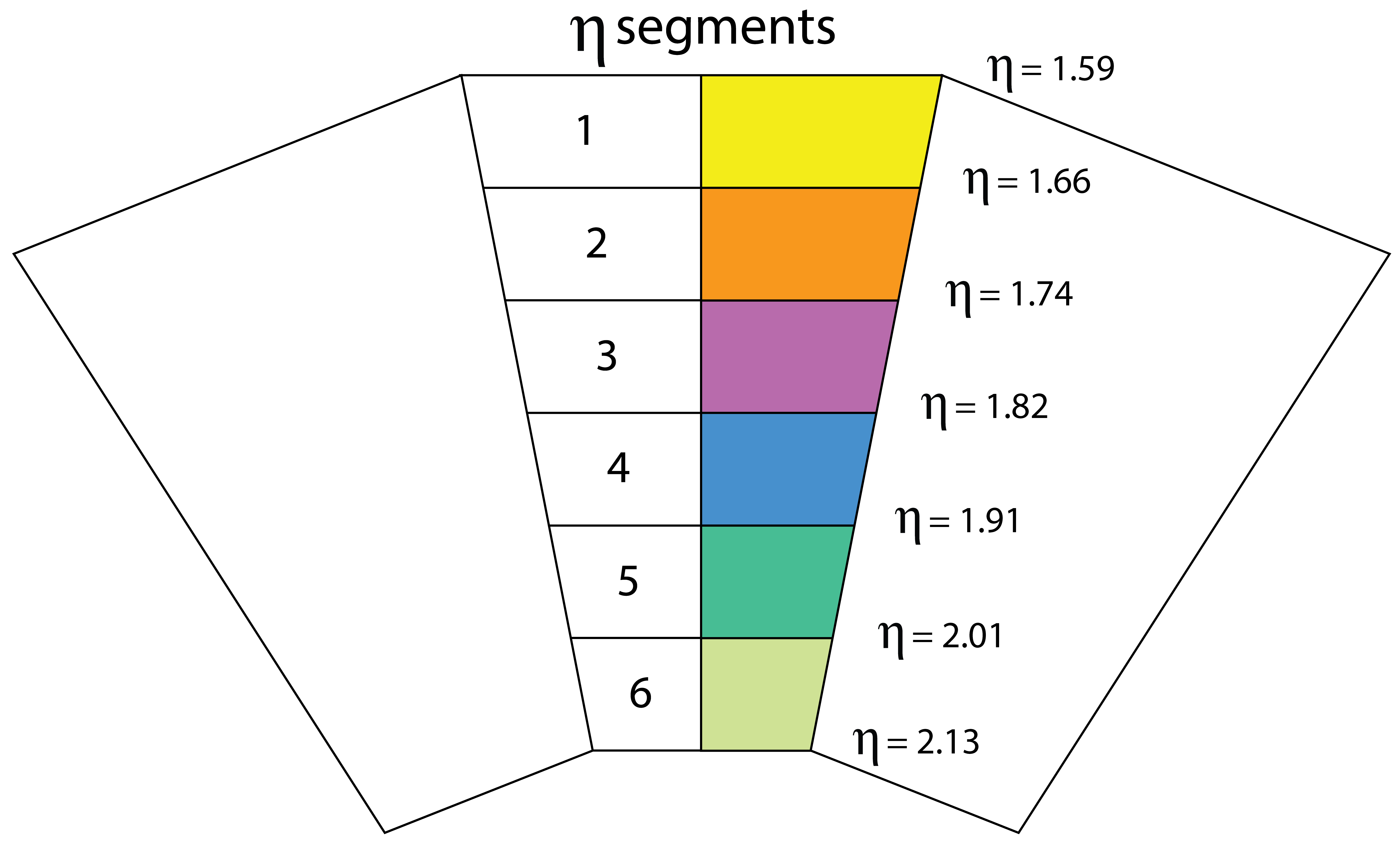}
			\caption{$ \eta $ partitioning of the GEM detectors.}
			\label{fig:simulation_environment__eta_partition}
		\end{figure}

		GEMs are divided into 6 segments in $ \eta $ and will be used at full granularity (128 strips by segment) for most studies. Figure \ref{fig:simulation_environment__eta_partition} depicts the $ \eta $ partitioning of the chambers and the naming convention: segment 6 has the highest $ \eta $ value and is the closest to the beam pipe, and segment 1 has the lowest $ \eta $ value and is the most outer segment. \\
		
		As reviewed in Section \ref{sec:gas_electron_multiplier_detectors__upgrade_of_the_cms_muon_system}, one of the interests of GEMs is to improve the CSCs' reconstruction capabilities, specifically ME1/1. Therefore, all algorithms will be tested using the setups listed in Table \ref{tab:simulation_environment__simulation_setups} which references the detector setups with the used detection layers and the resulting number of hits. When considering a particular setup, all chambers are required to be hit. If not, the event is dropped. \\

		\begin{table}[h!]
			\centering
			\begin{tabular}{c|c|c}
				Name & Used detection layers & Number of hits \\ \hline
				GE1 ME1 & GE1/1a, GE1/1b, and ME1/1 & 3 hits \\
				GE1 ME12 & GE1/1a, GE1/1b, ME1/1, and ME2/x & 4 hits \\
				GE1 ME123 & GE1/1a, GE1/1b, ME1/1, ME2/x, and ME3/x & 5 hits \\
				GE1 ME1234 & GE1/1a, GE1/1b, ME1/1, ME2/x, ME3/x, and ME4/x & 6 hits \\
				ME1234 & ME1/1, ME2/x, ME3/x, and ME4/x & 4 hits
			\end{tabular}
			\caption{List of detector setups with the used detection layers and the resulting number of hits.}
			\label{tab:simulation_environment__simulation_setups}
		\end{table}

		Studies will also be performed on the L1 acceptance rates, described in Section \ref{sec:trigger_system_and_reconstruction_algorithms__system_performances}, when reconstruction takes place with our algorithms. Furthermore, analysis will be done according to the hit segment of GE1/1 to compare the performances' evolution with $ \eta $. \\

\begin{figure}[h!]
			\centering
			\includegraphics[width = 0.7 \textwidth]{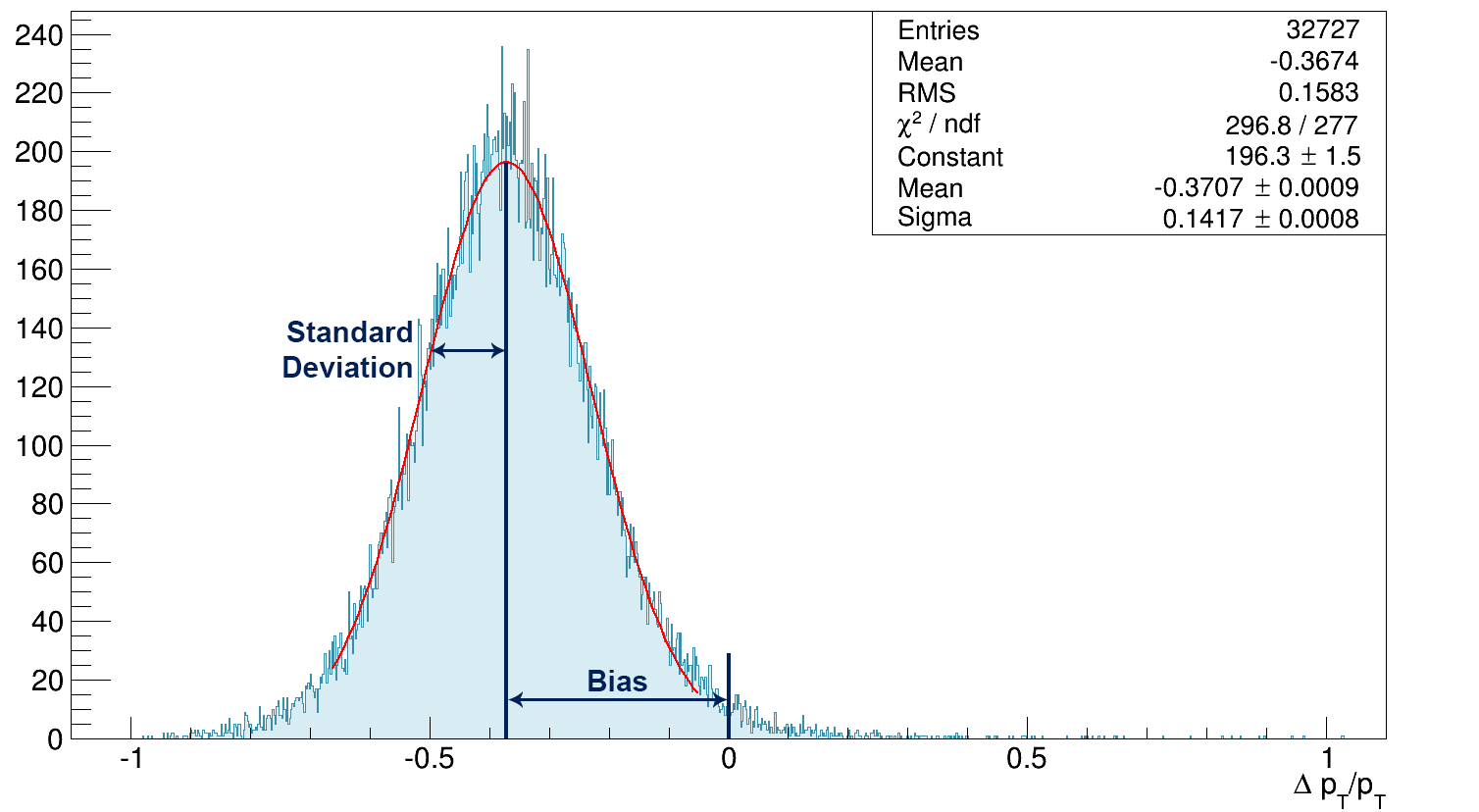}
			\caption{$ \frac{\Delta p_T}{p_T} $ for muon tracks generated with CMSSW with a simulated \pT{} of 60 \GeVc{}, and reconstructed with the Least Squares fit using the SimHits in GE1/1, ME1/1, and ME2/x.}
			\label{fig:simulation_environment__reco_pt_simhits_38T_100GeV_GEM_ME1234}
		\end{figure}

		To quantify the performance of the reconstruction algorithms, we use
		\begin{equation}
			\frac{\Delta p_T}{p_T} = \frac{\frac{1}{p_{T,reco}} - \frac{1}{p_{T,sim}}}{\frac{1}{p_{T,sim}}} \ ,
		\end{equation}
		where $ p_{T,reco} $ is the reconstructed transverse momentum and $ p_{T,sim} $ is the simulated transverse momentum. The bias and standard deviation of the results, both expressed in \%, are given by the parameters of a fitted gaussian distribution on the curve. These parameters are represented in Figure \ref{fig:simulation_environment__reco_pt_simhits_38T_100GeV_GEM_ME1234} which depicts $ \frac{\Delta p_T}{p_T} $ for muon tracks generated with CMSSW with a simulated \pT{} of 60 \GeVc{}, and reconstructed with the Least Squares fit using the SimHits in GE1/1, ME1/1, and ME2/x.

	\section{Topology of the Generated Muon Tracks in the CMS Muon Endcap System}
	\label{sec:simulation_environment__topology_of_the_generated_muon_tracks_in_the_cms_muon_endcap_system}

		To check the muon trajectory propagation by the 4th order Runge-Kutta method reviewed in Section \ref{sec:simulation_environment__fs_generation_and_propagation}, we have quantitatively observed the trajectories in three dimensions and in the transverse plane. In the following, we describe the topology of muon tracks generated with the FastSim in both the constant and the real magnetic field. \\ 

\begin{figure}[h!]
			\centering
			\includegraphics[width = 0.8 \textwidth]{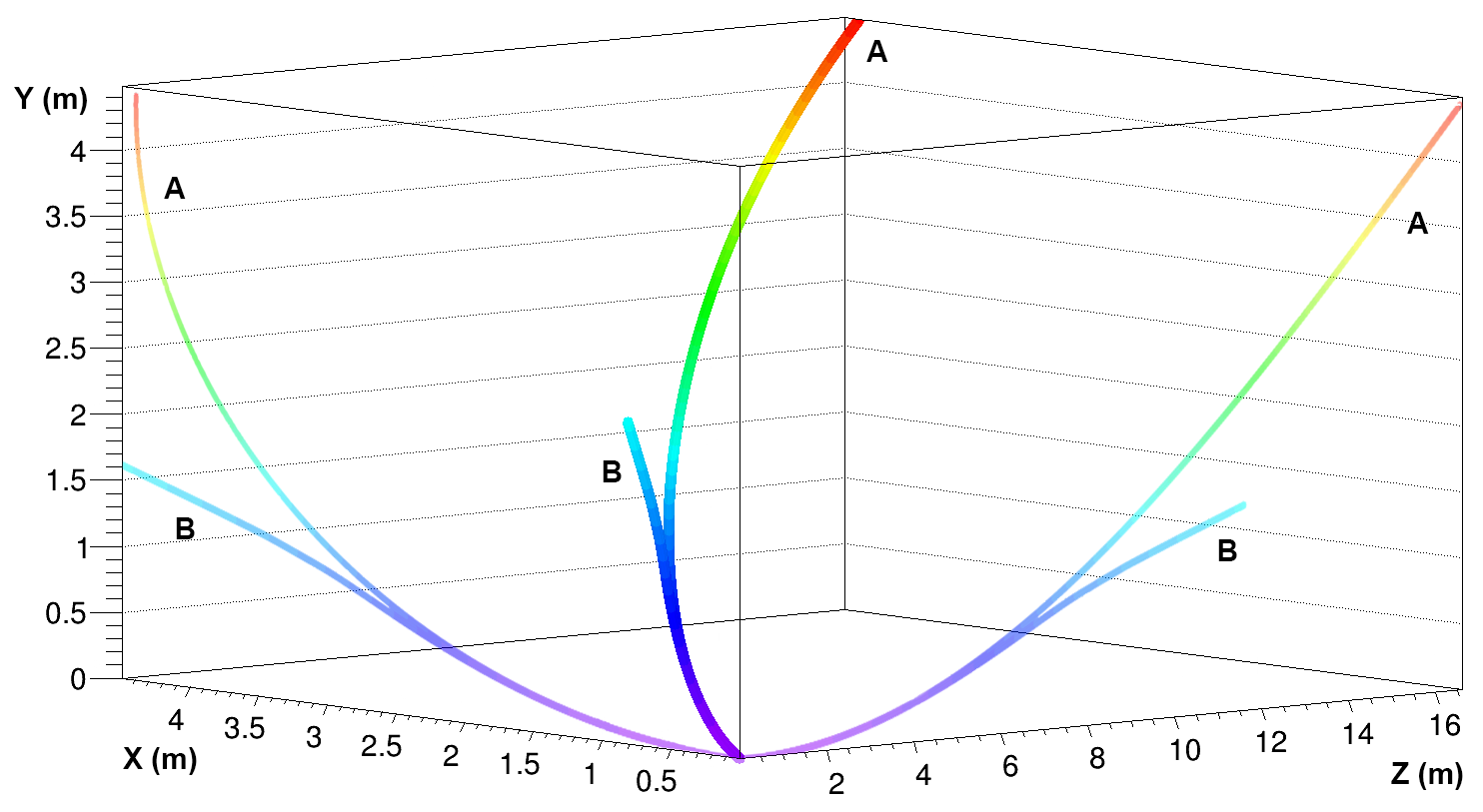}
			\caption{Three dimensional view of muon tracks generated with the FastSim in the constant magnetic field (blue arrows) and the real magnetic field (green arrows) with a simulated \pT{} of 5 \GeVc{}, an initial $ \eta $ of 1.6, and an initial $ \phi $ of 0. The full colored track represents the three dimensional propagation of the muons and th clearer tracks are the projections in the transverse plane and the longitudinal plane.}
			\label{fig:simulation_environment__3d_view}
		\end{figure}

		Figure \ref{fig:simulation_environment__3d_view} is a three dimensional view of two muon tracks generated with the FastSim in the constant magnetic field (A) and the real magnetic field (B) with a simulated \pT{} of 5 \GeVc{}, an initial $ \eta $ of 1.6, and an initial $ \phi $ of 0. At the beginning of the propagation, the two muon tracks follow the same trajectory as the magnetic fields are equal for both environments. However, once they reach a \axis{Z} around 5 m, they diverge as the real magnetic field starts to differ from the constant magnetic field. \\

		Figure \ref{fig:simulation_environment__3d_view} also depicts the projection of the tracks in the transverse and longitudinal plane. It is of importance to understand the impact of the track's parameters on the projection in the transverse plane as some of the hereafter developed algorithms rely only on this information to estimate the \pT{} of the particles.

		\subsection{Constant Magnetic Field}
		\label{sec:simulation_environment__constant_magnetic_field}

\begin{figure}[h!]
				\centering
				\includegraphics[width = 0.6 \textwidth]{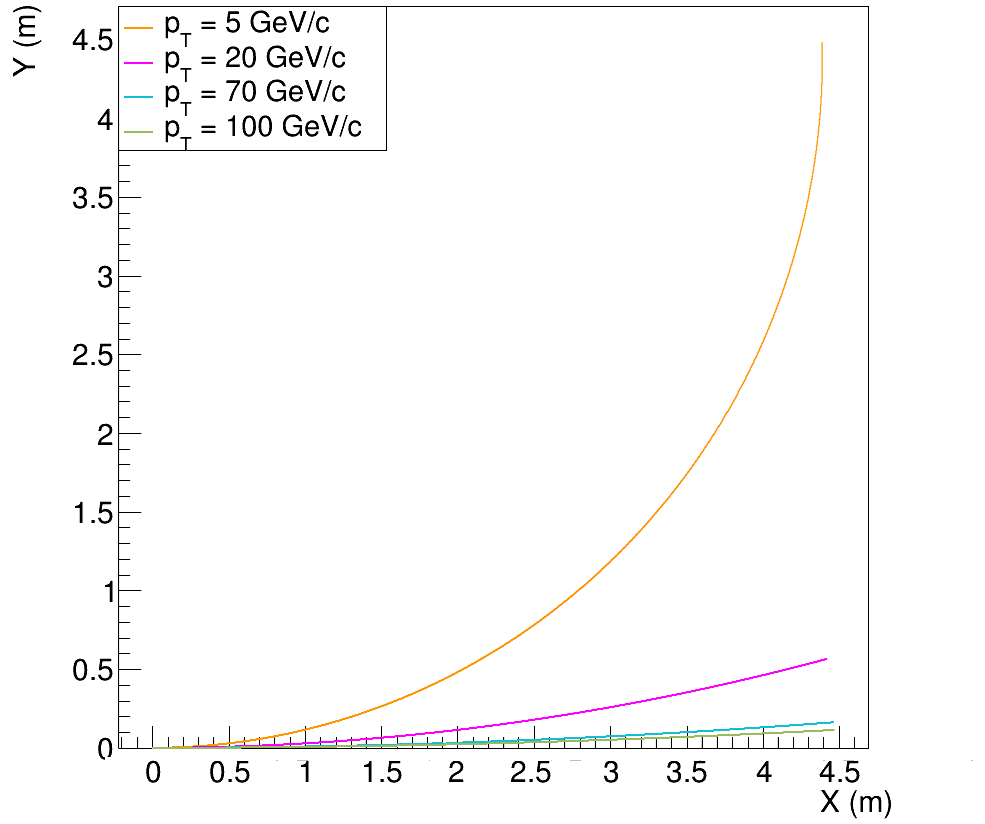}
				\caption{Projection in the transverse plane of muon tracks generated with the FastSim in the constant magnetic field with simulated \pT{} of 5, 20, 70, and 100 \GeVc{}, and an initial $ \eta $ of 1.6.}
				\label{fig:simulation_environment__xy_projection_38T_vs_pt}
			\end{figure}

			As shown in Equation \ref{eq:simulation_environment__helix_equation}, particles in the constant magnetic field describe helices along the axis of the field. The projection of these trajectories in the transverse plane are circles, which radius is given by Equation \ref{eq:lhc_and_cms__radius_to_momentum_relation} in Section \ref{sec:lhc_and_cms__tracker_system_performances}. Figure \ref{fig:simulation_environment__xy_projection_38T_vs_pt} represents the projection in the transverse plane of muon tracks generated with the FastSim in the constant magnetic field with simulated \pT{} of 5, 20, 70, and 100 \GeVc{}, and an initial $ \eta $ of 1.6. Particles with a small \pT{} (orange) have a smaller bending radius than those generated with a higher \pT{} (green). \\

\begin{figure}[h!]
				\centering
				\includegraphics[width = 0.6 \textwidth]{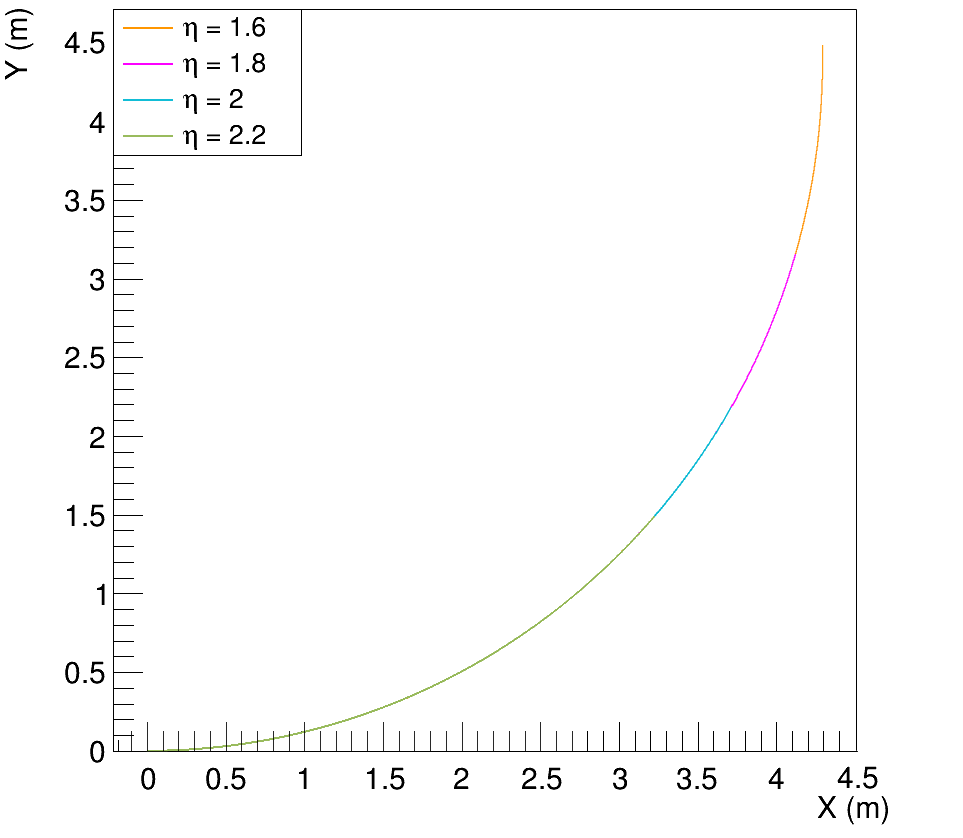}
				\caption{Projection in the transverse plane of muon tracks generated with the FastSim in the constant magnetic field with a simulated \pT{} of 5 \GeVc{} and initial $ \eta $ of 1.6, 1.8, 2, and 2.2 (the curves overlap but all start at 0).}
				\label{fig:simulation_environment__xy_projection_38T_vs_eta}
			\end{figure}

			Further, the projection of the particles' tracks in the transverse plane also varies with $ \eta $. Figure \ref{fig:simulation_environment__xy_projection_38T_vs_eta} depicts the projection in the transverse plane of muon tracks generated with the FastSim in the constant magnetic field with a simulated \pT{} of 5 \GeVc{} and initial $ \eta $ of 1.6, 1.8, 2, and 2.2 (the curves overlap but all start at 0). \\

			For the same simulated \pT{}, particles with an initial higher $ \eta $ (green) have a greater longitudinal momentum \pZ{} than particles generated at smaller $ \eta $
			\begin{equation}
				p_Z = p_T \sinh(\eta) \ .
			\end{equation}
			The higher \pZ{} results in shorter tracks in the transverse plane as particles propagate faster towards the detectors laid out along \axis{Z}.\\

\begin{figure}[h!]
				\centering
				\includegraphics[width = 0.7 \textwidth]{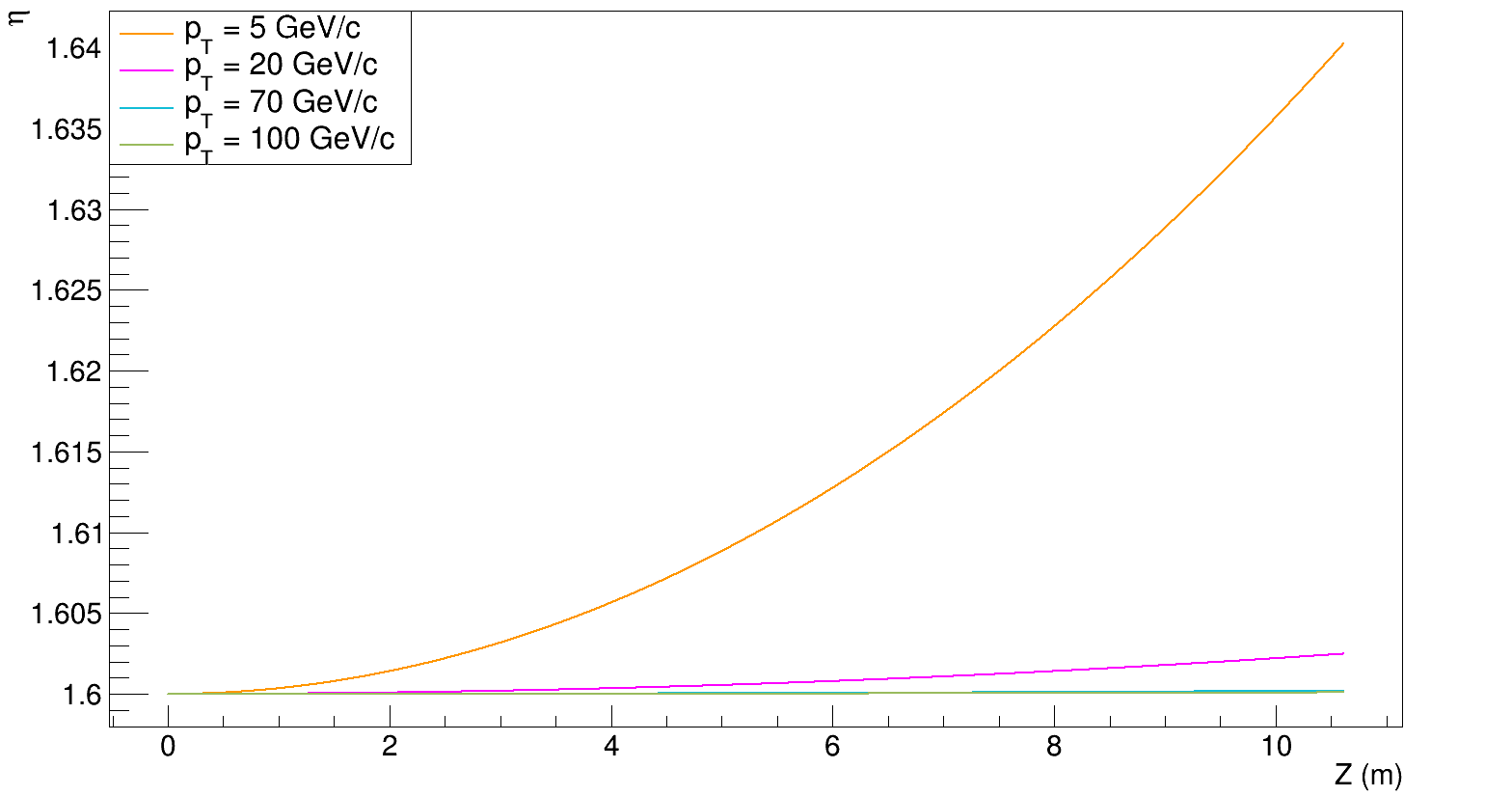}
				\caption{Variation of $ \eta $ as a function of $ Z $ for muon tracks generated with the FastSim in the constant magnetic field with simulated \pT{} of 5, 20, 70, and 100 \GeVc{}, and an initial $ \eta $ = 1.6.}
				\label{fig:simulation_environment__eta_evolution_38T_multi_pt}
			\end{figure}

			Moreover, $ \eta $ also plays another important role when constructing the RecHits from the SimHits. Figure \ref{fig:simulation_environment__eta_evolution_38T_multi_pt} shows the variation of $ \eta $ as a function of \axis{Z} for muon tracks generated with the FastSim in the constant magnetic field with simulated \pT{} of 5, 20, 70, and 100 \GeVc{}, and an initial $ \eta $ = 1.6. Tracks left by low \pT{} muons (orange) have a higher bending radius and lower \pZ{} compared to high \pT{} particles (green). The $ \eta $ difference $ \Delta \eta $ between the SimHits in the two GEM layers in GE1/1 is therefore greater ($ \Delta \eta \sim $ 10$ ^{-3} $ at 5 \GeVc{} and $ \Delta \eta \sim $ 2 10 $ ^{-7} $ at 100 \GeVc{}) as the propagation of the particle in the transverse plane is greater. \\

			When discretization occurs, the RecHits in the GEM detectors are arbitrarily assigned with a radial position corresponding to the middle of the hit segment, as GEMs do not measure the $ \eta $ coordinate. For low \pT{} particles, the error made will be larger than for particles of higher \pT{} as $ \Delta \eta $ is greater. Indeed, for 93.7\% of the events the hit segment in both chambers of GE1/1 are the same. Therefore, as the two chambers and their segmentations are identical, the radial coordinate of the RecHits in both chambers will be the same and equal the radial position of the middle of the segment. Tracks will locally appear to be parallel to the beam axis as the RecHits in the GEM detectors are located at the same radial distance from \axis{Z}. As high \pT{} tracks are straighter than low \pT{} tracks, they are closer to the locally reconstructed horizontal track.

		\subsection{Real Magnetic Field}
		\label{sec:simulation_environment__real_magnetic_field}

			The topology discussed in the previous section remains valid for tracks generated in the real magnetic field as long as its intensity does not significantly differ from the ideal case. Unfortunately, in CMS, the intensity of the field quickly decreases outside the solenoid. This effect is visible in Figure \ref{fig:simulation_environment__3d_view} where the track generated in the real magnetic field starts to deviate from the track generated in the constant magnetic field.

	\section{Muon Track Reconstruction Studies}
	\label{simulation_environment__muon_track_reconstruction_studies}

			In the following chapters, we will describe the study of several muon track reconstruction algorithms for the L1 Trigger of CMS using GEM data. The performance of each algorithm will first be analyzed with the constant magnetic field, then with the real magnetic field of CMS, and finally within the CMSSW framework. For each simulation environment, we will look at the performance of the reconstruction using the SimHits and the RecHits. The effect of different detector setups as well as the $ \eta $ segmentation of GEMs will also be investigated. 

	\cleardoublepage


\chapter{Least Squares Fit}
\label{chap:least_squares_fit}

	This chapter presents the first algorithm we developed to reconstruct the particles' trajectories, namely a Least Squared fit, which theory has been reviewed in Section \ref{sec:trigger_system_and_reconstruction_algorithms__least_squares}. We start by describing the algorithm before analyzing the results obtained with the FastSim and CMSSW.

	\section{Algorithm}
	\label{sec:least_squares_fit__algorithm}

		As reviewed in Section \ref{sec:simulation_environment__fs_generation_and_propagation}, charged particles traveling in a constant magnetic field will describe helices along the axis of the field, which projection in the transverse plane are circles. The radius of these circles is directly proportional to \pT{} as previously described in Equation \ref{eq:lhc_and_cms__radius_to_momentum_relation} in Section \ref{sec:lhc_and_cms__tracker_system_performances}. \\

	 	The value we try to minimize is 
		\begin{equation} 
			\chi^2(A, B, C) = \sum_i \left( (x_i - A)^2 + (y_i - B)^2 - C^2 \right)^2 \ ,
		\end{equation}
		where $ A $ is the \axis{X} coordinate of the center of the circle, $ B $ is the \axis{Y} coordinate of the center of the circle, and $ C $ is the radius of the circle. As is, the problem is non-linear and therefore harder and more time consuming to address, as we need to utilize numerical methods to solve a differential equation. In 1993, I. D. Coop proposed a solution to linearize the least squares method for a circle \Cite{Fit_Coop}. By defining $ D = 2A $, $ E = 2B $, and $ F = C^2 - A^2 - B^2 $, we find
		\begin{equation} 
			\chi^2(D, E, F) = \sum_i \left( x_i^2 + y_i^2 - D x_i - E y_i - F \right)^2 \ ,
		\end{equation}
		which is now linear in its parameters. We then derive this according to $ D $, $ E $, and $ F $
		\begin{equation} 
			\left\{  \begin{split}
				\frac{\partial \chi^2}{\partial D} & = \sum_i (-2 x_i) \left( x_i^2 + y_i^2 - D x_i - E y_i - F \right) = 0 \\
				\frac{\partial \chi^2}{\partial E} & = \sum_i (-2 y_i) \left( x_i^2 + y_i^2 - D x_i - E y_i - F \right) = 0 \\
				\frac{\partial \chi^2}{\partial F} & = \sum_i (-2) \left( x_i^2 + y_i^2 - D x_i - E y_i - F \right) = 0 \\
			\end{split} \right. \ ,
		\end{equation}
		giving us the following system to solve 
		\begin{equation} 
			\left( \begin{matrix} \sum_i x_i^2 & \sum_i x_i y_i & \sum_i x_i \\ \sum_i x_i y_i & \sum_i y_i^2 & \sum_i y_i \\ \sum_i x_i & \sum_i y_i & \sum_i 1 \end{matrix} \right) \left( \begin{matrix} D \\ E \\ F \end{matrix} \right) = \left( \begin{matrix} \sum_i x_i^3 + \sum_i x_i y_i^2 \\ \sum_i x_i^2 y_i + \sum_i y_i^3 \\ \sum_i x_i^2 + \sum_i y_i^2 \end{matrix} \right) \ .
		\end{equation}
		The small size of the matrix allows us to immediately compute the inverse without having to use matrix inversion algorithms. \\

		The downside of the linearizion is that we cannot use separated errors for $ \eta $ and $ \phi $. It is only possible to assign a weight $ w_i $ to each point 
		\begin{equation} 
			\chi^2(A, B, C) = \sum_i w_i \left( (x_i - A)^2 + (y_i - B)^2 - C^2 \right)^2 \ ,
		\end{equation}
		which is difficult in our case as GEMs have a poorer $ \eta $ resolution than CSCs but a higher $ \phi $ granularity. Therefore, we chose not to apply weights to the data. \\

		To reconstruct the particle's \pT{}, we invert Equation \ref{eq:lhc_and_cms__radius_to_momentum_relation} in Section \ref{sec:lhc_and_cms__tracker_system_performances} for a defined intensity of the magnetic field of 3.8 T and the previously found radius
		\begin{equation}
			p_T = 0.3 \ B[\mbox{T}] \ \sqrt{F + \frac{D^2}{4} + \frac{E^2}{4}} \ .
		\end{equation}

		We require a minimum of three hits in the detectors to fit the curve and then add the IP to the list of points.

	\section{FastSim Results: Constant Magnetic Field}
	\label{sec:least_squares_fit__fastsim_results_constant_magnetic_field}

		The algorithm is first applied to muon tracks obtained with the FastSim in the constant magnetic field in order to test its validity, as tracks describe a perfect circle in the transverse plane. This simulation environment gives us the opportunity to look at the effects of the $ \eta $ segmentation of GEMs without perturbations from the non-uniform field or physical processes.
		
		\subsection{Validation Using SimHits}
		\label{sec:least_squares_fit__constant_magnetic_field_validation_using_simhits}	

\begin{figure}[h!]
				\centering
				\includegraphics[width = 0.7 \textwidth]{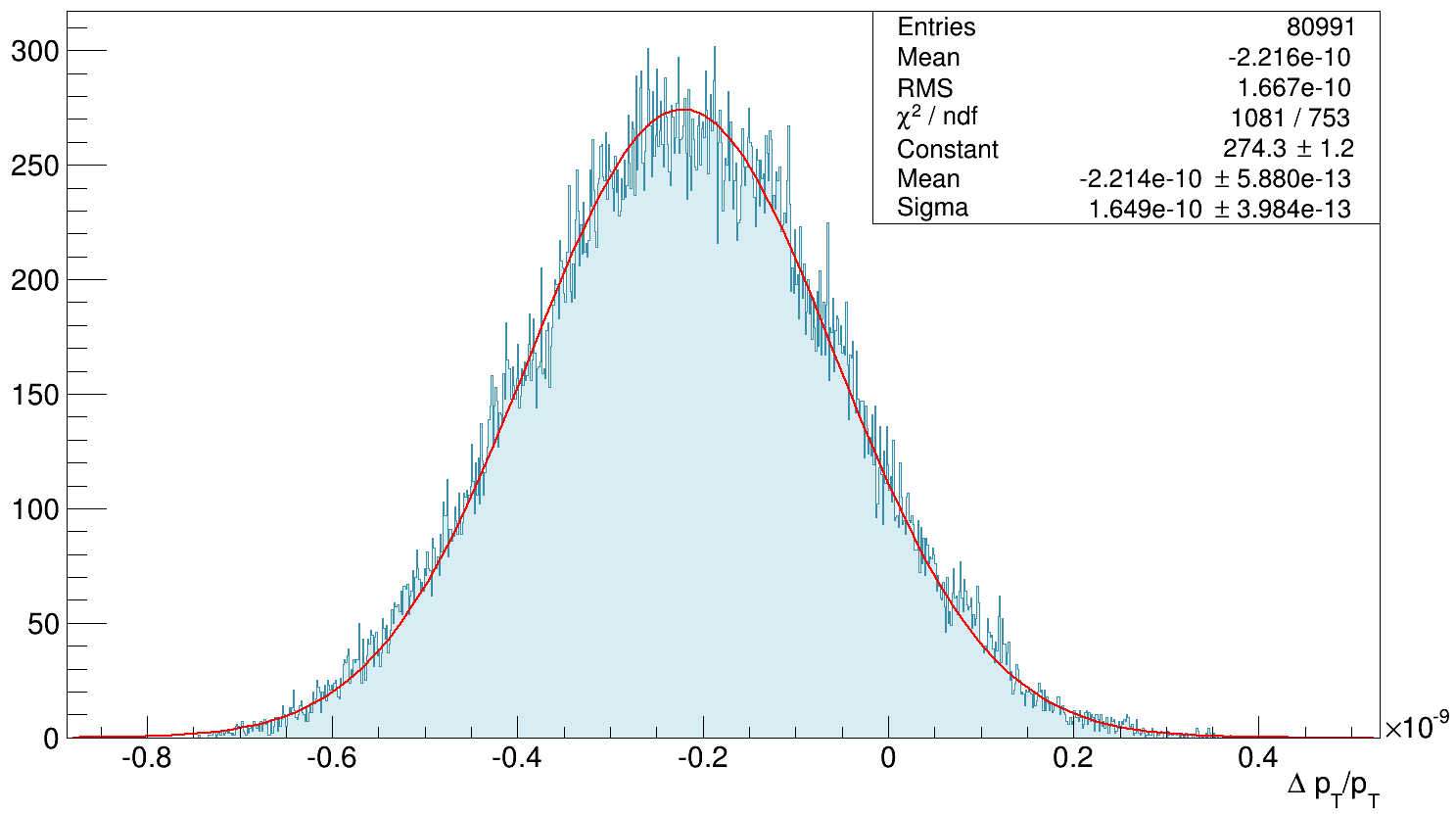}
				\caption{$ \frac{\Delta p_T}{p_T} $ for muon tracks generated with the FastSim in the constant magnetic field with a simulated \pT{} of 50 \GeVc{}, and reconstructed with the Least Squares fit using the SimHits in GE1/1, ME1/1, ME2/x, ME3/x, and ME4/x.}
				\label{fig:least_squares_fit__reco_pt_simhits_38T_50GeV_GEM_ME1234}
			\end{figure}

			We validate the fit by looking at the generated SimHits. Figure \ref{fig:least_squares_fit__reco_pt_simhits_38T_50GeV_GEM_ME1234} shows the fitted gaussian on $ \frac{\Delta p_T}{p_T} $ for muon tracks generated with the FastSim in the constant magnetic field with a simulated \pT{} of 50 \GeVc{}, and reconstructed with the Least Squares fit using the SimHits in GE1/1, ME1/1, ME2/x, ME3/x, and ME4/x. We emphasize the 10$ ^{-9} $ factor for the abscissa.

			 Both the bias and the standard deviation are of the order of 2 10$ ^{-8} $\% meaning the fit is successful. The Least Squares fit is applied to all the simulated \pT{} and all the detector setups. The bias is never higher than 3 10$ ^{-8} $\% and the standard deviation than 3 10$ ^{-7} $\%. Thus, we conclude that the algorithm works correctly and can be tested in more complex situations.
		
		\subsection{Impact of Segmentation}
		\label{sec:least_squares_fit__constant_magnetic_field_impact_of_segmentation}	

\begin{figure}[h!]
				\centering
				\includegraphics[width = 0.7 \textwidth]{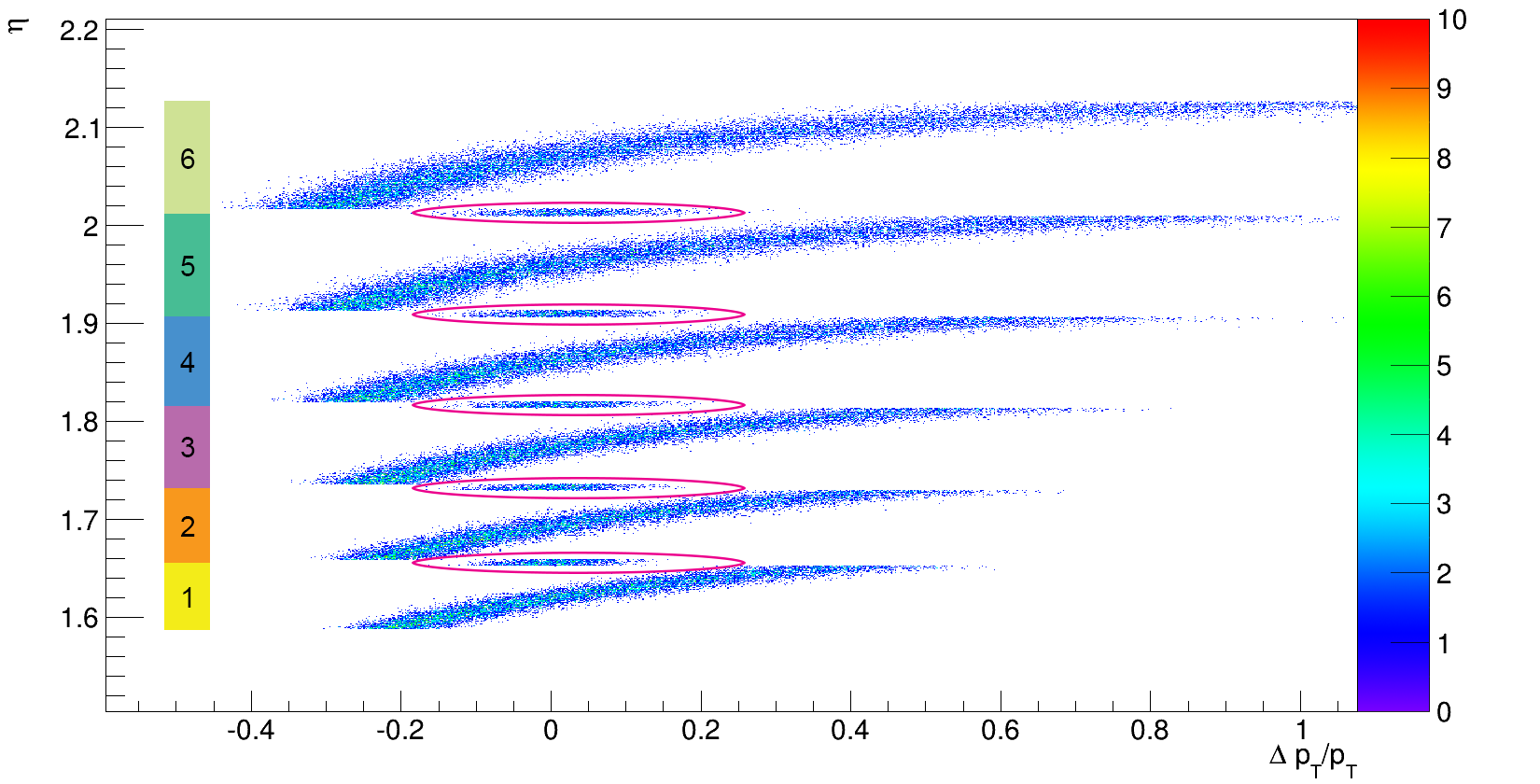}
				\caption{$ \frac{\Delta p_T}{p_T} $ as a function of the simulated $ \eta $ (before discretization) for muon tracks generated with the FastSim in the constant magnetic field with a simulated \pT{} of 5 \GeVc{}, and reconstructed with the Least Squares fit using the RecHits in GE1/1 and ME1/1.}
				\label{fig:least_squares_fit__reco_pt_rechits_38T_5GeV_GEM_ME11}
			\end{figure}

			The same analysis is done considering the RecHits generated with the FastSim in the constant magnetic field. The main challenge arising is the discrete $ \eta $ partitioning of GEM detectors. Figure \ref{fig:least_squares_fit__reco_pt_rechits_38T_5GeV_GEM_ME11} shows $ \frac{\Delta p_T}{p_T} $ as a function of the simulated $ \eta $ (before discretization) for muon tracks generated with the FastSim in the constant magnetic field with a simulated \pT{} of 5 \GeVc{}, and reconstructed with the Least Squares fit using the RecHits in GE1/1 and ME1/1. The influence of the digitization is clearly visible through the six different sectors, identified by their colors on the left, corresponding to the $ \eta $ segments in GE1/1. The surrounded points show a spread which is smaller than the other $ \eta $ region. Typically, the RMS of the $ \frac{\Delta p_T}{p_T} $ distribution for these points is of 0.05, while it is of the order of 0.2 elsewhere. The surrounded points correspond to muon tracks that hit different $ \eta $ segments in GE1/1a and in GE1/1b. This happens for approximately 6.3 \% of the events, and suggests that using two different $ \eta $ segmentations for the two GE1/1 chambers could improve the resolution on the $ \eta $ coordinate.

		\subsection{Standard Deviation for Different Detector Setups}
		\label{sec:least_squares_fit__constant_magnetic_field_standard_deviation_detectors_setups}					

\begin{figure}[h!]
				\centering
				\includegraphics[width = \textwidth]{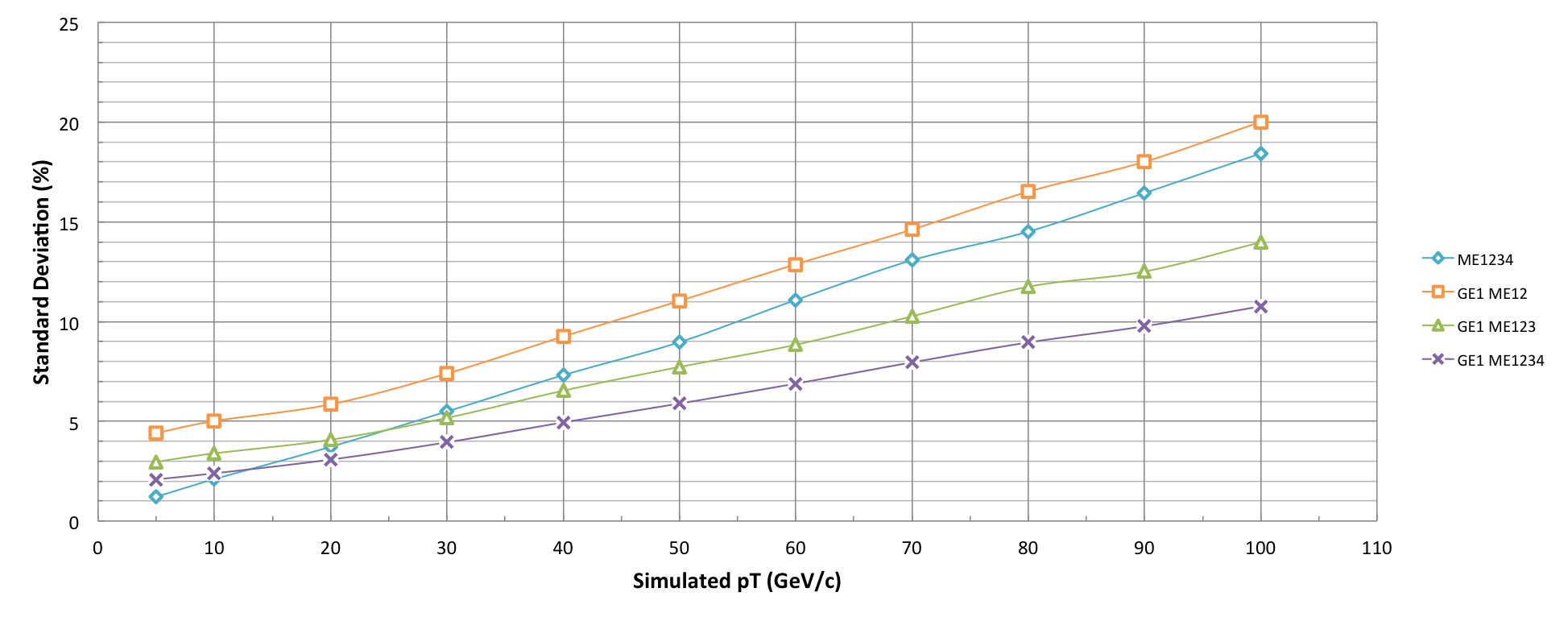}
				\caption{Standard deviation on $ \frac{\Delta p_T}{p_T} $ as a function of the simulated \pT{} for muon tracks generated with the FastSim in the constant magnetic field, and reconstructed with the Least Squares fit using the RecHits in multiple detector setups. The blue curve uses RecHits in ME1/1, ME2/x, ME3/x, and ME4/x, the orange curve in GE1/1, ME1/1, and ME2/x, the green curve in GE1/1, ME1/1, ME2/x, and ME3/x, and the purple curve in GE1/1, ME1/1, ME2/x, ME3/x, and ME4/x.}
				\label{fig:least_squares_fit__reco_sigma_rechits_38T_all_setups}
			\end{figure}

			In order to minimize the influence of the segmentation on the results, we must add other hits from CSC layers (ME2/x, ME3/x, and ME4/x) to reconstruct the muon tracks. We therefore analyze multiple detector setups, and more specifically, their standard deviation. Figure \ref{fig:least_squares_fit__reco_sigma_rechits_38T_all_setups} represents the standard deviation on $ \frac{\Delta p_T}{p_T} $ as a function of the simulated \pT{} for muon tracks generated with the FastSim in the constant magnetic field, and reconstructed with the Least Squares fit using the RecHits in multiple detector setups\footnote{The continuous lines are only there to ease the reading of the graph.}. Except below 10 \GeVc{}, we observe that adding the GEMs to the CSCs (purple compared to blue) improves the \pT{} resolution by more than 30\% for \pT{} $ \geq $ 50 \GeVc{}. We also note that using only the two first CSC stations, ME1/1 and ME2/x, together with GE1/1 (orange) degrades the resolution by less than 20\% at high \pT{} (\pT{} $ \geq $ 50 \GeVc{}) with respect to the CSC-only setup. \pagebreak

			As stated in Section \ref{sec:least_squares_fit__algorithm}, by linearizing the Least Squares fit for a circle, we are not able to account for the errors in $ \eta $ and $ \phi $ separately. This has an impact at lower \pT{} (\pT{} $ \leq $ 15 \GeVc{}) where the resolution on \pT{} of setups using GEM detectors is degraded due to the segmentation. As reviewed in Section \ref{sec:simulation_environment__constant_magnetic_field}, when the SimHits are discretized to yield the RecHits, an error on $ \eta $ is made. In most cases, the RecHits in the two GE1/1 chambers are assigned with the same radial coordinate $ R $, meaning that $ \Delta R $ = 0 between the two GEMs. This means that locally the track is parallel to \axis{Z}. At low \pT{}, tracks have a larger bending radius and therefore significantly differ from the locally straight tracks resulting in a larger error. \\	

\begin{figure}[h!]
				\centering
				\includegraphics[width = \textwidth]{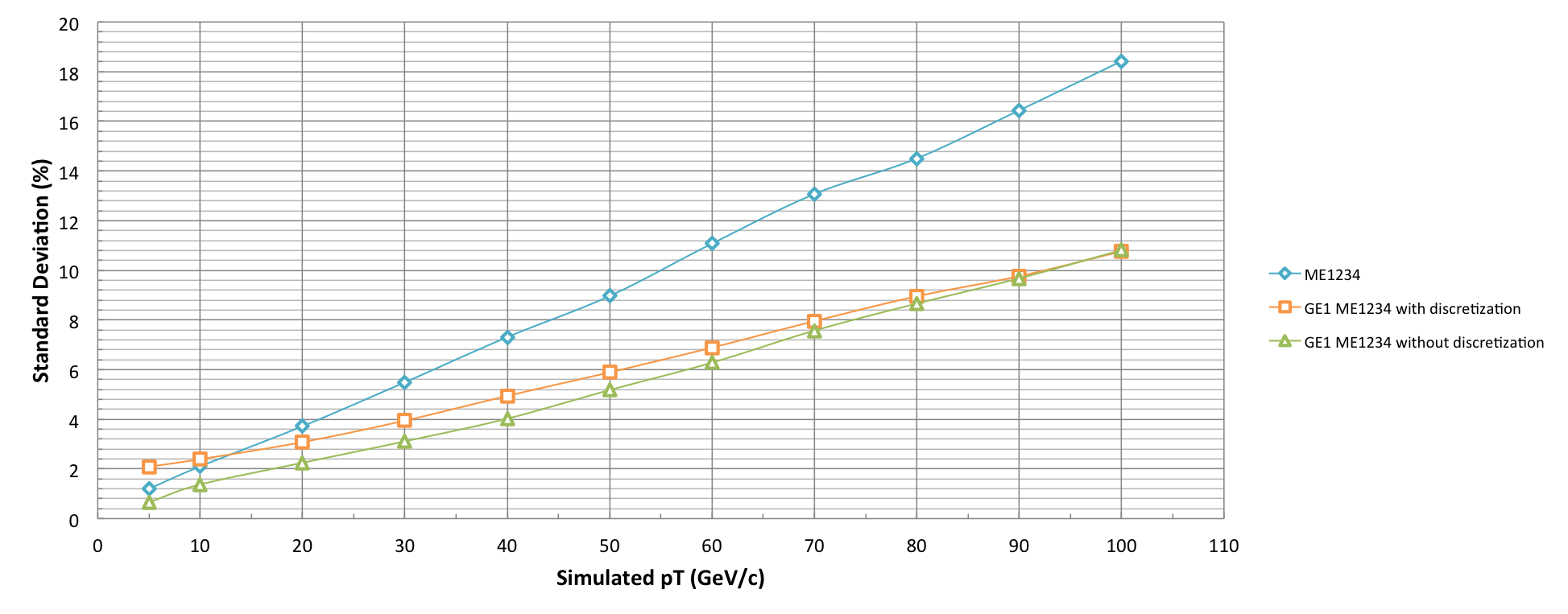}
				\caption{Standard deviation on $ \frac{\Delta p_T}{p_T} $ as a function of the simulated \pT{} for muon tracks generated with the FastSim in the constant magnetic field, and reconstructed with the Least Squares fit using the RecHits with and without discretization in multiple detector setups. The blue curve uses RecHits with discretization in ME1/1, ME2/x, ME3/x, and ME4/x, the purple curve uses RecHits with discretization in GE1/1, ME1/1, ME2/x, ME3/x, and ME4/x, and the orange curve uses RecHits without discretization in GE1/1, ME1/1, ME2/x, ME3/x, and ME4/x.}
				\label{fig:least_squares_fit__sigma_rechits_38T_setups_no_discretization}
			\end{figure}

			We verified the previous statement by testing the algorithm on datasets without $ \eta $ discretization (only noise on the $ \phi $ coordinate). Figure \ref{fig:least_squares_fit__sigma_rechits_38T_setups_no_discretization} depicts the standard deviation on $ \frac{\Delta p_T}{p_T} $ as a function of the simulated \pT{} for muon tracks generated with the FastSim in the constant magnetic field, and reconstructed with the Least Squares fit using the RecHits with and without discretization in multiple detector setups. 

			The obtained plots shows an improvement of 80\% to 14\% on the resolution between 5 \GeVc{} and 50 \GeVc{} when comparing the results with (purple) and without (orange) discretization, while the results remain mainly unchanged at higher \pT{}. \\

			Finally, we also observed, in Figure \ref{fig:least_squares_fit__reco_sigma_rechits_38T_all_setups}, an increase of the standard deviation with \pT{} due to straighter tracks in the transverse plane, as reviewed in Section \ref{sec:simulation_environment__constant_magnetic_field}. This makes it more difficult to estimate the bending radius, hence \pT{}.
		
		\subsection{Bias for Different Detector Setups}
		\label{sec:least_squares_fit__constant_magnetic_field_bias_detectors_setups}						

\begin{figure}[h!]
				\centering
				\includegraphics[width = \textwidth]{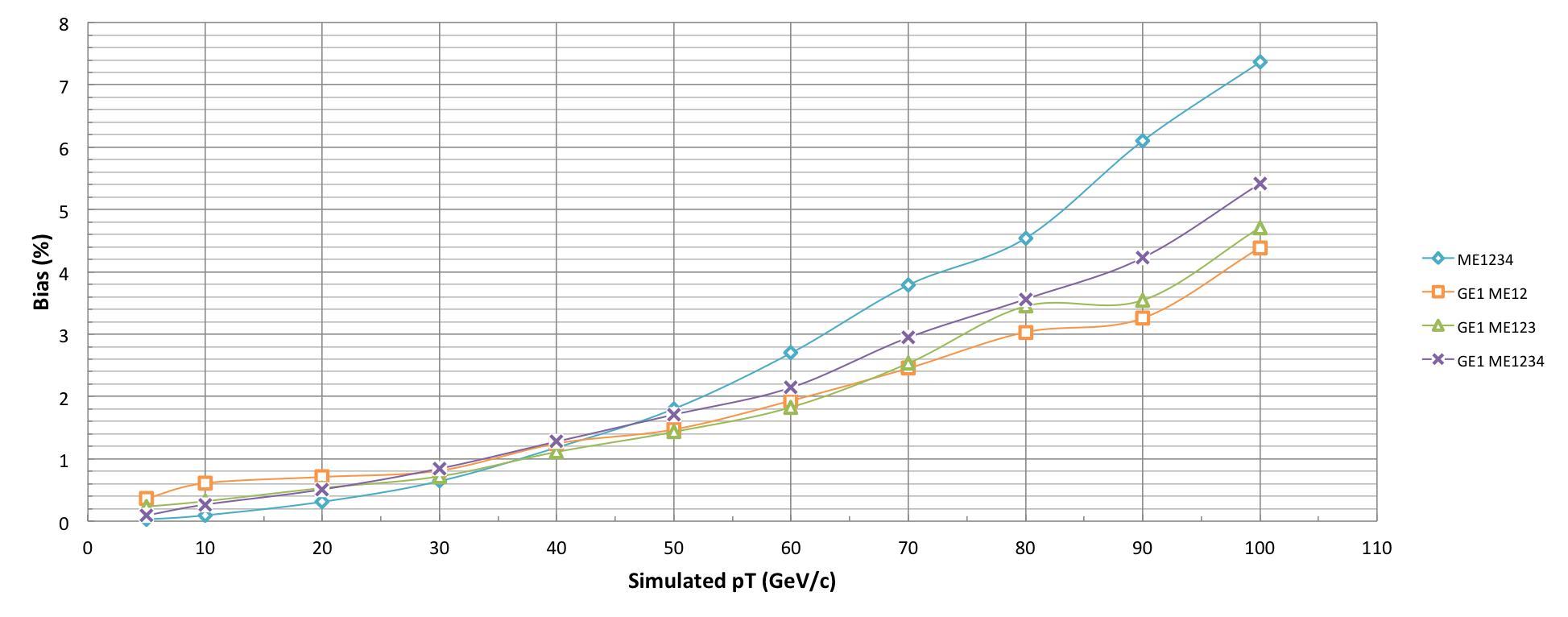}
				\caption{Bias on $ \frac{\Delta p_T}{p_T} $ as a function of the simulated \pT{} for muon tracks generated with the FastSim in the constant magnetic field, and reconstructed with the Least Squares fit using the RecHits in multiple detector setups. The blue curve uses RecHits in ME1/1, ME2/x, ME3/x, and ME4/x, the orange curve in GE1/1, ME1/1, and ME2/x, the green curve in GE1/1, ME1/1, ME2/x, and ME3/x, and the purple curve in GE1/1, ME1/1, ME2/x, ME3/x, and ME4/x.}
				\label{fig:least_squares_fit__mu_rechits_38T_all_setups}
			\end{figure}

			We now consider Figure \ref{fig:least_squares_fit__mu_rechits_38T_all_setups} which represents the bias on $ \frac{\Delta p_T}{p_T} $ as a function of the simulated \pT{} for muon tracks generated with the FastSim in the constant magnetic field, and reconstructed with the Least Squares fit using the RecHits in multiple detector setups. Below 40 \GeVc, all detector setups yield a bias smaller than 1\%. Above 40 \GeVc, we observe that GEM detectors reduce the bias by up to 40\% at 100 \GeVc.

		\subsection{Evolution with $ \eta $}
		\label{sec:least_squares_fit__constant_magnetic_field_evolution_eta}		

\begin{figure}[h!]
				\centering
				\includegraphics[width = \textwidth]{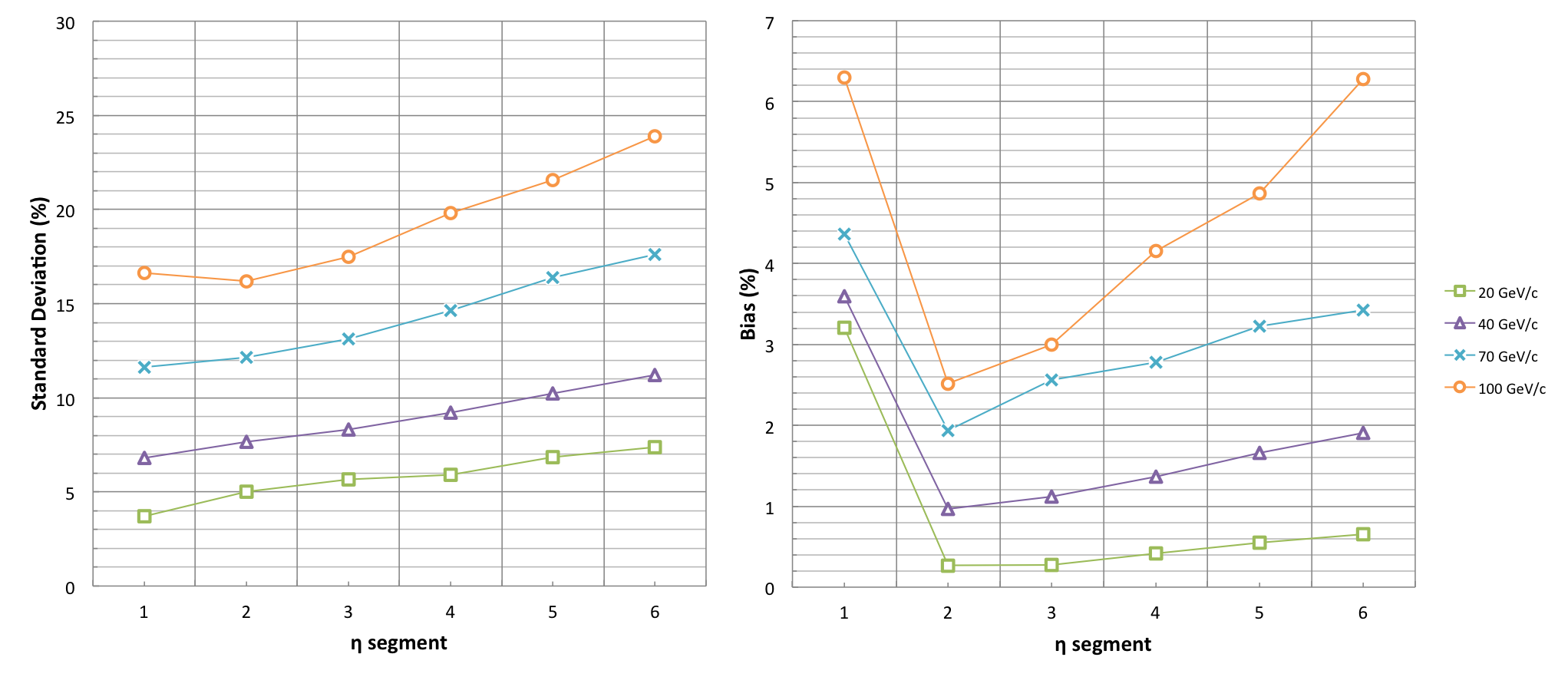}
				\caption{Standard deviation (left) and bias (right) on $ \frac{\Delta p_T}{p_T} $ as a function of the hit $ \eta $ segment in GE1/1 for muon tracks generated with the FastSim in the constant magnetic field with simulated \pT{} of 20, 40, 70, and 100 \GeVc{}, and reconstructed with the Least Squares fit using the RecHits in GE1/1, ME1/1, and ME2/x.}
				\label{fig:least_squares_fit__sigma_mu_rechits_38T_all_setups_evolution_eta}
			\end{figure}

			Finally, we look in Figure \ref{fig:least_squares_fit__sigma_mu_rechits_38T_all_setups_evolution_eta} at the evolution of the standard deviation (left) and bias (right) on $ \frac{\Delta p_T}{p_T} $ as a function of the hit $ \eta $ segment in GE1/1 for muon tracks generated with the FastSim in the constant magnetic field with simulated \pT{} of 20, 40, 70, and 100 \GeVc{}, and reconstructed with the Least Squares fit using the RecHits in GE1/1, ME1/1, and ME2/x. Both parameters improve when considering segments with lower $ \eta $. \\

			The large bias observed in the right plot for segment 1 is due to the lack of hits in ME2/x for 1.6 < $ | \eta | $ < 1.63, resulting in the reconstruction not being performed. Indeed, Figure \ref{fig:muon_chambers__placement} in Section \ref{sec:muon_chambers__disposition_of_the_detectors} shows that this region corresponds to the transition between ME2/1 and ME2/2. As the rings do not overlap, some events are rejected due to missing hits. We observed in Figure \ref{fig:least_squares_fit__reco_pt_rechits_38T_5GeV_GEM_ME11} that the bias increases in each segment at higher $ \eta $. As part of the hits at lower $ \eta $, hence lower bias, are rejected, the average bias for the segment 1 is overestimated. \\

			Furthermore, to understand the performances' deterioration at higher $ \eta $, we refer to Section \ref{sec:simulation_environment__constant_magnetic_field}. Particles with an initial high $ \eta $ have a greater \pZ{} thus describe smaller tracks in the transverse plane. Therefore, the fitted circle arc is smaller and more sensitive to errors, as the lever arm between measurements decreases.
		
		\subsection{Summary}
		\label{sec:least_squares_fit__constant_magnetic_field_summary}		

			From the study of the SimHits generated with the FastSim in the constant magnetic field, we conclude that our algorithm successfully reconstructs tracks due to the small standard deviation and bias, both of the order of 10$ ^{-7} $ to 10$ ^{-8} $\%. \\

			In general, when considering the RecHits, the resolution on the muons \pT{} improves when we add the hits from GE1/1 to the ones from the CSCs. An improvement of more than 30\% for \pT{} $ \geq $ 50 \GeVc{} is observed when comparing the CSC-only setup and the GEMs-plus-CSCs setup. The bias also improves when we add the GEM detectors. However, at low \pT{}, less than 10 \GeVc{}, the $ \eta $ segmentation of the GEMs degrades the precision. This can be explained by the fact that the linearized Least Squares fit does not allow to weight independently the $ \eta $ and $ \phi $ measurements from the GEM detectors. \\

			Moreover, as expected, we notice a deterioration of the performances at higher $ \eta $, hence when looking at segments closer to the beam pipe, due to the shorter tracks left in the transverse plane. 
	
	\section{FastSim Results: Real Magnetic Field}
	\label{sec:least_squares_fit__fastsim_results_real_magnetic_field}

		After analyzing the results in the constant magnetic field, we look at the FastSim with the real field of CMS to see its influence on the results.

		\subsection{Least Squares Fit with FastSim SimHits in the Real Magnetic Field}
		\label{sec:least_squares_fit__real_magnetic_field_validation_using_simhits}	

\begin{figure}[h!]
				\centering
				\includegraphics[width = 0.7 \textwidth]{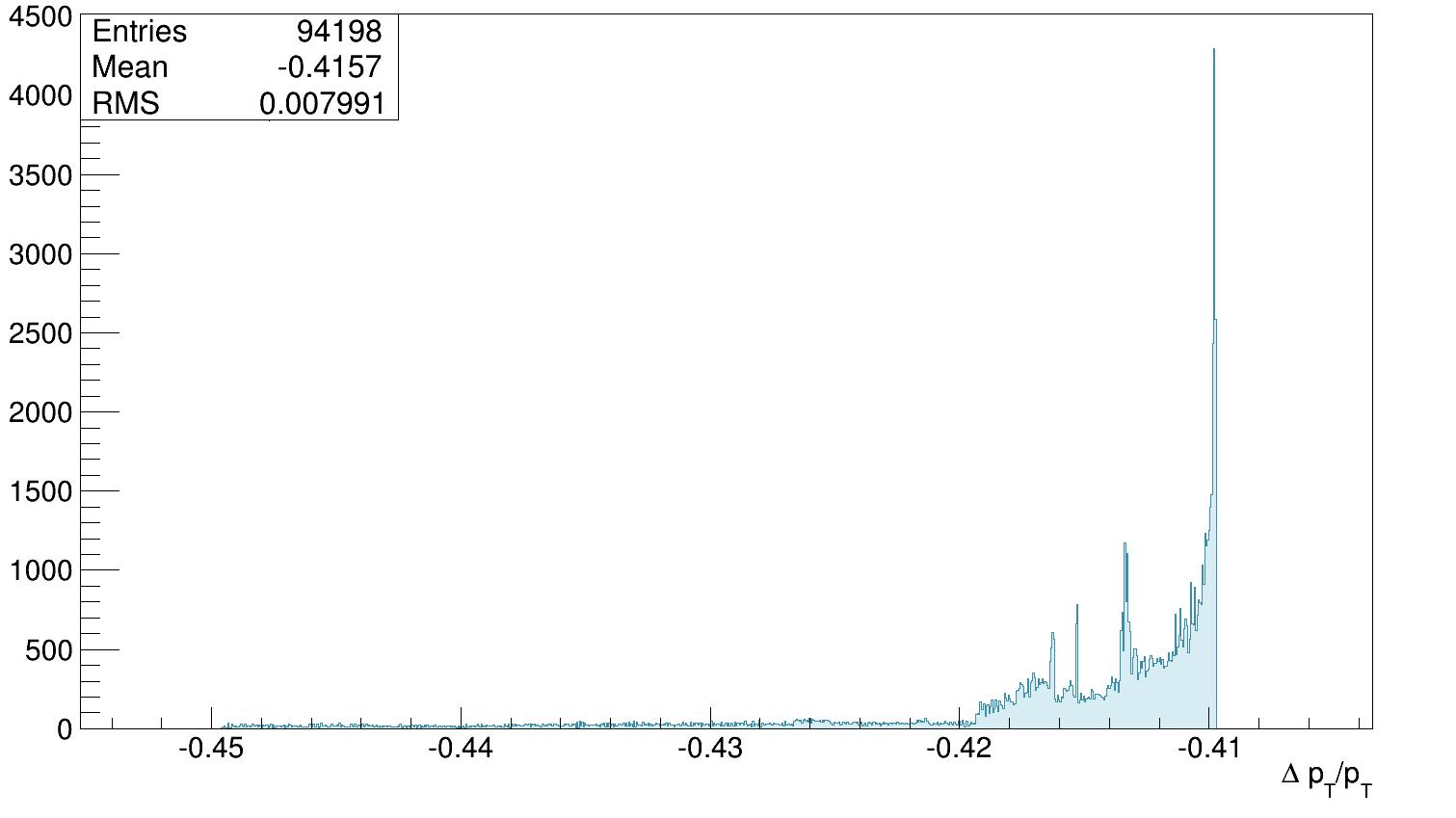}
				\caption{$ \frac{\Delta p_T}{p_T} $ for muon tracks generated with the FastSim in the real magnetic field with a simulated \pT{} of 10 \GeVc{}, and reconstructed with the Least Squares fit using the SimHits in GE1/1, ME1/1, and ME2/x.}
				\label{fig:least_squares_fit__reco_pt_simhits_RT_10GeV_GEM_ME12}
			\end{figure}

			First, we look at the results using the SimHits. Figure \ref{fig:least_squares_fit__reco_pt_simhits_RT_10GeV_GEM_ME12} depicts $ \frac{\Delta p_T}{p_T} $ for muon tracks generated with the FastSim in the real magnetic field with a simulated \pT{} of 10 \GeVc{}, and reconstructed with the Least Squares fit using the SimHits in GE1/1, ME1/1, and ME2/x. We emphasize that the RMS of the distribution is small (0.008) and that the abscissa has been enlarged in order to distinguish the variations of the distribution. A bias of the order of -40\% is observed on the reconstructed \pT{} due to the non-uniformity of the magnetic field. The deviation being negative, muons of lower \pT{} are reconstructed like particles of higher \pT{}. Correcting this effect is important, as this is one of the issues we want to address for the L1 Trigger, as reviewed in Section \ref{sec:trigger_system_and_reconstruction_algorithms__system_performances}.  \\

\begin{figure}[h!]
				\centering
				\includegraphics[width = 0.7 \textwidth]{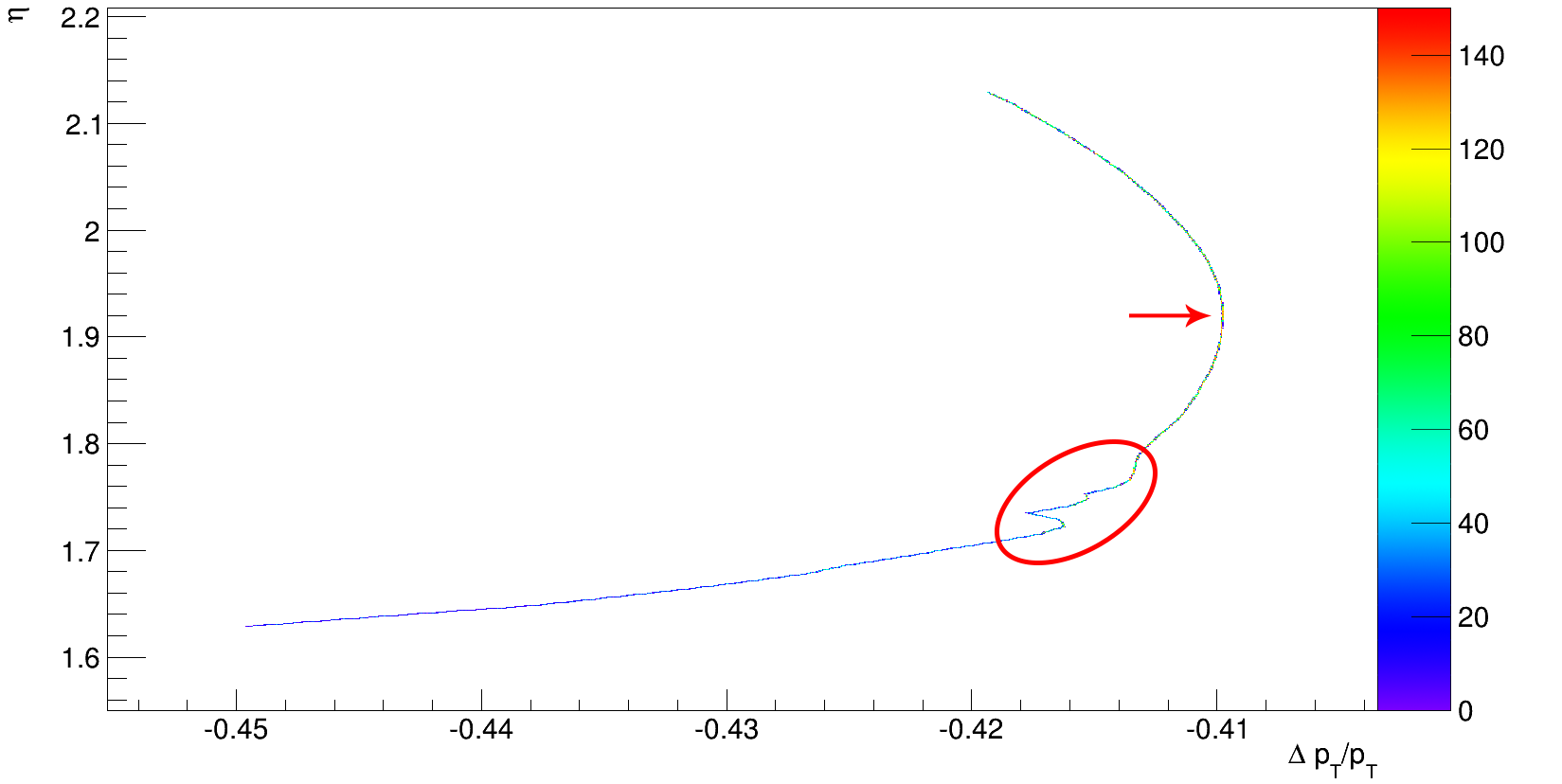}
				\caption{$ \frac{\Delta p_T}{p_T} $ as a function of the simulated $ \eta $ for muon tracks generated with the FastSim in the real magnetic field with a simulated \pT{} of 10 \GeVc{}, and reconstructed with the Least Squares fit using the SimHits in GE1/1, ME1/1, and ME2/x.}
				\label{fig:least_squares_fit__reco_pt_simhits_RT_10GeV_GEM_ME12_vs_eta}
			\end{figure}

\begin{figure}[h!]
				\centering
				\includegraphics[width = 0.7 \textwidth]{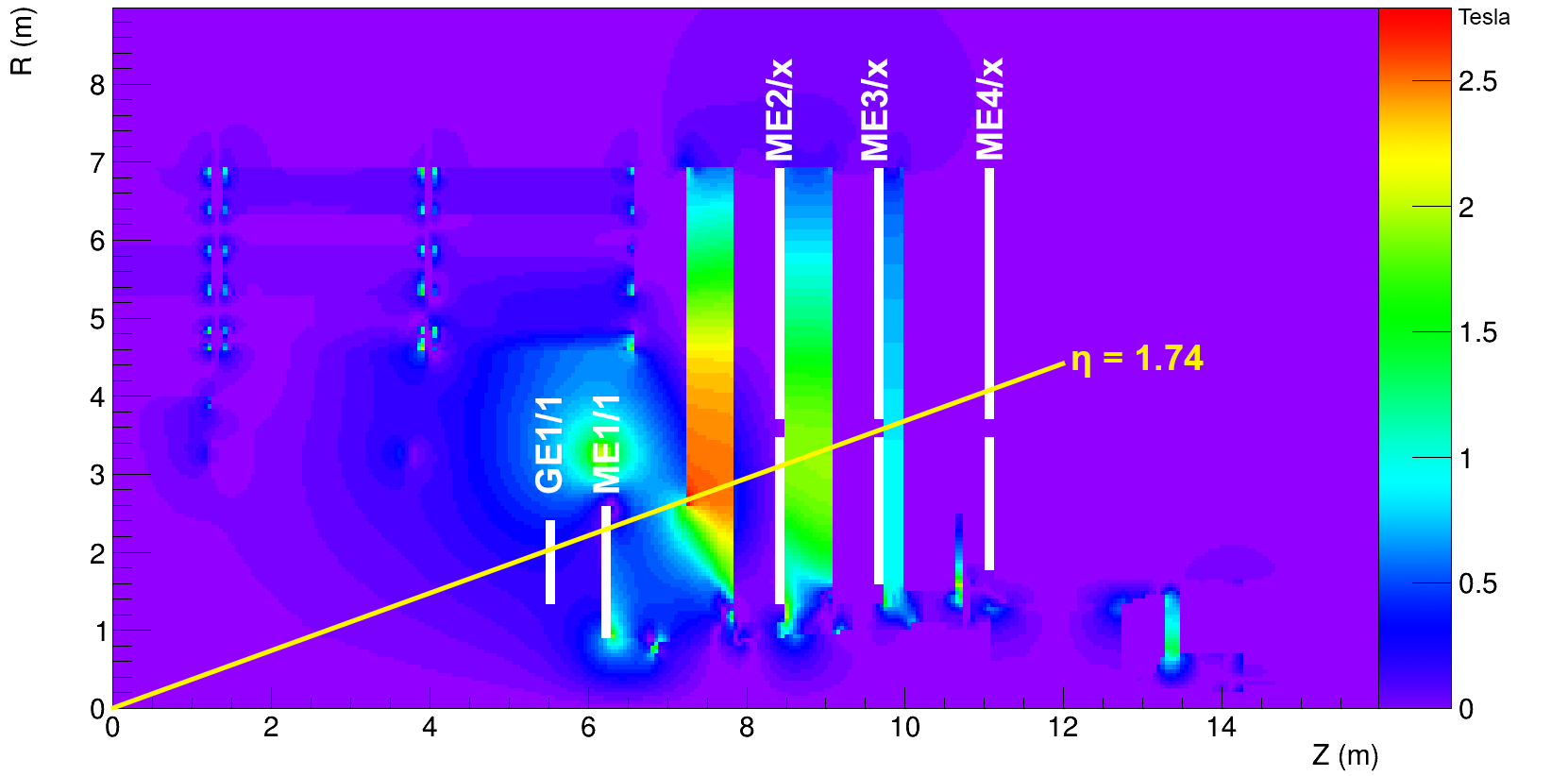}
				\includegraphics[width = 0.7 \textwidth]{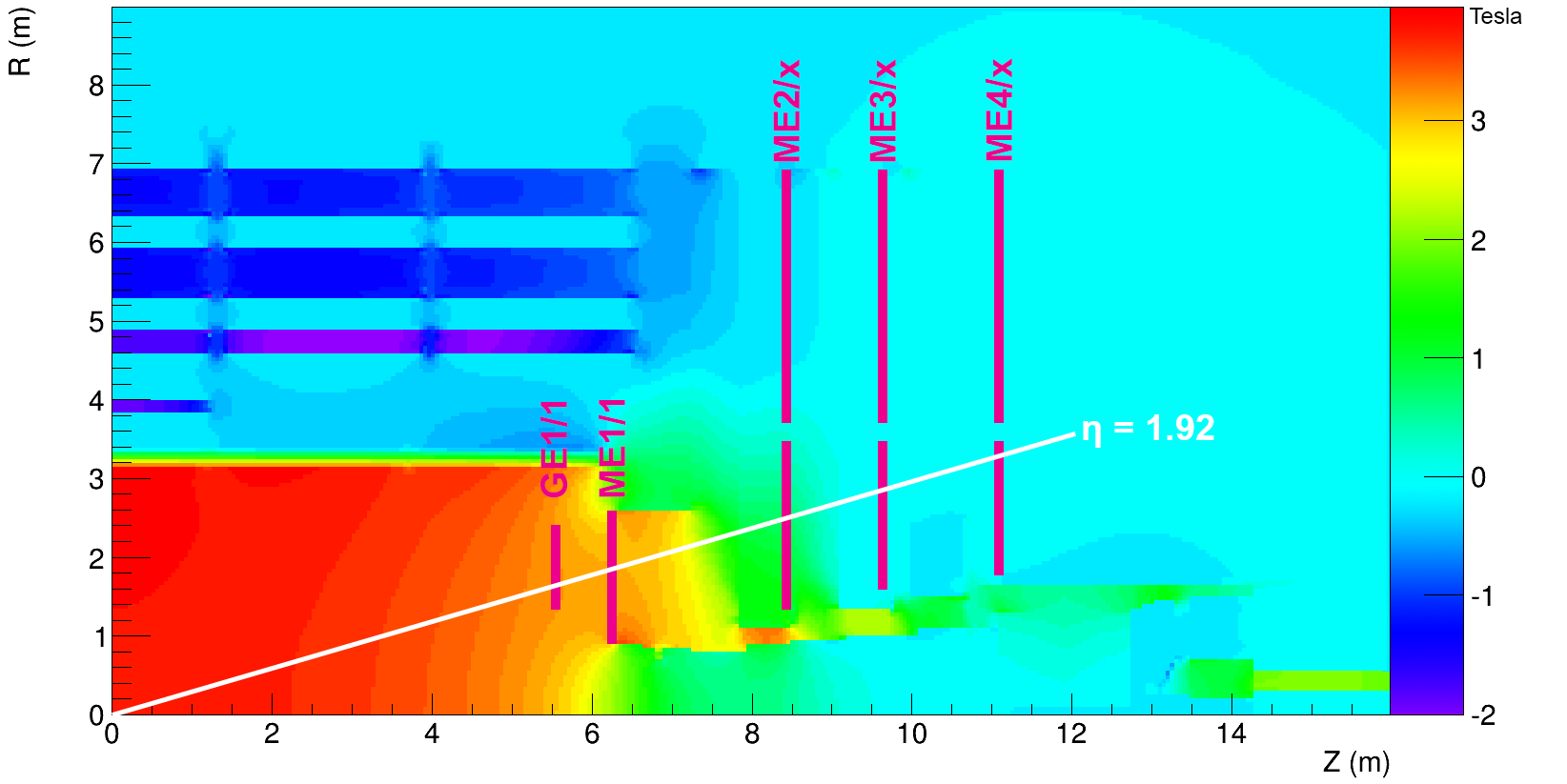}
				\caption{Intensity of the transverse (top) and longitudinal (bottom) component of the CMS magnetic field.}
				\label{fig:least_squares_fit__RT_magnetic_field}
			\end{figure}

			To fully understand the impact of the magnetic field on the bias, Figure \ref{fig:least_squares_fit__reco_pt_simhits_RT_10GeV_GEM_ME12_vs_eta} represents $ \frac{\Delta p_T}{p_T} $ as a function of the simulated $ \eta $ for muon tracks generated with the FastSim in the real magnetic field with a simulated \pT{} of 10 \GeVc{}, and reconstructed with the Least Squares fit using the SimHits in GE1/1, ME1/1, and ME2/x. We observe two anomalies: a strong fluctuation around $ \eta $ = 1.74 (surrounded), and a change in direction at $ \eta $ = 1.92 (arrow). Both are due to strong variations in the intensity of the magnetic field. Figure \ref{fig:least_squares_fit__RT_magnetic_field} shows the intensity of the transverse (top) and longitudinal (bottom) component of the CMS magnetic field. \\

			The first anomaly can be explained using the top image in Figure \ref{fig:least_squares_fit__RT_magnetic_field} which illustrates the intensity of the transverse component of the CMS magnetic field. We observe a large variation of the magnetic field in the iron yoke between ME1/1 and ME2/x, with a peak around $ \eta $ = 1.74, as indicated by the yellow line. The presence of a radial component in the field disturbs the particles' trajectories and result in variations of the reconstructed \pT{}. \\

\begin{figure}[h!]
				\centering
				\includegraphics[width = \textwidth]{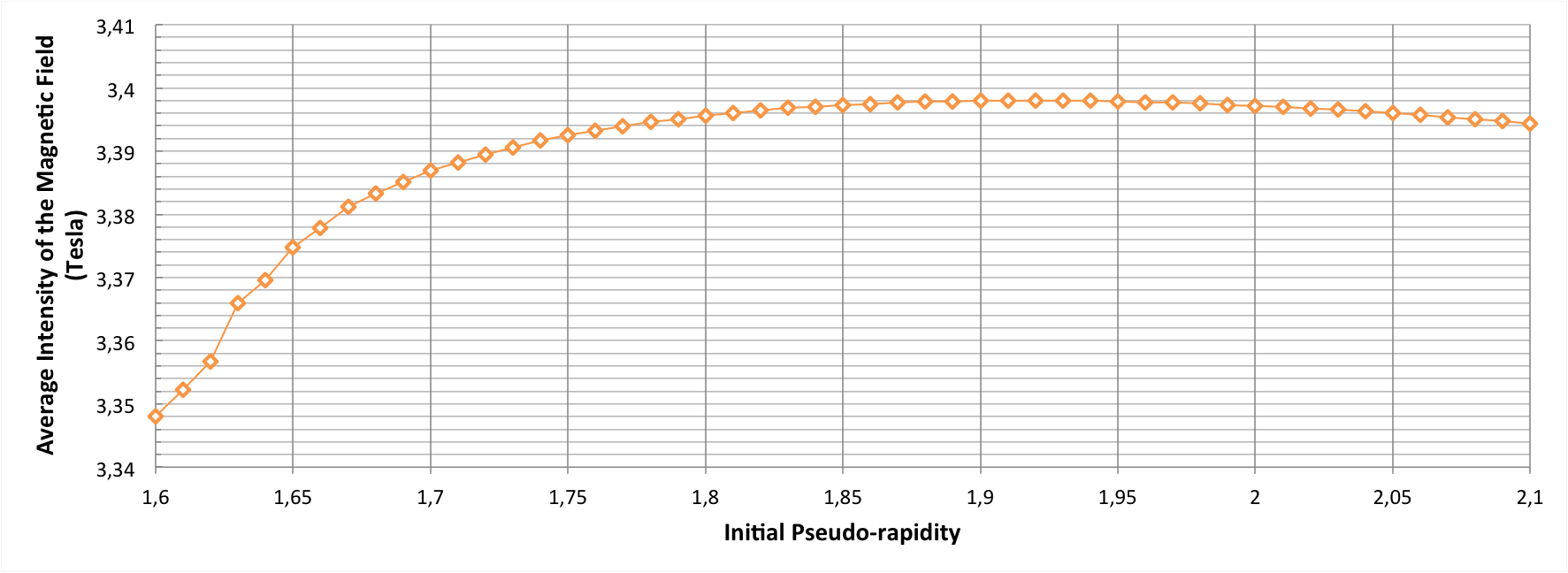}
				\caption{Average intensity of the magnetic field particles travel through as a function of the simulated $ \eta $ for muon tracks generated with the FastSim in the real magnetic field with a simulated \pT{} of 10 \GeVc.}
				\label{fig:least_squares_fit__RT_average_magnetic_field}
			\end{figure}

			The second anomaly has to be explained using the bottom image in Figure \ref{fig:least_squares_fit__RT_magnetic_field} which depicts the intensity of the longitudinal component of the CMS magnetic field. When muons pass through a region where the magnetic field is less intense along \axis{Z}, their bending radius increases confusing the algorithm into assigning them with a higher \pT{}. More specifically, the reconstruction of \pT{} with the Least Squares fit, assuming a constant intensity of 3.8 T, overestimates the particle's \pT{} as if the greater bending radius was due to a higher \pT{} and not a smaller intensity of the magnetic field. To understand the feature arising at $ \eta $ = 1.92, we have computed the average intensity of the magnetic field as a function of $ \eta $ at which the muon is emitted from the IP. This is shown in Figure \ref{fig:least_squares_fit__RT_average_magnetic_field}. We observe a steep slope between 1.6 < $ | \eta | $ < 1.7 which translates in Figure \ref{fig:least_squares_fit__reco_pt_simhits_RT_10GeV_GEM_ME12_vs_eta} in a quickly decreasing bias. Between $ \eta $ = 1.6 and $ \eta $ = 1.92, the average intensity of the magnetic field increases, meaning the overall bending radius is closer to the one generated in the constant magnetic field. The bias will therefore be reduced. Between $ \eta $ = 1.92 and $ \eta $ = 2.1, the opposite effect arises as the average intensity decreases. \\

			We cannot perform a study on the standard deviations of the SimHits generated with the FastSim in the real magnetic field due to the non-gaussian shape of the $ \frac{\Delta p_T}{p_T} $ distributions, as represented in Figure \ref{fig:least_squares_fit__reco_pt_simhits_RT_10GeV_GEM_ME12}. The shape of this distribution is the consequence of the non-uniform magnetic field and results from the anomalies observed in Figure \ref{fig:least_squares_fit__reco_pt_simhits_RT_10GeV_GEM_ME12_vs_eta}. However, the obtained RMS is of 8 10$ ^{-3} $, meaning the spread of these results is small. \\

			Due to the error that is applied on the SimHits to yield the RecHits, the distributions of $ \frac{\Delta p_T}{p_T} $ become gaussian but with a much larger spread, allowing us to further analyze the results.
		
		\subsection{Impact of Segmentation}
		\label{sec:least_squares_fit__real_magnetic_field_impact_of_segmentation}	

			The impact of segmentation on the results is the same as those reviewed in Section \ref{sec:least_squares_fit__constant_magnetic_field_impact_of_segmentation}	for the constant magnetic field. The GE1/1 and ME1/1 setup is not sufficient to correctly reconstruct tracks as the larger errors on $ \eta $ of the GEM detectors significantly biases the results.

		\subsection{Standard Deviation for Different Detector Setups}
		\label{sec:least_squares_fit__real_magnetic_field_standard_deviation_detectors_setups}		

\begin{figure}[h!]
				\centering
				\includegraphics[width = \textwidth]{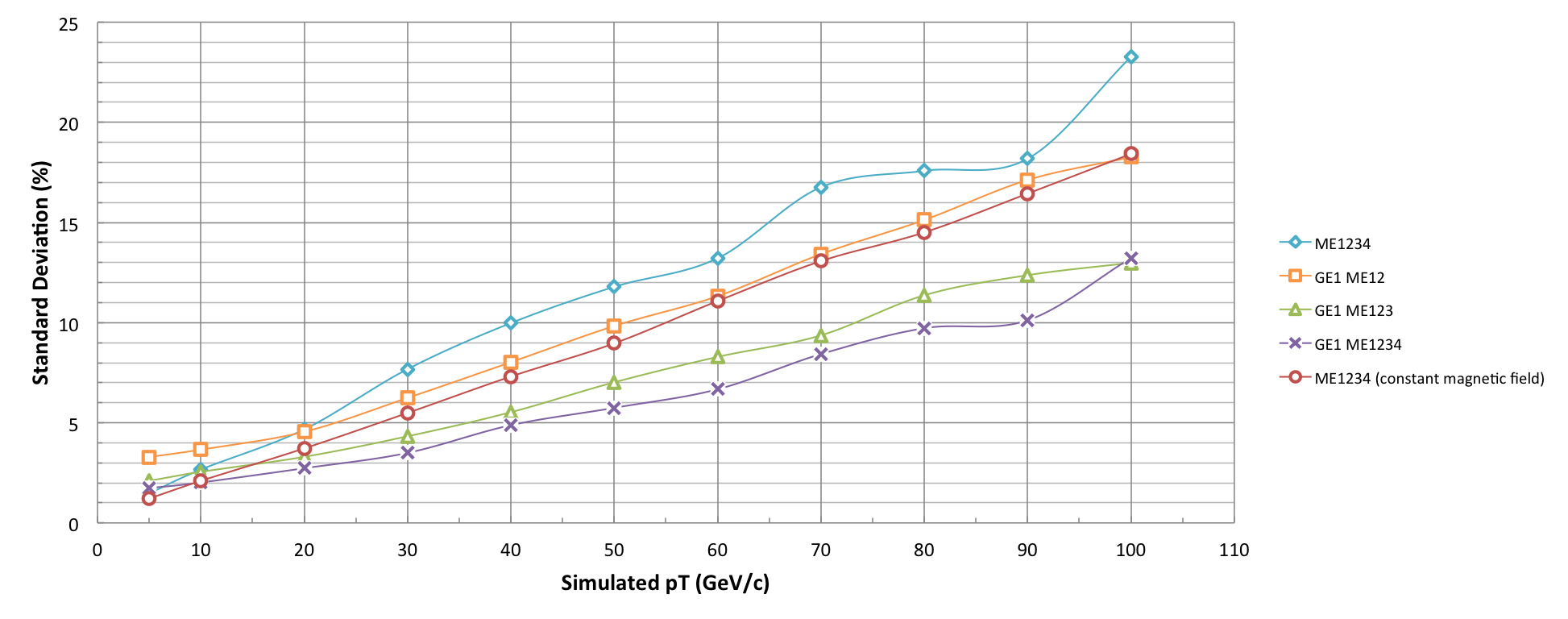}
				\caption{Standard deviation on $ \frac{\Delta p_T}{p_T} $ as a function of the simulated \pT{} for muon tracks generated with the FastSim in the real magnetic field, and reconstructed with the Least Squares fit using the RecHits in multiple detector setups. The blue curve uses RecHits in ME1/1, ME2/x, ME3/x, and ME4/x, the orange curve in GE1/1, ME1/1, and ME2/x, the green curve in GE1/1, ME1/1, ME2/x, and ME3/x, the purple curve in GE1/1, ME1/1, ME2/x, ME3/x, and ME4/x, and the red curve in ME1/1, ME2/x, ME3/x, ME4/x generated with the FastSim in the constant magnetic field.}
				\label{fig:least_squares_fit__reco_sigma_rechits_RT_all_setups}
			\end{figure}

			To use the Least Squares fit in the real magnetic field, we need to add more CSC layers. Therefore, we consider the standard deviation of different setups. Figure \ref{fig:least_squares_fit__reco_sigma_rechits_RT_all_setups} shows the standard deviation on $ \frac{\Delta p_T}{p_T} $ as a function of the simulated \pT{} for muon tracks generated with the FastSim in the real magnetic field, and reconstructed with the Least Squares fit using the RecHits in multiple detector setups. As for the constant magnetic field, the GEMs' $ \eta $ segmentation deteriorates the resolution at low \pT{} while having little to no effect at high \pT{}. However, considering RecHits in GE1/1, ME1/1, and ME2/x setup (orange) offers a better resolution at higher \pT{} (20\% at 100 \GeVc{}) than the ME1/1, ME2/x, ME3/x, and ME4/x setup (blue). This is due to the fact that GEMs are placed in a region of CMS where the magnetic field is still uniform while CSCs, especially ME3/x, and ME4/x, lie in regions where the field is more complex. Moreover, the GE1/1, ME1/1, ME2/x, ME3/x, and ME4/x setup (purple) improves the CSC-only setup (blue) by 40 to 55\% above 20 \GeVc{}. \\

			To illustrate this, we refer to Figures \ref{fig:least_squares_fit__RT_magnetic_field} for a visualization of the magnetic field of CMS. These emphasize that considering hits only occurring in CSCs is more difficult. Particles may significantly be deviated by the transverse component of the field, whose intensity is stronger in the yokes between the CSC layers. This is a strong argument in favor of the installation of GEMs which would be placed near ME1/1, where the field is still uniform. \\

			To compare the results with the previous simulations, we added the results obtained with the RecHits in ME1/1, ME2/x, ME3/x, and ME4/x generated with the FastSim in the constant magnetic field (red). We observe a deterioration of the resolution (between 30 and 20\%) which is expected due to the non-uniform magnetic field.
		
		\subsection{Bias for Different Detector Setups}
		\label{sec:least_squares_fit__real_magnetic_field_bias_detectors_setups}	

\begin{figure}[h!]
				\centering
				\includegraphics[width = \textwidth]{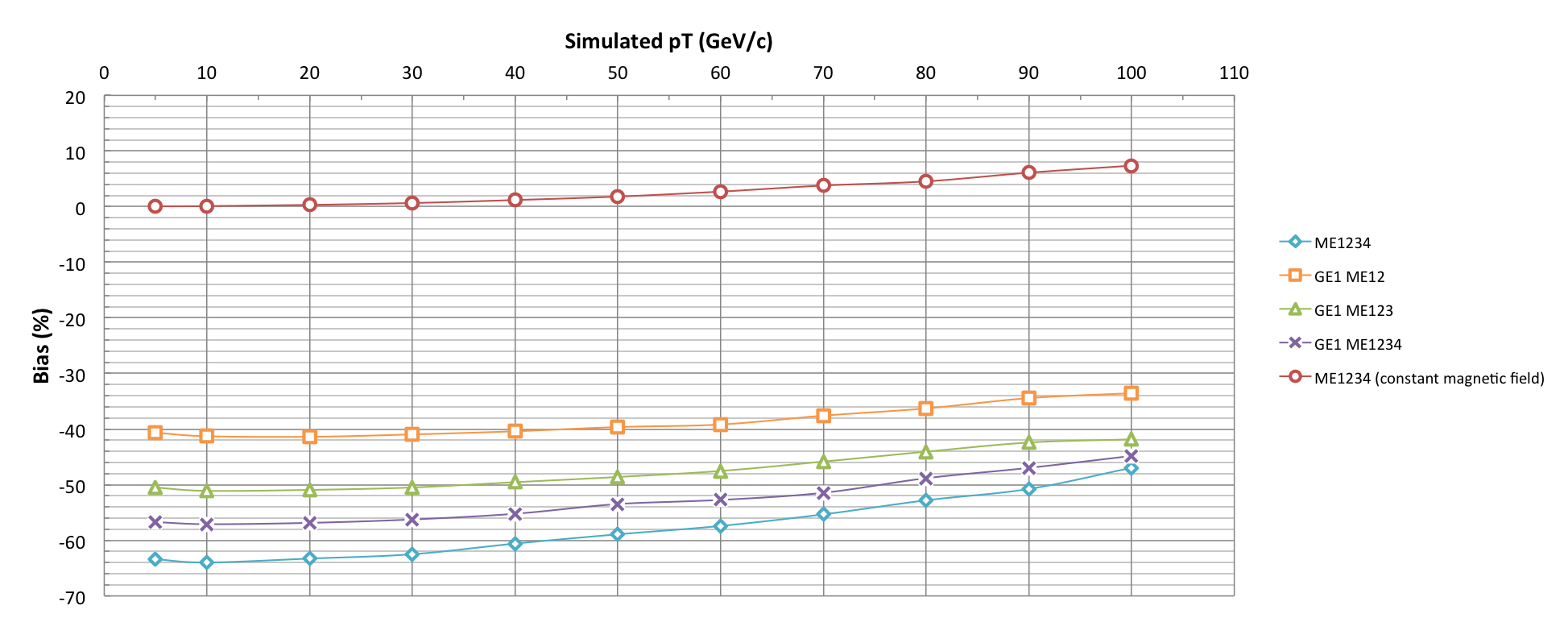}
				\caption{Bias on $ \frac{\Delta p_T}{p_T} $ as a function of the simulated \pT{} for muon tracks generated with the FastSim in the real magnetic field, and reconstructed with the Least Squares fit using the RecHits in multiple detector setups. The blue curve uses RecHits in ME1/1, ME2/x, ME3/x, and ME4/x, the orange curve in GE1/1, ME1/1, and ME2/x, the green curve in GE1/1, ME1/1, ME2/x, and ME3/x, the purple curve in GE1/1, ME1/1, ME2/x, ME3/x, and ME4/x, and the red curve in ME1/1, ME2/x, ME3/x, ME4/x generated with the FastSim in the constant magnetic field.}
				\label{fig:least_squares_fit__mu_rechits_RT_all_setups}
			\end{figure}

			For the same set of detector setups, we look at the impact of the real magnetic field on the bias. Figure \ref{fig:least_squares_fit__mu_rechits_RT_all_setups} shows the bias on $ \frac{\Delta p_T}{p_T} $ as a function of the simulated \pT{} for muon tracks generated with the FastSim in the real magnetic field, and reconstructed with the Least Squares fit using the RecHits in multiple detector setups. Adding GEM detectors to the CSCs-only setup improves the results by an average of 9\%. The less CSC layers we consider, the more the bias diminishes. We observe 18\% improvement between the GE1/1, ME1/1, and ME2/x setup (orange) and the GE1/1, ME1/1, ME2/x, and ME3/x setup (green), and 25\% improvement compared to GE1/1, ME1/1, ME2/x, ME3/x, and ME4/x (purple). Finally, if we compare the GE1/1, ME1/1, and ME2/x setup (orange) to the CSCs is standalone (blue), an average improvement of 33\% is obtained.

\begin{figure}[h!]
				\centering
				\includegraphics[width = 0.5 \textwidth]{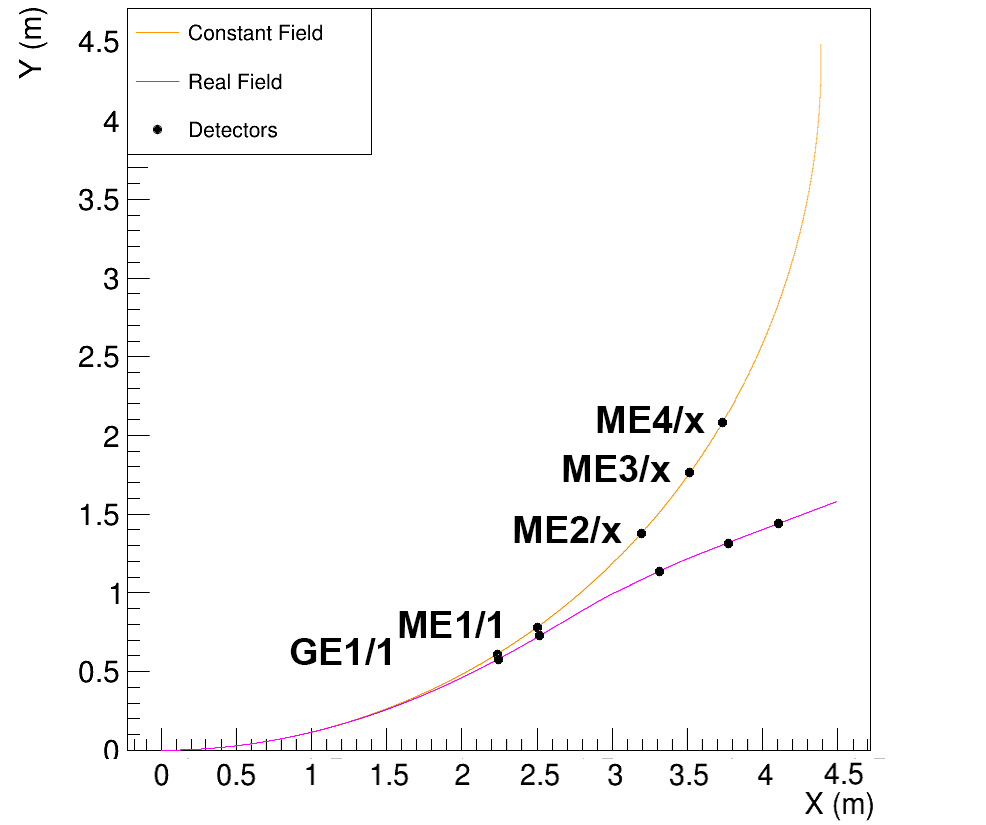}
				\caption{Projection in the transverse plane of muon tracks generated with the FastSin in the constant and real magnetic field with a simulated \pT{} of 5 \GeVc{} and an initial $ \eta $ of 1.6, with the placement of the detectors on both trajectories.}
				\label{fig:least_squares_fit__xy_projection_RT_vs_38T}
			\end{figure}

			To explain this, we look at Figure \ref{fig:least_squares_fit__xy_projection_RT_vs_38T} that represents the projection in the transverse plane of muon tracks generated with the FastSim in the constant and real magnetic field with a simulated \pT{} of 5 \GeVc{} and an initial $ \eta $ of 1.6, with the placement of the detectors on both trajectories. We note that the track at the GE1/1 and ME1/1 locations is still close to the ideal track, while at the ME2/x, ME3/x, and ME4/x locations the curvature varies. If we start by looking at the GE1/1, ME1/1, and ME2/x setup (orange), we have a bias of 40\%. When we add the RecHit in ME3/x, the fitted circle will have a greater bending radius and be reconstructed with a higher \pT{}, yielding a deteriorating bias. The more CSC layers we add, the further away we pull the track from its ideal trajectory.

\begin{figure}[h!]
				\centering
				\includegraphics[width = 0.5 \textwidth]{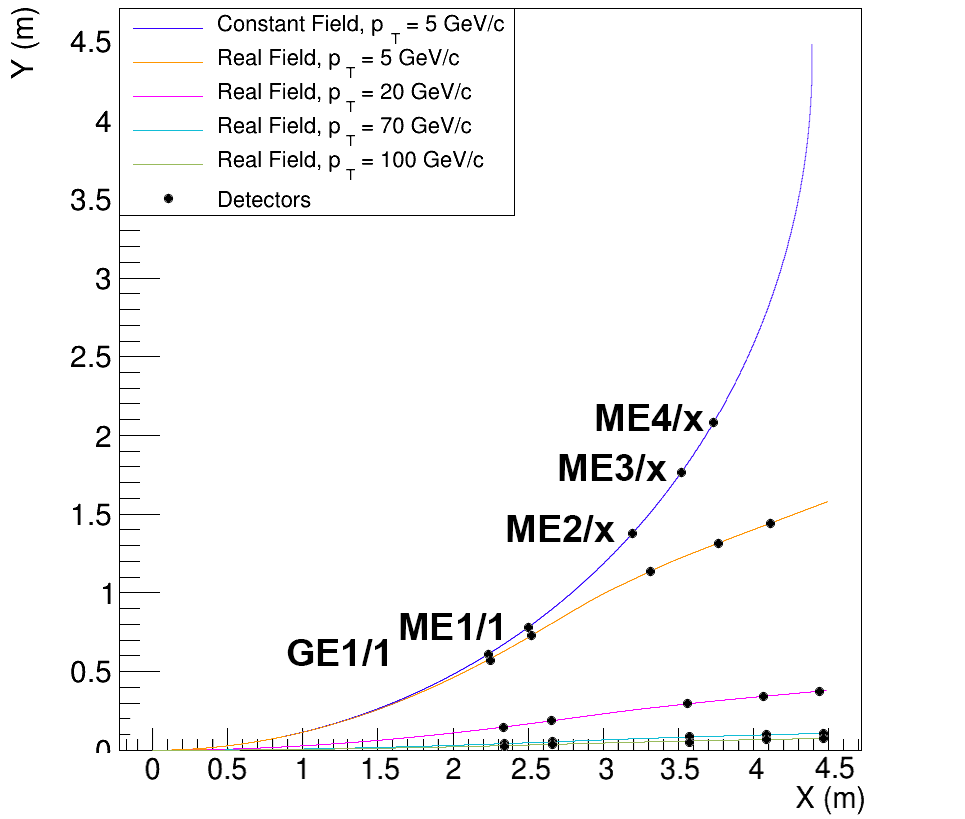}
				\caption{Projection in the transverse plane of muon tracks generated with the FastSim in the constant and real magnetic field with simulated \pT{} of 5, 20, 70, and 100 \GeVc{}, and an initial $ \eta $ of 1.6 with the placement of the detectors on the trajectories.}
				\label{fig:least_squares_fit__xy_projection_RT_vs_pt}
			\end{figure}

			Paradoxically, the bias improves at higher \pT{}. This can be understood from Figure \ref{fig:least_squares_fit__xy_projection_RT_vs_pt} which depicts the projection in the transverse plane of muon tracks generated with the FastSim in the constant and real magnetic field with simulated \pT{} of 5, 20, 70, and 100 \GeVc{}, and an initial $ \eta $ of 1.6 with the placement of the detectors on the trajectories. At higher \pT{} (green), muon tracks are straighter and therefore less sensible to the magnetic field's non-uniformity.

		\subsection{Evolution with $ \eta $}
		\label{sec:least_squares_fit__real_magnetic_field_evolution_eta}		

\begin{figure}[h!]
				\centering
				\includegraphics[width = \textwidth]{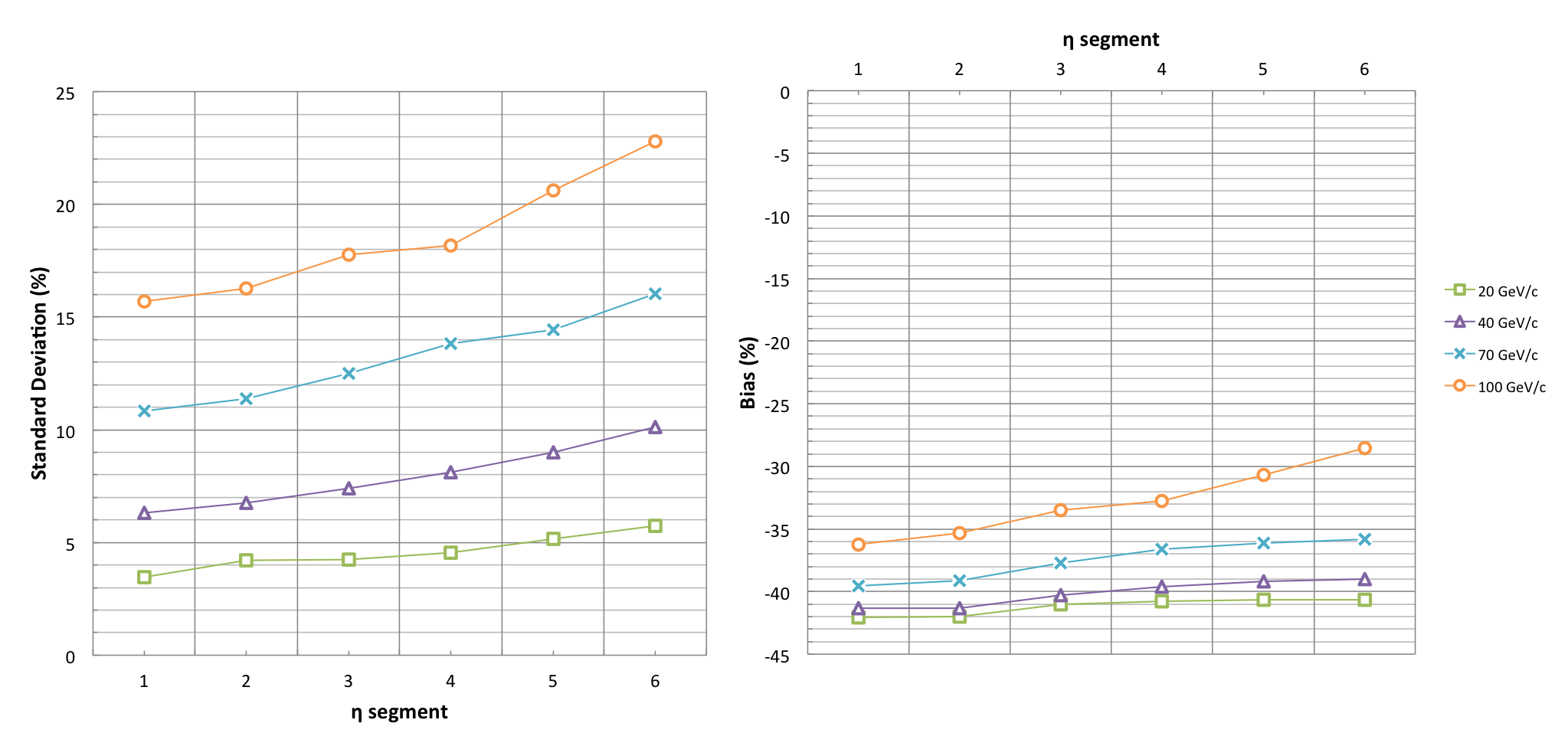}
				\caption{Standard deviation (left) and bias (right) on $ \frac{\Delta p_T}{p_T} $ as a function of the hit $ \eta $ segment in GE1/1 for muon tracks generated with the FastSim in the real magnetic field with simulated \pT{} of 20, 40, 70, and 100 \GeVc{}, and reconstructed with the Least Squares fit using the RecHits in GE1/1, ME1/1, and ME2/x.}
				\label{fig:least_squares_fit__sigma_mu_rechits_RT_all_setups_evolution_eta}
			\end{figure}

			Finally, we analyze the evolution of the standard deviation and the bias with $ \eta $. Figure \ref{fig:least_squares_fit__sigma_mu_rechits_RT_all_setups_evolution_eta} shows the standard deviation (left) and bias (right) on $ \frac{\Delta p_T}{p_T} $ as a function of the hit $ \eta $ segment in GE1/1 for muon tracks generated with the FastSim in the real magnetic field with simulated \pT{} of 20, 40, 70, and 100 \GeVc{}, and reconstructed with the Least Squares fit using the RecHits in GE1/1, ME1/1, and ME2/x. We observe the degradation of the standard deviation with $ \eta $ while the bias is improved. \\

			The observed degradation in the standard deviation has the same origin as with the constant magnetic field, as reviewed in Section \ref{sec:least_squares_fit__constant_magnetic_field_evolution_eta}. At high $ \eta $, the projected tracks in the transverse plane are shorter and thus leave more room for error when fitting them. \\

\begin{figure}[h!]
				\centering
				\includegraphics[width = 0.5 \textwidth]{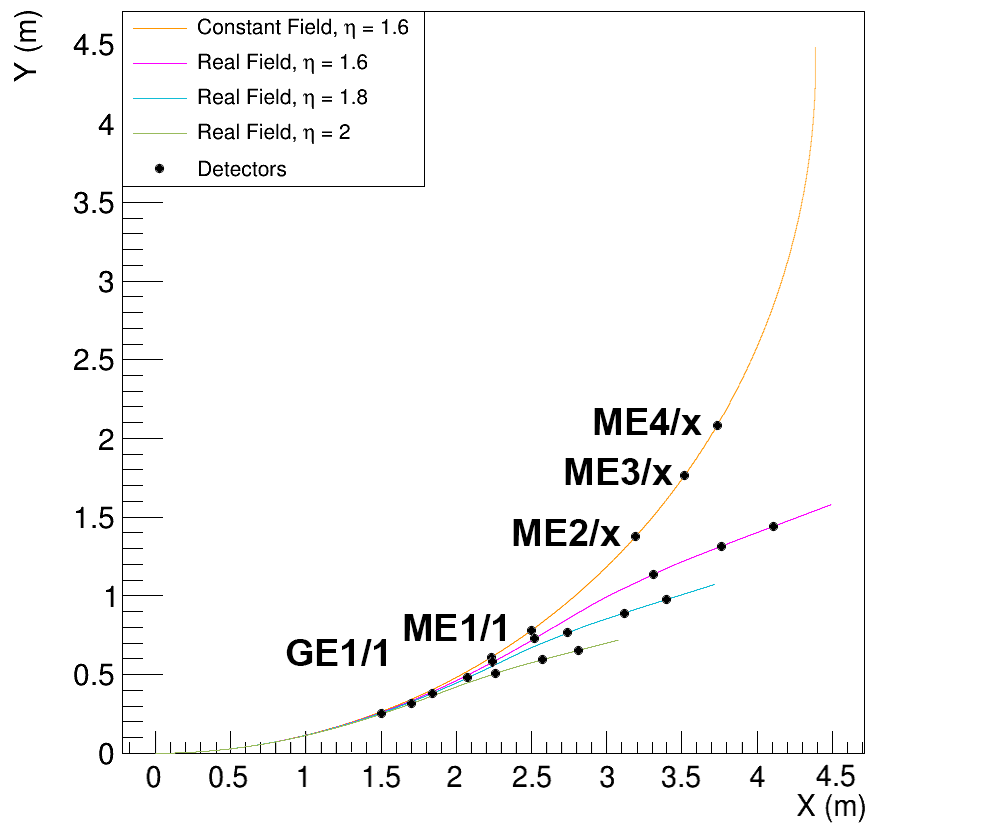}
				\caption{Projection in the transverse plane of muon tracks generated with the FastSim in the constant and real magnetic field for a simulated \pT{} of 5 \GeVc{} and initial $ \eta $ of 1.6, 1.8, and 2, with the placement of the detectors on the trajectories.}
				\label{fig:least_squares_fit__xy_projection_vs_eta}
			\end{figure}

			Regarding the improvement of the bias with $ \eta $, we previously observed this effect in Section \ref{sec:least_squares_fit__real_magnetic_field_validation_using_simhits}. At high $ \eta $, the average intensity of the magnetic field is higher, meaning tracks are closer to those in the constant magnetic field, hence the bias is smaller. Moreover, another effect adds up to this. Figure \ref{fig:least_squares_fit__xy_projection_vs_eta} represents the projection in the transverse plane of muon tracks generated with the FastSim in the constant and real magnetic field for a simulated \pT{} of 5 \GeVc{} and initial $ \eta $ of 1.6, 1.8, and 2, with the placement of the detectors on the trajectories. At higher $ \eta $ (green), the projections in the transverse plane become shorter and the hits in the detectors are closer to the ideal track yielding a better reconstruction. This effect is even more significant at higher \pT{} where tracks are even shorter.

		\subsection{Summary}
		\label{sec:least_squares_fit__real_magnetic_field_summary}	

			Using the SimHits generated with the FastSim in the real magnetic field, we see that the results are significantly biased by the non-uniformity of the field (up to -40\% bias at 10 \GeVc{}). This behavior of the Least Squares fit is expected since the algorithm assumes a constant magnetic field. Therefore, the bias is minimal for the GE1/1, ME1/1, and ME2/x setup which benefits from the uniform magnetic field in GE1/1. This setups improves the results of the CSCs in standalone by an average of 33\%. As tracks significantly differ from the ideal trajectory in the outer CSC layers (ME3/x and ME4/x), setups using them deteriorate the results. An improvement at higher \pT{} is observed due to the straighter tracks the particles leave, for which the deviation caused by the CSC layers is less important. \\

			Furthermore, in the real magnetic field, detector setups using GEM detectors offer a better resolution due to the uniformity of the field near GE1/1. Additionally, considering more hits constrains the fit and improves the standard deviation. An improvement between 40 and 55\% of the standard deviation is noted when adding the GEMs to the CSCs. \\

			Although the standard deviation degrades at higher $ \eta $ due to straighter tracks, the bias is improved as tracks are shorter and hits in the detectors are closer to the ideal trajectory. This effect is more visible at higher \pT{}, where tracks are shorter. 
	
	\section{CMSSW Results}
	\label{sec:least_squares_fit__cmssw_results}	

		Finally, we consider hits generated using CMSSW. This simulation environment takes into account multiple scattering and energy losses, which adds a level of difficulty to the reconstruction.
		
		\subsection{Least Squares Fit Using SimHits in CMSSW}
		\label{sec:least_squares_fit__cmssw_validation_using_simhits}	

\begin{figure}[h!]
				\centering
				\includegraphics[width = 0.7 \textwidth]{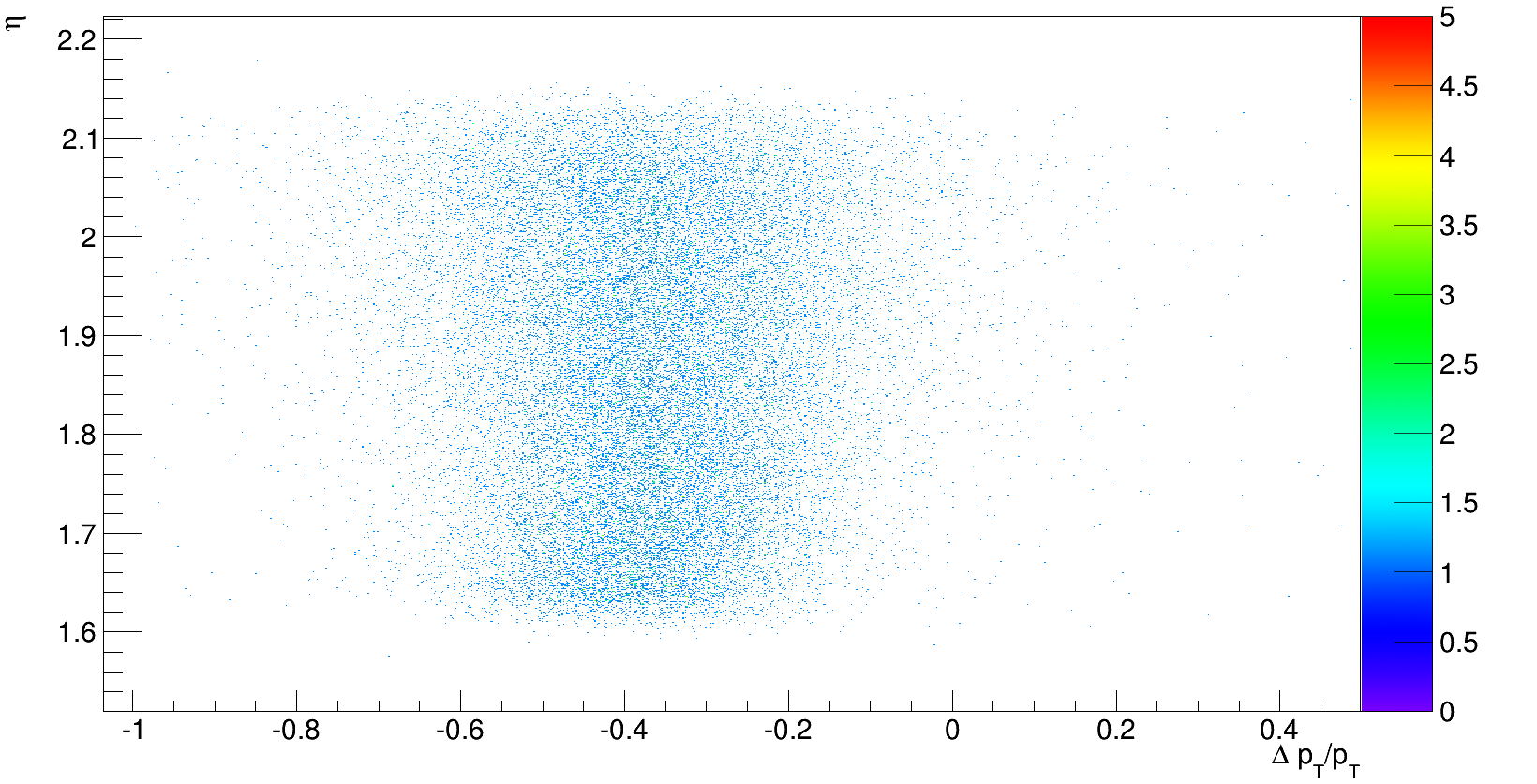}
				\caption{$ \frac{\Delta p_T}{p_T} $ as a function of the simulated $ \eta $ for muon tracks generated with CMSSW with a simulated \pT{} of 10 \GeVc{}, and reconstructed with the Least Squares fit using the SimHits in GE1/1, ME1/1, and ME2/x.}
				\label{fig:least_squares_fit__reco_pt_simhits_CMSSW_10GeV_GEM_ME12_vs_eta}
			\end{figure}

			Figure \ref{fig:least_squares_fit__reco_pt_simhits_CMSSW_10GeV_GEM_ME12_vs_eta} is a density plot of $ \frac{\Delta p_T}{p_T} $ as a function of the simulated $ \eta $ for muon tracks generated with CMSSW with a simulated \pT{} of 10 \GeVc{}, and reconstructed with the Least Squares fit using the SimHits in GE1/1, ME1/1, and ME2/x. Contrary to the SimHits generated with the FastSim in the real magnetic field reviewed in Section \ref{sec:least_squares_fit__real_magnetic_field_validation_using_simhits}, the points do not form a well defined line along $ \eta $. This is due to multiple scattering that dominates the small variations due to the magnetic field's non-uniformity and deviates particles. However, a bias of the order of -40\% is still visible.

		\subsection{Impact of Segmentation}
		\label{sec:least_squares_fit__cmssw_impact_of_segmentation}	

			One of the consequences of multiple scattering and energy losses is that they mask the segmentation of the GEMs. We are not able to distinguish the different $ \eta $ segments of GE1/1. The percentage of tracks that hit a different segment in GE1/1a and GE1/1b increased from 6.3\% to 7.8\% but does not yield better results.

		\subsection{Standard Deviation for Different Detector Setups}
		\label{sec:least_squares_fit__cmssw_standard_deviation_detectors_setups}		

\begin{figure}[h!]
				\centering
				\includegraphics[width = \textwidth]{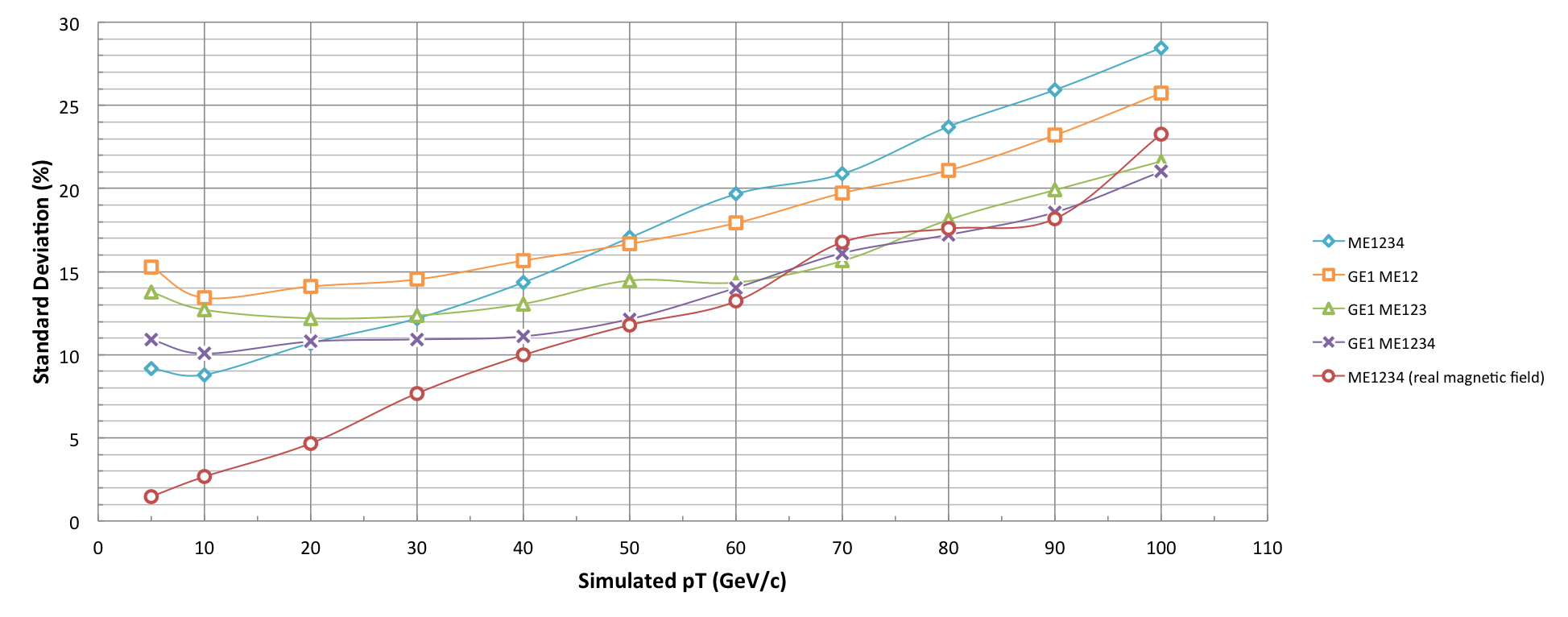}
				\caption{Standard deviation on $ \frac{\Delta p_T}{p_T} $ as a function of the simulated \pT{} for muon tracks generated with CMSSW, and reconstructed with the Least Squares fit using the RecHits in multiple detector setups. The blue curve uses RecHits in ME1/1, ME2/x, ME3/x, and ME4/x, the orange curve in GE1/1, ME1/1, and ME2/x, the green curve in GE1/1, ME1/1, ME2/x, and ME3/x, the purple curve in GE1/1, ME1/1, ME2/x, ME3/x, and ME4/x, and the red curve in ME1/1, ME2/x, ME3/x, ME4/x generated with the FastSim in the real magnetic field.}
				\label{fig:least_squares_fit__reco_sigma_rechits_CMSSW_all_setups}
			\end{figure}

			Using the RecHits, we analyze the standard deviation of the results for different detector setups. Figure \ref{fig:least_squares_fit__reco_sigma_rechits_CMSSW_all_setups} shows the standard deviation on $ \frac{\Delta p_T}{p_T} $ as a function of the simulated \pT{} for muon tracks generated with CMSSW, and reconstructed with the Least Squares fit using the RecHits in multiple detector setups. If we compare these results to those yielded by ME1/1, ME2/x, ME3/x, and ME4/x generated with the FastSim in the real magnetic field (red), we observe that the trend remains unchanged, except at low \pT{} where multiple scattering and energy losses degrade the resolution. \\

			With the FastSim in the real magnetic field, the GE1/1, ME1/1, and ME2/x setup offered better resolution than the CSC layers alone at $ p_T $ > 20 \GeVc{}, while in CMSSW, this value is raised up to $ p_T $ > 50 \GeVc{}. The same yields for the GE1/1, ME1/1, ME2/x, and ME3/x setup that goes from an observed improvement above $ p_T $ > 10 \GeVc{} to $ p_T $ > 30 \GeVc{}. The differences in the results are due to both multiple scattering and the change in simulation environment. As stated in Section \ref{sec:simulation_environment__fast_simulation}, several differences exist between the FastSim and CMSSW. \\

			Finally, using the RecHits in GE1/1 and all the CSCs (purple) significantly improves the resolution above 20 \GeVc{} (around 28\% improvement at \pT{} > 50 \GeVc) compared to the ME1/1, ME2/x, ME3/x, and ME4/x setup (blue), while only slightly degrading it below 20 \GeVc{} (18\% degradation at 5 \GeVc). Considering only GE1/1, ME1/1, and ME2/x (orange) also improves the results at higher \pT{} (10\% improvement at 100 \GeVc{}). 
		
		\subsection{Bias for Different Detector Setups}
		\label{sec:least_squares_fit__cmssw_bias_detectors_setups}	

\begin{figure}[h!]
				\centering
				\includegraphics[width = \textwidth]{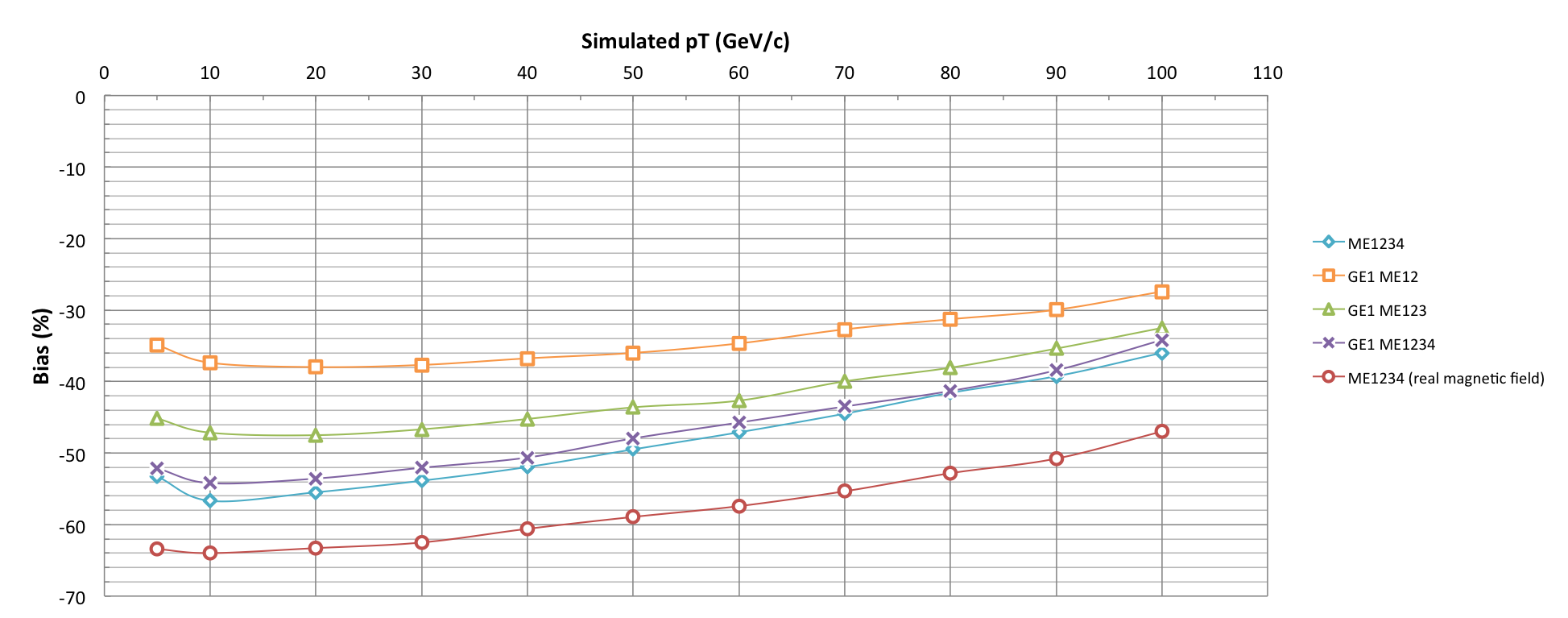}
				\caption{Bias on $ \frac{\Delta p_T}{p_T} $ as a function of the simulated \pT{} for muon tracks generated with CMSSW, and reconstructed with the Least Squares fit using the RecHits in multiple detector setups. The blue curve uses RecHits in ME1/1, ME2/x, ME3/x, and ME4/x, the orange curve in GE1/1, ME1/1, and ME2/x, the green curve in GE1/1, ME1/1, ME2/x, and ME3/x, the purple curve in GE1/1, ME1/1, ME2/x, ME3/x, and ME4/x, and the red curve in ME1/1, ME2/x, ME3/x, ME4/x generated with the FastSim in the real magnetic field.}
				\label{fig:least_squares_fit__mu_rechits_CMSSW_all_setups}
			\end{figure}

			After looking at the standard deviation, we consider Figure \ref{fig:least_squares_fit__mu_rechits_CMSSW_all_setups} depicting the bias on $ \frac{\Delta p_T}{p_T} $ as a function of the simulated \pT{} for muon tracks generated with CMSSW, and reconstructed with the Least Squares fit using the RecHits in multiple detector setups. The observed trends remain the same as for the RecHits generated with the FastSim in the real magnetic field reviewed in Section \ref{sec:least_squares_fit__real_magnetic_field_bias_detectors_setups}. As before, GEM detectors diminish the bias as they are placed in a region of CMS where the magnetic field is still uniform (improvement between 24 and 34\% if we compare GE1/1, ME1/1, and ME2/x (orange) to ME1/1, ME2/x, ME3/x, and ME4/x (blue)). Therefore hits are closer to the ideal track. Adding CSC layers degrades the results as they lie in a greatly non-uniform field. 

		\subsection{Evolution with $ \eta $}
		\label{sec:least_squares_fit__cmssw_evolution_eta}		

\begin{figure}[h!]
				\centering
				\includegraphics[width = \textwidth]{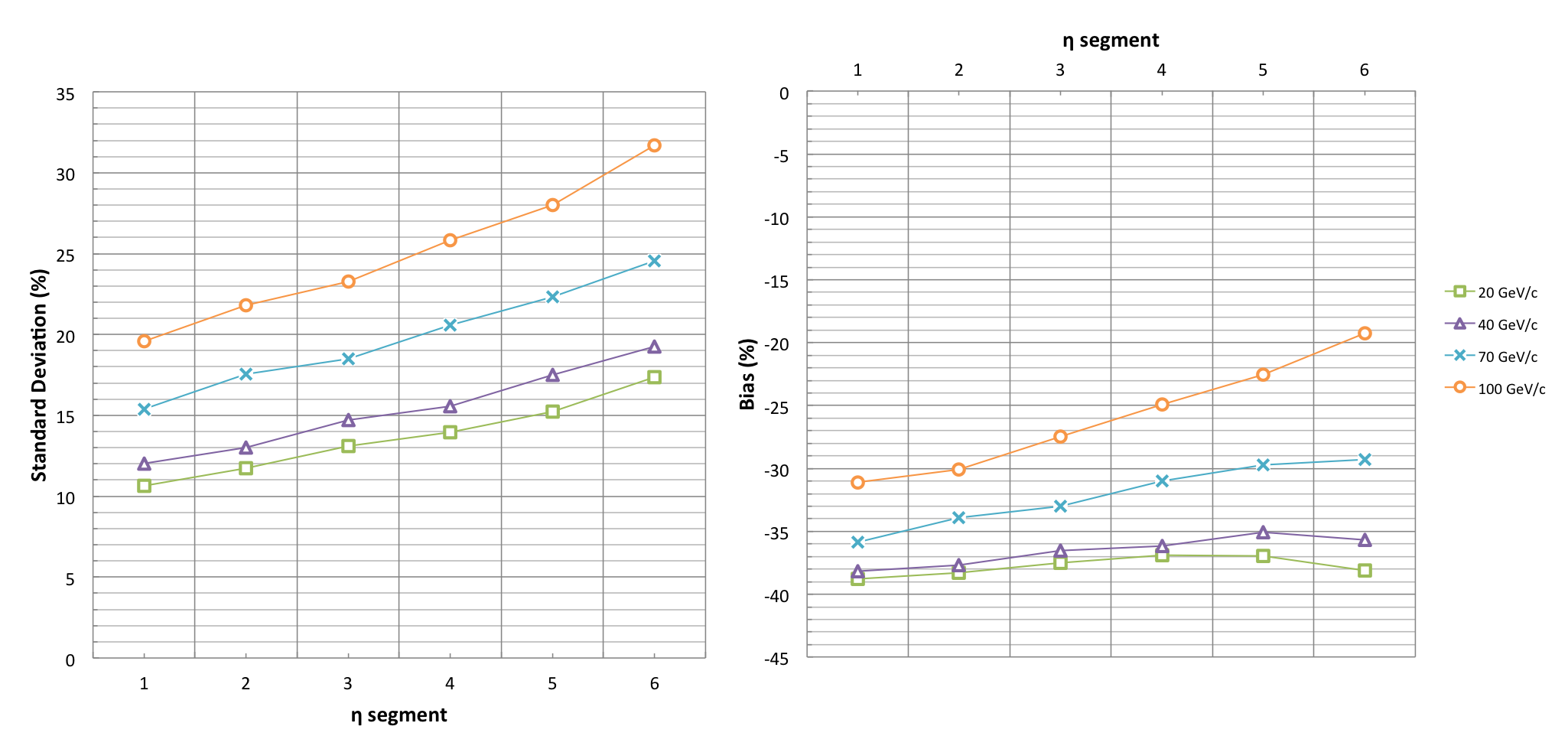}
				\caption{Standard deviation (left) and bias (right) on $ \frac{\Delta p_T}{p_T} $ as a function of the hit $ \eta $ segment in GE1/1 for muon tracks generated with CMSSW with simulated \pT{} of 20, 40, 70, and 100 \GeVc{}, and reconstructed with the Least Squares fit using the RecHits in GE1/1, ME1/1, and ME2/x.}
				\label{fig:least_squares_fit__sigma_mu_rechits_CMSSW_all_setups_evolution_eta}
			\end{figure}

			Finally, we look at Figure \ref{fig:least_squares_fit__sigma_mu_rechits_CMSSW_all_setups_evolution_eta} which represents the standard deviation (left) and bias (right) on $ \frac{\Delta p_T}{p_T} $ as a function of the hit $ \eta $ segment in GE1/1 for muon tracks generated with CMSSW with simulated \pT{} of 20, 40, 70, and 100 \GeVc{}, and reconstructed with the Least Squares fit using the RecHits in GE1/1, ME1/1, and ME2/x. A deterioration of the standard deviation with $ \eta $ is observed, while the bias improves. The first effect being due to shorter tracks in the transverse plane which leave more room for error. The second one resulting from the more constant magnetic field inside CMS at higher $ \eta $. This has been previously noticed in the FastSim in the real magnetic field. \\

		\subsection{Summary}
		\label{sec:least_squares_fit__cmssw_summary}

			Physical processes have multiple effects on the reconstruction: they even out the $ \frac{\Delta p_T}{p_T} $ distribution along $ \eta $, dominate all the effects of the non-uniform magnetic field expect for the strong bias, and degrade the standard deviation at lower \pT{}. However, the trends observed for the FastSim with the real magnetic field remain valid. \\

			At \pT{} < 20 \GeVc{}, the CSCs in standalone offer a better resolution than setups using GEMs due to the $ \eta $ segmentation of the latter (18\% degradation when we compared the setup using the GEMs and all the CSCs and the CSCs in standalone), while at \pT{} > 20 \GeVc{}, this effect becomes negligible, and the higher $ \phi $ resolution of GEMs improves the results of the CSCs-only by up to 28\%. \\

			Moreover, setups using GEMs have a smaller bias, as they are placed in a region of CMS where the magnetic field is constant. An improvement between 24 and 34\% is observed when comparing GEM detectors and two layers of CSCs against the CSCs in standalone. Adding more CSC layers to the GEMs degrades the results as tracks diverge from the ideal trajectory, but always yields a smaller bias than CSCs in standalone. \\

			Finally, we observe an improvement of the bias at higher $ \eta $, along with a degradation of the standard deviation.
	
	\section{Conclusion}
	\label{sec:least_squares_fit__conclusion}	

		This analysis shows that we cannot consider GE1/1 and ME1/1 alone as it was initially foreseen due to the $ \eta $ segmentation of GEM detectors. However, using GE1/1, ME1/1, and ME2/x is a viable option. Even if the consequences of segmentation are still visible at low \pT{} by degrading the standard deviation, this setup minimizes the bias for all \pT{} and proves to have a greater resolution than CSCs alone at higher \pT{}. Within the FastSim with the real magnetic field using GEM detectors and only two layers of CSCs improves the standard deviation by up to 20\% at high \pT{} and the bias by an average of 33\% over the entire \pT{} range. \\

		Setups with more CSC layers, such as GE1/1, ME1/1, ME2/x, and ME3/x or GE1/1, ME1/1, ME2/x, ME3/x, and ME4/x improve the standard deviation, due to higher constraints on the fit, but increase the bias because of the non-uniform magnetic field in which the CSCs are placed. In CMSSW, using GEMs and all the CSCs improve the standard deviation of the CSCs in standalone by an average of 28\% at \pT{} > 50 \GeVc. \\

		Moreover, we observed an improvement in the bias with $ \eta $, while, on the other hand, a degradation of the standard deviation was noticed as the projected trajectories in the transverse plane become shorter and leave more room for error. \\

		Those conclusion are supported and verified by the different simulation environments. \\

		Regarding the different simulation environments, we observed that using the real magnetic field, either in the FastSim or in CMSSW, brings a large bias on the reconstructed tracks which cannot be erased using this particular method. Therefore, we implemented a Kalman Filter which function and results are explained in the next chapter.

	\cleardoublepage


\chapter{Kalman Filter}
\label{chap:kalman_filter}

	In the perspective of improving the results, especially the bias, of the previously reviewed Least Squares fit, we implemented a Kalman filter. The theory behind this method has been reviewed in Section \ref{sec:trigger_system_and_reconstruction_algorithms__kalman_filter} and will be applied to helical tracks in this chapter.
	
	\section{Helix Parameterization}
	\label{sec:kalman_filter__helix_parameterization}
 	
\begin{figure}[h!]
			\centering
			\includegraphics[width = 6cm]{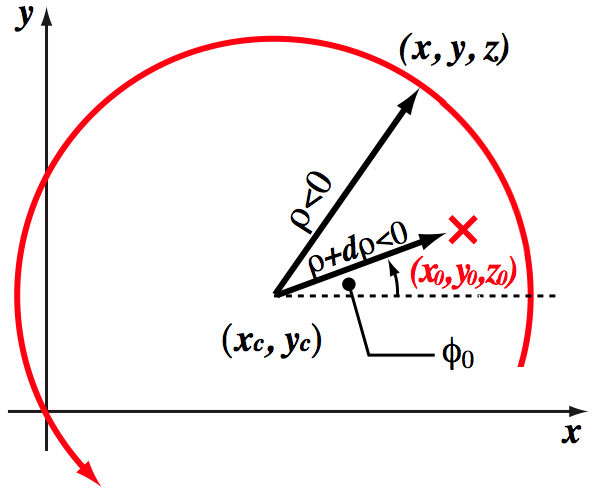}
			\caption{Helical track parametrization for the Kalman Filter \Cite{Fit_Extended_Kalman}.}
			\label{fig:kalman_filter__fit_parameters}
		\end{figure}

	 	One of the parameterizations of helices for the Kalman filter \Cite{Fit_Extended_Kalman} uses 5 parameters to describe the track locally with respect to a reference point, or pivot
	 	\begin{equation} 
	 		\mathbf{x}_0 = \left( \begin{array}{ccc} x_0 & y_0 & z_0 \end{array} \right)^\intercal \ .
	 	\end{equation} 
		The state vector of the system, $ \mathbf{a} $, describes the helix close to the pivot and is parametrized by
		\begin{equation}
			\mathbf{a} = \left( \begin{array}{ccccc} d_\rho & \phi_0 & \kappa & d_z & \tan \lambda \end{array} \right)^\intercal
		\end{equation}
		where $ d_\rho $ is the radial distance between the helix and the pivot in the transverse plane, $ \phi_0 $ is the azimuthal angle of the pivot, $ \kappa = \frac{Q}{p_T} $ is the charge sign divided by \pT{}, $ d_z $ is the distance between the helix and the pivot along \axis{Z}, and $ \tan \lambda $ is the tangent of the polar angle. The $ d_\rho $ and $ \phi_0 $ parameters are represented in Figure \ref{fig:kalman_filter__fit_parameters}. In all future calculations, the hits in the detectors will be used as pivots. \\

		The propagation from one site to another is done by changing the pivot. If at site $ (k - 1) $ the parameters describe the helix for $ \mathbf{x}_0 $, by changing the pivot to $ \mathbf{x}_0' $, we obtain the parameters at site $ (k) $. The extrapolated state vector is given by
		\begin{equation}
			\mathbf{a}^k_{k-1} \left\{ \begin{array}{ll}
				d_\rho' & = \left( X_c - x_0' \right) \cos \phi_0' + \left( Y_c - y_0' \right) \sin \phi_0' - \frac{\alpha}{\kappa} \\
				\phi_0' & = \left\{ \begin{split} 
					\tan^{-1}\left( \frac{Y_c - y_0'}{X_c - x_0'} \right) & & (\kappa > 0) \\
					\tan^{-1}\left( \frac{y_0' - Y_c}{x_0' - X_c} \right) & & (\kappa < 0)
				\end{split} \right. \\
				\kappa' & = \kappa \\
				d_z' & = z_0 - z_0' + d_z - \frac{\alpha}{\kappa} \left( \phi_0' - \phi_0 \right) \tan \lambda \\
				\tan \lambda' & = \tan \lambda \\
			\end{array} \right. \ ,
			\label{eq:kalman_filter__parameter_extrapolation}
		\end{equation} 
		where $ X_c $ and $ Y_c $ are the coordinates of the center of the helix in the transverse plane
		\begin{equation}
			\left\{ \begin{split}
				X_c & = x_0 + \left( d_\rho + \frac{\alpha}{\kappa} \right) \cos \phi_0 \\
				Y_c & = y_0 + \left( d_\rho + \frac{\alpha}{\kappa} \right) \sin \phi_0
			\end{split} \right. \ ,
		\end{equation} and 
		\begin{equation}
			\alpha = \frac{1}{c B} \ ,
		\end{equation}	
		where $ c $ is the speed of light and $ B $ is the intensity of the magnetic field along the axis of the helix. \\

		The derivation of $ \mathbf{F} $ is not explicitly done here but can be calculated using Equation \ref{eq:trigger_system_and_reconstruction_algorithms__kalman_propagator} in Section \ref{sec:trigger_system_and_reconstruction_algorithms__kalman_filter_extended_kalman_filter}	, yielding
		\begin{equation}
			\mathbf{F}_{k-1} = \left( \begin{array}{ccccc} 
				\cos \Phi & \left( d_\rho + \frac{\alpha}{\kappa} \right) \sin \Phi & \frac{\alpha}{\kappa^2} \left( 1 - \cos \Phi \right) & 0 & 0 \\
				- \frac{\sin \Phi}{d_\rho' + \frac{\alpha}{\kappa}} & \frac{d_\rho + \frac{\alpha}{\kappa}}{d_\rho' + \frac{\alpha}{\kappa}} \cos \Phi & \frac{\alpha}{\kappa^2} \frac{\sin \Phi}{d_\rho' + \frac{\alpha}{\kappa}} & 0 & 0 \\
				0 & 0 & 1 & 0 & 0 \\
				\frac{\alpha}{\kappa} \frac{\tan \lambda \sin \Phi}{d_\rho' + \frac{\alpha}{\kappa}} & \frac{\alpha}{\kappa} \tan \lambda \left( 1 - \frac{d_\rho + \frac{\alpha}{\kappa}}{d_\rho' + \frac{\alpha}{\kappa}} \cos \Phi \right) & \frac{\alpha}{\kappa^2} \tan \lambda \left( \Phi - \frac{\alpha}{\kappa} \frac{\sin \Phi}{d_\rho' + \frac{\alpha}{\kappa}} \right) & 1 & - \frac{\alpha}{\kappa} \left( \Phi \right) \\
				0 & 0 & 0 & 0 & 1 
			\end{array} \right) \ , 
		\end{equation}
		where $ \Phi = \phi_0' - \phi_0 $. In order to compute the propagation noise between two pivots, we have to divide the interval into small iterations during which the matter density remains unchanged. The propagation noise matrix for each step is given by
		\begin{equation}
			\mathbf{Q}_m = \theta_{RMS} \left( \begin{array}{ccccc} 
				0 & 0 & 0 & 0 & 0 \\ 0 & 1 + \tan^2 \lambda & 0 & 0 & 0 \\ 
				0 & 0 & \left( \kappa \tan \lambda \right)^2 & 0 & \kappa \tan \lambda \left( 1 + \tan^2 \lambda \right) \\ 
				0 & 0 & 0 & 0 & 0 \\
				0 & 0 & \kappa \tan \lambda \left( 1 + \tan^2 \lambda \right) & 0 & \left( 1 + \tan^2 \lambda \right)^2 \\ 
			\end{array} \right) \ ,
		\end{equation}
		where $ \theta_{RMS} $ is defined by Equation \ref{eq:muon_chambers__multiple_scattering_RMS} in Section \ref{sec:muon_chambers__multiple_scattering}. To calculate the cumulated noise, we use the propagation matrix for each iteration 
		\begin{equation}
			\mathbf{Q}_{k-1} = \sum_m \mathbf{F}_m \mathbf{Q}_m \mathbf{F}^\intercal_m \ .
		\end{equation}	
		Due to time constraints, we ignored propagation noise in our implementation of the Kalman filter. Therefore, the results in CMSSW will not be optimal due to the presence of physical processes such as multiple scattering and energy losses. \\

		The extrapolated state is then compared to the hit by collapsing it onto the measurement vector
		\begin{equation}
			\mathbf{m}_k = \left( \begin{array}{cc} \phi & \eta \end{array} \right)^\intercal \ .
		\end{equation}
		We obtain this by first computing the intersection between the helix and the measurement plane 
		\begin{equation}
			\mathbf{x}_k = \left( \begin{array}{c}
				x_0 + d_\rho \cos \phi_0 + \frac{\alpha}{\kappa} \left( \cos \phi_0 - \cos\left( \phi_0 + \phi_k \right) \right) \\
				y_0 + d_\rho \sin \phi_0 + \frac{\alpha}{\kappa} \left( \sin \phi_0 - \sin\left( \phi_0 + \phi_k \right) \right) \\
				z_0 + d_z - \frac{\alpha}{\kappa} \phi_k \tan \lambda
			\end{array} \right) \ ,
			\label{eq:kalman_filter__propagated_position}
		\end{equation}
		where $ \phi_k $ is the deflection angle, the local angle between the helix and the measurement plane. \\

		To calculate $ \phi_k $, we consider the equation of a detection layer $ S_k $
		\begin{equation}
			S_k \equiv S_k \left( \mathbf{x}_k \right) = 0 \ .
		\end{equation}	
		In our case, it becomes
		\begin{equation}
			S_k \equiv z = z_k \ ,
		\end{equation}	
		as detectors are placed at constant $ z $, which yields
		\begin{equation}
			S_k \equiv z_0 + d_z - \frac{\alpha}{\kappa} \phi_k \tan \lambda - z_k = 0 
		\end{equation}
		or
		\begin{equation}
			S_k \equiv d_z - \frac{\alpha}{\kappa} \phi_k \tan \lambda = 0 \ ,
		\end{equation}
		as the hit is used as pivot so that $ z_0 = z_k $. We expand this relation around an approximation $ \phi_n $ of $ \phi_k $
		\begin{equation}
			S_k \left( \mathbf{x} \left( \mathbf{a}_k, \phi_n \right) \right) + \frac{\partial S_k}{\partial \mathbf{x}_k} \frac{\partial \mathbf{x}_k}{\partial \phi_n} \left( \phi - \phi_n \right) \approx 0 \ . 
		\end{equation}		
		Using a recurrence method, we find that $ \phi_k $ can be obtained by iteration
		\begin{equation}
			\phi_{n+1} = \phi_n - \frac{S_k}{\left( \frac{\partial S_k}{\partial \mathbf{x}_k} \right) \left( \frac{\partial \mathbf{x}_k}{\partial \phi_k} \right)} \ ,
		\end{equation}	
		until $ S_k \approx 0 $, hence the found position is on the measurement layer, with
		\begin{equation}
			\frac{\partial S_k}{\partial \mathbf{x}_k} = \left( \begin{array}{ccc} 0 & 0 & 1 \end{array} \right)^\intercal
		\end{equation}
		and
		\begin{equation}
			\frac{\partial \mathbf{x}_k}{\partial \phi_k} = \left( \begin{array}{ccc} \frac{\alpha}{\kappa} \sin \left( \phi_0 + \phi_k \right) & - \frac{\alpha}{\kappa} \cos \left( \phi_0 + \phi_k \right) & - \frac{\alpha}{\kappa} \tan \lambda \end{array} \right) \ .
		\end{equation} \\

		Finally, we can derive the projection matrix using the previously defined quantities
		\begin{equation}
			\mathbf{H}_k = \frac{\partial \mathbf{m}_k}{\partial \mathbf{x}_k} \frac{\partial \mathbf{x}_k}{\partial \mathbf{a}^k_{k-1}} = \frac{\partial \mathbf{m}_k}{\partial \mathbf{x}_k} \left( \frac{\partial \mathbf{x}_k}{\partial \phi_k} \frac{\partial \phi_k}{\partial \mathbf{a}^k_{k-1}} + \frac{\partial \mathbf{x}_k }{\partial \mathbf{a}^k_{k-1}} \right) \ ,
		\end{equation}
		where
		\begin{equation} 
			\frac{\partial \phi_k}{\partial \mathbf{a}^k_{k-1}} = - \frac{\left( \frac{\partial S_k}{\partial \mathbf{x}_k} \right) \left( \frac{\partial \mathbf{x}_k}{\partial \mathbf{a}^k_{k-1}} \right)}{\left( \frac{\partial S_k}{\partial \mathbf{x}_k} \right) \left( \frac{\partial \mathbf{x}_k}{\partial \phi_k} \right)} \ ,
			\label{eq:kalman_filter__dphi_da}
		\end{equation}
		\begin{equation}
			\frac{\partial \mathbf{x}_k}{\partial \mathbf{a}^k_{k-1}} = \left( \begin{array}{ccccc} 
				\cos \phi_0 & - d_\rho \sin \phi_0 - \frac{\alpha}{\kappa} \left( \sin \phi_0 - \sin \Phi_k \right) & - \frac{\alpha}{\kappa^2} \left( \cos \phi_0 - \cos \Phi_k \right) & 0 & 0 \\
				\sin \phi_0 & d_\rho \cos \phi_0 + \frac{\alpha}{\kappa} \left( \cos \phi_0 - \cos \Phi_k \right) & - \frac{\alpha}{\kappa^2} \left( \sin \phi_0 - \sin \Phi_k \right) & 0 & 0 \\
				0 & 0 & \frac{\alpha}{\kappa^2} \tan \lambda \phi_k & 1 & - \frac{\alpha}{\kappa} \phi_k
			\end{array} \right)
		\end{equation}
		with $ \Phi_k = \phi_0 + \phi_k $, and
		\begin{equation} 
			\frac{\partial \mathbf{m}_k}{\partial \mathbf{x}_k} = \left( \begin{array}{ccc} 
				- \frac{y}{x^2 + y^2} & \frac{x}{x^2 + y^2} & 0 \\
				- \frac{x z}{\left( x^2 + y^2 \right) \sqrt{x^2 + y^2 + z^2}} & - \frac{y z}{\left( x^2 + y^2 \right) \sqrt{x^2 + y^2 + z^2}} & \frac{1}{\sqrt{x^2 + y^2 + z^2}}
			 \end{array} \right) \ .
		\end{equation} \\

		\pT{} is obtained from $ \kappa $ 
		\begin{equation}
			p_T = \left| \frac{1}{\kappa} \right| \ .
		\end{equation} \\

		We require a minimum of three hits in the detectors and then add the IP to the list of points.

	\section{Standard Kalman Filter}
	\label{sec:kalman_filter__standard_kalman_filter}

		We implemented the hereabove described Kalman filter and tested it in the various simulation environments by performing the same analysis as those reviewed in Chapter \ref{chap:least_squares_fit}. 
		
		\subsection{FastSim Results: Constant Magnetic Field}
		\label{sec:kalman_filter__standard_fastsim_results_constant_magnetic_field}

			We validate the algorithm by using the hits generated with the FastSim in the constant magnetic field. This simulation environment allows us to look at the impact of the $ \eta $ segmentation without having to take into account the non-uniform magnetic field or physical processes. 
			
			\subsubsection{Validation Using SimHits}
			\label{sec:kalman_filter__standard_constant_field_validation_using_simhits}	

				The results given by the SimHits generated with the FastSim in the constant magnetic field yield a standard deviation and a bias on $ \frac{\Delta p_T}{p_T} $ never greater than 10$ ^{-6} $\%. We notice that these are larger than the results obtained for the Least Squares fit (of the order of 2 10$ ^{-8} $\%) discussed in Section \ref{sec:least_squares_fit__constant_magnetic_field_validation_using_simhits}. This is due to the iterative process and is explained in more details in the next section.
			
			\subsubsection{Impact of Segmentation}
			\label{sec:kalman_filter__standard_constant_field_impact_of_segmentation}	

\begin{figure}[h!]
					\centering
					\includegraphics[width = 0.7 \textwidth]{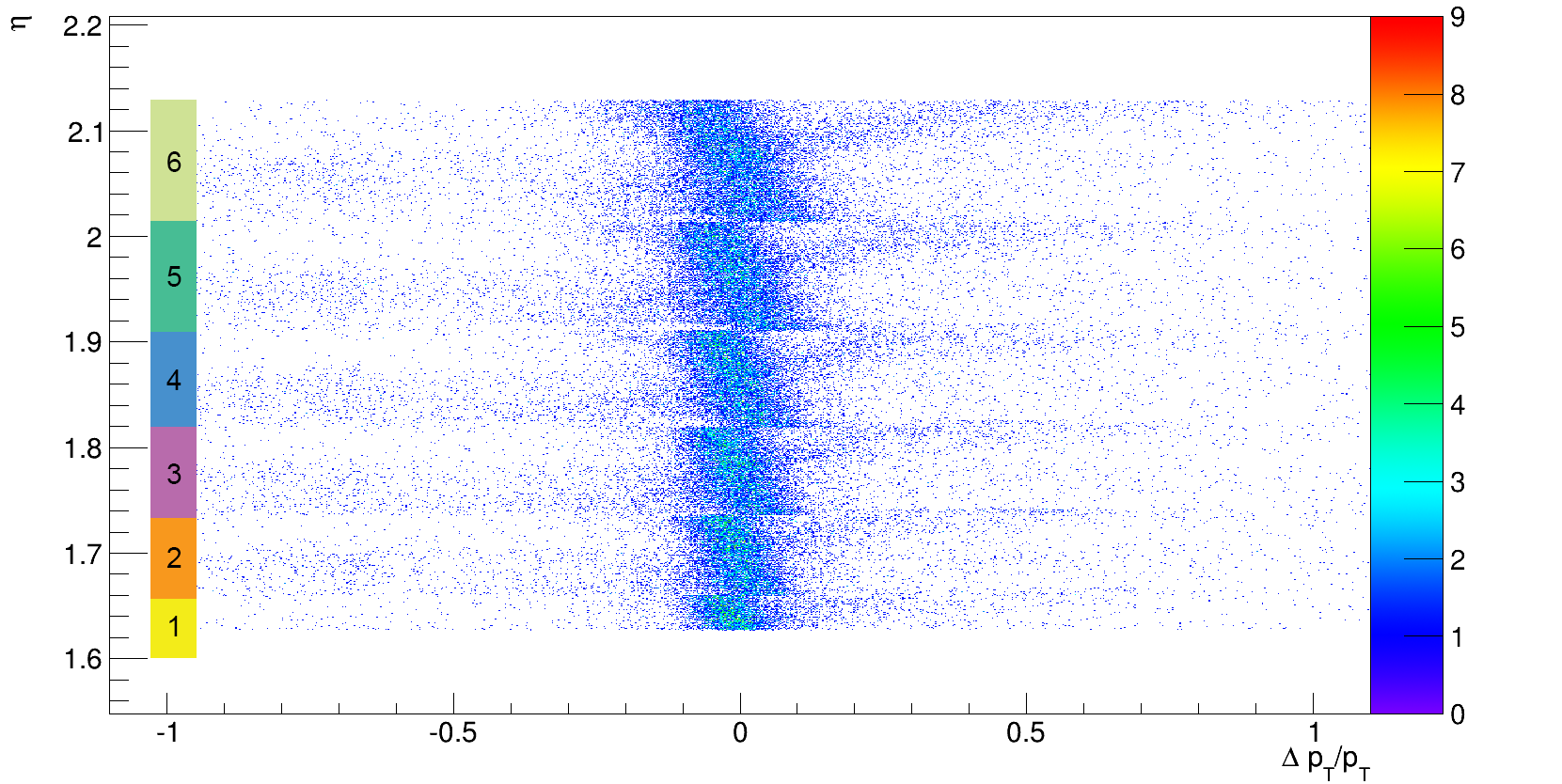}
					\caption{$ \frac{\Delta p_T}{p_T} $ as a function of the simulated $ \eta $ for muon tracks generated with the FastSim in the constant magnetic field with a \pT{} of 10 \GeVc{}, and reconstructed with the standard Kalman filter using the RecHits in GE1/1, ME1/1, and ME2/x.}
					\label{fig:kalman_filter__standard_reco_pt_rechits_38T_5GeV_GEM_ME11}
				\end{figure}

				We use the RecHits generated with the FastSim in the constant magnetic field to look at the impact of segmentation. Figure \ref{fig:kalman_filter__standard_reco_pt_rechits_38T_5GeV_GEM_ME11} shows $ \frac{\Delta p_T}{p_T} $ as a function of the simulated $ \eta $ for muon tracks generated with the FastSim in the constant magnetic field with a \pT{} of 10 \GeVc{}, and reconstructed with the standard Kalman filter using the RecHits in GE1/1, ME1/1, and ME2/x. The impact of segmentation is visible through the 6 sectors that appear in the image. We also observe that some events are badly reconstructed and spread over the entire range -1 < $ \frac{\Delta p_T}{p_T} $ < 1. 

				This is caused by both the iterative process and the direction in which the iteration is done. The Kalman filter should return better results than the Least Squares fit, as it can take into account the errors of the detectors. Unfortunately, due to the complexity of the implementation of the helix, reconstruction sometimes fails causing the large visible spread. We also made the choice to start with ME4/x, the most outer detection layer, and reconstruct the track backwards to the IP. By doing so, the final parameters of the helix are computed in the region where the field is the most uniform: at the IP. This is not important in this simulation environment, but becomes critical when considering the non-uniform magnetic field. On the other hand, the last used detectors are GE1/1b and GE1/1a where the error on $ \eta $ is greater, therefore inefficiently improving the resolution. 
			
			\subsubsection{Standard Deviation for Different Detector Setups}
			\label{sec:kalman_filter__standard_constant_field_standard_deviation_detectors_setup}	

\begin{figure}[h!]
					\centering
					\includegraphics[width = \textwidth]{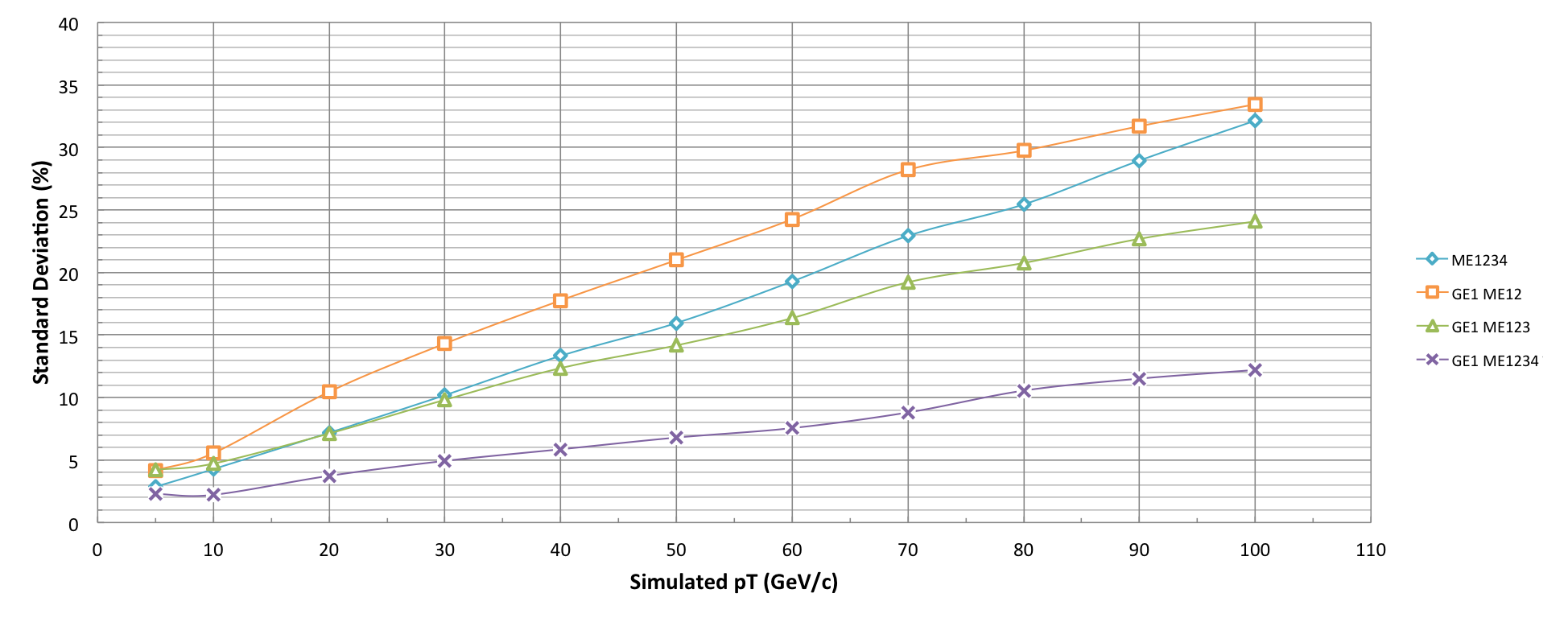}
					\caption{Standard deviation on $ \frac{\Delta p_T}{p_T} $ as a function of the simulated \pT{} for muon tracks generated with the FastSim in the constant magnetic field, and reconstructed with the standard Kalman filter using the RecHits in multiple detector setups. The blue curve uses RecHits in ME1/1, ME2/x, ME3/x, and ME4/x, the orange curve in GE1/1, ME1/1, and ME2/x, the green curve in GE1/1, ME1/1, ME2/x, and ME3/x, and the purple curve in GE1/1, ME1/1, ME2/x, ME3/x, and ME4/x.}
					\label{fig:kalman_filter__standard_reco_sigma_rechits_38T_all_setups}
				\end{figure}

				Figure \ref{fig:kalman_filter__standard_reco_sigma_rechits_38T_all_setups} shows the standard deviation on $ \frac{\Delta p_T}{p_T} $ as a function of the simulated \pT{} for muon tracks generated with the FastSim in the constant magnetic field, and reconstructed with the standard Kalman filter using the RecHits in multiple detector setups. The observed trends are the same as those obtained in Section \ref{sec:least_squares_fit__constant_magnetic_field_standard_deviation_detectors_setups} for the Least Squares fit. \\

				Adding the GEMs to the CSCs (purple) significantly improves the resolution over the entire range of simulated \pT{} (62\% improvement at 100 \GeVc{}) compared to the CSCs in standalone (blue). Considering the GEMs and only three layers of CSCs (green) offers a poorer resolution (20\% degradation at 5 \GeVc{}) below 20 \GeVc{} compared to the CSCs in standalone (blue) but brings improvements to the results (up to 26\% improvement) at higher \pT{}. This is due to the impact of segmentation which is important at small \pT{}. 

			\subsubsection{Bias for Different Detector Setups}
			\label{sec:kalman_filter__standard_constant_field_bias_detectors_setup}	

\begin{figure}[h!]
					\centering
					\includegraphics[width = \textwidth]{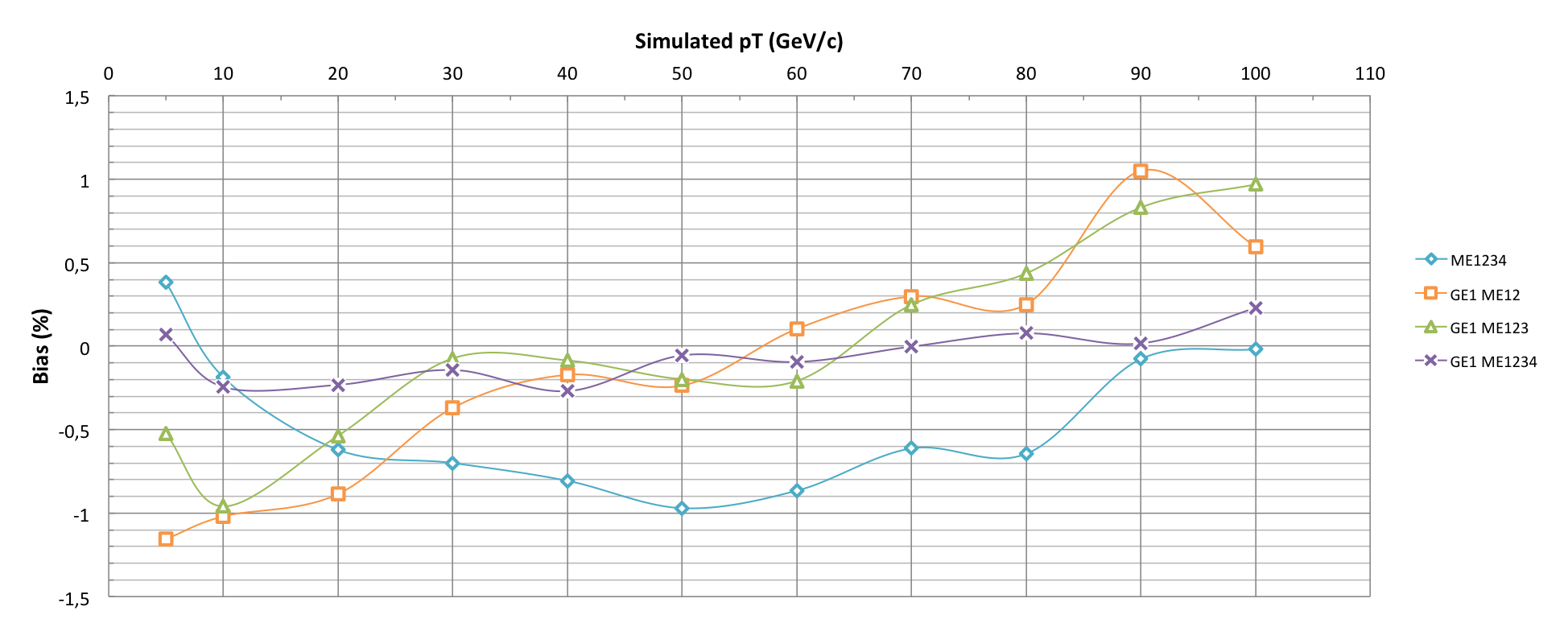}
					\caption{Bias on $ \frac{\Delta p_T}{p_T} $ as a function of the simulated \pT{} for muon tracks generated with the FastSim in the constant magnetic field, and reconstructed with the standard Kalman filter using the RecHits in multiple detector setups. The blue curve uses RecHits in ME1/1, ME2/x, ME3/x, and ME4/x, the orange curve in GE1/1, ME1/1, and ME2/x, the green curve in GE1/1, ME1/1, ME2/x, and ME3/x, and the purple curve in GE1/1, ME1/1, ME2/x, ME3/x, and ME4/x.}
					\label{fig:kalman_filter__standard_mu_rechits_38T_all_setups}
				\end{figure}

				After looking at the standard deviation, we analyze the bias. Figure \ref{fig:kalman_filter__standard_mu_rechits_38T_all_setups} represents the bias on $ \frac{\Delta p_T}{p_T} $ as a function of the simulated \pT{} for muon tracks generated with the FastSim in the constant magnetic field, and reconstructed with the standard Kalman filter using the RecHits in multiple detector setups. The results fluctuate between -1 and 1, meaning the bias is rather small. We will not further discuss the curves' trends as the variations are too small to extract global tendencies.

			\subsubsection{Evolution with $ \eta $}
			\label{sec:kalman_filter__standard_constant_field_evolution_eta}		

\begin{figure}[h!]
					\centering
					\includegraphics[width = \textwidth]{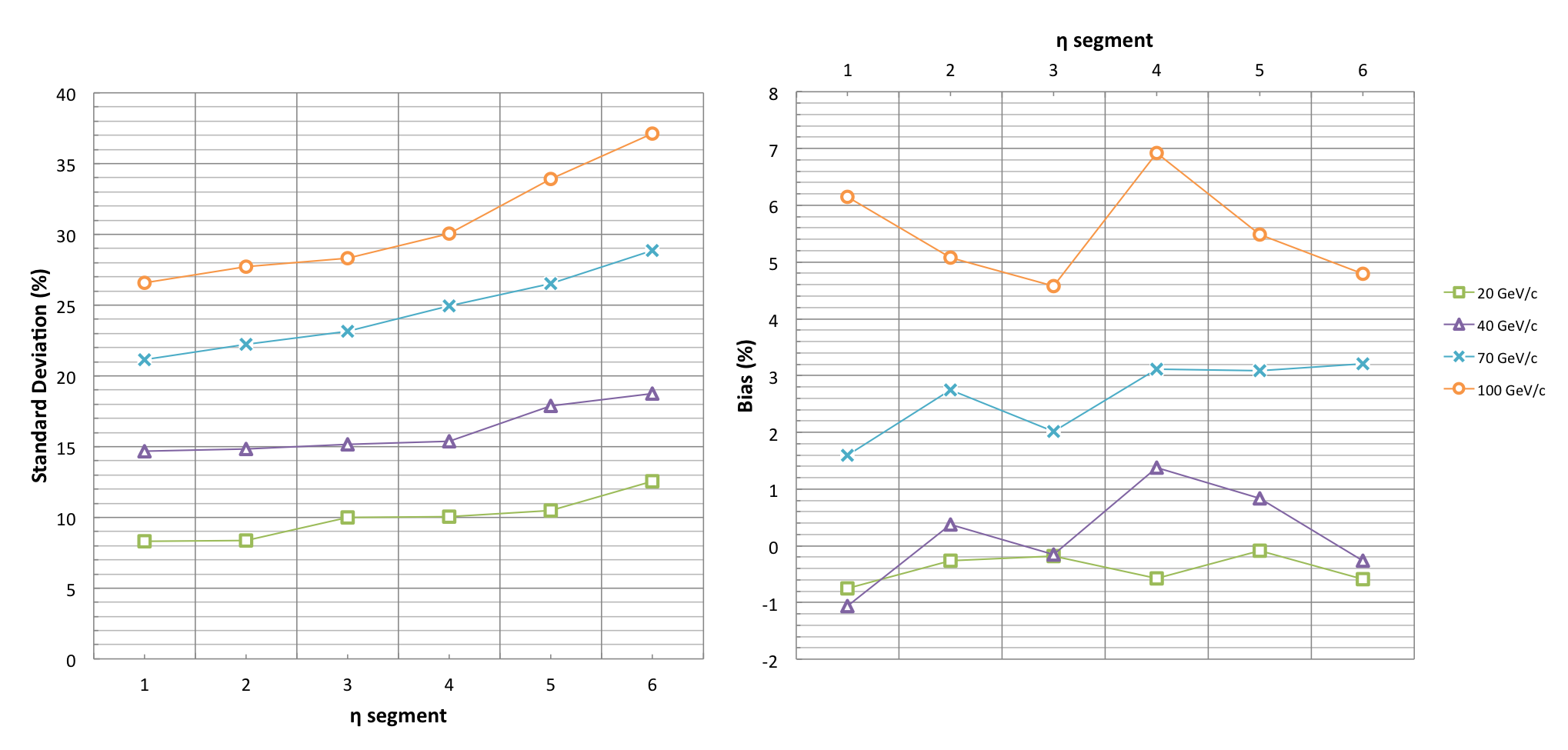}
					\caption{Standard deviation (left) and bias (right) on $ \frac{\Delta p_T}{p_T} $ as a function of the hit $ \eta $ segment in GE1/1 for muon tracks generated with the FastSim in the constant magnetic field with simulated \pT{} of 20, 40, 70, and 100 \GeVc{}, and reconstructed with the standard Kalman filter using the RecHits in GE1/1, ME1/1, and ME2/x.}
					\label{fig:kalman_filter__modified_sigma_mu_rechits_38T_all_setups_evolution_eta}
				\end{figure}

				Finally, we analyze the evolution of the parameters with $ \eta $. Figure \ref{fig:kalman_filter__modified_sigma_mu_rechits_38T_all_setups_evolution_eta} depicts the standard deviation (left) and bias (right) on $ \frac{\Delta p_T}{p_T} $ as a function of the hit $ \eta $ segment in GE1/1 for muon tracks generated with the FastSim in the constant magnetic field with simulated \pT{} of 20, 40, 70, and 100 \GeVc{}, and reconstructed with the standard Kalman filter using the RecHits in GE1/1, ME1/1, and ME2/x. As for the Least Squares fit, the standard deviation increases with $ \eta $ due to shorter projections of the tracks in the transverse plane. The bias varies along $ \eta $ and does not follow a visible trend, as with the evolution along \pT{}.

			\subsubsection{Summary}
			\label{sec:kalman_filter__standard_constant_field_summary}	

				From the SimHits generated with the FastSim in the constant magnetic field, we conclude that the algorithm is working as expected as the standard deviation and bias are never greater than 10$ ^{-6} $\%. \\ 

				Even thought, the impact of segmentation is still visible through 6 discontinuous sectors in $ \eta $, considering the GEM detectors and all the CSC layers improves the resolution over the entire range of simulated \pT{} by up to 62\% at 100 \GeVc{}. The bias is quasi null and varies between -1 and 1\% for both the evolution along the \pT{} and along $ \eta $.

		\subsection{FastSim Results: Real Magnetic Field}
		\label{sec:kalman_filter__standard_fastsim_results_real_magnetic_field}

			We move on to the tracks generated with the FastSim in the real magnetic field and look at its influence on the results. This implementation of the Kalman filter is not designed to support non-uniform field and will perform reconstruction as if the intensity of the field is of 3.8 T everywhere. 

			\subsubsection{Standard Kalman Filter with FastSim SimHits in the Real Magnetic Field}
			\label{sec:kalman_filter__standard_real_field_validation_using_simhits}	

\begin{figure}[h!]
					\centering
					\includegraphics[width = 0.7 \textwidth]{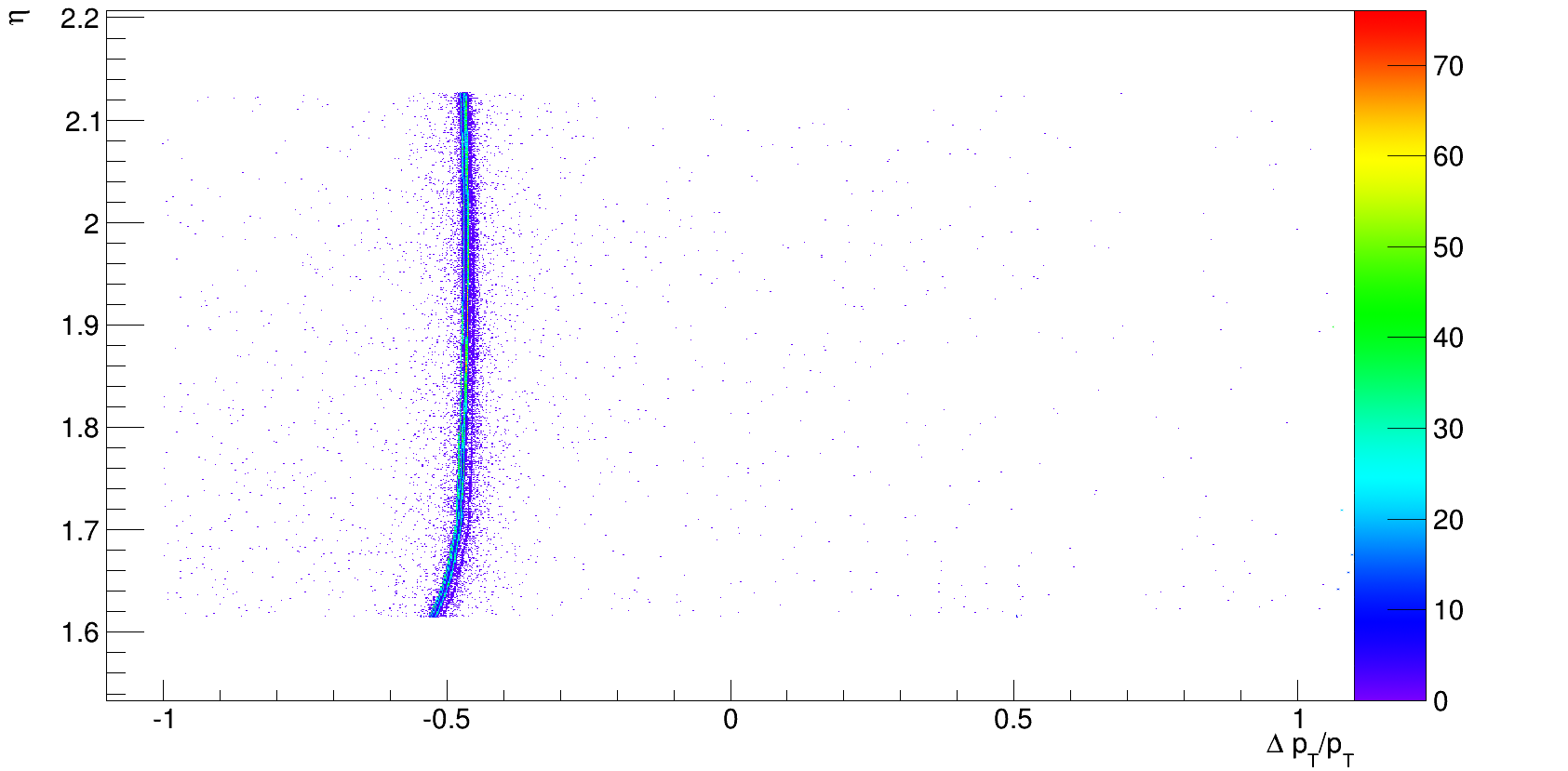}
					\caption{$ \frac{\Delta p_T}{p_T} $ as a function of the simulated $ \eta $ for muon tracks generated with the FastSim in the real magnetic field with a simulated \pT{} of 5 \GeVc{}, and reconstructed with the standard Kalman filter using the SimHits in GE1/1, ME1/1, and ME2/x.}
					\label{fig:kalman_filter__standard_reco_pt_simhits_RT_10GeV_GEM_ME12_vs_eta}
				\end{figure}

				By looking at the SimHits we can gauge the impact of the magnetic field without perturbations from the segmentation. Figure \ref{fig:kalman_filter__standard_reco_pt_simhits_RT_10GeV_GEM_ME12_vs_eta} is a density plot of $ \frac{\Delta p_T}{p_T} $ as a function of the simulated $ \eta $ for muon tracks generated with the FastSim in the real magnetic field with a simulated \pT{} of 5 \GeVc{}, and reconstructed with the standard Kalman filter using the SimHits in GE1/1, ME1/1, and ME2/x. We observe that the results are centered around $ \frac{\Delta p_T}{p_T} = $ -0.5. A similar deviation around -40\% at 10 \GeVc{} has been noted in Section \ref{sec:least_squares_fit__real_magnetic_field_bias_detectors_setups} for the Least Squares fit using the SimHits generated with the FastSim in the real magnetic field.

			\subsubsection{Impact of Segmentation}
			\label{sec:kalman_filter__standard_real_field_impact_of_segmentation}	

				The impact of the segmentation on the results is similar to the one observed in the constant magnetic field. Although less visible than for the Least Squares fit, 6 sectors are still present, corresponding to the different $ \eta $ segments of GE1/1. 
			
			\subsubsection{Standard Deviation for Different Detector Setups}
			\label{sec:kalman_filter__standard_real_field_standard_deviation_detectors_setup}	

\begin{figure}[h!]
					\centering
					\includegraphics[width = \textwidth]{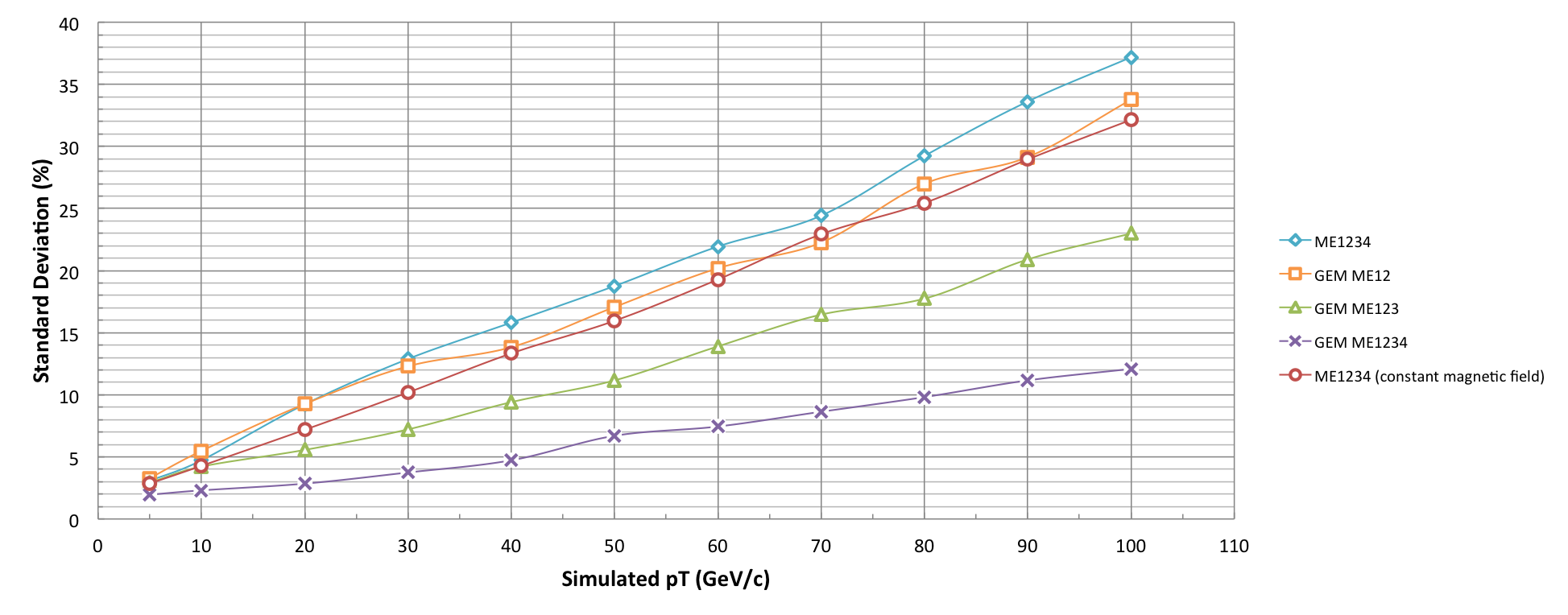}
					\caption{Standard deviation on $ \frac{\Delta p_T}{p_T} $ as a function of the simulated \pT{} for muon tracks generated with the FastSim in the real magnetic field, and reconstructed with the standard Kalman filter using the RecHits in multiple detector setups. The blue curve uses RecHits in ME1/1, ME2/x, ME3/x, and ME4/x, the orange curve in GE1/1, ME1/1, and ME2/x, the green curve in GE1/1, ME1/1, ME2/x, and ME3/x, the purple curve in GE1/1, ME1/1, ME2/x, ME3/x, and ME4/x, and the red curve in ME1/1, ME2/x, ME3/x, and ME4/x generated with the FastSim in the constant magnetic field.}
					\label{fig:kalman_filter__standard_reco_sigma_rechits_RT_all_setups}
				\end{figure}

				Figure \ref{fig:kalman_filter__standard_reco_sigma_rechits_RT_all_setups} shows the standard deviation on $ \frac{\Delta p_T}{p_T} $ as a function of the simulated \pT{} for muon tracks generated with the FastSim in the real magnetic field, and reconstructed with the standard Kalman filter using the RecHits in multiple detector setups. We observe that setups using GEM detectors (orange, green, and purple) offer an improved resolution than CSCs-only (blue) (between 69\% improvement at 30 \GeVc{} and 70\% at 100 \GeVc{}), even at lower \pT{}, where the impact of segmentation degrades the results yielded by GEM detectors. \\

				Moreover, as previously noted, the resolution degrades at higher \pT{} where the trajectories in the transverse plane are shorter and tracks straighter, and setups with more hits yield a better standard deviation. \\

				Finally, a degradation of the standard deviation is also observed when comparing the results obtained with the RecHits generated with the FastSim in the constant magnetic field (red), which is expected due to the complexer magnetic field.
			
			\subsubsection{Bias for Different Detector Setups}
			\label{sec:kalman_filter__standard_real_field_bias_detectors_setup}	

\begin{figure}[h!]
					\centering
					\includegraphics[width = \textwidth]{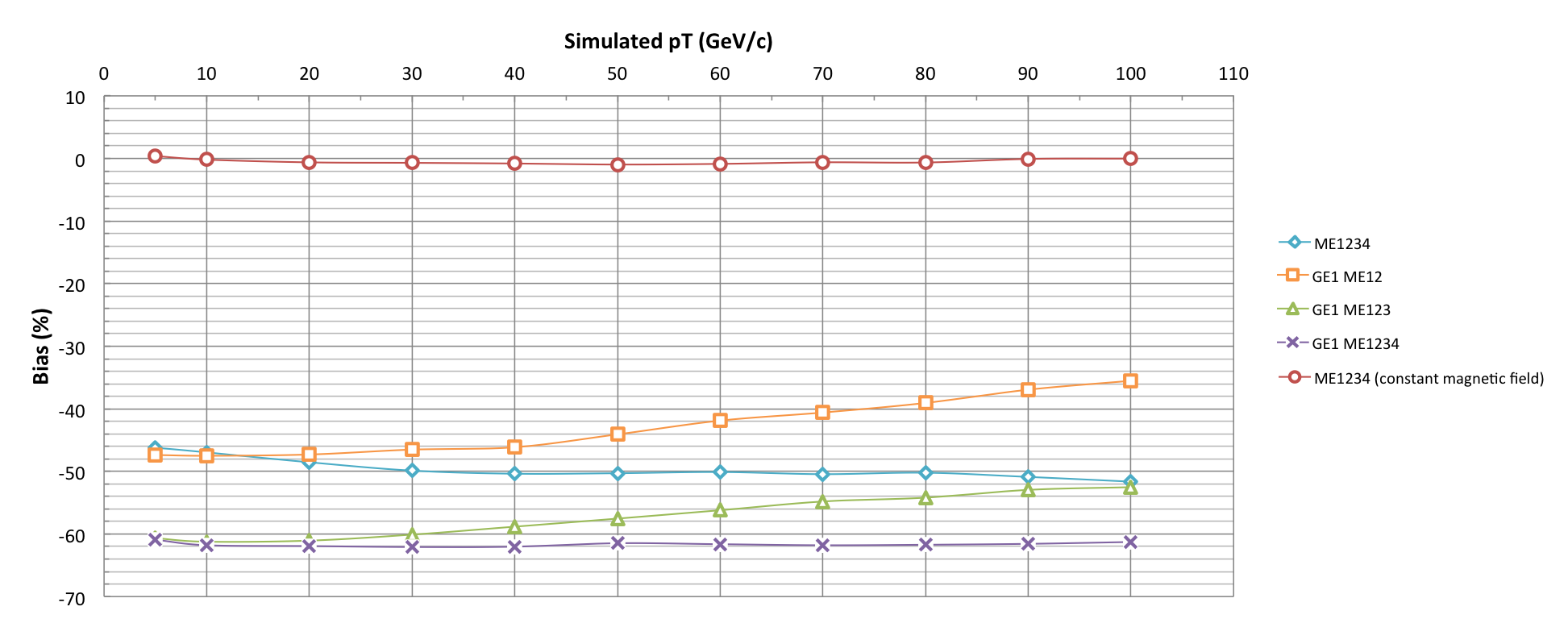}
					\caption{Bias on $ \frac{\Delta p_T}{p_T} $ as a function of the simulated \pT{} for muon tracks generated with the FastSim in the real magnetic field, and reconstructed with the standard Kalman filter using the RecHits in multiple detector setups. The blue curve uses RecHits in ME1/1, ME2/x, ME3/x, and ME4/x, the orange curve in GE1/1, ME1/1, and ME2/x, the green curve in GE1/1, ME1/1, ME2/x, and ME3/x, and the purple curve in GE1/1, ME1/1, ME2/x, ME3/x, and ME4/x, and the red curve in ME1/1, ME2/x, ME3/x, and ME4/x generated with the FastSim in the constant magnetic field.}
					\label{fig:kalman_filter__standard_mu_rechits_RT_all_setups}
				\end{figure}

				The standard Kalman filter uses a constant magnetic field intensity to reconstruct particles' tracks. Therefore, we expect a large bias, as previously seen with the Least Squares fit. Figure \ref{fig:kalman_filter__standard_mu_rechits_RT_all_setups} depicts the bias on $ \frac{\Delta p_T}{p_T} $ as a function of the simulated \pT{} for muon tracks generated with the FastSim in the real magnetic field, and reconstructed with the standard Kalman filter using the RecHits in multiple detector setups. As anticipated, we observe a bias between -35\% and -60\%. However, using GE1/1 and only two layers of CSCs (orange) offers better results than the CSCs in standalone (blue) (8 to 31\% improvement between 40 and 100 \GeVc), as GE1/1 lies in a region where the magnetic field is still uniform. \\

				To explain the behavior of the bias of the different detector setups, we have to understand the seeding process. To obtain a rough estimation of the state vector for the Kalman filter at the first measurement site, we use three hits to estimate $ \phi_0 $, $ \kappa $, and $ \tan \lambda $ (the two other parameters are set to zero). This initial estimation is of importance as the algorithm encounters difficulties in the non-uniform magnetic field to update the state and improve the track parameters. The initial estimation is then propagated backwards from the most outer stations to the IP. Because of the larger error on $ \eta $ in the GEMs, GE1/1 is not used for the initial approximation of the parameters. Then the two most natural choices for the initial parameter estimation are: either to consider the hits in the three most outer stations, or to use the hit in the most outer station, the hit in ME1/1, and the IP. The first choice yields an estimation that is closer to the trajectory in the outer stations. The algorithm can thus efficiently reconstruct the track backwards. However, the used detectors are all placed in a non-uniform field increasing the bias on \pT{} when propagating the state assuming a constant magnetic field. The second choice takes advantage of the fact that the IP and ME1/1 lay in a region where the field is still uniform. On the other hand, the initial state vector does not match very well the track during the first iteration happening in the non-uniform magnetic field. This confuses the filter and increases the standard deviation on \pT{}. \\

				In the GE1/1, ME1/1, and ME2/x setup, there is no other choice than using the IP, ME1/1, and ME2/x for the initial parameter estimation. For the other setups, we chose to use the hits in the three most outer stations to compute the initial parameters as it yields the best standard deviation. 
			
			\subsubsection{Evolution with $ \eta $}
			\label{sec:kalman_filter__standard_real_field_evolution_eta}		

\begin{figure}[h!]
					\centering
					\includegraphics[width = \textwidth]{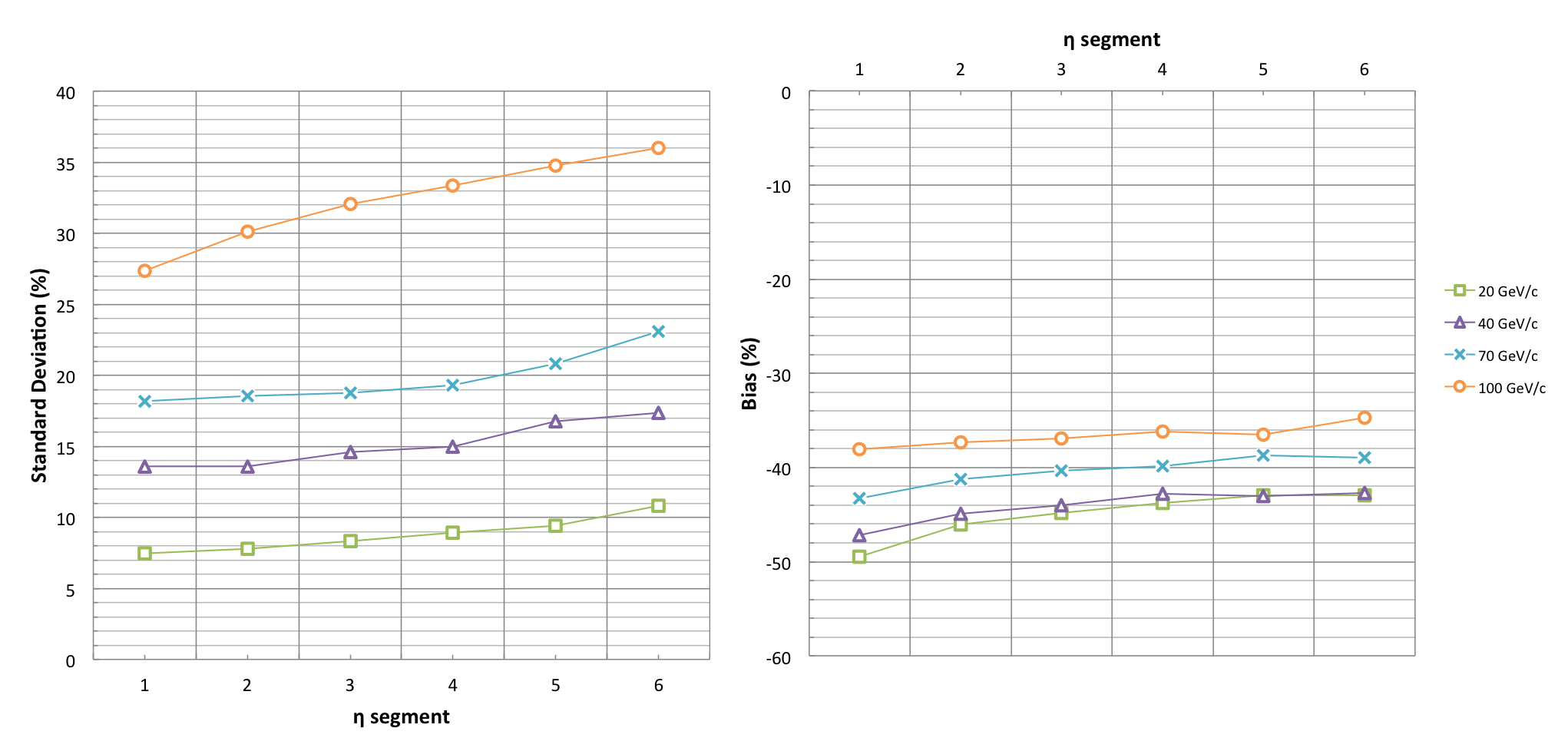}
					\caption{Standard deviation (left) and bias (right) on $ \frac{\Delta p_T}{p_T} $ as a function of the hit $ \eta $ segment in GE1/1 for muon tracks generated with the FastSim in the real magnetic field with simulated \pT{} of 20, 40, 70, and 100 \GeVc{}, and reconstructed with the standard Kalman filter using the RecHits in GE1/1, ME1/1, and ME2/x.}
					\label{fig:kalman_filter__standard_sigma_mu_rechits_RT_all_setups_evolution_eta}
				\end{figure}

				The evolution of the standard deviation and the bias with $ \eta $ follows the same trends as previously observed for the Least Squared fit in Section \ref{sec:least_squares_fit__real_magnetic_field_evolution_eta}. Figure \ref{fig:kalman_filter__standard_sigma_mu_rechits_RT_all_setups_evolution_eta} depicts the standard deviation (left) and bias (right) on $ \frac{\Delta p_T}{p_T} $ as a function of the hit $ \eta $ segment in GE1/1 for muon tracks generated with the FastSim in the real magnetic field with simulated \pT{} of 20, 40, 70, and 100 \GeVc{}, and reconstructed with the standard Kalman filter using the RecHits in GE1/1, ME1/1, and ME2/x. The standard deviation increases with $ \eta $ due to shorter tracks in the transverse plane, while the bias diminishes because hits are placed on a trajectory that is closer to the ideal track.

			\subsubsection{Summary}
			\label{sec:kalman_filter__standard_real_field_summary}		

				As for the Least Squares fit, using the real magnetic field biases the results (bias of the order of -50\%) as the reconstruction is performed supposing the field remains constant along all the trajectory. Moreover, the segmentation remains visible when looking at the RecHits generated with the FastSim in the real magnetic field. \\

				Further, setups using GEM detectors offer a better resolution than CSCs in standalone over the entire range of simulated \pT{}. An improvement of the order of 70\% above 30 \GeVc{} is observed when adding GEM detectors to the CSCs. \\

				Finally, we also observe that the first estimation of the parameters is of importance and explains the poorer results of the GE1/1, ME1/1, ME2/x, and ME3/x, and the GE1/1, ME1/1, ME2/x, ME3/x, and ME4/x setups. Finally, the GE1/1, ME1/1, and ME2/x setup yields the smallest bias which is still around -40\% but brings an improvement of 31\% at 100 \GeVc{} when compared to the CSC-only setup.
		
		\subsection{CMSSW Results}
		\label{sec:kalman_filter__standard_cmssw_results}	

			As stated in the description of the algorithm, this implementation of the Kalman filter does not take into account physical processes. This will play an important role at lower \pT{} where multiple scattering is significant.

			\subsubsection{Standard Kalman Filter with CMSSW SimHits}
			\label{sec:kalman_filter__standard_cmssw_validation_using_simhits}	

\begin{figure}[h!]
					\centering
					\includegraphics[width = 0.7 \textwidth]{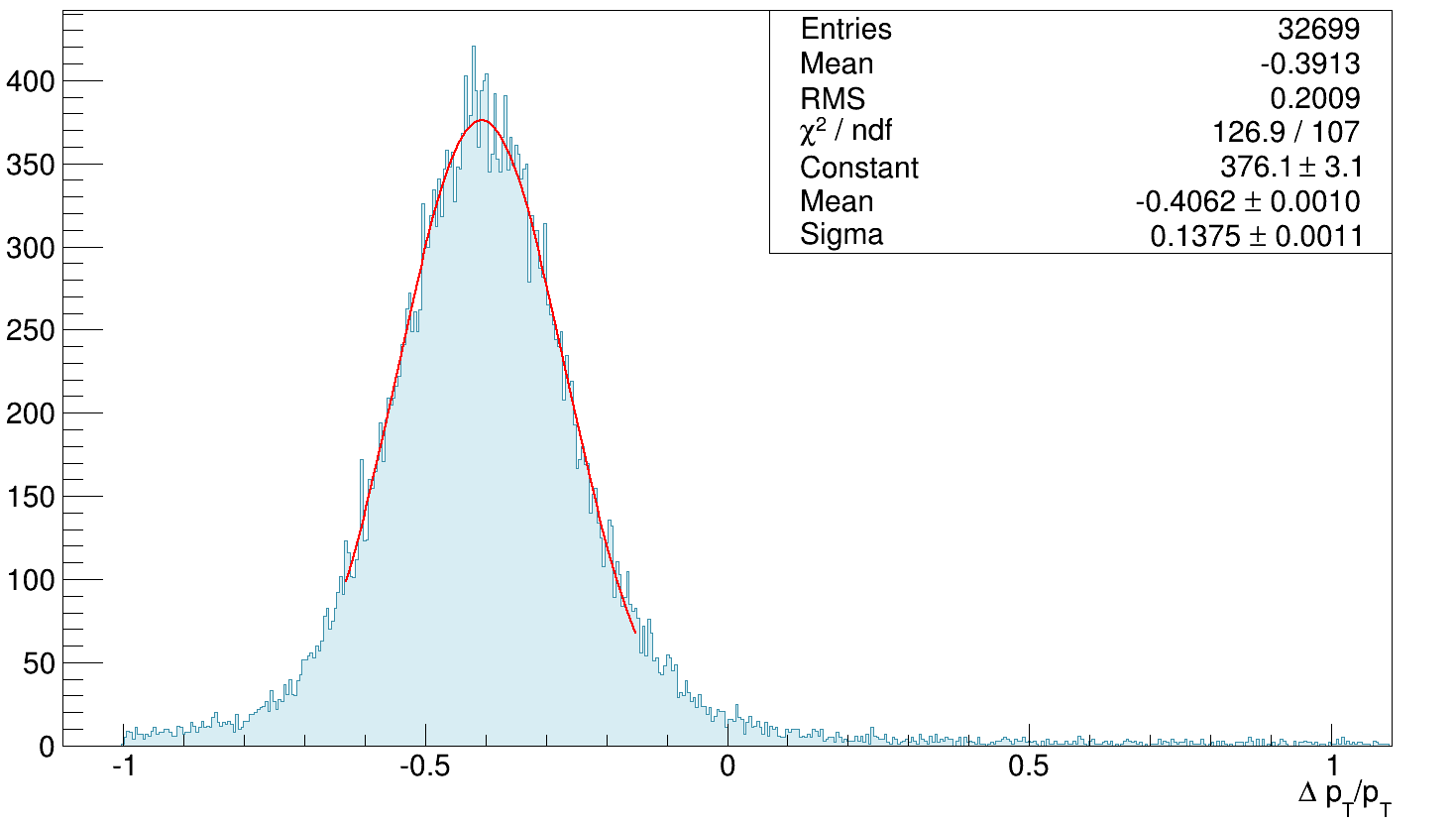}
					\caption{$ \frac{\Delta p_T}{p_T} $ for muon tracks generated with CMSSW with all a simulated \pT{} of 30 \GeVc{}, and reconstructed with the standard Kalman filter using the SimHits in GE1/1, ME1/1, and ME2/x.}
					\label{fig:kalman_filter__standard_reco_pt_simhits_CMSSW_30GeV_GEM_ME12}
				\end{figure}

				Figure \ref{fig:kalman_filter__standard_reco_pt_simhits_CMSSW_30GeV_GEM_ME12} represents $ \frac{\Delta p_T}{p_T} $ for muon tracks generated with CMSSW with all a simulated \pT{} of 30 \GeVc{}, and reconstructed with the standard Kalman filter using the SimHits in GE1/1, ME1/1, and ME2/x. We observe that the results remain usable, hence the algorithm is capable of reconstructing tracks, and notice the presence of the expected bias due to the non-uniform magnetic field.

			\subsubsection{Impact of Segmentation}
			\label{sec:kalman_filter__standard_cmssw_impact_of_segmentation}	

				As for the Least Squares fit, multiple scattering and energy losses completely dominate the impact of the $ \eta $ segmentation on the results. 

			\subsubsection{Standard Deviation for Different Detector Setups}
			\label{sec:kalman_filter__standard_cmssw_standard_deviation_detectors_setup}	

\begin{figure}[h!]
					\centering
					\includegraphics[width = \textwidth]{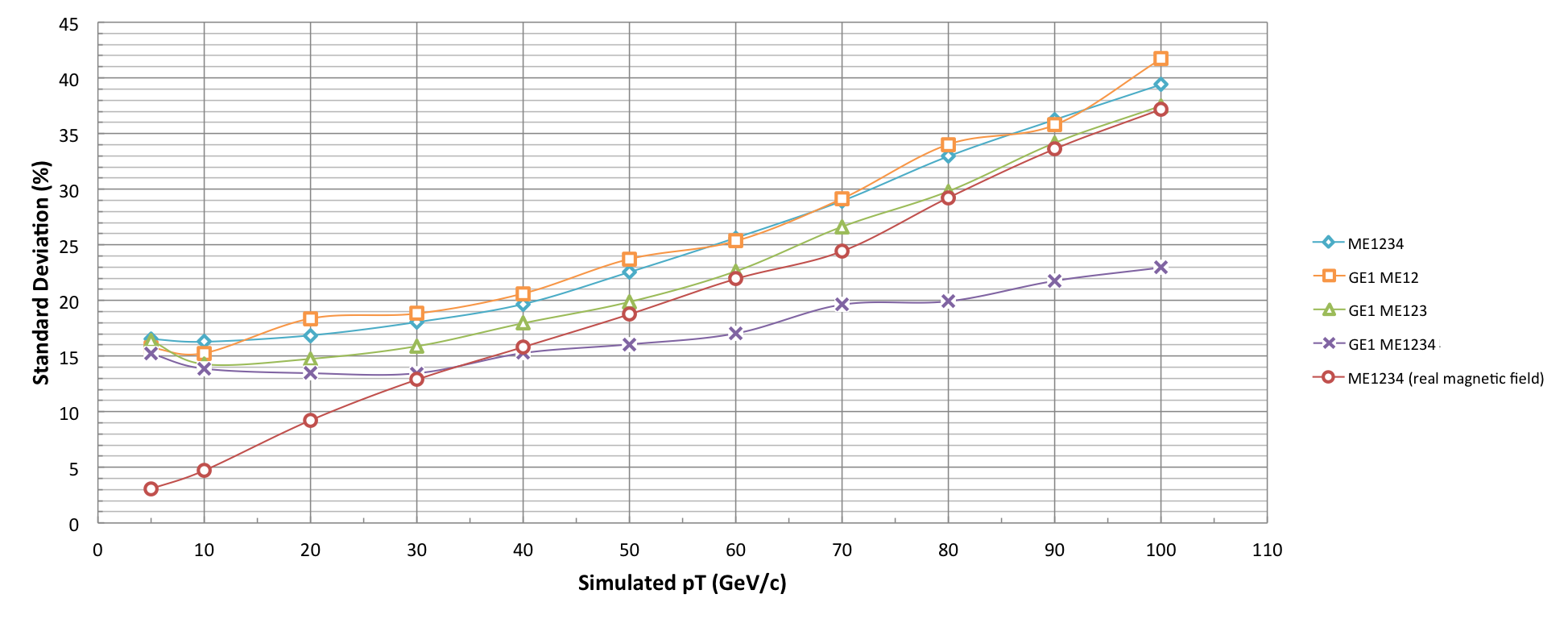}
					\caption{Standard deviation on $ \frac{\Delta p_T}{p_T} $ as a function of the simulated \pT{} for muon tracks generated with CMSSW, and reconstructed with the standard Kalman filter using the RecHits in multiple detector setups. The blue curve uses RecHits in ME1/1, ME2/x, ME3/x, and ME4/x, the orange curve in GE1/1, ME1/1, and ME2/x, the green curve in GE1/1, ME1/1, ME2/x, and ME3/x, and the purple curve in GE1/1, ME1/1, ME2/x, ME3/x, and ME4/x, and the red curve in ME1/1, ME2/x, ME3/x, and ME4/x generated with the FastSim in the real magnetic field.}
					\label{fig:kalman_filter__standard_reco_sigma_rechits_CMSSW_all_setups}
				\end{figure}

				We consider the standard deviation. Figure \ref{fig:kalman_filter__standard_reco_sigma_rechits_CMSSW_all_setups} shows the standard deviation on $ \frac{\Delta p_T}{p_T} $ as a function of the simulated \pT{} for muon tracks generated with CMSSW, and reconstructed with the standard Kalman filter using the RecHits in multiple detector setups. As mentioned in Section \ref{sec:kalman_filter__standard_real_field_bias_detectors_setup}, multiple choices for the estimations of the initial parameters are available. We chose to use the seed yielding the best standard deviation.

 				\pagebreak

				Considering GEM detectors and all the CSC layers (purple) improves the results compared to the CSCs in standalone (blue) over the entire range of simulated \pT{}, and by up to 40\% at 100 \GeVc{}. While the trends remain the same as those observed for the FastSim RecHits in the real magnetic field, we notice that the CSCs' resolution (blue) is poorer by a few \% than the one obtained with the RecHits generated with the FastSim in the real magnetic field (red). The resolution degrades even more at low \pT{}, below 30 \GeVc{}. This is due to multiple scattering and energy losses, and is also present for the other setups.
			
			\subsubsection{Bias for Different Detector Setups}
			\label{sec:kalman_filter__standard_cmssw_bias_detectors_setup}	

\begin{figure}[h!]
					\centering
					\includegraphics[width = \textwidth]{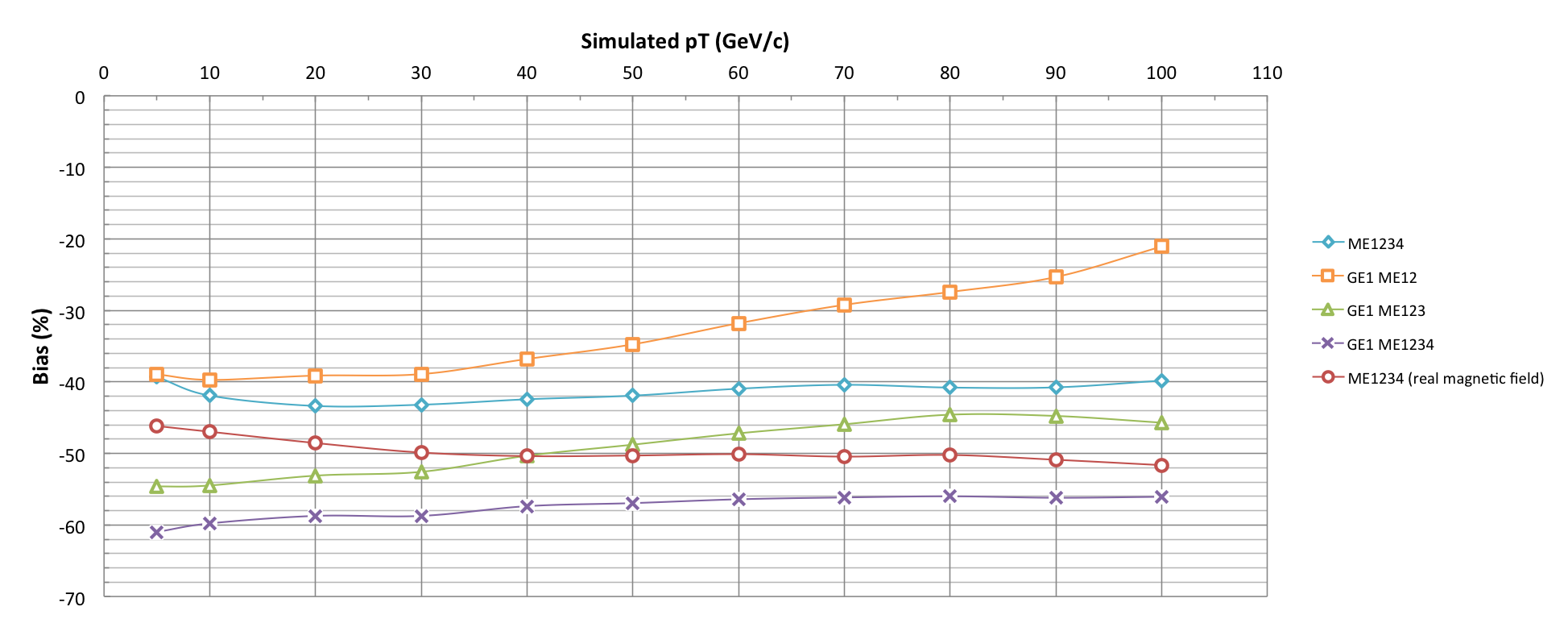}
					\caption{Bias on $ \frac{\Delta p_T}{p_T} $ as a function of the simulated \pT{} for muon tracks generated with CMSSW, and reconstructed with the standard Kalman filter using the RecHits in multiple detector setups. The blue curve uses RecHits in ME1/1, ME2/x, ME3/x, and ME4/x, the orange curve in GE1/1, ME1/1, and ME2/x, the green curve in GE1/1, ME1/1, ME2/x, and ME3/x, and the purple curve in GE1/1, ME1/1, ME2/x, ME3/x, and ME4/x, and the red curve in ME1/1, ME2/x, ME3/x, and ME4/x generated with the FastSim in the real magnetic field.}
					\label{fig:kalman_filter__standard_mu_rechits_CMSSW_all_setups}
				\end{figure}

				After looking at the standard deviation, we study the bias on the results. Figure \ref{fig:kalman_filter__standard_mu_rechits_CMSSW_all_setups} represents the bias on $ \frac{\Delta p_T}{p_T} $ as a function of the simulated \pT{} for muon tracks generated with CMSSW, and reconstructed with the standard Kalman filter using the RecHits in multiple detector setups. Once again, the trends are comparable with those observed with the RecHits generated with the FastSim in the real magnetic field, namely that the bias degrades when adding more CSC layers to the setups. The GE1/1, ME1/1, and ME2/x setup (orange) still yields the best results and improves the bias when compared against the actual system of CSCs (blue) (17\% improvement at 50 \GeVc{} and 47\% at 100 \GeVc{}). \\

				We also observe a slight difference between the CSCs in standalone in CMSSW (blue) and in the FastSim in the real magnetic field (red) due to the transition from one simulation environment to another. The differences between the two frameworks have been reviewed in Section \ref{sec:simulation_environment__fast_simulation}.

			\subsubsection{Evolution with $ \eta $}
			\label{sec:kalman_filter__standard_cmssw_evolution_eta}		

\begin{figure}[h!]
					\centering
					\includegraphics[width = \textwidth]{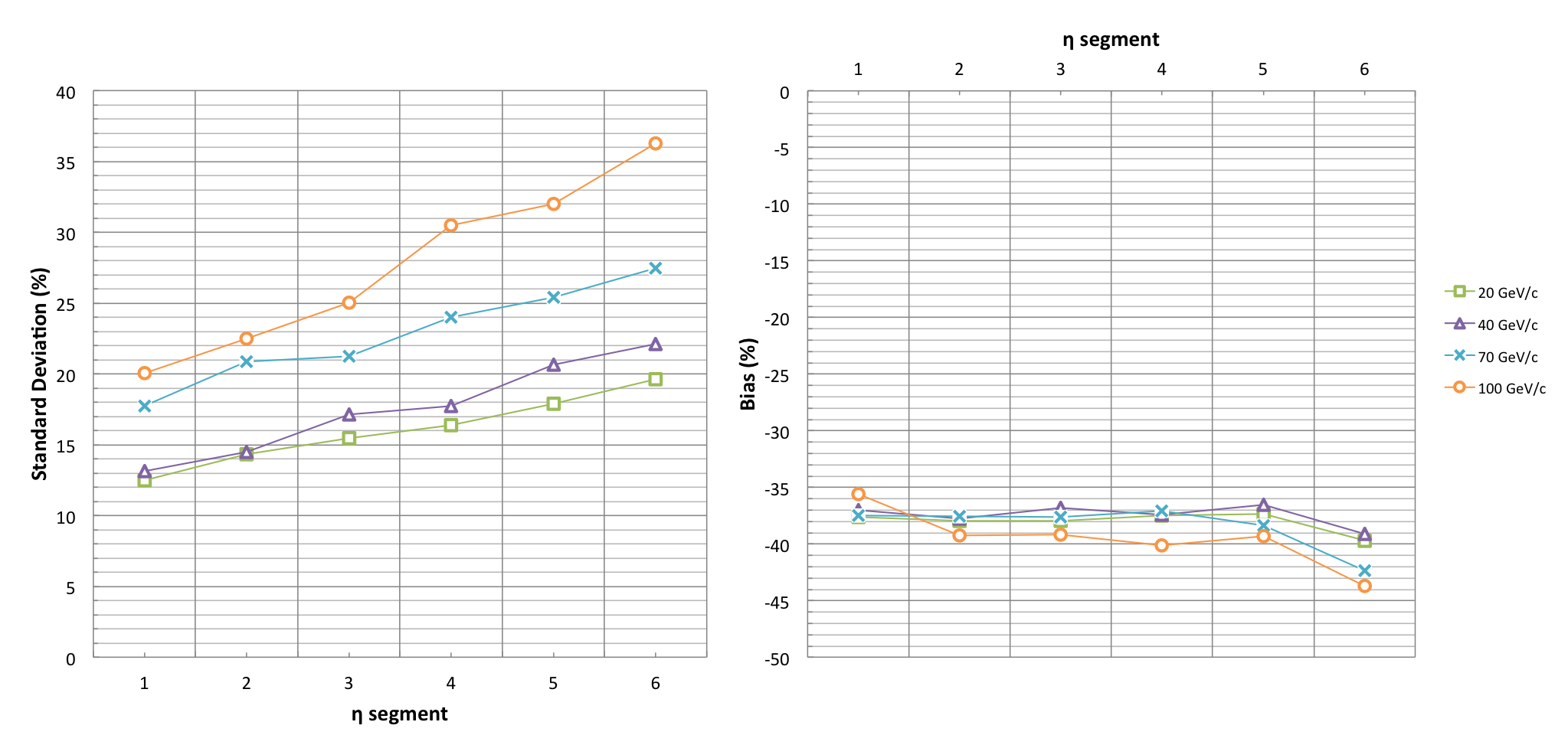}
					\caption{Standard deviation (left) and bias (right) on $ \frac{\Delta p_T}{p_T} $ as a function of the hit $ \eta $ segment in GE1/1 for muon tracks generated with CMSSW with simulated \pT{} of 20, 40, 70, and 100 \GeVc{}, and reconstructed with the standard Kalman filter using the RecHits in GE1/1, ME1/1, and ME2/x.}
					\label{fig:kalman_filter__standard_rechits_CMSSW_GEM_ME12_evolution_eta}
				\end{figure}

				Finally, we analyze the evolution of the parameters with $ \eta $. Figure \ref{fig:kalman_filter__standard_rechits_CMSSW_GEM_ME12_evolution_eta} depicts the standard deviation (left) and bias (right) on $ \frac{\Delta p_T}{p_T} $ as a function of the hit $ \eta $ segment in GE1/1 for muon tracks generated with CMSSW with simulated \pT{} of 20, 40, 70, and 100 \GeVc{}, and reconstructed with the standard Kalman filter using the RecHits in GE1/1, ME1/1, and ME2/x. As with the other simulation environments, we observe a deterioration of the standard deviation at higher $ \eta $, while the bias diminishes. This improvement is more significant at higher \pT{} where tracks are even closer to the trajectories described in the constant magnetic field.

			\subsubsection{Summary}
			\label{sec:kalman_filter__standard_real_field_summary}	

				The standard Kalman filter returns valid results in CMSSW with SimHits even though a large bias of the order of -40\% is observed. Multiple scattering and energy losses completely dominate the effect of the $ \eta $ segmentation on the results. \\

				Adding the GEM detectors to the current system of CSCs brings a large improvement on the resolution, by up to 40\% at 100 \GeVc{} even though multiple scattering and energy losses degrade the performances at lower \pT{} where they are significant.  \\

				While the GE1/1, ME1/1, and ME2/x setup yields a standard deviation of the order of the one returned by the CSCs in standalone, it offers a smaller bias which is improved by 17 to 47\% between 50 and 100 \GeVc{}. \\

				Finally, as with the other simulation environments, a degradation of the standard deviation is observed at higher $ \eta $, while the bias improves.
	
	\section{Runge-Kutta Propagator}
	\label{sec:kalman_filter__runge_kutta_propagator}	

		To take into account the non-uniformity of the magnetic field, we developed a modified Kalman Filter that uses a Runge-Kutta propagator instead of Equation \ref{eq:kalman_filter__parameter_extrapolation}. \\

		From an initial approximation of the parameters, we calculate the position of the particle using Equation \ref{eq:kalman_filter__propagated_position}, and the momentum using the following relation
		\begin{equation}
			\mathbf{p} = \left( \begin{array}{ccc} - \frac{Q}{| \kappa |} \cos \phi_0 & \frac{Q}{| \kappa |} \cos \phi_0 & \frac{Q}{| \kappa |} \tan \lambda \end{array} \right)^\intercal \ ,
		\end{equation}
		where $ Q $ is the charge number. Those are propagated towards the next detectors using the Runge-Kutta equations describes in Section \ref{sec:simulation_environment__fs_generation_and_propagation}. The extrapolated state vector at the new measurement site is given, in function of the old parameters and the extrapolated momentum $ \mathbf{p}' $, by
		\begin{equation}
			\mathbf{a}^k_{k-1} = \left( \begin{array}{ccccc} d_\rho & \tan^{-1}\left( - \frac{p_x'}{p_y'} \right) & \frac{\sign \kappa}{p_T'} & d_z & \frac{p_z'}{p_T'} \end{array} \right)^\intercal \ .
		\end{equation} \\

		The $ \alpha $ parameter is also updated to match the \axis{Z} component of the magnetic field at the given position.
		\begin{equation}
			\alpha = \frac{1}{c B_z(\mathbf{x})}
		\end{equation} \\

		This modified version of the Kalman filter does not compute the propagated covariance matrix using the Runge-Kutta propagator as this tasks is complex and therefore not implemented in this first version. Instead, we use Equation \ref{eq:trigger_system_and_reconstruction_algorithms__propagated_covariance} in Section \ref{sec:trigger_system_and_reconstruction_algorithms__kalman_filter_classical_kalman_filter} to propagate the covariance, which means the method is not yet used at maximum capacity.
	
	\section{Modified Kalman Filter}
	\label{sec:kalman_filter__modified_kalman_filter}

		Developed to take into account the non-uniformity of the magnetic field of CMS, this algorithm has been tested only in the FastSim in the real magnetic field and in CMSSW.
		
		\subsection{FastSim Results: Real Magnetic Field}
		\label{sec:kalman_filter__modified_fastsim_results_real_magnetic_field}
			
			We start by analyzing the results of the RecHits generated with the FastSim in the real magnetic field to see how reconstruction performs without the physical processes of multiple scattering and energy losses.
			
			\subsubsection{Modified Kalman Filter with FastSim SimHits in the Real Magnetic Field}
			\label{sec:kalman_filter__modified_real_field_validation_using_simhits}	

\begin{figure}[h!]
					\centering
					\includegraphics[width = 0.7 \textwidth]{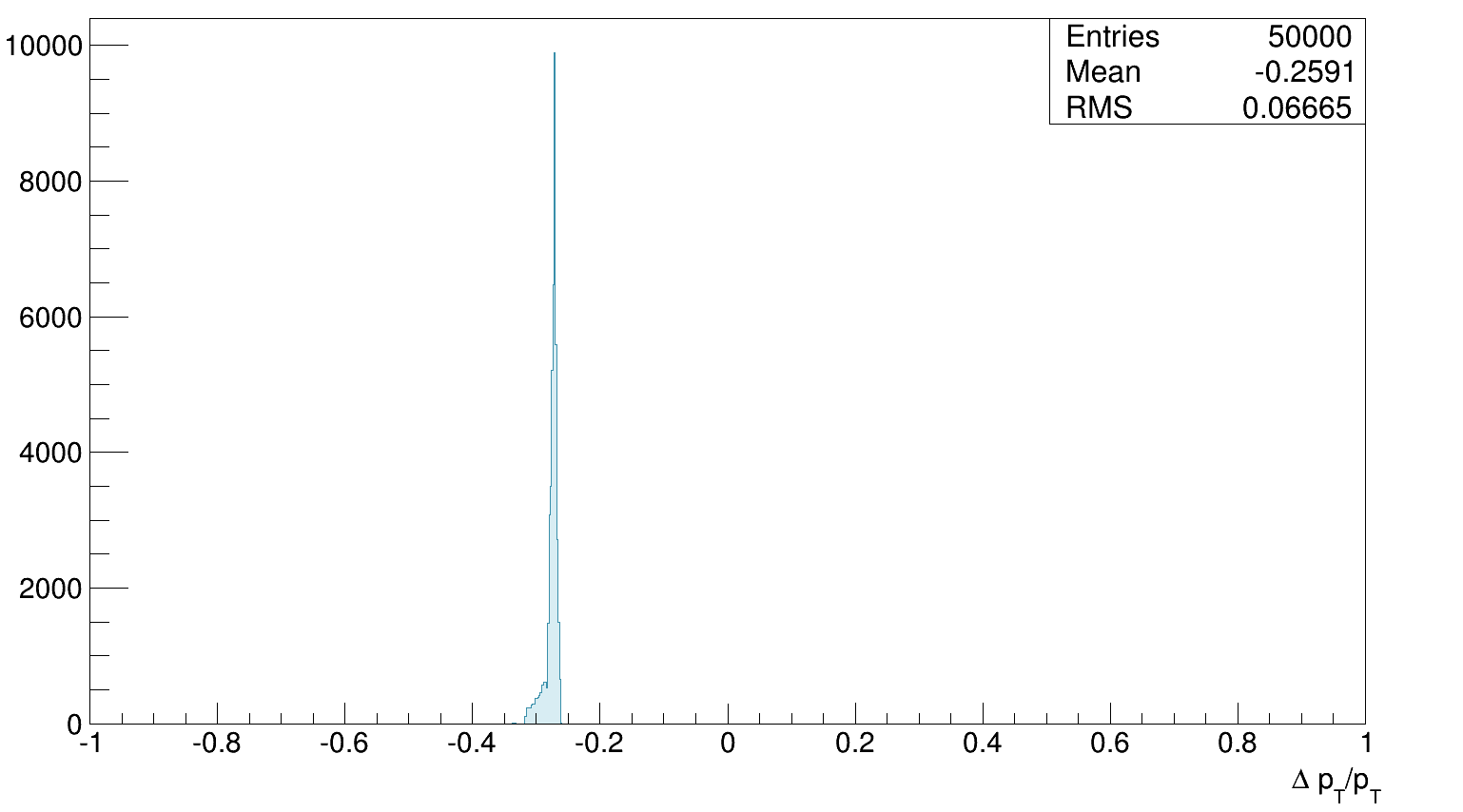}
					\caption{$ \frac{\Delta p_T}{p_T} $ for muon tracks generated with the FastSim in the real magnetic field with a simulated \pT{} of 60 \GeVc{}, and reconstructed with the modified Kalman filter using the SimHits in GE1/1, ME1/1, and ME2/x.}
					\label{fig:kalman_filter__modified_reco_pt_simhits_CMSSW_60GeV_GEM_ME12}
				\end{figure}

				We first consider the SimHits by looking at Figure \ref{fig:kalman_filter__modified_reco_pt_simhits_CMSSW_60GeV_GEM_ME12} which shows $ \frac{\Delta p_T}{p_T} $ for muon tracks generated with the FastSim in the real magnetic field with a simulated \pT{} of 60 \GeVc{}, and reconstructed with the modified Kalman filter using the SimHits in GE1/1, ME1/1, and ME2/x. We immediately notice that the bias (of the order of -28\%) is divided by a factor of two in comparison to the standard Kalman filter (of the order of -50\%) reviewed in Section \ref{sec:kalman_filter__standard_real_field_validation_using_simhits}.

			\subsubsection{Impact of Segmentation}
			\label{sec:kalman_filter__modified_real_field_impact_of_segmentation}	

				The modifications made on the Kalman filter do not mask the impact of segmentation on the RecHits generated with the FastSim in the real magnetic field. As before, the distribution of $ \frac{p_T}{p_T} $ along $ \eta $ is divided in six sectors corresponding to the six $ \eta $ segments of the GEM detectors.

			\subsubsection{Standard Deviation for Different Detector Setups}
			\label{sec:kalman_filter__modified_real_field_standard_deviation_detectors_setup}	

\begin{figure}[h!]
					\centering
					\includegraphics[width = \textwidth]{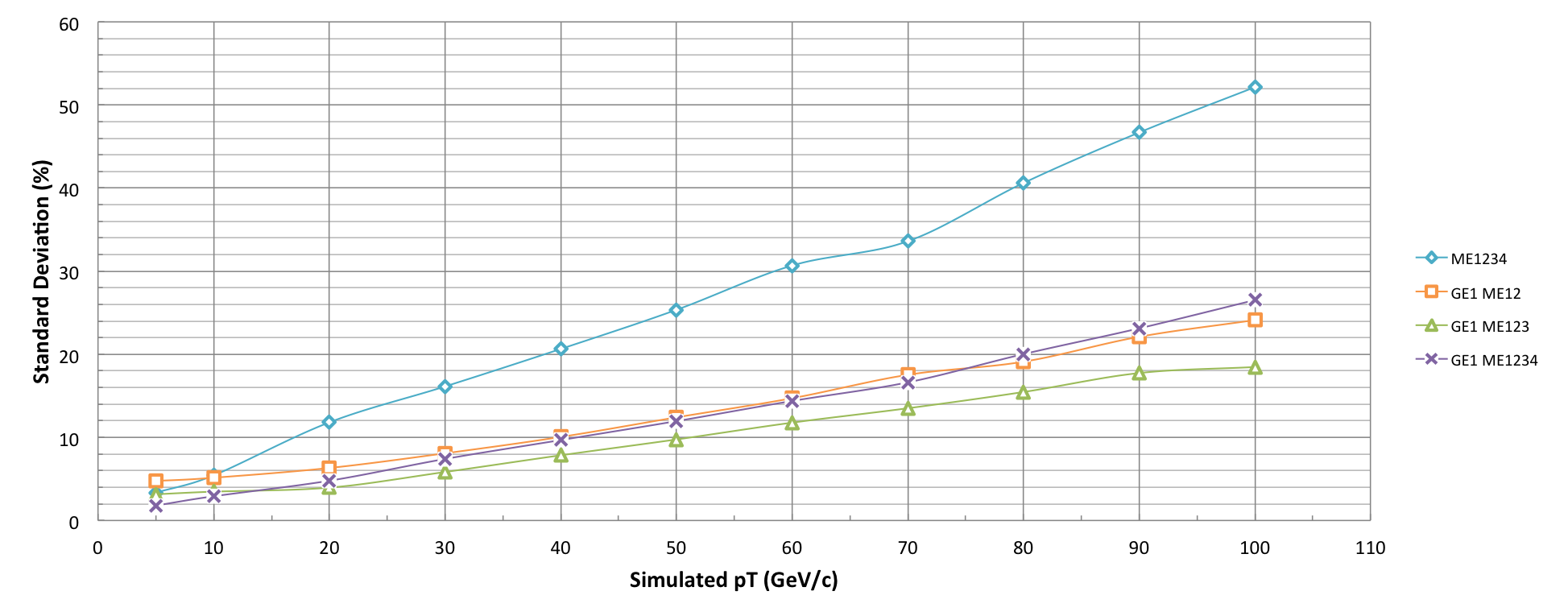}
					\caption{Standard deviation on $ \frac{\Delta p_T}{p_T} $ as a function of the simulated \pT{} for muon tracks generated with the FastSim in the real magnetic field, and reconstructed with the modified Kalman filter using the RecHits in multiple detector setups. The blue curve uses RecHits in ME1/1, ME2/x, ME3/x, and ME4/x, the orange curve in GE1/1, ME1/1, and ME2/x, the green curve in GE1/1, ME1/1, ME2/x, and ME3/x, and the purple curve in GE1/1, ME1/1, ME2/x, ME3/x, and ME4/x.}
					\label{fig:kalman_filter__modified_reco_sigma_rechits_RT_all_setups}
				\end{figure}

				Figure \ref{fig:kalman_filter__modified_reco_sigma_rechits_RT_all_setups} represents the standard deviation on $ \frac{\Delta p_T}{p_T} $ as a function of the simulated \pT{} for muon tracks generated with the FastSim in the real magnetic field, and reconstructed with the modified Kalman filter using the RecHits in multiple detector setups. The impact of segmentation at low \pT{} is almost negligible, allowing setups equipped with GEM detectors (orange, green, and purple) to yield a better resolution than the CSCs in standalone (blue). The GE1/1, ME1/1, and ME2/x setup (orange) brings an improvement of 52\% at 100 \GeVc{} compared to the actual setup (blue). \\ 

				If we compare those results to the standard Kalman filter for the RecHits generated with the FastSim in the real magnetic field studied in Figure \ref{fig:kalman_filter__standard_reco_sigma_rechits_RT_all_setups} in Section \ref{sec:kalman_filter__standard_real_field_standard_deviation_detectors_setup}, we notice that the standard deviation of the setups using ME4/x increases. This is due to the modified propagation process. The error on the initial estimation of the parameters of the helix used by the Kalman filter is large due to the fact that it is done in a region where the magnetic field is non-uniform. Therefore, both the initial position and momentum are badly estimated. These are then propagated using an iterative process which further increases the uncertainties. The initial estimations of the parameters done using ME1/1 which lies in a region where the field is still constant is therefore closer to the ones of the ideal track, which is the case for the GE1/1, ME1/1, and ME2/x setup (orange).
			
			\subsubsection{Bias for Different Detector Setups}
			\label{sec:kalman_filter__modified_real_field_bias_detectors_setup}	

\begin{figure}[h!]
					\centering
					\includegraphics[width = \textwidth]{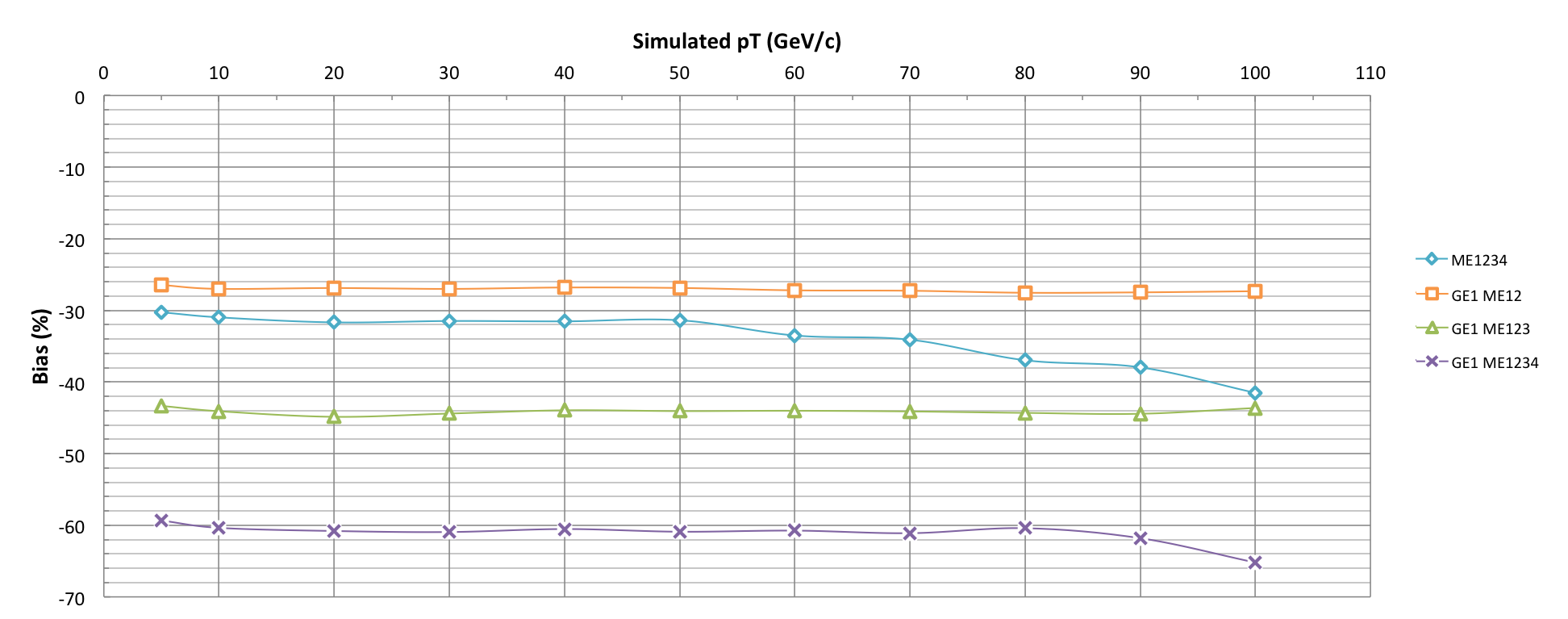}
					\caption{Bias on $ \frac{\Delta p_T}{p_T} $ as a function of the simulated \pT{} for muon tracks generated with the FastSim in the real magnetic field, and reconstructed with the modified Kalman filter using the RecHits in multiple detector setups. The blue curve uses RecHits in ME1/1, ME2/x, ME3/x, and ME4/x, the orange curve in GE1/1, ME1/1, and ME2/x, the green curve in GE1/1, ME1/1, ME2/x, and ME3/x, and the purple curve in GE1/1, ME1/1, ME2/x, ME3/x, and ME4/x.}
					\label{fig:kalman_filter__modified_mu_rechits_RT_all_setups}
				\end{figure}

				We then consider the bias on the results. Figure \ref{fig:kalman_filter__modified_mu_rechits_RT_all_setups} depicts the bias on $ \frac{\Delta p_T}{p_T} $ as a function of the simulated \pT{} for muon tracks generated with the FastSim in the real magnetic field, and reconstructed with the modified Kalman filter using the RecHits in multiple detector setups. As we stated in Section \ref{sec:kalman_filter__standard_real_field_bias_detectors_setup} regarding the bias for the standard Kalman filter in the real magnetic field, the larger observed biases for the GE1/1, ME1/1, ME2/x, and ME3/x (green), and the GE1/1, ME1/1, ME2/x, ME3/x, and ME4/x setups (purple) are caused by the detectors used to perform the initial estimation of the parameters. We decided to use the same seeds as with the standard Kalman filter which yields the best standard deviation but a poorer bias. However, the bias of the GE1/1, ME1/1, and ME2/x setup (orange) reconstructed by the modified Kalman filter is improved by 24\% compared to the standard Kalman filter shown in Figure \ref{fig:kalman_filter__standard_mu_rechits_RT_all_setups} in Section \ref{sec:kalman_filter__standard_real_field_bias_detectors_setup}. \\

				Furthermore, we notice that the GE1/1, ME1/1, and ME2/x setup (orange) performs better than the CSCs in standalone (blue), especially at high \pT{} (improvement between 12\% at 5 \GeVc and 34\% at 100 \GeVc). The degradation of the performances for the CSC setup is, as for the other setups, related to the initial estimation of the parameters. 	
			
			\subsubsection{Evolution with $ \eta $}
			\label{sec:kalman_filter__modified_real_field_evolution_eta}	

\begin{figure}[h!]
					\centering
					\includegraphics[width = \textwidth]{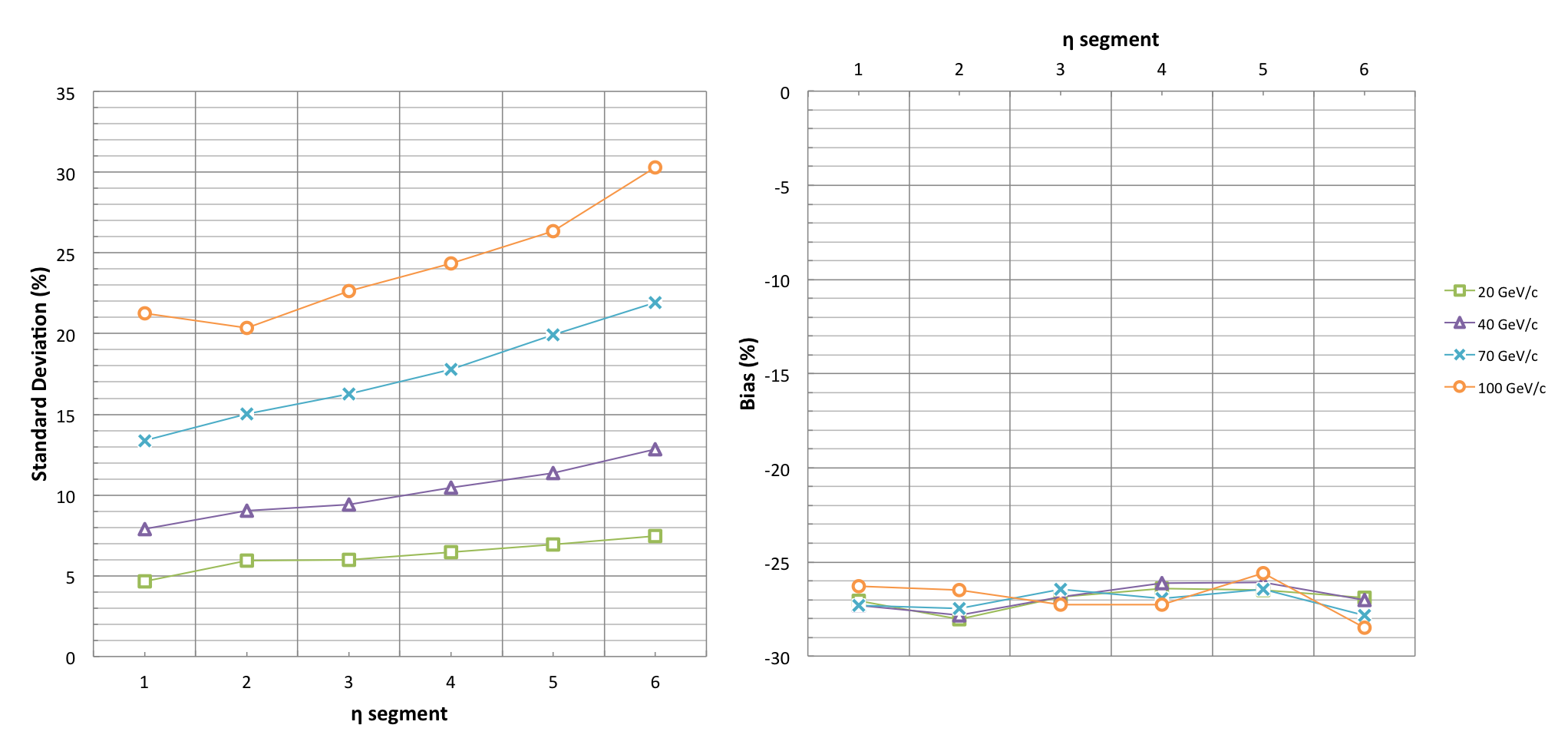}
					\caption{Standard deviation (left) and bias (right) on $ \frac{\Delta p_T}{p_T} $ as a function of the hit $ \eta $ segment in GE1/1 for muon tracks generated with the FastSim in the real magnetic field with simulated \pT{} of 20, 40, 70, and 100 \GeVc{}, and reconstructed with the modified Kalman filter using the RecHits in GE1/1, ME1/1, and ME2/x.}
					\label{fig:kalman_filter__modified_sigma_mu_rechits_RT_all_setups_evolution_eta}
				\end{figure}

				Finally, we look at Figure \ref{fig:kalman_filter__modified_sigma_mu_rechits_RT_all_setups_evolution_eta} which represents the standard deviation (left) and bias (right) on $ \frac{\Delta p_T}{p_T} $ as a function of the hit $ \eta $ segment in GE1/1 for muon tracks generated with the FastSim in the real magnetic field with simulated \pT{} of 20, 40, 70, and 100 \GeVc{}, and reconstructed with the modified Kalman filter using the RecHits in GE1/1, ME1/1, and ME2/x. As always, the standard deviation decreases at higher $ \eta $ due to shorter projected tracks in the transverse plane. However, contrary to the Least Squares fit and the standard Kalman filter, the bias remains unchanged along $ \eta $.

			\subsubsection{Summary}
			\label{sec:kalman_filter__modified_real_field_summary}		

				The modified Kalman filter yields a smaller bias than the standard Kalman filter for all the setups, although it does not significantly improves the GE1/1, ME1/1, ME2/x, and ME3/x and GE1/1, ME1/1, ME2/x, ME3/x, and ME4/x setups. For example, the bias of the GE1/1, ME1/1, and ME2/x setup is improved by an average of 24\% by the modified Kalman filter compared to the standard Kalman filter. \\ 

				The standard deviation of setups using station ME4/x is degraded. This is due to the errors made at the initial estimation of the track parameters which are propagated by the iterative Runge-Kutta propagator. However, the GEM detectors and only two stations of CSCs decrease the bias by up to 52\% at 100 \GeVc{} compared to the CSCs in standalone. \\ 

				The evolution of the standard deviation along $ \eta $ remains unchanged, namely it deteriorates due to smaller tracks in the transverse plane, while the bias stays constant.

		\subsection{CMSSW Results}
		\label{sec:kalman_filter__modified_cmssw_results}	

			We now test our algorithm with the CMSSW environment that includes the non-uniform magnetic field and physical processes.

			\subsubsection{Modified Kalman Filter with CMSSW SimHits}
			\label{sec:kalman_filter__modified_CMSSW_validation_using_simhits}	

				By looking at the reconstruction performed using the SimHits generated with CMSSW, we observe results similar to those encountered with the FastSim with the real magnetic field reviewed in Section \ref{sec:kalman_filter__modified_real_field_validation_using_simhits}. Namely, that the bias is still present (around -20\%) although smaller than with the standard Kalman filter (around -35\%).

			\subsubsection{Impact of Segmentation}
			\label{sec:kalman_filter__modified_CMSSW_impact_of_segmentation}	

				As with the other algorithms tested in CMSSW, multiple scattering and energy losses mask the impact of the $ \eta $ segmentation on the results as these processes largely dominate.
			
			\subsubsection{Standard Deviation for Different Detector Setups}
			\label{sec:kalman_filter__modified_CMSSW_standard_deviation_detectors_setup}	

\begin{figure}[h!]
					\centering
					\includegraphics[width = \textwidth]{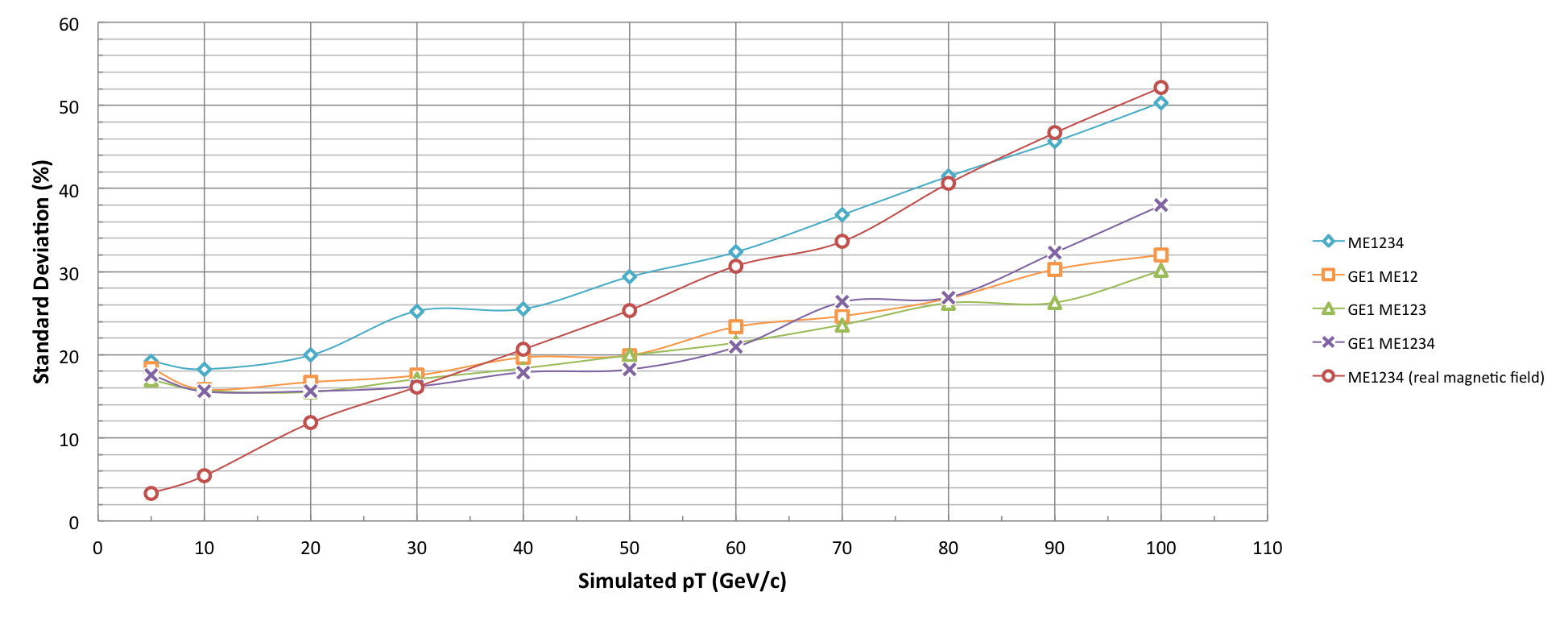}
					\caption{Standard deviation on $ \frac{\Delta p_T}{p_T} $ as a function of the simulated \pT{} for muon tracks generated with CMSSW, and reconstructed with the modified Kalman filter using the RecHits in multiple detector setups. The blue curve uses RecHits in ME1/1, ME2/x, ME3/x, and ME4/x, the orange curve in GE1/1, ME1/1, and ME2/x, the green curve in GE1/1, ME1/1, ME2/x, and ME3/x, and the purple curve in GE1/1, ME1/1, ME2/x, ME3/x, and ME4/x, and the red curve in ME1/1, ME2/x, ME3/x, and ME4/x generated with the FastSim in the real magnetic field.}
					\label{fig:kalman_filter__modified_reco_sigma_rechits_CMSSW_all_setups}
				\end{figure}

				The standard deviation on $ \frac{\Delta p_T}{p_T} $ as a function of the simulated \pT{} for muon tracks generated with CMSSW, and reconstructed with the modified Kalman filter using the RecHits in multiple detector setups is represented in Figure \ref{fig:kalman_filter__modified_reco_sigma_rechits_CMSSW_all_setups}. As with the standard Kalman filter, we observe a degradation of the performances at low \pT{} compared to the RecHits generated with the FastSim in the real magnetic field (red) due to multiple scattering and energy losses. We also notice that all setups using GEM detectors (orange, green, and purple) are more efficient than CSCs in standalone (blue). An improvement of 36\% at 100 \GeVc{} is observed between the GE1/1, ME1/1, and ME2/x setup (orange) and the actual system of CSCs (blue).
				
			\subsubsection{Bias for Different Detector Setups}
			\label{sec:kalman_filter__modified_CMSSW_bias_detectors_setup}	

\begin{figure}[h!]
					\centering
					\includegraphics[width = \textwidth]{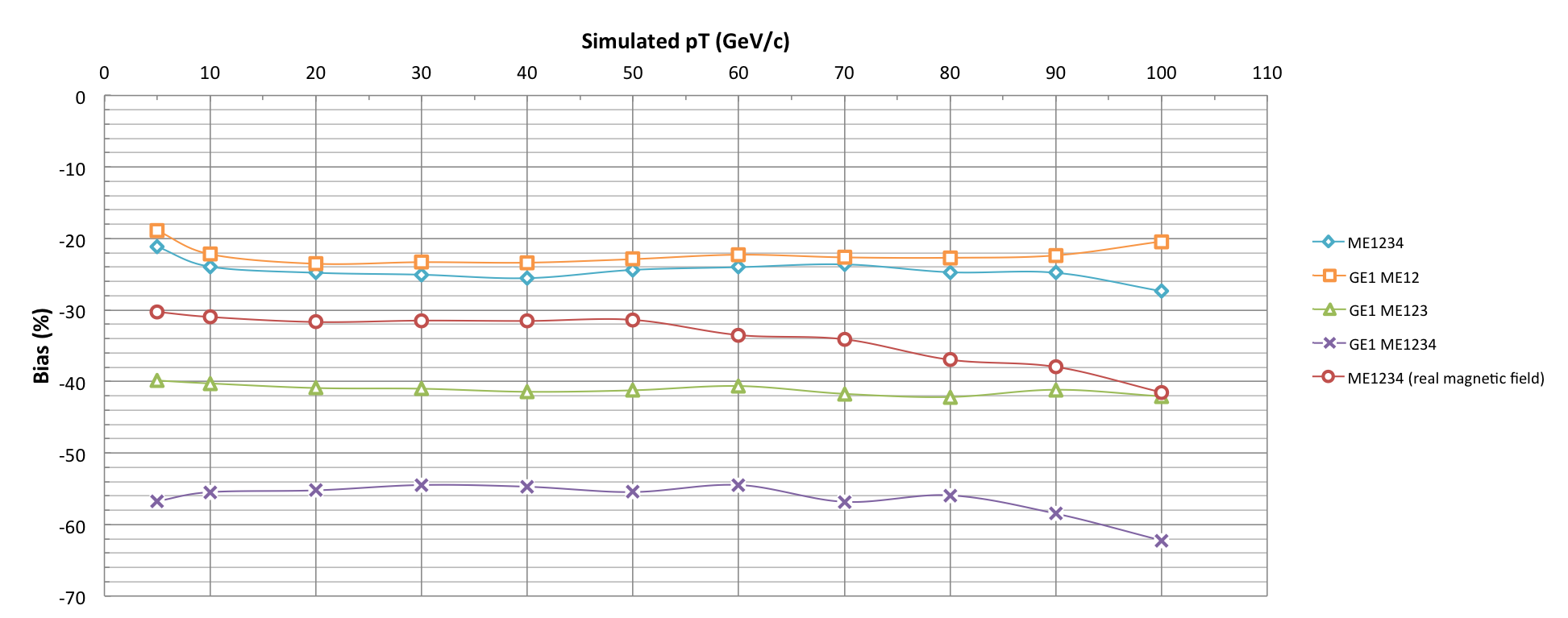}
					\caption{Bias on $ \frac{\Delta p_T}{p_T} $ as a function of the simulated \pT{} for muon tracks generated with CMSSW, and reconstructed with the modified Kalman filter using the RecHits in multiple detector setups. The blue curve uses RecHits in ME1/1, ME2/x, ME3/x, and ME4/x, the orange curve in GE1/1, ME1/1, and ME2/x, the green curve in GE1/1, ME1/1, ME2/x, and ME3/x, and the purple curve in GE1/1, ME1/1, ME2/x, ME3/x, and ME4/x, and the red curve in ME1/1, ME2/x, ME3/x, and ME4/x generated with the FastSim in the real magnetic field.}
					\label{fig:kalman_filter__modified_mu_rechits_CMSSW_all_setups}
				\end{figure}

				The observations that we can do on the bias also remain the same as with the standard Kalman filter. Figure \ref{fig:kalman_filter__modified_mu_rechits_CMSSW_all_setups} depicts the bias on $ \frac{\Delta p_T}{p_T} $ as a function of the simulated \pT{} for muon tracks generated with CMSSW, and reconstructed with the modified Kalman filter using the RecHits in multiple detector setups. We notice a slight difference between the CSCs in CMSSW (blue) and in the FastSim with the real magnetic field (red) due to the differences in the simulation environments as reviewed in Section \ref{sec:simulation_environment__fast_simulation}. \\

				We also observe that the GE1/1, ME1/1, and ME2/x setup (orange) yields better results than the CSCs in standalone (blue) (10 to 25\% improvement between 5 and 100 \GeVc) or than the two other setups using GEMs. Once again, this is due to the initial estimation of the parameters of the helix for the Kalman filter, as reviewed in Section \ref{sec:kalman_filter__standard_real_field_bias_detectors_setup}. 
			
			\subsubsection{Evolution with $ \eta $}
			\label{sec:kalman_filter__modified_CMSSW_evolution_eta}	

\begin{figure}[h!]
					\centering
					\includegraphics[width = \textwidth]{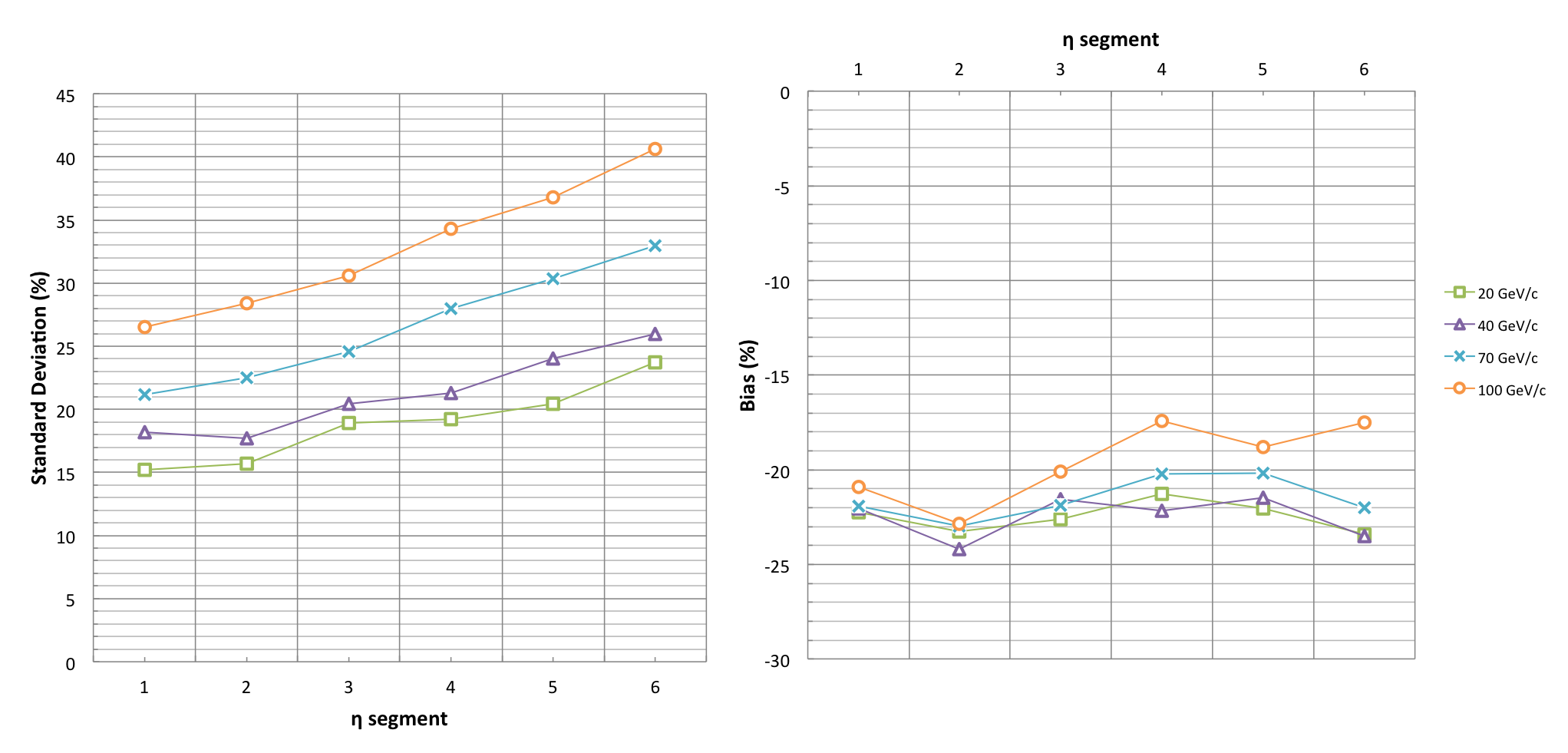}
					\caption{Standard deviation (left) and bias (right) on $ \frac{\Delta p_T}{p_T} $ as a function of the hit $ \eta $ segment in GE1/1 for muon tracks generated with CMSSW with all the simulated \pT{}, and reconstructed with the modified Kalman filter using the RecHits in GE1/1, ME1/1, and ME2/x.}
					\label{fig:kalman_filter__modified_sigma_mu_rechits_CMSSW_all_setups_evolution_eta}
				\end{figure}

				Finally, we look at the evolution of the parameters with $ \eta $. Figure \ref{fig:kalman_filter__modified_sigma_mu_rechits_CMSSW_all_setups_evolution_eta} represents the standard deviation (left) and bias (right) on $ \frac{\Delta p_T}{p_T} $ as a function of the hit $ \eta $ segment in GE1/1 for muon tracks generated with CMSSW with all the simulated \pT{}, and reconstructed with the modified Kalman filter using the RecHits in GE1/1, ME1/1, and ME2/x. As for most of the previously discussed simulation environments, the standard deviation degrades at higher $ \eta $ where the projection of the tracks in the transverse plane are shorter. On the other hand, the bias stays the same or slightly improves. We do not observe a clear improvement as with the standard Kalman filter. 

			\subsubsection{Summary}
			\label{sec:kalman_filter__modified_cmssw_summary}	

				From the SimHits, we observe that the bias on the reconstructed events using the modified Kalman filter is smaller than with the standard Kalman filter, respectively -20 and -35\%, improving the results by 43\%. The digitization process is dominated by multiple scattering and energy losses, which also degrades the resolution at lower \pT{}. \\

				The results produced using the GE1/1, ME1/1, and ME2/x setup yield a better resolution and bias than the CSCs in standalone, respectively improving these parameters by 30 and 25\% at 100 \GeVc{}. \\ 

				Finally, both the standard deviation and the bias vary with $ \eta $, respectively degrading and slightly improving at higher $ \eta $.

	\section{Conclusion}
	\label{sec:kalman_filter__conclusion}	

		We successfully implemented a Kalman filter to reconstruct helical tracks and to correct the bias returned by the Least Squares fit. \\

		The first version of the Kalman filter we used reconstructs tracks without considering the non-uniform magnetic field. This induces a large bias on the results for both the FastSim in the real magnetic field and the CMSSW environments of the order of -50\% when considering the SimHits. With this algorithm, the setup yielding the smallest bias is GE1/1, ME1/1, and ME2/x, which brings an improvement between 8 and 31\%  above 40 \GeVc{} compared to the CSCs in standalone in the FastSim with the real magnetic field. \\

		Setups using GEM detectors and more than two CSCs, even though they offered a poorer bias, yield an improved standard deviation. For example, in the FastSim with the real field, adding the RecHits in GE1/1 to the CSCs improves the resolution by 30\% above 30 \GeVc{}. The large bias these setups yield are due to the initial estimations of the track's parameters which are crucial for this algorithm. Before iterating on the various hits, the Kalman filter makes a rough estimation of the track's parameters by considering a small number of hits. In the non-uniform magnetic field, this estimation is significantly biased and the error propagates throughout the entire iterative process.  \\

		To improve those results, we developed our own modified Kalman filter which yielded a smaller bias (divided by a factor of two) than the standard Kalman filter. \\

		With this modified algorithm, we observe an improvement of the standard deviation of 36\% at 100 \GeVc{} when comparing the GE1/1, ME1/1, and ME2/x setup and the CSCs in standalone in CMSSW. Moreover, comparing the same setups, we observe an improvement between 10 and 25\% of the bias over the entire range of simulated \pT{} when using GEM detectors instead of CSCs. \\

		For both algorithms, the GE1/1, ME1/1, and ME2/x setup returns better results than the CSCs in standalone for both the bias and the standard deviation. \\

		In this version of the modified Kalman filter, we did not take into account multiple scattering or energy losses. By implementing these features we should be able to improve the standard deviation, especially in CMSSW. \\

		In the next chapter, we compare the different algorithms' results, their impact on the L1 Trigger, and their timing.

	\cleardoublepage


\chapter{Algorithms Performances, Impact on the L1 Trigger, and Prospects}
\label{chap:algorithms_performances_timing_prospects}

	In this chapter, we compare the results obtained using the algorithms described in the previous chapters. Furthermore, to give a reference point of the performances one could expect at the L1 Trigger, we describe one last algorithm which resembles to what is currently used to reconstruct tracks in the CSC L1 Trigger. Finally, we describe the guidelines for the implementations of the modified Kalman filter on programmable electronics.
	
	\section{Comparison of the Algorithms}
	\label{sec:algorithms_performances_prospects__comparison_of_the_algorithms}

		As the Kalman filters we implemented do not take into account the physical processes of multiple scattering and energy losses, we will compare the standard deviation and the bias for the RecHits generated with both the FastSim in the real magnetic field and CMSSW. The comparison will be done using the GE1/1, ME1/1, and ME2/x setup as we demonstrated that this setup is the most efficient.
		
		\subsection{FastSim Results: Real Magnetic Field}
		\label{sec:algorithms_performances_prospects__fastsim_results_real_magnetic_field}	

\begin{figure}[h!]
				\centering
				\includegraphics[width = \textwidth]{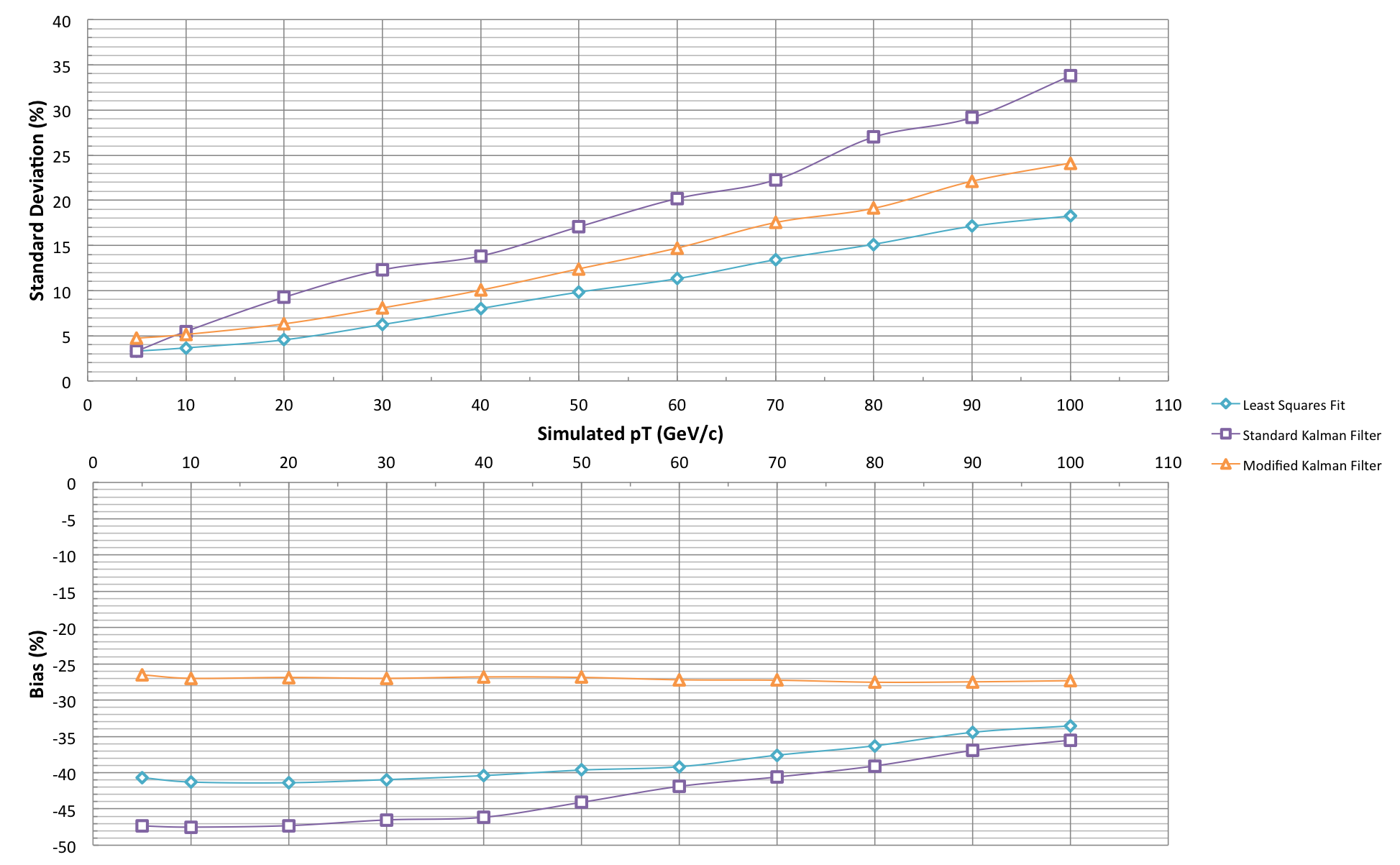}
				\caption{Standard deviation (top) ans bias (bottom) on $ \frac{\Delta p_T}{p_T} $ as a function of the simulated \pT{} for muon tracks generated with the FastSim in the real magnetic field, and reconstructed with each tested algorithm using the RecHits in GE1/1, ME1/1, and ME2/x.}
				\label{fig:algorithms_performances_prospects__RT_GEM_ME12}
			\end{figure}

			We start by comparing the results in Figure \ref{fig:algorithms_performances_prospects__RT_GEM_ME12} which depicts the standard deviation (top) and bias (bottom) on $ \frac{\Delta p_T}{p_T} $ as a function of the simulated \pT{} for muon tracks generated with the FastSim in the real magnetic field, and reconstructed with each tested algorithm using the RecHits in GE1/1, ME1/1, and ME2/x. \\

			We observe that the Least Squares fit yields a better standard deviation (blue; 5 to 20\%) than the modified Kalman filter (orange; 5 to 25\%) and the standard Kalman filter (purple; 5 to 35\%). As reviewed in Section \ref{sec:kalman_filter__helix_parameterization} and Section \ref{sec:kalman_filter__standard_real_field_bias_detectors_setup}, the degradation in the standard deviation for these algorithms is due to their complexity. For the standard Kalman filter, the propagation between two measurement sites is done as if the magnetic field was constant, and for the modified Kalman filter, the Runge-Kutta propagator does not take into account the evolution of the covariance matrix. Moreover, for the latter, a small deviation at the first site will be amplified by the iterative process used for the propagation. This yields a larger spread of the results than for the Least Squares fit, which matches a track to all the points at the same time. The standard deviation is a measure of the constraints that we impose on the fit, which are less important for the Kalman filter than for the Least Squares fit. \\

			However, when considering the bias on $ \frac{\Delta p_T}{p_T} $, the tendency is inverted. The modified Kalman filter yields a smaller bias (around -27\%) than the Least Squares fit (between -41 and -33\%) which brings an improvement of 18 to 35\% to the results.
		
		\subsection{CMSSW Results}
		\label{sec:algorithms_performances_prospects__cmssw_results}	

\begin{figure}[h!]
				\centering
				\includegraphics[width = \textwidth]{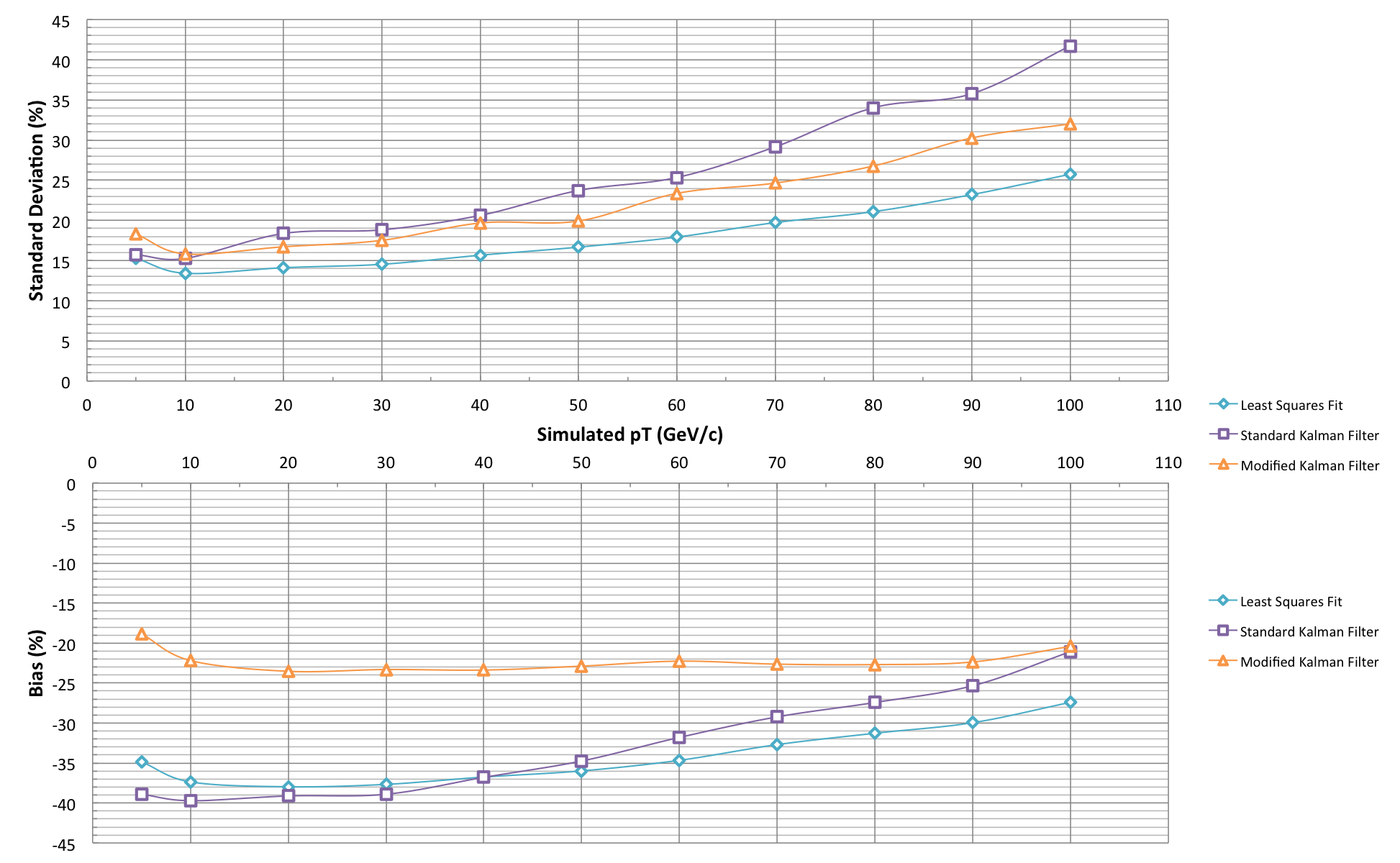}
				\caption{Standard deviation (top) and bias (bottom) on $ \frac{\Delta p_T}{p_T} $ as a function of the simulated \pT{} for muon tracks generated with CMSSW, and reconstructed with each tested algorithm using the RecHits in GE1/1, ME1/1, and ME2/x.}
				\label{fig:algorithms_performances_timing_prospects_CMSSW_GEM_ME12}
			\end{figure}

			We now compare the results with the RecHits generated with CMSSW. Figure \ref{fig:algorithms_performances_timing_prospects_CMSSW_GEM_ME12} depicts the standard deviation (top) and bias (bottom) on $ \frac{\Delta p_T}{p_T} $ as a function of the simulated \pT{} for muon tracks generated with CMSSW, and reconstructed with each tested algorithm using the RecHits in GE1/1, ME1/1, and ME2/x. The same trends are observed as with the FastSim in the real magnetic field, except that the resolution is poorer at lower \pT{} where physical processes play the most significant role. None of the algorithms were implemented to take into account these effects which explains this degradation. \\

			The trends also remain the same with the bias on $ \frac{\Delta p_T}{p_T} $. The only difference is that the standard Kalman filter yields a smaller bias than the Least Squares fit above 40 \GeVc{}. As before, the modified Kalman filter offers a substantial improvement (bias around -23\%) of the bias compared to the other algorithms. An improvement between 25 and 46\% is observed when compared to the results of the Least Squares fit. On the other hand, the standard deviation is degraded by an average of 24\% over the entire range of simulated \pT{}.
	
		\subsection{Summary}
		\label{sec:algorithms_performances_prospects__comp_summary}	

			In both simulation environments, the modified Kalman filter yields the smallest bias, which is one of the issues we want to address at the L1 Trigger. \\

			Between the two frameworks, the FastSim in the real magnetic field is the one that allows the best comparison as multiple scattering and energy losses have not been implemented in the Kalman filters. The modified Kalman filter improves the bias of the Least Squares fit by 18 to 35\% over the entire range of simulated \pT{}, while the standard deviation is degraded by an average of 32\%. 

	\section{Impact on the L1 Trigger}
	\label{sec:algorithms_performances_prospects__impact_on_the_l1_trigger}

		Using the algorithms we developed, we compute the turn-on curves and the rate of accepted events for a threshold of 14 \GeVc{}. To compute the latter, we first calculate the percentage of event that are over the threshold for a given simulated \pT{}. We then multiply this value by the corresponding generated rate inside CMS represented in Figure \ref{fig:trigger_system_and_reconstruction_algorithms__acceptance} in Section \ref{sec:trigger_system_and_reconstruction_algorithms__system_performances}, yielding the rate of accepted events for a given \pT{}. Finally, we sum the obtained rates to obtain the total rate of accepted events. \\

		The current rate of accepted events for single muons for the DT/CSC trigger is of the order of 8 kHz.

		\subsection{FastSim Results: Real Magnetic Field}
		\label{sec:algorithms_performances_prospects__rates_fastsim_results_real_magnetic_field}	

\begin{figure}[h!]
				\centering
				\includegraphics[width = \textwidth]{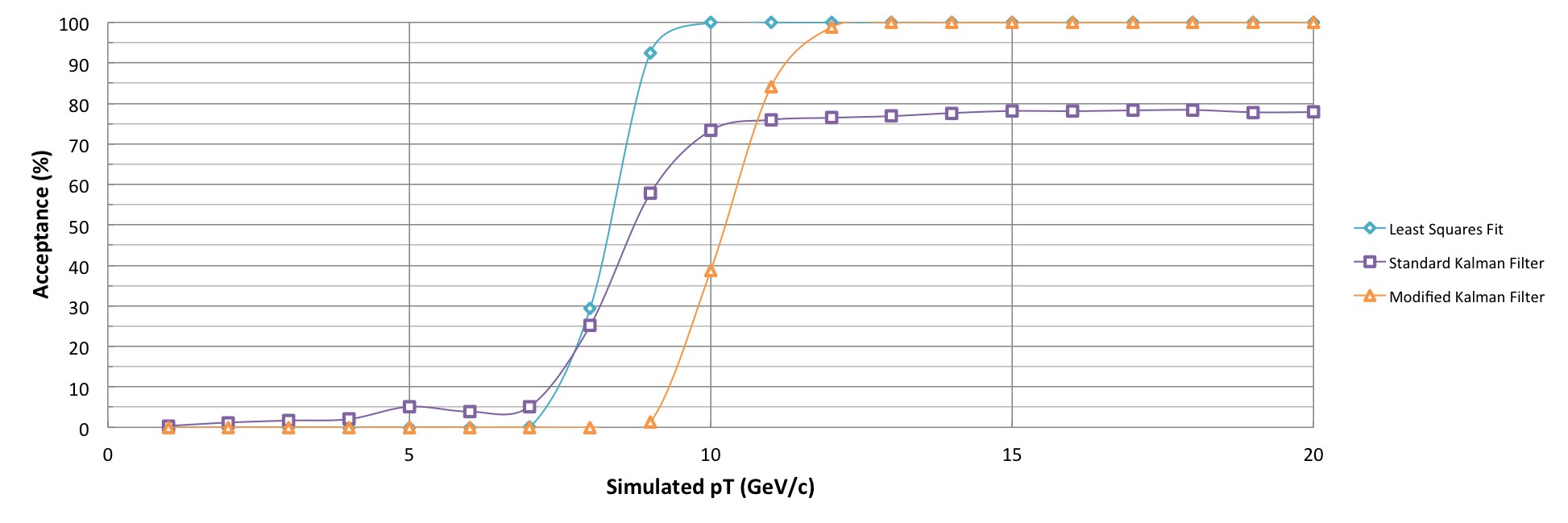}
				\caption{Turn-on curve for a threshold of 14 \GeVc{} as a function of the simulated \pT{} for single muon tracks generated with the FastSim in the real magnetic field, and reconstructed with each tested algorithm using the RecHits in GE1/1, ME1/1, and ME2/x.}
				\label{fig:algorithms_performances_prospects__turn_on_RT_GEM_ME12}
			\end{figure}

			Figure \ref{fig:algorithms_performances_prospects__turn_on_RT_GEM_ME12} represents the turn-on curve for a threshold of 14 \GeVc{} as a function of the simulated \pT{} for single muon tracks generated with the FastSim in the real magnetic field, and reconstructed with each tested algorithm using the RecHits in GE1/1, ME1/1, and ME2/x. \\

			Ideally, the curves should be at 0\% before 14 \GeVc{}, meaning we reject every event below the threshold, and 100\% after, meaning we accept every event beyond the threshold. We notice that the greater bias for the Least Squares fit results in a shift of the curve to lower \pT{} with respect to the modified Kalman filter. The algorithm starts to accept events around 7 \GeVc{}, while the modified Kalman filter is effectively rejecting events up to 9 \GeVc{}. Both algorithms rapidly reach 100\% acceptance, which is not the case for the standard Kalman filter which only goes up to approximately 80\%. This is due to long distribution tails where events are reconstructed with a much higher or much smaller \pT{}. The standard Kalman filter is therefore not a viable option for the L1 Trigger. \\

			The obtained rate for a threshold of 14 \GeVc{} for muon tracks generated with the FastSim in the real magnetic field, and reconstructed with the modified Kalman filter using the RecHits in GE1/1, ME1/1, and ME2/x is of 5.9 kHz. The improvement compared to the actual value is of 26\%. We also computed this rate for the CSCs in standalone and found 7.24 kHz. We observe that GEM detectors diminish the rate by 19\% and therefore have a positive effect on the L1 Trigger. \\

			Moreover, the Least Squares fit in the same conditions for RecHits in GE1/1, ME1/1, and ME2/x, and in ME1/1, ME2/x, ME3/x, and ME4/x respectively yields rates of 9.42 and 27.8 kHz. The modified Kalman filter improves these results by 37 and 74\% respectively. 

		\subsection{CMSSW Results}
		\label{sec:algorithms_performances_prospects__rates_cmssw_results}

\begin{figure}[h!]
				\centering
				\includegraphics[width = \textwidth]{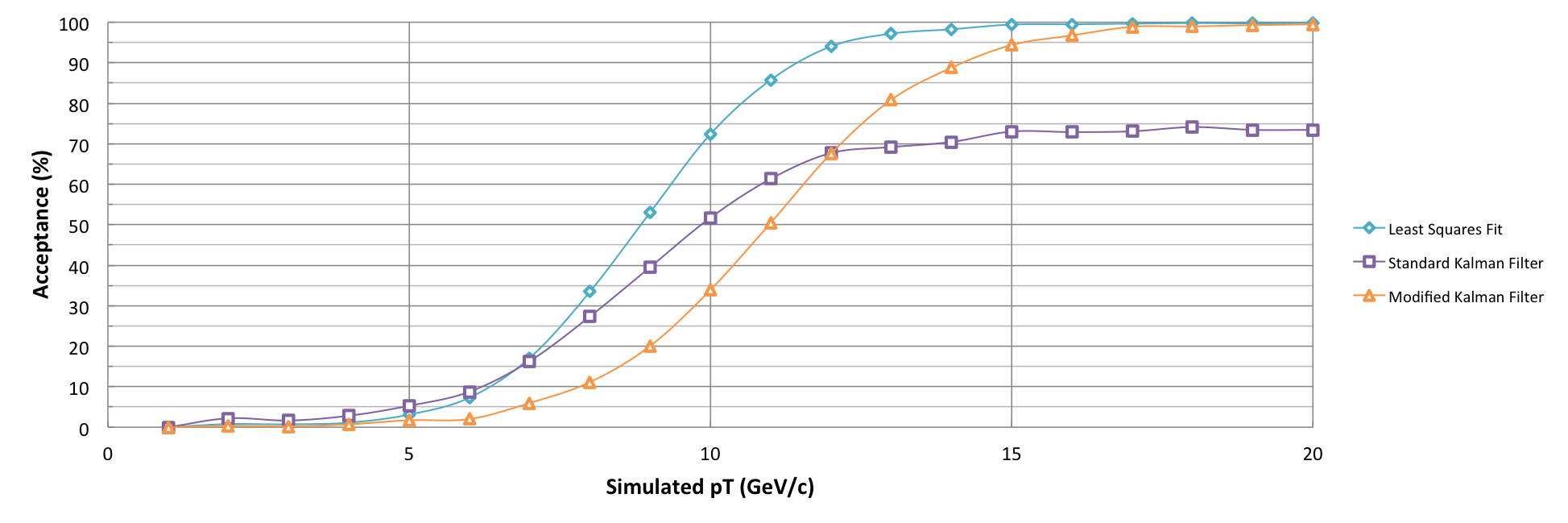}
				\caption{Turn-on curve for a threshold of 14 \GeVc{} as a function of the simulated \pT{} for single muon tracks generated with CMSSW, and reconstructed with each tested algorithm using the RecHits in GE1/1, ME1/1, and ME2/x.}
				\label{fig:algorithms_performances_prospects__turn_on_CMSSW_GEM_ME12}
			\end{figure}

			The same analysis is performed in CMSSW. Figure \ref{fig:algorithms_performances_prospects__turn_on_CMSSW_GEM_ME12} depicts the turn-on curve for a threshold of 14 \GeVc{} as a function of the simulated \pT{} for single muon tracks generated with CMSSW, and reconstructed with each tested algorithm using the RecHits in GE1/1, ME1/1, and ME2/x. \\

			We notice that the Least Squares fit starts to accept event around 5 \GeVc{} and the modified Kalman filter around 7 \GeVc{} which is at lower \pT{} than for the FastSim with the real magnetic field. Paradoxically, the bias for both algorithms is smaller in CMSSW. To compare the impact of the bias, we have to look at the point where the curve reaches 50\% and not the value at which the algorithms start to accept events. For the Least Squares fit in CMSSW, an acceptance of 50\% is reached around 9 \GeVc{}, while in the FastSim with the real magnetic field, it is reached around 8 \GeVc{}. \\

			To understand why the algorithms in CMSSW start to accept events at lower \pT{}, we have to consider the slope of the curves. We observe that the slope is less steep than for the FastSim with the real magnetic field which is due to the higher standard deviation, hence the longer distributions' tails. While the bias influences the position of the middle of the curve, the standard deviation modifies its slope. The longer the tails of the distributions, the more events will be accepted at lower \pT{}, which explains why the curves start to accept events at lower \pT{} than in the FastSim with the real magnetic field, even though their bias is smaller. \\

			The resulting rate for a threshold of 14 \GeVc{} for muon tracks generated with CMSSW, and reconstructed with each tested algorithm using the RecHits in GE1/1, ME1/1, and ME2/x is of 7.39 kHz. The modified Kalman filter brings an improvement of 8\% compared to the current rate. Moreover, using the modified Kalman filter, we computed the rate yielded by the CSCs in standalone and found 7.87 kHz, which is of the order of the current rate for the CSCs at the L1 Trigger. \\

			Finally, the Least Squares fit in the same conditions for the GE1/1, ME1/1, and ME2/x setup, and the ME1/1, ME2/x, ME3/x, and ME4/x setup respectively yields rates of 11.6 and 18 kHz. These are respectively improved by 37 and 56\% when reconstructing the tracks with the modified Kalman filter.

		\subsection{Summary}
		\label{sec:algorithms_performances_prospects__rates_summary}	

			The GE1/1, ME1/1, and ME2/x setup with the modified Kalman filter improves the acceptance rate by 26\% when compared to the CSCs-only in the FastSim with the real magnetic field. \\

			Moreover, the modified Kalman filter yields the best results of the three algorithms in both simulation frameworks. An improvement of 37\% of the accepted rate for single muons reconstructed with the modified Kalman filter using RecHits in GE1/1, ME1/1, and ME2/x is observed in both the FastSim with the real magnetic field and CMSSW compared to the Least Squares fit. \\

			This confirms that GEM detectors improve the reconstruction process as well as the efficiency of the L1 Trigger. \\

			Future improvements of the modified Kalman filter to take into account physical processes will reduce this rate even more.			

	\section{$ \Delta \phi $ Method}
	\label{sec:algorithms_performances_prospects__delta_phi_method}

\begin{figure}[h!]
			\centering
			\includegraphics[width = 0.7 \textwidth]{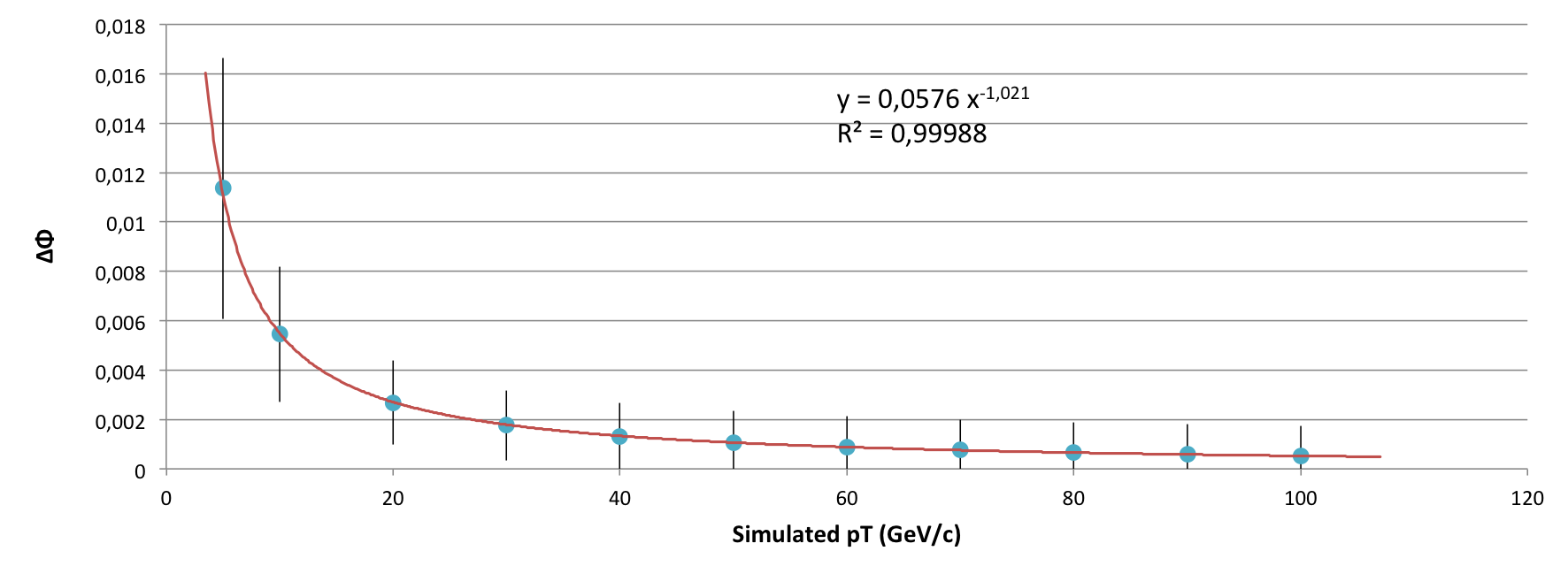}
			\caption{$ \Delta \phi $ between RecHits in GE1/1a and ME1/1 as a function of the simulated \pT{} for muon tracks generated with CMSSW.}
			\label{fig:algorithms_performances_prospects__dphi_vs_pt_CMSSW_GEM_ME1}
		\end{figure}

		We present one last method to estimate the \pT{} of the particles which is similar to what is currently done in the L1 Trigger. The distance between GE1/1 and ME1/1 is large enough (roughly 30 cm) so that 
		\begin{equation}
			\Delta \phi = \phi_{GE1/1a} - \phi_{ME1/1} 
		\end{equation}
		becomes significant and can be used to yield the transverse momentum of the particles. Figure \ref{fig:algorithms_performances_prospects__dphi_vs_pt_CMSSW_GEM_ME1} presents $ \Delta \phi $ between RecHits in GE1/1a and ME1/1 as a function of the simulated \pT{} for muon tracks generated with CMSSW. We notice that the points are almost perfectly fit by the trend curve (R$ ^2 $ = 0.99988). \\

		By inverting this relation, we are able to determine the \pT{} of the particles for each event. We observe that below 20 \GeVc{} points are separated and even though their errors overlap, can still be identified. However, above 20 \GeVc{}, it becomes more difficult, to near impossible, to isolate the points. Furthermore, the higher errors at low \pT{} are due to the impact of multiple scattering which is significant. When reaching higher \pT{}, the particles are less affected and errors diminish. \\

\begin{figure}[h!]
			\centering
			\includegraphics[width = \textwidth]{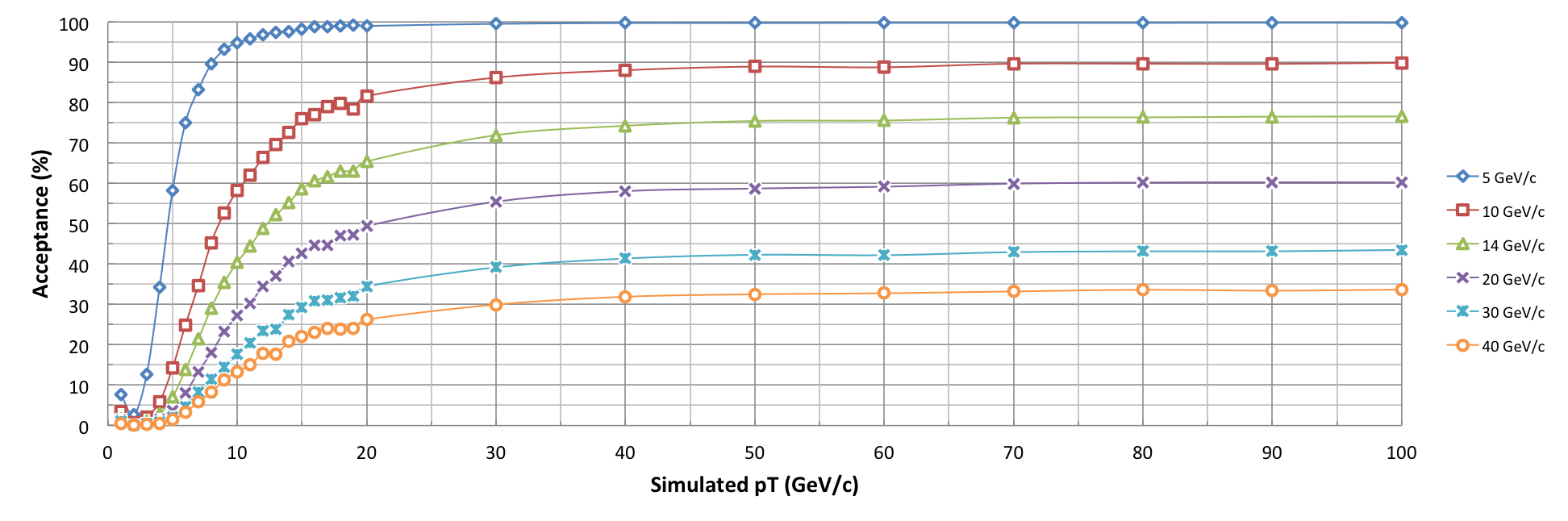}
			\caption{Turn-on curves for threshold of 5, 10, 14, 20, 30, and 40 \GeVc{} as a function of the simulated \pT{} for muon tracks generated with CMSSW, and reconstructed with the $ \Delta \phi $ method using the RecHits in GE1/1, ME1/1, and ME2/x.}
			\label{fig:algorithms_performances_prospects__dphi_turn_on_CMSSW}
		\end{figure}

		To prove that the algorithms we developed are more powerful than those actually used and improve the system, we analyze the turn-on curves of the methods for various thresholds. Figure \ref{fig:algorithms_performances_prospects__dphi_turn_on_CMSSW} presents the turn-on curves for threshold of 5, 10, 14, 20, 30, and 40 \GeVc{} as a function of the simulated \pT{} for muon tracks generated with CMSSW, and reconstructed with the $ \Delta \phi $ method using the RecHits in GE1/1, ME1/1, and ME2/x. \\

		We observe that for a threshold of 5 \GeVc{}, an acceptance of 100\% is quickly reached, while the other curves converge to different plateaus. To understand this effect, we look back at Figure \ref{fig:algorithms_performances_prospects__dphi_vs_pt_CMSSW_GEM_ME1}. At higher \pT{}, the points cover the same $ \Delta \phi $ range resulting in numerous faulty reconstruction. For example, a muon of 50 \GeVc{} is often reconstructed with a \pT{} of 30 \GeVc{} or less and is therefore not accepted for certain threshold. \\

\begin{figure}[h!]
			\centering
			\includegraphics[width = \textwidth]{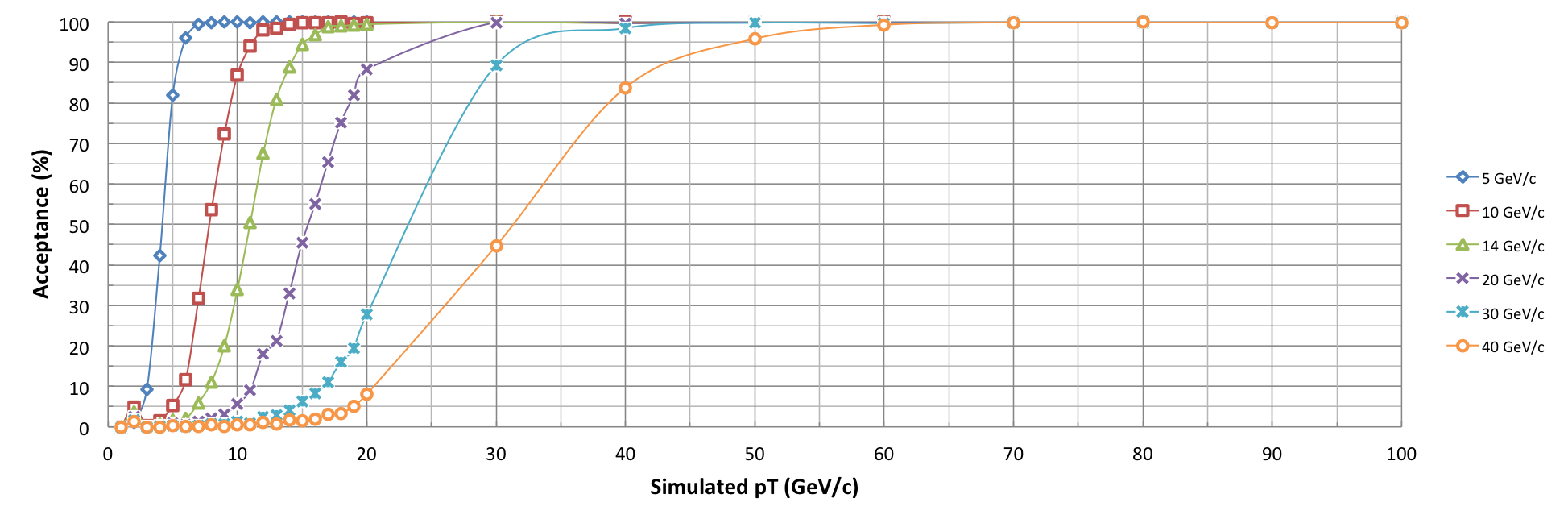}
			\caption{Turn-on curves for threshold of 5, 10, 14, 20, 30, and 40 \GeVc{} as a function of the simulated \pT{} for muon tracks generated with CMSSW, and reconstructed with the modified Kalman filter using the RecHits in GE1/1, ME1/1, and ME2/x.}
			\label{fig:algorithms_performances_prospects__rkal_turn_on_CMSSW_GEM_ME12}
		\end{figure}

		In comparison, Figure \ref{fig:algorithms_performances_prospects__rkal_turn_on_CMSSW_GEM_ME12} depicts the turn-on curves for threshold of 5, 10, 14, 20, 30, and 40 \GeVc{} as a function of the simulated \pT{} for muon tracks generated with CMSSW, and reconstructed with the modified Kalman filter using the RecHits in GE1/1, ME1/1, and ME2/x. We immediately notice that for every threshold the curves quickly reach 100\% acceptance. Moreover, the filter starts to accept events later than the $ \Delta \phi $ method. With a threshold of 40 \GeVc{}, the $ \Delta \phi $ method reaches 10\% acceptance around 9 \GeVc{} while the modified Kalman filter reaches 10\% acceptance a little above 20 \GeVc{}. \\ 

		The actual algorithm for the L1 Trigger using $ \Delta \phi $ is more complex than what has been presented here. It relies on the multiple readout layers inside a CSC module to measure the angle of the particle with the chamber and uses that information to discriminate between multiple possible \pT{} given by $ \Delta \phi $. Using this information allows the system to reach an acceptance's plateau of 95-98\% for all the thresholds.
	
	\section{Prospects}
	\label{sec:algorithms_performances_prospects__prospects}

		Until now, we ignored the execution time of the various algorithms, and analyzed the results yielded by the programs using computer simulation. To ensure that the methods meet the L1 Trigger's timing requirement for which these reconstruction methods are intended, we have to closely look at their time consumption. Moreover, the L1 Trigger, as stated in Section \ref{sec:trigger_system_and_reconstruction_algorithms__level_1_trigger}, is composed of dedicated electronics which functions differently than computer \emph{Central Processing Units} (CPU). This section addresses both points and provides the basis for further development.

		\subsection{Timing}
		\label{sec:algorithms_performances_prospects__timing}

\begin{figure}[h!]
				\centering
				\includegraphics[width = \textwidth]{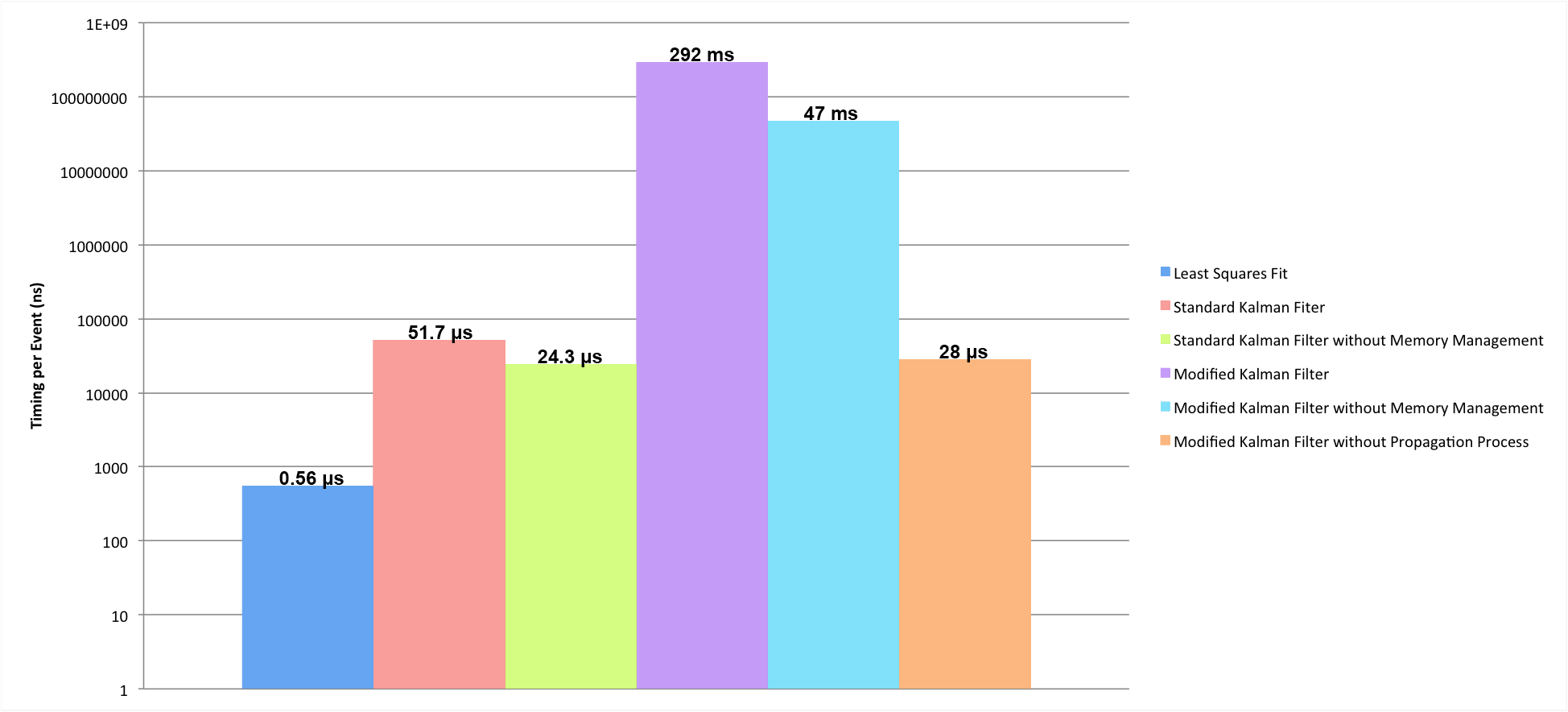}
				\caption{Execution time per event for each tested algorithm.}
				\label{fig:algorithms_performances_prospects__execution_time}
			\end{figure}

			We performed timing analysis on the different algorithms using Mac OS X's Instruments to analyze the programs performances, memory management, system calls, etc. The given results were obtained on a Macbook Pro mid 2012 equipped with a 2.3 GHz Intel Core i7 and 16 Go of RAM (1600 MHz DDR3) running OS X version 10.8.3. The graphics card, although not used, is a NVIDIA GeForce GT 650M with 512MB of GDDR5 memory. The execution time per event for each tested algorithm is shown in Figure \ref{fig:algorithms_performances_prospects__execution_time}. The dark blue, red, and purple columns respectively correspond to the Least Squares fit, the standard Kalman filter, and the modified Kalman filter. We observe that the Least Squares fit runs in less that 3.2 \us{}, the maximum execution time for algorithms in the L1 Trigger. \\

			Using Instruments, we were able to subtract the time needed to initialize the program (allocations of memory, calls of the constructor/destructor of classes, etc) from the total execution time. The Least Squares fit algorithm being simple, was not affected by this. The Kalman filters, on the other hand, perform a lot of calls that slow down the processes. The green and light blue columns in Figure \ref{fig:algorithms_performances_prospects__execution_time} respectively show the execution time per event without, what we call, the Memory Management. We notice that for both algorithms, timing is divided by at least a factor of two. This analysis is performed as it will be useful in the next section when discussing the implementation of the algorithms on electronics circuits. \\

			Furthermore, for the modified Kalman filter which is the slowest algorithm (47 ms per event), we analyzed the time used by the Runge-Kutta propagator. 99.99\% of the total execution time is spent propagating the particle from one detection layer to another. The execution time drops to 28 \us{} if we ignore that process as represented by the orange column in Figure \ref{fig:algorithms_performances_prospects__execution_time}. \\
		
			To effectively eliminate the time spent to propagate the particle between the measurements sites, we intend to use a LUT. For a defined set of ($ \mathbf{x} $, $ \mathbf{p} $) couples at measurement site $ (k - 1) $, the LUT would hold the resulting ($ \mathbf{x}' $, $ \mathbf{p}' $) couple at site $ (k) $, hereby eliminating the Runge-Kutta propagation. This has not yet been implemented, but we simulated this behavior by applying a discretization on $ \mathbf{x} $ and $ \mathbf{p} $ before and after being propagated. The positions are rounded with a 20 cm interval, and the momenta are matched to the closest $ \mathbf{p}_{Code} $ which can be found in Table \ref{tab:algorithms_performances_prospects__lut_momenta}. For each measurement site, there would be of the order of 256 $ \times $ 32,768 entries corresponding to 8 bits for the position and 5 bits per component of the momentum. Those choices are preliminary and have not been refined to reduce the required space. 

			\begin{table}[h!]
				\centering
				\begin{tabular}{c|c||c|c||c|c}
					$ \mathbf{p}_{Code} $ & $ \mathbf{p} $ (\GeVc{}) & $ \mathbf{p}_{Code} $ & $ \mathbf{p} $ (\GeVc{}) & $ \mathbf{p}_{Code} $ & $ \mathbf{p} $ (\GeVc{}) \\ \hline
					0 & 1 	& 11 & 20 & 22 & 100 \\
					1 & 2 	& 12 & 25 & 23 & 110 \\
					2 & 3 	& 13 & 30 & 24 & 120 \\
					3 & 4 	& 14 & 35 & 25 & 130 \\
					4 & 5 	& 15 & 40 & 26 & 140 \\
					5 & 6 	& 16 & 45 & 27 & 150 \\
					6 & 7 	& 17 & 50 & 28 & 160 \\
					7 & 8 	& 18 & 60 & 29 & 170 \\
					8 & 9 	& 19 & 70 & 30 & 180 \\
					9 & 10 	& 20 & 80 & 31 & 190 \\
					10 & 15 & 21 & 90 &    & 
				\end{tabular}
				\caption{List of momenta that would be used to generate a LUT for the modified Kalman Filter}
				\label{tab:algorithms_performances_prospects__lut_momenta}
			\end{table}

\begin{figure}[h!]
				\centering
				\includegraphics[width = \textwidth]{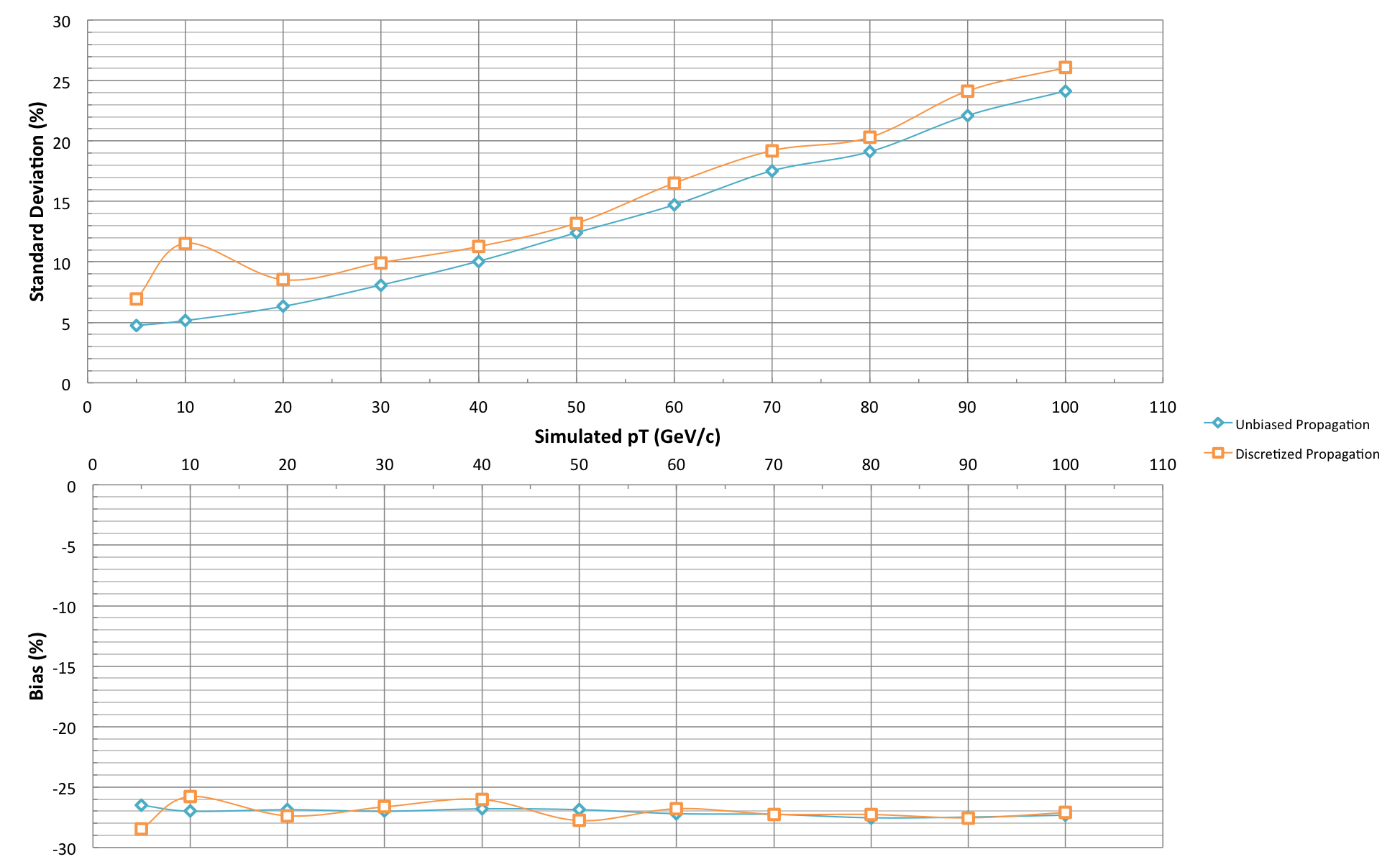}
				\caption{Standard deviation (top) and bias (bottom) on $ \frac{\Delta p_T}{p_T} $ as a function of the simulated \pT{} for muon tracks generated with CMSSW, and reconstructed with the modified Kalman filter with and without a LUT using RecHits in GE1/1, ME1/1, and ME2/x.}
				\label{fig:algorithms_performances_prospects__discrete_RT_GEM_ME12}
			\end{figure}

			As mentioned, we simulated the use of a LUT for the propagation between two detection layers. The obtained standard deviation (top) and bias (bottom) on $ \frac{\Delta p_T}{p_T} $ as a function of the simulated \pT{} for muon tracks generated with CMSSW, and reconstructed with the modified Kalman filter with and without a LUT using RecHits in GE1/1, ME1/1, and ME2/x are shown in Figure \ref{fig:algorithms_performances_prospects__discrete_RT_GEM_ME12}. We observe a degradation of the standard deviation, especially at lower \pT{}, while the bias remains the same. \\

			The achieved 28 \us{} for the modified Kalman filter using a LUT are still too long for the actual L1 Trigger. However, it has been proposed \Cite{Trigger_Update} to increase the latency of the trigger after LS2, allowing the algorithms to run for 20 \us{}. As the installation of GEM detectors in GE1/1 is foreseen to take place during LS2, we can focus on reducing the timing below 20 \us{} instead of 3.2 \us{} 

		\subsection{Field Programmable Gate Array}
		\label{sec:algorithms_performances_prospects__field_programmable_gate_array}

			The L1 Trigger's algorithms are programmed on dedicated electronics, such as \emph{Field Programmable Gate Arrays} (FPGA). FPGAs are digital electronic components which behavior can be programmed using a \emph{Hardware Description Language} (HDL). They are made out of hundreds of thousands of \emph{Configurable Logic Blocks} (CLBs) which response is programmable, of logic blocks which perform a dedicated task (addition, multiplication, etc) and of \emph{Input/Output} (I/O) pins. All of these can be connected to each other to form a complex system. Each CLB is composed of a LUT with, typically, four input and one output pins, and of a latching register which synchronizes the output with the system clock. For each possible combination at the entries (2$ ^4 $), the LUT holds the corresponding output. When the FPGA is programmed, the HDL code is translated into a map with all the values of the CLBs' LUT and the routes between the blocks. Most of the operations performed by a FPGA are synchronous and driven by the system's clock. \\

			FPGA development is different than computer programing. In a C++ program, statements are executed sequentially, while FPGAs, by design, offer the possibility to execute multiple statements in parallel. In FPGAs, the execution of an algorithm has to be seen like the functioning of an electronic circuit. Each block runs separately and is presented with multiple inputs, comparable to wires. After a short delay, it returns and holds the result on the output pin. \\

			Table \ref{tab:algorithms_performances_prospects__c_hdl_operation} compares the operations performed by an algorithm that sums up four numbers (A, B, C, and D), executed by a computer CPU and by a FPGA. Due to the fact that the FPGA is specifically designed to perform that task, it can break down the process and improve efficiency. Moreover, as variables can be seen like wires that hold the information, no memory calls are needed. However, the clock speed used to drive a FPGA is much slower (typically of the order of 100 MHz for FPGAs compared to 1 GHz for computer CPUs). \\

			\begin{table}[h!]
				\centering
				\begin{tabular}{c|c}
					Computer CPU & FPGA \\ \hline
					Retrieve A from memory and load the adder & Perform A + B = E, and C + D = F \\
					Retrieve B from memory & Perform E + F = G \\
					Perform A + B = E & \\
					Retrieve C from memory & \\
					Perform E + C = F & \\
					Retrieve D from memory & \\
					Perform F + D = G & \\
					Place G in memory & 
				\end{tabular}
				\caption{Comparison of the operations performed by an algorithm that sums up four numbers (A, B, C, and D), executed by a computer CPU and by a FPGA}
				\label{tab:algorithms_performances_prospects__c_hdl_operation}
			\end{table}		 

\begin{figure}[h!]
				\centering
				\includegraphics[width = \textwidth]{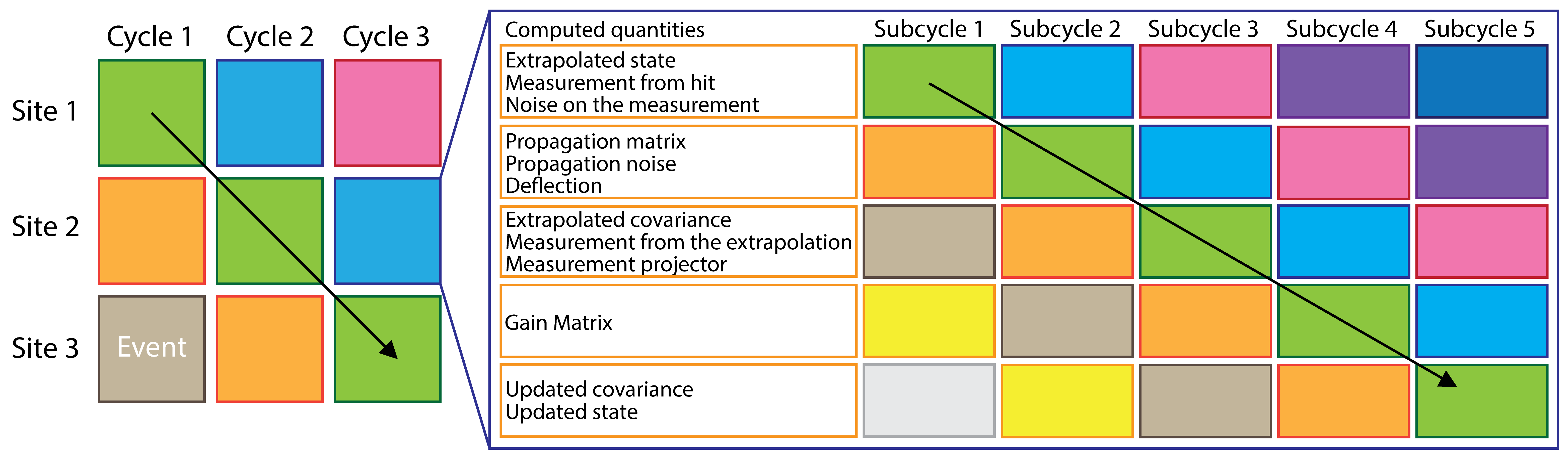}
				\caption{Schematic view of the implementation of the Kalman filter on a FPGA which enables pipelining for the events (colored block).}
				\label{fig:algorithms_performances_prospects__fpga_kalman}
			\end{figure}

			We previously measured the execution time without what we called Memory Management as it is not present in FPGAs. The development of the previously discussed algorithms will be a complex task, but not an impossible one as the capabilities currently offered by FPGAs are considerable and evolve fast. Figure \ref{fig:algorithms_performances_prospects__fpga_kalman} gives the ideas behind the implementation of the modified Kalman filter in a FPGA and how it will be implemented. \\

			For each measurement site (left block), we will use a separated execution loop performing the iterative process resulting in the update of the state. This allows the system to use pipelining on the events (colored blocks): when the update of event A is done at measurement site 1, it moves on to site 2 and event B is updated at site 1. At each cycle, the events are shifted from one site to the other. Moreover, at each site, the iteration is furthermore broken down into five operations (right block) which cannot be run in parallel and depend upon one another. With this design, for $ n $ measurement sites, we can have up to $ 5 \times n $ events in the system which significantly reduces the number of FPGAs we need to use to keep up with the rate of generated events. \\

\begin{figure}[h!]
				\centering
				\includegraphics[width = 0.7 \textwidth]{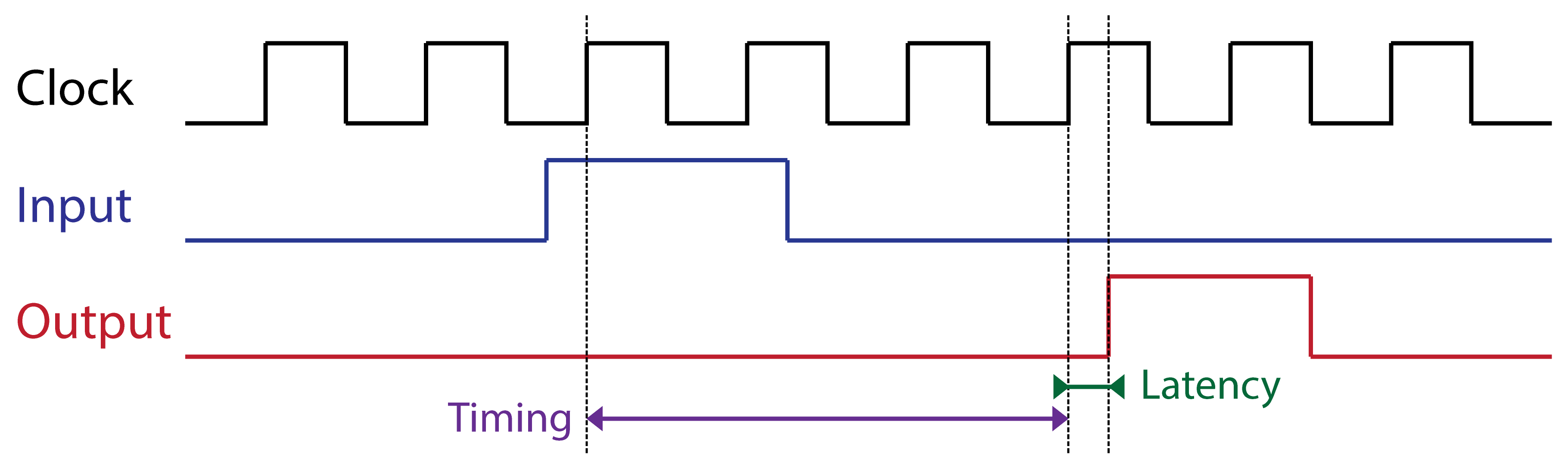}
				\caption{Representation of the timing (purple) and the latency (green) of the output signals (red) relative to the input signal (blue).}
				\label{fig:algorithms_performances_prospects__fpga_timing}
			\end{figure}

			In order to have an idea of the execution speed of statements in a FPGA, we used Xilinx's simulation tools to time some basic function we will need. The timing is defined as the number of clock cycles from rising edge to rising edge required to complete the statement, and the latency as the delay between the closest rising edge and the modification of the output signal as shown in Figure \ref{fig:algorithms_performances_prospects__fpga_timing} which represents the timing (purple) and latency (green) for a given clock (black), input signal (blue), and the resulting output signal (red). \\

			Table \ref{tab:algorithms_performances_prospects__fpga_timing} lists the timing for multiple operations simulated using ISim on a Xilinx Virtex7 (device: xc7vx330t; package: ffg1157; speed grade: -2) FPGA. We notice that divisions are the most time consuming operation and that on the other hand, additions and subtractions calculations are straightforward. The sine, cosine, square root and arctangent functions are computed using the CORDIC algorithm \Cite{FPGA_Cordic} which is fast, but subject to errors as the values have to be coded using a limited number of bits. \\

			\begin{table}[h!]
				\centering
				\begin{tabular}{c|c|c}
					Operation & Timing (clock cycles) & Latency \\ \hline
					Addition and Subtraction & 0 & 359 ps \\
					Multiplication & 0 & 1.716 ns \\
					Division (16 bits by 16 bits) & 35 & < 10 ps \\
					Cosine and Sine (16 bits) & 0 & 8.613 ns \\
					Square root (16 bits) & 0 & 7.806 ns \\
					Arctangent (16 bits) & 0 & 10.750 ns
				\end{tabular}
				\caption{Timing of multiple operations simulated using ISim for a Xilinx Virtex7 (device: xc7vx330t; package: ffg1157; speed grade: -2) FPGA.}
				\label{tab:algorithms_performances_prospects__fpga_timing}
			\end{table}

	\section{Conclusion}
	\label{sec:algorithms_performances_prospects__prospects_conclusion}

		The modified Kalman filter is from the three implemented algorithms the one that yields the best results and brings the largest improvements to the L1 Trigger. In the FastSim in the real magnetic field using RecHits in GE1/1, ME1/1, and ME2/x, the modified Kalman filter improves the bias of the Least Squares fit by 18 to 35\% over the entire range of simulated \pT{}, while the standard deviation is degraded by an average of 32\%. \\

		Moreover, in CMSSW with the same setup, we observe an improvement of 37\% of the rate of accepted events at a threshold of 14 \GeVc{} for single muons when we compare the modified Kalman filter to the Least Squares fit. \\

		Further development has to be performed in order to take into account physical processes and improve the results in CMSSW. Moreover, a greater integration of the magnetic field inside the iterative process will have to be put in place to even more reduce the bias. \\

		Finally, to be used in the L1 Trigger, the modified Kalman filter will have to be implemented on a FPGA. Earlier work \Cite{FPGA_Kalman} showed that a simple Kalman filter can run in less than 0.4013 \us{} per iteration on a FPGA. Therefore, using the ideas we developed in the previous section, namely the usage of a LUT and nested pipelining, we believe it is possible to execute a modified Kalman filter for helical tracks on a FPGA and reach timings that meet the L1 Trigger's requirements. 

	\cleardoublepage


\chapter*[Conclusion]{Conclusion}
\addcontentsline{toc}{chapter}{Conclusion}
\label{chap:conclusion}

	This work aimed at developing and characterizing new algorithms for the L1 Trigger for the upgrade of the CMS muon spectrometer with Triple-GEM detectors. \\

	After the LS2 upgrade, The LHC will run at luminosities beyond its nominal value of 10$ ^{34} $ cm$ ^{-2} $ s$ ^{-1} $. CMS will suffer from higher particle fluxes and background rates that will degrade the performances of the L1 Trigger. The CMS GEM collaboration proposes to instrument the forward region of the CMS muon spectrometer, 1.6 < $ | \eta | $ < 2.1, with Triple-GEM detectors to increase redundancy and make use of the vacant space initially foreseen to host RPCs. Previous studies demonstrated that Triple-GEM detectors are able to sustain rates as high as 10 MHz cm$ ^{-2} $ while yielding a spatial resolution between 170 and 340 \um{}, and a detection efficiency of 98\%. Moreover, the installation of Triple-GEM detectors in CMS could bring a significant improvement to the L1 Trigger. To study their impact, we developed simulation tools and track reconstruction algorithms, and analyzed the influence of Triple-GEM detectors on the current system. \\

	As GEM detectors were not yet present in the official simulation software of CMS when we started, we developed our own simulation frameworks allowing us to test and debug the algorithms. Later on, when a first description of GEM detectors became available, we continued our studies with both environments. Using these, we implemented and studied three algorithms: a Least Squares fit, a standard Kalman filter, and a modified Kalman filter. Currently, none of these methods accounts for multiple scattering and energy losses. \\

	The major difficulty we faced was the non-uniformity of the magnetic field of CMS, which deviates muon tracks from their ideal trajectory in a constant magnetic field and confuses the reconstruction algorithms. Both the Least Squares fit and the standard Kalman filter were significantly affected as they perform reconstruction assuming a constant the magnetic field. For both methods, the reconstructed transverse momentum is overestimated, having a negative impact on the trigger which then starts to accept events below a defined threshold. \\

	To reduce the bias on the reconstructed transverse momentum, we developed a modified Kalman filter which uses a Runge-Kutta propagator to perform the reconstruction. Using this algorithm we were able to drastically diminish the bias compared to the two other methods and significantly improve the transverse momentum's resolution compared to the standard Kalman filter. \\	

	For each algorithm, we observe that Triple-GEM detectors improve the performances of the current system composed of CSCs. Using data from CMSSW and performing reconstruction using the Least Squares fit, an improvement between 10 and 29\% on the standard deviation of the $ \frac{\Delta p_T}{p_T} $ distribution is noticed above 20 \GeVc{} when adding the Triple-GEM detectors to the CSCs and the bias is slightly reduced by a few \%. However the bias is significantly reduced, by 24 to 35\%, when comparing the Triple-GEM detectors with only the two first CSC stations, ME1/1 and ME2/x, against the CSCs in standalone. This behavior is explained by the fact that CSCs of ME3/x and ME4/x are located in a very non-uniform magnetic field, not taken into account by the Least Squares fit. \\

	We observe that the standard Kalman filter in CMSSW yields an improvement of up to 25\% on the bias of the reconstructed transverse momentum compared to the Least Squares fit, but that the bias remains large, of the order of -40\% for a transverse momentum less than 40 \GeVc{}. The bias could only be improved by using the modified Kalman filter. In the FastSim, the modified Kalman filter improves the bias by 25 to 46\% in comparison to the Least Squares fit, while the standard deviation is degraded by an average of 24\%. \\

	Future studies are needed to further improve the results obtained with the Kalman filters by taking into account multiple scattering and energy losses. \\

	Using the algorithms we developed, we have also computed the rates of accepted events with a muon in the final state for a defined threshold of 14 \GeVc{} at the L1 Trigger. With the modified Kalman filter, Triple-GEM detectors and the two first CSC stations yield the smallest rates of 5.9 and 7.39 kHz for the FastSim in the real magnetic field and CMSSW respectively. The CSCs in standalone respectively result in rates of 7.24 and 7.87 kHz. Triple-GEM detectors therefore improve the results of the L1 Trigger by respectively 19 and 6\% for both simulation frameworks compared to the current system of CSCs in standalone. \\

	Further, we compared the rates between the different algorithms. The modified Kalman filter which yields the smallest bias also results in the smallest rates. In CMSSW and for the setup using Triple-GEM detectors and only the two first CSC stations, the Least Squares fit returns a rate of 11.6 kHz compared to 7.39 kHz for the modified Kalman filter. The improvement is of 36\% on the rate at the L1 Trigger. \\

	The here-above presented results are only a part of the studies that were done in this work. We also measured the execution time of the algorithms. The Least Squares fit runs in less than 1 \us{} per event, while both Kalman filters are more time consuming with an execution time superior to 1 ms. To allow these to run in the L1 Trigger, we present solutions to speed up the processes, among which the use of Look Up Tables, parallel processing and pipelining. We also give guidelines to implement the modified Kalman filter on a Field Programmable Gate Array that can be integrated in the L1 Trigger of CMS.
	\cleardoublepage

	\listoffigures
	\cleardoublepage	

	\listoftables
	\cleardoublepage

\chapter*[List of Abbreviations]{List of Abbreviations}
\addcontentsline{toc}{chapter}{List of Abbreviations}
\label{chap:list_of_abbreviations}

	\begin{description}

		\item[Booster]		\dotfill	Proton Synchrotron Booster
		\item[BX]			\dotfill	Bunch Crossing
		\item[CERN]			\dotfill	European Organization for Nuclear Research
		\item[CLB]			\dotfill 	Configurable Logic Block
		\item[CMS] 			\dotfill 	Compact Muon Solenoid
		\item[CMSSW] 		\dotfill	CMS Software
		\item[COG] 			\dotfill 	Center of Gravity
		\item[CPU]			\dotfill 	Central Processing Unit
		\item[CSC]			\dotfill	Cathode Strip Chambers
		\item[DAQ]			\dotfill 	Data Acquisition
		\item[Digis] 		\dotfill	Digitized Hits
		\item[DT] 			\dotfill	Drift Tube
		\item[ECAL]			\dotfill 	Electromagnetic Calorimeter
		\item[ENC]			\dotfill 	Equivalent Noise Charge
		\item[eV]			\dotfill	Electron-Volt
		\item[FastSim] 		\dotfill	Fast Simulation
		\item[FPGA]			\dotfill 	Field Programmable Gate Array
		\item[GCT]			\dotfill 	Global Calorimeter Trigger
		\item[GEM]			\dotfill	Gas Electron Multiplier
		\item[GMT]			\dotfill 	Global Muon Trigger
		\item[GT]			\dotfill 	Global Trigger
		\item[HCAL]			\dotfill 	Hadronic Calorimeter
		\item[HDL]			\dotfill	Hardware Description Language
		\item[HL-LHC]		\dotfill 	High Luminosity Large Hadron Collider
		\item[HLT]			\dotfill	High Level Trigger
		\item[HV]			\dotfill 	High Voltage
		\item[I/O]			\dotfill 	Input/Output
		\item[IP]			\dotfill	Interaction Point		
		\item[L1 Trigger] 	\dotfill 	Level-1 Trigger
		\item[LEP]			\dotfill	Large Electron Positron
		\item[LHC] 			\dotfill 	Large Hadron Collider
		\item[LLSQ]			\dotfill 	Linear Least Squares
		\item[LS] 			\dotfill 	Long Shutdown
		\item[LUT]			\dotfill 	Look Up Table
		\item[MIP]			\dotfill	Minimum Ionizing Particle
		\item[NLLSQ]		\dotfill 	Non-Linear Least Squares
		\item[PAC]			\dotfill 	Pattern Comparator
		\item[PS]			\dotfill	Proton Synchrotron
		\item[RecHits] 		\dotfill	Reconstructed Hits
		\item[RPC]			\dotfill	Resistive Plate Chambers	
		\item[SimHits] 		\dotfill	Simulated Hits
		\item[SL]			\dotfill 	Super Layer
		\item[SPS]			\dotfill	Super Proton Synchrotron
		\item[TF]			\dotfill 	Track-Finder
		\item[TK]			\dotfill 	Tracker
		\item[TOF]			\dotfill 	Time of Flight
		\item[TOT]			\dotfill 	Time Over Threshold

	\end{description}
	\cleardoublepage


	\printbibliography
	\cleardoublepage

\appendix

\chapter{Impact of the Granularity on the Standard Deviation and the Bias}
\label{chapter:impact_of_the_granularity}

	One complementary study that can be performed using the datasets we produced is the evolution of the standard deviation and the bias as a function of the granularity of the GEM detectors in GE1/1. The lower the granularity, the poorer the resolution in $ \phi $ will be for the GEMs, as reviewed in Section \ref{sec:gas_electron_multiplier_detectors__spatial_resolution}. The use of a smaller granularity may be required at the L1 Trigger in order to reduce the processing time of the events. Reading-out, shaping, and digitizing 2304 electronic channels (128 per chamber, 18 chambers per detector) for each GEM introduces a non-negligible delay. Therefore, channels may be regrouped to decrease the granularity and speed up the readout process for the L1 Trigger, while the full granularity remains accessible at the HLT. In the following, we compare the results yielded by the Least Squares fit and by the modified Kalman filter at granularities of: 128, 64, 32, 16, and 8. \\

\begin{figure}[h!]
		\centering
		\includegraphics[width = \textwidth]{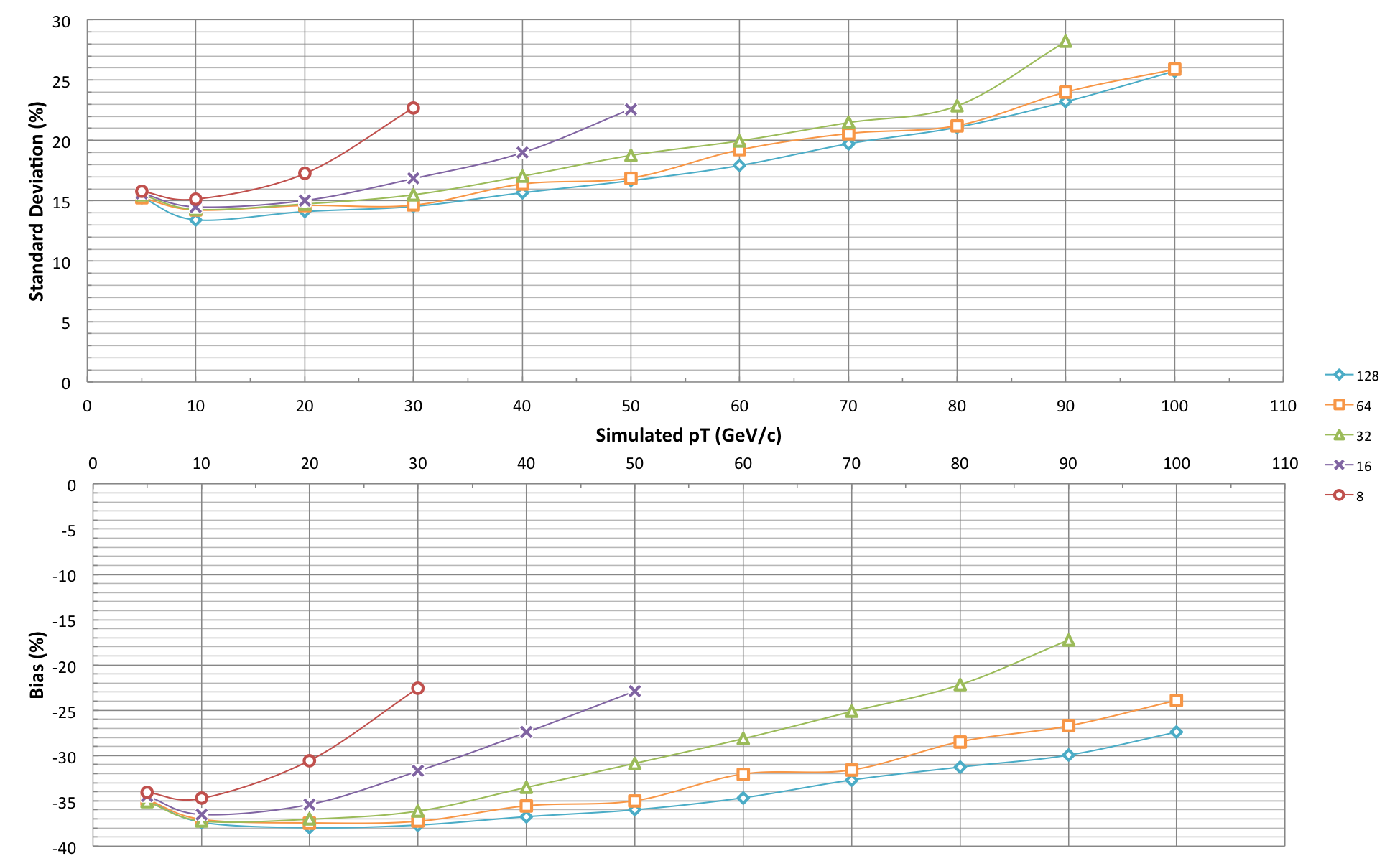}
		\caption{Standard deviation (top) and bias (bottom) on $ \frac{\Delta p_T}{p_T} $ as a function of the simulated \pT{} for muon tracks generated with CMSSW, and reconstructed with the Least Squares fit using the RecHits in GE1/1, ME1/1, and ME2/x with granularities of 128, 64, 32, 16, and 8.}
		\label{fig:impact_of_the_granularity__ls_granularity_CMSSW_GEM_ME12}
	\end{figure}

	We start by looking at the standard deviation (top) and bias (bottom) on $ \frac{\Delta p_T}{p_T} $ as a function of the simulated \pT{} for muon tracks generated with CMSSW, and reconstructed with the Least Squares fit using the RecHits in GE1/1, ME1/1, and ME2/x with granularities of 128, 64, 32, 16, and 8, represented in Figure \ref{fig:impact_of_the_granularity__ls_granularity_CMSSW_GEM_ME12}. When a line is discontinued, it means that the reconstruction fails. We observe that granularities of 8 (red), 16 (purple), and 32 (green) fail before reaching the highest simulated \pT{} of 100 \GeVc{}. Even a granularity of 64 (orange) becomes difficult to use when reaching the higher \pT{}. \\ 

	The observable smaller bias at lower granularity is caused by the failure of the reconstruction and is not an actual improvement of the results. This means that for the Least Squares fit, we cannot use a granularity other than 64 or 128, which may be of concern for the L1 Trigger as we plan to use a granularity of 16 or 32. We also notice a degradation of the standard deviation due to the larger error on the $ \phi $ coordinate. \\

\begin{figure}[h!]
		\centering
		\includegraphics[width = \textwidth]{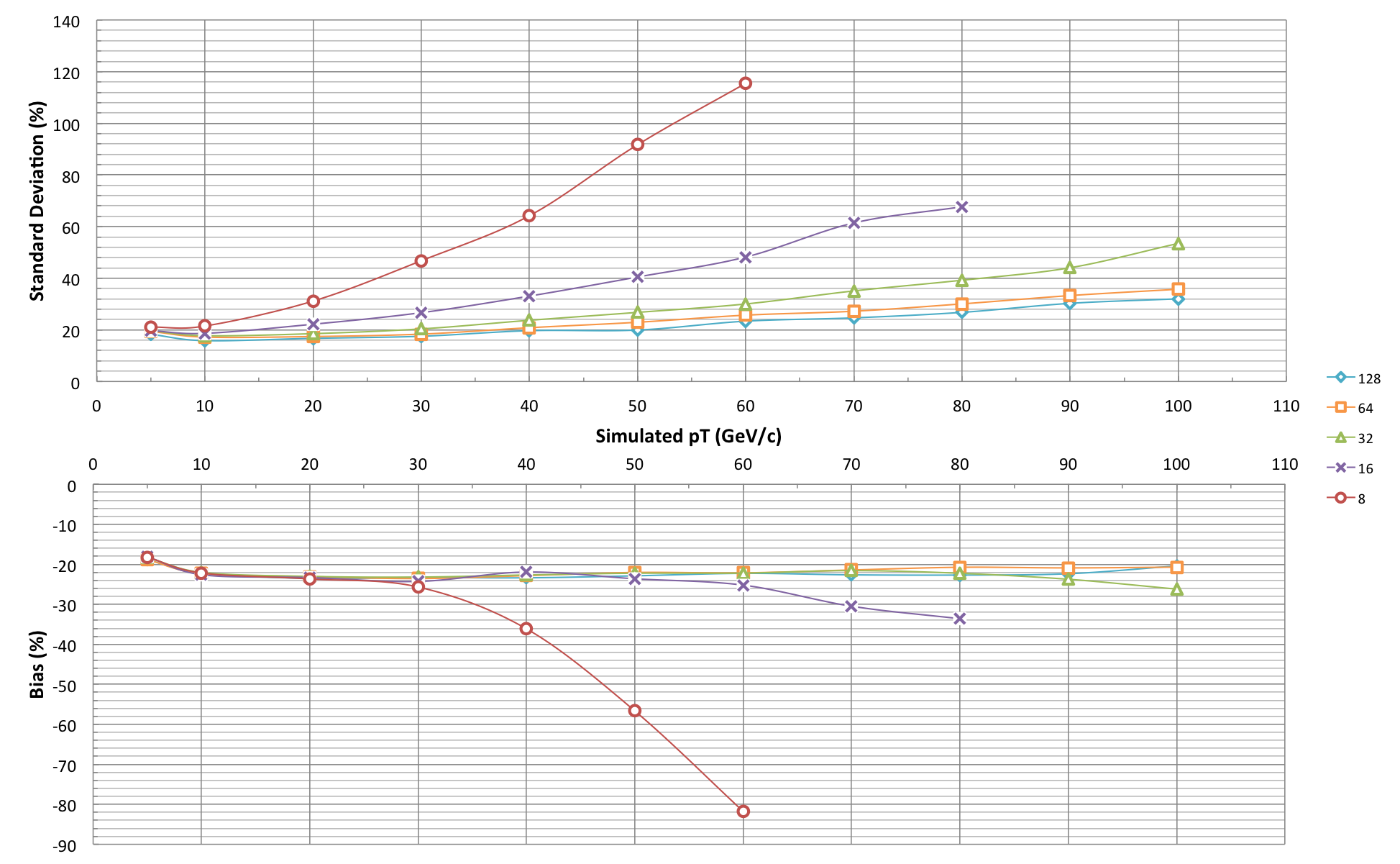}
		\caption{Standard deviation (top) and bias (bottom) on $ \frac{\Delta p_T}{p_T} $ as a function of the simulated \pT{} for muon tracks generated with CMSSW, and reconstructed with the modified Kalman filter using the RecHits in GE1/1, ME1/1, and ME2/x with granularities of 128, 64, 32, 16, and 8.}
		\label{fig:impact_of_the_granularity__rkal_granularity_CMSSW_GEM_ME12}
	\end{figure}

	We perform the same analysis for the modified Kalman filter. Figure \ref{fig:impact_of_the_granularity__rkal_granularity_CMSSW_GEM_ME12} depicts the standard deviation (top) and bias (bottom) on $ \frac{\Delta p_T}{p_T} $ as a function of the simulated \pT{} for muon tracks generated with CMSSW, and reconstructed with the modified Kalman filter using the RecHits in GE1/1, ME1/1, and ME2/x with granularities of 128, 64, 32, 16, and 8. First of all, we notice that granularities 8 (red) and 16 (purple) remain usable longer than for the Least Squares fit (until 60 \GeVc{} for a granularity of 8 with the modified Kalman filter against a limit of 30 \GeVc{} with the Least Squares fit). Moreover, the 32 (green) granularity is effective for all the simulated \pT{}. The degradation of the standard deviation with the granularity is observed between the granularity of 32 (green) and of 64 (orange). \\

	The previous discussion reveals one more advantage of the modified Kalman filter over the Least Squares fit, namely the fact that setups with lower granularities remain usable for a wider range of \pT{}.
	\cleardoublepage

\chapter{Why Unbiasing the Results is not Straightforward}
\label{chapter:unbiasing_the_results}

	In this short annex, we explain why we cannot simply subtract the computed bias represented in Figure \ref{fig:kalman_filter__modified_mu_rechits_CMSSW_all_setups} in Section \ref{sec:kalman_filter__modified_CMSSW_bias_detectors_setup} from the results in order to unbias them. \\

	We first have to understand how physicists measure the cross-section of a process. Let us for example consider 
	\begin{equation}
		q + \bar{q} \rightarrow Z
	\end{equation}
	which cross-section $ \sigma_{q + \bar{q} \rightarrow Z} $ we want to measure through its decay into two muons
	\begin{equation}
		Z \rightarrow \mu + \mu 
	\end{equation}
	using the clear signature it produces in the detector. Modifying Equation \ref{eq:lhc_and_cms__luminosity_to_N} in Section \ref{sec:lhc_and_cms__beam_structure_and_luminosity} to account for the decay channel, we find that
	\begin{equation}
		\sigma_{q + \bar{q} \rightarrow Z} = \frac{N_{Z \rightarrow \mu + \mu}}{\Gamma_{Z \rightarrow \mu + \mu}} \frac{1}{L} \ ,
	\end{equation}
	where $ \Gamma_{Z \rightarrow \mu + \mu} $ is the branching ration of $ Z $ into two muons, $ N_{Z \rightarrow \mu + \mu} $ is the number of found events, and $ L $ is the integrated luminosity. \\

	Physicists performing this study cannot analyze all the data volume produced by CMS as it is considerable. Therefore, they focus on datasets that are created according to the filters the events passed in the triggers, under which the di-muons filter. A powerful reconstruction algorithm is then ran on the events in the di-muons datasets and the invariant mass of the system is computed. If the latter is compatible with the mass of the $ Z $ boson, the event is accepted for this analysis. \\

	When using a specific datasets, physicists have to compute the efficiency of the filters to know how many correct events were rejected. If the system discards $ \epsilon $ percent of valid event, the total number of events will be
	\begin{equation}
		N_{Z \rightarrow \mu + \mu;\ total} = \frac{N_{Z \rightarrow \mu + \mu;\ accepted}}{1 - \epsilon} \ .
	\end{equation}	 
	This efficiency is not constant but varies with the parameters of the run, under which the L1 Trigger's threshold on the \pT{} reviewed in Section \ref{sec:trigger_system_and_reconstruction_algorithms__system_performances}. Changing the threshold complicates the analysis and requires the physicists to recompute the efficiency for each change. \\

\begin{figure}[h!]
		\centering
		\includegraphics[width = 0.6 \textwidth]{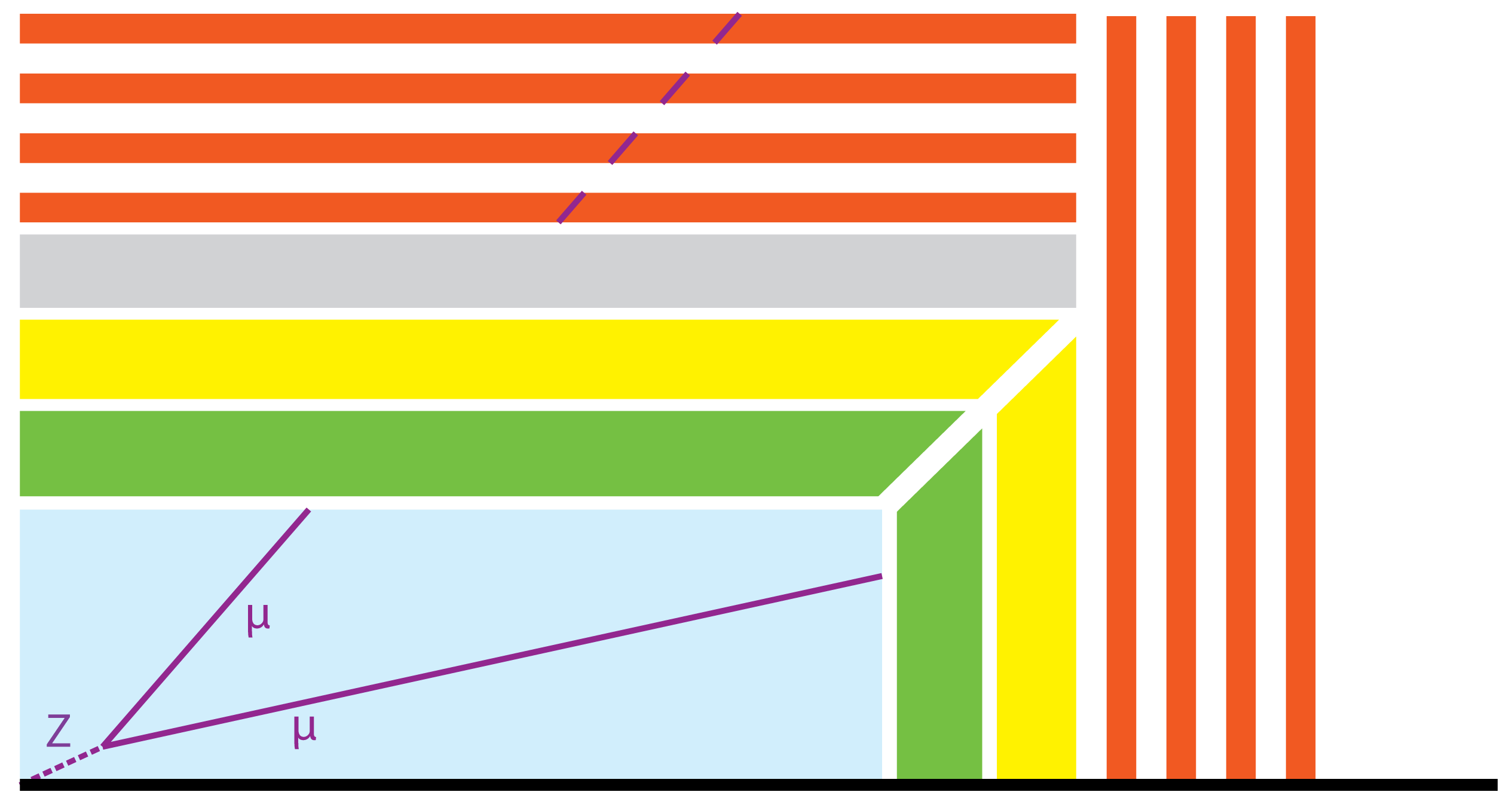}
		\caption{Schematic representation of a $ Z $ boson decaying in two muons inside CMS.}
		\label{fig:unbiasing_the_results__z_decay}
	\end{figure}

	Moreover, the efficiency depends upon the performances of the local triggers of the CMS muon spectrometer that test the events against a first set of filters. To measure the impact of the triggers on the efficiency, we look at events that produced two muons: one in the barrel and one in the endcap. We require the muon in the endcap not to have been identified as a muon in the muon chambers but to have left a track in the tracker. The other muon has to have left a track in both the tracker and the muon chambers and therefore be identified as a single-muon event. This event is represented in Figure \ref{fig:unbiasing_the_results__z_decay}. As the muon in the endcap did not pass the filters, the event will not be flagged as di-muons but as single-muons. However, using the invariant mass of the particles and the tracks left in the tracker, it is possible to reconstruct it correctly. By analyzing the single-muon datasets, physicists can compute the efficiency of the triggers, in this case the endcaps trigger, by dividing the number of events that are correctly flagged by the total number of events. \\

	Furthermore, we observed in Figure \ref{fig:kalman_filter__modified_sigma_mu_rechits_CMSSW_all_setups_evolution_eta} in Section \ref{sec:kalman_filter__modified_CMSSW_evolution_eta} that the standard deviation and the bias on $ \frac{\Delta p_T}{p_T} $ varies with $ \eta $. If we unbias the results according to $ \eta $, we will bring one more degree of complexity to the computation of the efficiency which is not an option. \\

	Finally, we reviewed in Section \ref{sec:kalman_filter__modified_CMSSW_bias_detectors_setup} that the bias changes with the \pT{}. If we obtain a bias of 50\% for a simulated \pT{} of 60 \GeVc{} and a bias of 28\% for a simulated \pT{} of 70 \GeVc{}, both will return a reconstructed \pT{} of 90 \GeVc{}. At the L1 Trigger, when the algorithm has reconstructed a \pT{} of 90 \GeVc{}, an ambiguity arises when we try to unbias it. \\

	For all the previously stated reasons, the algorithms themselves must provide the smallest possible bias.

	\cleardoublepage	


\end{document}